\documentclass[apj,twocolumn]{openjournal}
\usepackage{newtxtext,newtxmath,enumitem,multirow,ctable}
\usepackage[T1]{fontenc}
\usepackage[breaklinks,colorlinks,citecolor=blue,urlcolor=blue]{hyperref}

\newcommand{\Alf}{{Alfv\'en}}

\newcommand{\paperone}{Paper {\small I}}
\newcommand{\papertwo}{Paper {\small II}}
\newcommand{\paperthree}{Paper {\small III}}

\newcommand{\orcidauthor}[3]{\author{\href{http://orcid.org/#1}{#2$^{#3}$}}}

\shorttitle{Multi-Phase, Hyper-Magnetized Disks}
\shortauthors{Hopkins}

\begin{document}

\title{\vspace{-0.8cm}Multi-Phase Thermal Structure \&\ The Origin of the Broad-Line Region, Torus, and Corona in Magnetically-Dominated Accretion Disks\vspace{-1.5cm}}
\orcidauthor{0000-0003-3729-1684}{Philip F. Hopkins}{1,*}
\affiliation{$^{1}$TAPIR, Mailcode 350-17, California Institute of Technology, Pasadena, CA 91125, USA}
\thanks{$^*$E-mail: \href{mailto:phopkins@caltech.edu}{phopkins@caltech.edu}},

\begin{abstract}
Recent simulations have demonstrated the formation of ``flux-frozen'' and hyper-magnetized disks, qualitatively distinct from both classical $\alpha$ disks and magnetically-arrested disks, as a natural consequence of fueling gas to supermassive black holes in galactic nuclei. We previously showed that the dynamical structure of said disks can be approximated by simple analytic similarity models. Here we study the thermal properties of these models over a wide range of physical scales and accretion rates (from highly sub-critical to super-critical). 
%The states can be maintained even for super-critical accretion (while luminosity saturates near-Eddington) but transition to a hot, optically-thin flow below $\sim 1-5\%$ of Eddington. Above that accretion rate, 
We show there are several characteristic zones: a dusty ``torus''-like region, a multi-phase neutral and then multi-phase ionized, broad line-emitting region interior to the sublimation radius, before finally a transition to a thermal accretion disk with a warm Comptonizing layer. The disks are strongly-flared with large scale heights, and reprocess and/or scatter an order-one fraction of the central disk emission. As a result, this simple accretion disk model predicts phenomena including the existence of a dusty torus and its covering factor, geometry, clumpiness, and dust temperatures; a broad-line-region (BLR) with its characteristic sizes and luminosities and ionization properties; extended scattering/reprocessing surfaces producing cooler disk continuum and apparently large observed disk sizes; and existence of warm Comptonizing layers and hard coronal gas. Remarkably, these properties emerge without our having to introduce new ``components'' or parameters: they are all {\em part of} the accretion flow if the disks are in the hyper-magnetized limit.
\end{abstract}


\maketitle

\section{Introduction}
\label{sec:intro}

\begin{table*}
\begin{center}
\begin{footnotesize}
\caption{Common variables in this manuscript (others are defined throughout where relevant).\label{tbl:variables}}
\begin{tabular}{cl}
\hline\hline
$M_{\rm BH}$, $m_{7}$ & BH mass $M_{\rm BH} \equiv m_{7} 10^{7} M_{\odot}$ \\
$\dot{M}$, $\dot{m}$ & Accretion rate $\dot{M}$ and accretion relative to critical $\dot{m} \equiv \dot{M}/\dot{M}_{\rm crit}$ with $\dot{M}_{\rm crit} \equiv L_{\rm Edd}/0.1 c^{2}$ \\
$\epsilon_{r}$, $\epsilon_{r,\,0.1}$ & Integrated (bolometric) radiative efficiency $L \equiv \epsilon_{r} \dot{M} c^{2}$ with $\epsilon_{r} \equiv \epsilon_{r,\,0.1}\,0.1$ \\
\hline
$R_{g}$, $x_{g}$, $R_{\rm ISCO}$ & BH Schwarzschild/gravitational radius $R_{g}\equiv 2\,G M_{\rm BH}/c^{2}$, $x_{g} \equiv R/R_{g}$, and ISCO radius (for non-spinning BH $R_{\rm ISCO}=3 R_{g}$) \\
$R_{\rm BHROI}$ & BH radius of influence $R_{\rm BHROI} \equiv G M_{\rm BH}/\sigma_{\rm gal}^{2}$ (with $\sigma_{\rm gal}$ the galactic velocity dispersion) \\
$r_{\rm ff}$, $r_{{\rm ff},\,5}$ & Outer boundary or ``freefall'' radius into Keplerian potential $r_{\rm ff} \approx R_{\rm BHROI}$, defined by $r_{\rm ff} \equiv r_{{\rm ff},\,5} 5\,{\rm pc}$ (with $r_{{\rm ff},\,5} \sim m_{7}^{1/2}$ expected) \\
\hline
$R$, $r$, $\phi$, $\theta$, $z$ & Cylindrical $R$ (spherical $r$) radii, azimuthal $\phi$ (polar $\theta$) angles, and vertical height $z$ (disk-aligned and centered on the BH) \\
${\bf B}$, $B_{i}$ & Magnetic field ${\bf B}$ and components $B_{i}$ (e.g.\ radial, toroidal, poloidal components $B_{R}$, $B_{\phi}$, $B_{z}$) \\
${\bf v}$, $v_{i}$ & Gas velocity ${\bf v}$ and components $v_{i}$ (e.g.\ radial, azimuthal, vertical components $v_{R}$, $v_{\phi}$, $v_{z}$) \\
\hline
$\rho$, $n$, $\Sigma_{\rm gas}$ & Gas 3D density $\rho$ or number density $n_{\rm gas}$ (in ${\rm particles\,cm^{-3}}$), and projected surface density $\Sigma_{\rm gas}$ \\
$T$, $c_{s}$ & Gas temperature $T=T_{\rm gas}$ ($T_{\rm rad}$, $T_{\rm dust}$ denote radiation and dust temperatures) and thermal sound speed $c_{s} \equiv \sqrt{k_{B}\,T/\mu\,m_{p}}$ \\
$v_{A}$, $v_{\rm turb}$ & \Alf\ speed $v_{A} \equiv |{\bf B}|/\sqrt{4\pi\rho}$, and typical turbulent velocity $v_{\rm turb}$ \\
\hline
$\beta$, $\mathcal{M}_{s}$, $\mathcal{M}_{A}$ & Plasma $\beta \equiv c_{s}^{2}/v_{A}^{2}$ parameter, sonic $\mathcal{M}_{s} \equiv |\delta {\bf v}|/c_{s}$ and \Alf\ $\mathcal{M}_{A} \equiv |\delta {\bf v}|/v_{A}$ Mach numbers \\
$H$ & Gas disk vertical scale-height $H$ (defined within a given annulus $R$) \\
$v_{\rm K}$, $\Omega$ & Keplerian circular velocity $v_{\rm K} \equiv G\,M_{\rm BH}/r$ and frequency $\Omega \equiv v_{\rm K}/R$ \\
\hline
$t_{\rm dyn}$, $t_{\rm cool}$ & Disk dynamical time $t_{\rm dyn} \equiv \Omega^{-1}$, and gas cooling time $t_{\rm cool}$ (at a given radius and temperature, etc.) \\
$\kappa$, $\kappa_{s}$, $\kappa_{a}$, $\kappa^{\ast}$ & Total, scattering, absorption and effective $\kappa^{\ast} \equiv \sqrt{\kappa_{a}(\kappa_{a}+\kappa_{s})}$ opacities, with corresponding optical depths $\tau$, $\tau_{s}$, $\tau_{a}$, $\tau^{\ast}$ \\
$Z$, $x_{e}$, $f_{\rm mol}$, $\xi$ & gas metallicity $Z\equiv \tilde{Z} Z_{\odot}$, free electron \&\ molecular fractions $x_{e} \equiv n_{e}/n$,  $f_{\rm mol}$, ionization parameter $\xi \equiv ({\rm d}\dot{N}_{\rm ion}/{\rm dA})/(n c)$ \\
\hline\hline
\end{tabular}
\end{footnotesize}
\end{center}
\end{table*}

Active galactic nuclei (AGN) and quasars are powered by accretion disks around supermassive BHs \citep{schmidt:1963.qso.redshift,soltan82}, with accretion rates exceeding $\gtrsim 10\,{\rm M_{\odot}\,yr^{-1}}$ in the most luminous sources. It has been known for decades that there must be multi-phase gas structure around such BHs, with large covering factors or vertical extent above the midplane \citep[e.g.][and references therein]{davidson.netzer:1979.qso.emission.lines.reviews,peterson:1997.agn.book,krolik:1999.agn.book}, in order to explain many observational features of their spectra and variability. This includes, for example (1) optical narrow line regions (NLR), probably from more ``typical'' interstellar medium (ISM) gas in the AGN host galaxy; (2) the infrared dusty ``torii'' of clumpy gas with $\mathcal{O}(1)$ covering factor, cool and well-shielded enough to host dust at a range of temperatures in ``clumpy'' structures at $\sim 0.01-10\,$pc \citep{antonucci:agn.unification.review,urry:radio.unification.review,burtscher:2013.agn.torii.compilation}; (3) the optical/UV broad emission line region (BLR), believed to come from partially-ionized gas at $\sim 10^{4}$\,K with a range of densities but relatively modest range of ionization parameters at $\sim 1-100\,$ld distances, reprocessing $\sim 10-20\%$ of the light in a thick-disk type geometry \citep{kaspi:2005.blr.size.reverb.mapping,Peterson2006:BLR.review,gravity:2018.sturm.blr.rotating.thick.disk}; (4) an extended, relatively cool central thermal optical/UV continuum region perhaps dominated by scattering (potentially related to any extended scattering surface from a thick disk, or outflows/``warm absorbers,'' or patchy cloud-covering) to explain both the weak dependence of AGN SEDs on BH mass and microlensing observations indicating emission or reprocessing of continuum at large radii \citep{laor:warm.absorber,dai:2010.agn.microlensing.xray.optical.larger.than.expected,giustini.proga:2019.summary.acc.states.winds.qual.phenomenology}; (5) ``warm'' ($\sim 1\,$keV) Comptonizing structures covering the thermal continuum source needed to explain the EUV/soft X-ray excess \citep{kubota:2018.soft.excess.comptonizing.layers,liu.qiao:2022.agn.acc.disk.review.w.focus.on.coronae.disk.states} and a more diffuse/extended ``hard'' ($\sim 10-100\,$keV) X-ray corona, plus again large covering-factor scattering structures needed to explain the X-ray reflection spectra \citep{george:1991.agn.xray.corona,haardt:1991.coronal.heating.model,marinucci:2018.agn.coronae.review}. 

Many theoretical papers have been written about the potential origins of these various multi-phase structures. But notably, the vast majority assume as a starting point that the accretion disk itself is thermal and/or radiation-pressure dominated, and so is qualitatively something akin to a \citet{shakurasunyaev73}-like $\alpha$-disk (hereafter SS73), whether geometrically ``thin'' or supercritical/``slim.'' While there are many variant accretion disk models in the literature, (including radiatively inefficient, advection-dominated, magnetically-arrested, magnetically-elevated, ``slim,'' and gravito-turbulent disks; for reviews see \citealt{frank:2002.accretion.book,abramowicz:accretion.theory.review}), for luminous quasars the fundamental assumption of SS73, that magnetic pressure is small compared to thermal pressure ($\beta \gg 1$), is still most often the ``baseline'' for both analytic studies, observational forward-modeling, or setting up initial conditions for idealized accretion-disk simulations. 
Importantly, in this specific category of thermal-pressure dominated $\alpha$-disk models, the multi-phase structure observed must arise ``outside'' of the accretion disk -- SS73 essentially require (by assumption) that the disk be single-phase, and the predicted densities, temperatures, and covering factors within standard thermal-pressure-dominated $\alpha$ disks are wildly different than these observed multi-phase structures. Moreover, it has been known for decades that thermal-pressure dominated disks are violently gravitationally unstable at most of these (larger) radii while simultaneously unstable to catastrophic thermal-viscous instability at smaller radii, and predict gas densities and temperatures many orders-of-magnitude different (at the same distances from the BH) to those inferred for the BLR/torus/Comptonizing and scattering surfaces/coronae \citep{goodman:qso.disk.selfgrav}. This in turn has led to other many other models for the origins of this multi-phase structure, including popular ideas such as the BLR being a part of an outflow or ``failed wind''/fountain-flow \citep{murray:1995.acc.disk.rad.winds,krolik:clumpy.torii,elitzur:torus.wind,naddaf:2021.blr.structure.from.failed.winds}. But these have their own challenges, and some variants of such models may even be ruled out by recent observations finding a rotating, thick-disk geometry \citep{gravity:2018.sturm.blr.rotating.thick.disk,gravity:2020.resolved.blr.size.disk.inside.dust.sub,gravity:2021.resolved.blr.disk.hot.dust.coronal.regions}.

Recently, \citet[][\paperone]{hopkins:superzoom.overview} and \citet[][\papertwo]{hopkins:superzoom.disk} presented the first simulations to self-consistently follow gas in a cosmological simulation from $>$\,Mpc to $<100\,$au scales (a few hundred gravitational radii) around an accreting SMBH, including the physics of magnetic fields (seeded from trace cosmological values), multi-band radiation-hydrodynamics, non-equilibrium multi-phase radiative thermo-chemistry and cooling, self-gravity, star formation, and stellar evolution/feedback (jets, stellar mass-loss, radiation, supernovae). In these simulations, gas around the black hole radius of influence (BHROI)\footnote{Defined as the radius interior to which the BH dominates the potential over its host galaxy of characteristic velocity dispersion $\sigma_{\rm gal}$, or $R_{\rm BHROI} \sim G\,M_{\rm bh}/\sigma_{\rm gal}^{2}$ (about $\sim 5\,{\rm pc}$ in the reference simulations).} is tidally captured by the SMBH of mass $M_{\rm bh} \sim 10^{7}\,M_{\odot}$ from larger-scale ISM gas complexes in the galaxy, and free-falls briefly before circularizing to form an accretion disk with $Q\gg 1$ and little to no star formation or fragmentation on sub-pc scales. This disk evolves in quasi-steady-state and sustains super-critical accretion (up to $\dot{M} \sim 20-30\,{\rm M_{\odot}\,yr^{-1}}$) onto the SMBH for at least $\sim 10^{5}$ disk dynamical times (the simulation duration). Crucially, in \papertwo\ where the disk properties were studied in detail, it was shown that these ``hyper-magnetized'' and ``flux-frozen'' disks have $\beta \sim 10^{-4}-10^{-2}$ in the midplane, in the form of primarily toroidal magnetic field (but with mean radial fields and quasi-isotropic turbulent fields only a factor of a few less strong) owing to amplification of magnetic flux accreted from the ISM. These stabilize the disk against {\em both} thermal-viscous instability in the inner regions as well as catastrophic fragmentation and star formation in the outer regions: without magnetic fields, the disks were shown to be orders-of-magnitude less massive and support factor of $\sim 1000$ lower accretion rates and higher star formation rates. The disks also have a flared structure ($H/R \sim 0.1-1$ at large radii) with weak vertical stratification owing to trans-\Alf{ic},  highly super-sonic turbulence, which is sustained by rapid cooling ($\mathcal{M}_{A} \sim v_{\rm turb} / v_{A} \sim 1$, with $\mathcal{M}_{s}^{2} \sim 1/\beta \sim 1/t_{\rm cool}\,\Omega \gg 1$). With it now possible to capture these multi-scale ISM-to-disk conditions, since \paperone\ these flux-frozen disks have been seen in a number of other simulations of distinct parameter spaces including sub-Eddington accretion of galactic hot gas onto much more massive BHs \citep{guo:2024.fluxfrozen.disks.lowmdot.ellipticals}, accretion onto intermediate-mass to $\sim 10^{6}\,{\rm M}_{\odot}$ BHs in dense star clusters \citep{shi:2024.imbh.growth.feedback.survey,shi:2024.seed.to.smbh.case.study.subcluster.merging.pairing.fluxfrozen.disk}, horizon-scale idealized and/or radiation-GRMHD idealized simulations \citep{kaaz:2024.hamr.forged.fire.zoom.to.grmhd.magnetized.disks}, ideal-MHD simulations of ``quasi-star'' formation \citep{luo:2024.magnetically.dominated.disk.like.our.zoomins.zoomin.on.first.supermassive.star.situation}, and circumbinary AGN disks \citep{most:2024.bh.circumbinary.acc.disk.decoupling.when.mad}, as well as more idealized global (\citealt{gaburov:2012.public.moving.mesh.code} as well as Guo et al., Tomar et al., in preparation) and local/shearing-box \citep{squire:2024.mri.shearing.box.strongly.magnetized.different.beta.states} simulations with the appropriate initial conditions.

Here, we show that such disks could present a natural solution to the puzzle of the origin of the various multi-phase structures or components of the AGN ecosystem reviewed above. We take the simple analytic flux-frozen disk model in \citet{hopkins:superzoom.analytic} (\paperthree; which was shown therein to reasonably reproduce the simulation properties from \paperone-\papertwo), and explore the opacity and thermal structure of the disk in more detail.\footnote{Note that \papertwo-\paperthree\ briefly discussed an extremely simple estimate of the characteristic disk temperatures, but this largely ignored the details of the real opacity and thermo-chemistry in the disk, as it was only intended to demonstrate that for even the most extreme plausible opacity structure producing the highest possible disk temperatures, the disk models above would still be self-consistent in that they would still maintain $\beta \ll 1$ in the midplane down to horizon scales. Moreover their discussion was primarily focused on the largest radii in the disk, and ignored much of the phenomenology we discuss here.} We find that -- without introducing any new ``components'' or parameters to the model -- this naturally predicts both the existence and properties (including densities, temperatures, scale heights/covering factors, reprocessed light fractions, ionization parameters, and clumping/density variations) of many different AGN ecosystem components above including the dusty ``torus''; the broad-line region; a cooler, more spatially extended scattering effective central emitter (compared to SS73); and the warm Comptonizing and hard coronae. The gravitational, thermal, and buoyant stability of this gas, as well as its location/height above the midplane, are naturally explained by the disk itself -- in brief, they are ``held up'' by magnetic fields. Crucially, these flux-frozen disks are (1) vastly geometrically thicker and more strongly-flared at large radii, (2) much lower density {\em and} lower mass/surface density, and (3) much more strongly turbulent (in terms of the {\em sonic} Mach number $\mathcal{M}_{s} \equiv v_{\rm turb}/c_{s}$) than thermal-pressure-dominated disks. These completely change the predictions for their thermal structure.

\begin{figure}
	\centering\includegraphics[width=0.95\columnwidth]{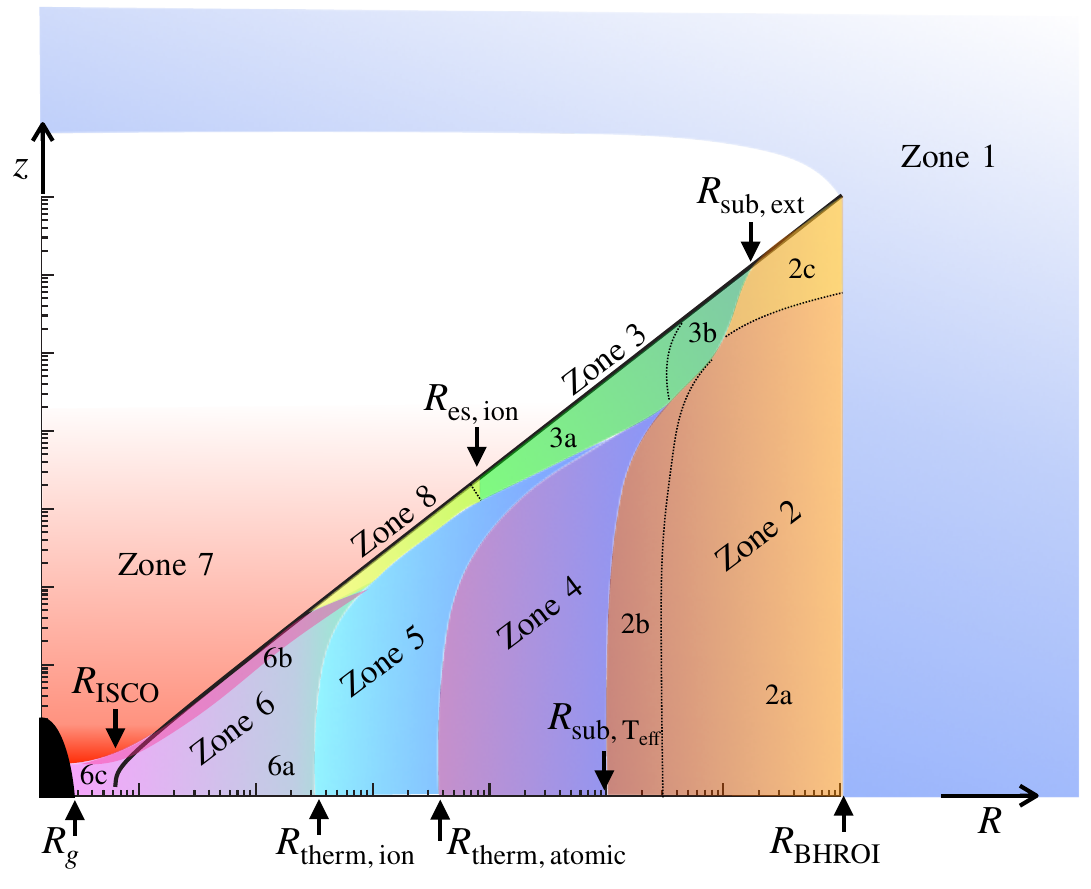} \\
	\caption{Heuristic illustration of the accretion disk properties for a magnetically-dominated, flux-frozen disk, with distinct ``zones'' with different thermo-chemical properties. Black line shows the disk scale-height $z=H$ versus cylindrical radius $R$. We label critical radii including: the gravitational radius/horizon $R_{g}$, ISCO $R_{\rm ISCO}$, radius interior to which the midplane is thermalized and self-ionized $R_{\rm therm,\,ion}$, radius at which the atomic disk is thermalized $R_{\rm therm,\,atomic}$, radius of dust sublimation in the shielded midplane $R_{{\rm sub},\,T_{\rm eff}}$, radius interior to which there is a Thompson-thick electron-scattering intercepting surface illumination $R_{\rm es,\,ion}$, radius interior to which the surface/illuminated layers have dust sublimated $R_{\rm sub,\,ext}$, and BH radius of influence $R_{\rm BHROI}$. We divide the system into distinct ``zones'' labeled (\S~\ref{sec:zone1.ism}-\ref{sec:zone8.warm.scattering}), described in Fig.~\ref{fig:cartoon.quadrants}.
	\label{fig:cartoon.definitions}}
\end{figure}

In \S~\ref{sec:model}, we outline some fundamental assumptions and scalings we will use throughout (common variables are defined in Table~\ref{tbl:variables} for reference). In subsequent sections, we divide the disk into ``zones'' where different gas phases or sources of illumination dominate, outlined in Fig.~\ref{fig:cartoon.definitions}, which we discuss in subsequent sections, including the galactic ISM (\S~\ref{sec:zone1.ism}), dust torus (\S~\ref{sec:zone2.torus}), broad-line region (\S~\ref{sec:zone3.blr}), neutral (\S~\ref{sec:zone4.neutral}) and multi-phase (\S~\ref{sec:zone5.multiphase.disk}) optically-thin disks, central thermalized and blackbody-emitting disk (\S~\ref{sec:zone6.disk}), corona (\S~\ref{sec:zone7.corona}), and extended scattering layers and surfaces (\S~\ref{sec:zone8.warm.scattering}). We describe how behavior should change at both super-critical and highly subcritical accretion rates (\S~\ref{sec:extreme.mdot}), and briefly comment on outflows and jets there and in \S~\ref{sec:jets.winds}. We discuss the key differences from thermal or radiation-pressure dominated disks in \S~\ref{sec:discussion}, and conclude in \S~\ref{sec:conclusions}.

\begin{figure*}
	\centering\includegraphics[width=0.95\textwidth]{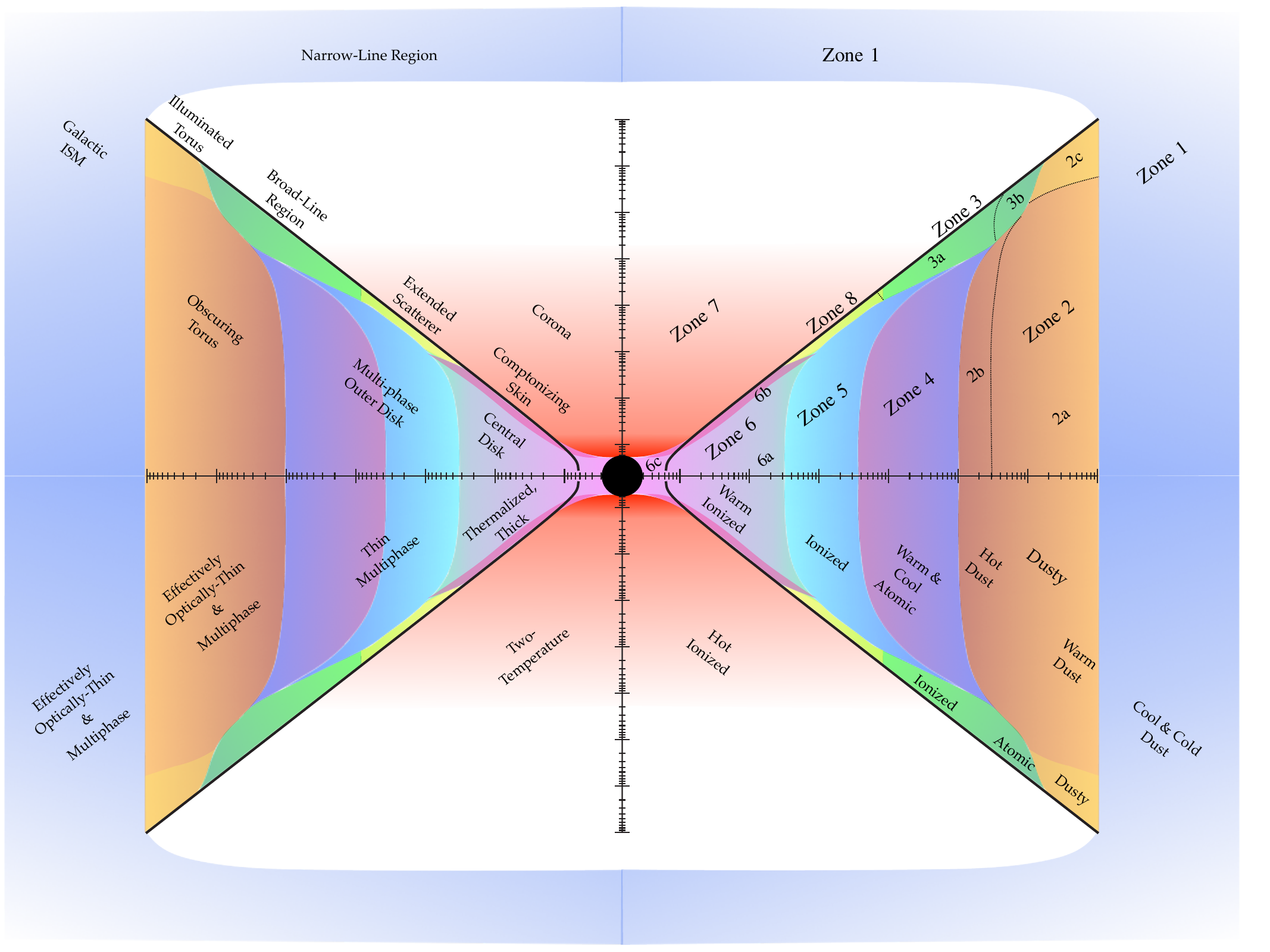} \\
	\caption{Illustration as Fig.~\ref{fig:cartoon.definitions} of a flux-frozen disk, with descriptions of each zone. {\em Top right:} Zone label (\S~\ref{sec:zone1.ism}-\ref{sec:zone8.warm.scattering}; as Fig.~\ref{fig:cartoon.definitions}). Black line shows disk scale-height $z=H$, versus cylindrical radius $R$. {\em Bottom right:} Phase of the gas in each zone (e.g.\ ``dusty'' outside of the sublimation radii, or atomic, or ionized). {\em Bottom left:} Whether each phase is effectively optically-thin or thick to absorption, and multiphase or thermalized. {\em Top left:} Phenomenological structures corresponding to each zone. Zones include the galactic ISM (1; \S~\ref{sec:zone1.ism}), dust torus (2; \S~\ref{sec:zone2.torus}), broad-line region (3; \S~\ref{sec:zone3.blr}), neutral (4; \S~\ref{sec:zone4.neutral}) and multi-phase (5; \S~\ref{sec:zone5.multiphase.disk}) optically-thin disks, central thermalized and blackbody-emitting disk (6; \S~\ref{sec:zone6.disk}), corona (7; \S~\ref{sec:zone7.corona}), and extended scattering layers and surfaces (8; \S~\ref{sec:zone8.warm.scattering}).
	\label{fig:cartoon.quadrants}}
\end{figure*}

\begin{figure*}
	\centering\includegraphics[width=0.97\textwidth]{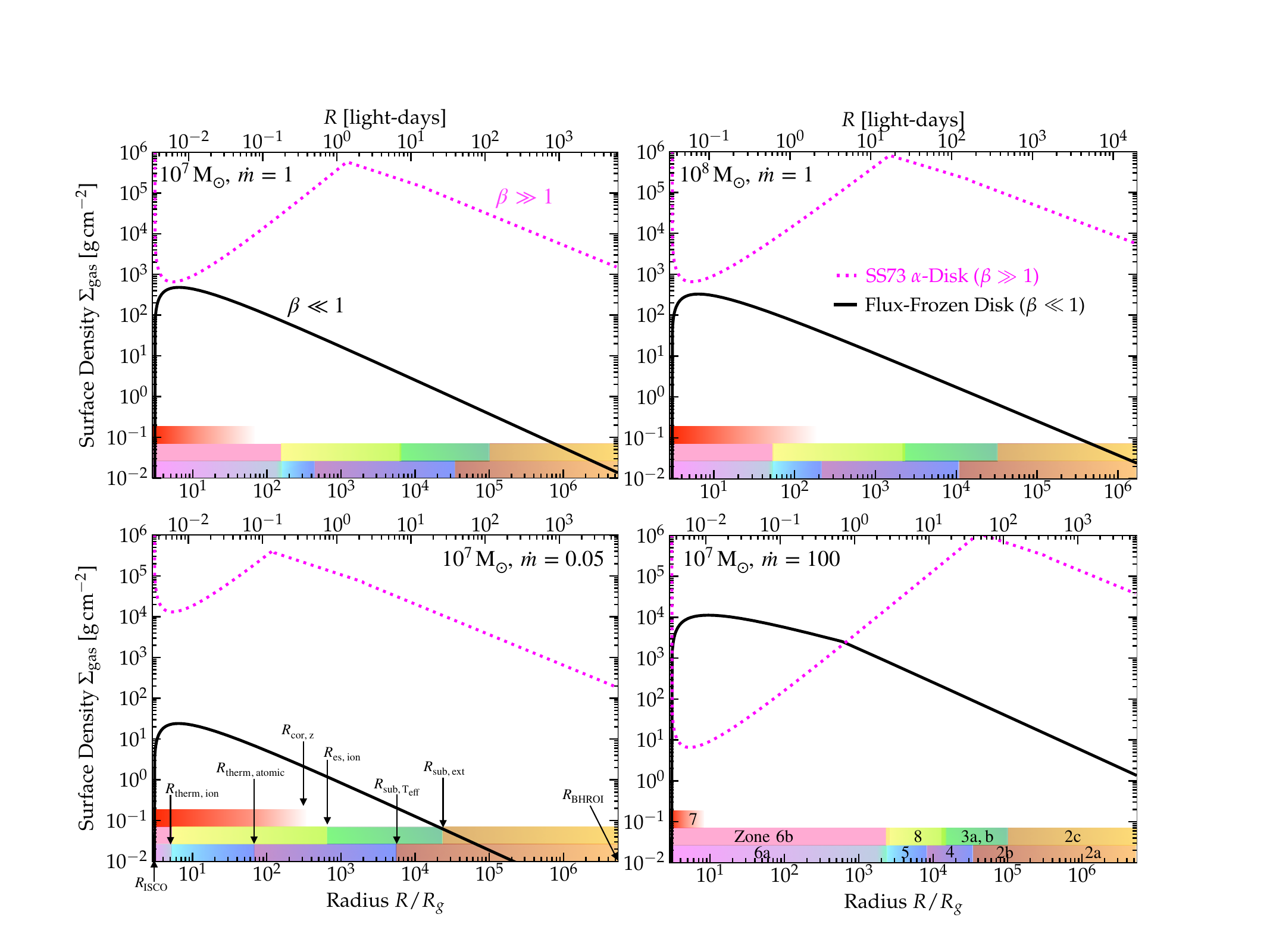} 
	\vspace{-0.2cm}
	\caption{Predicted disk surface densities $\Sigma_{\rm gas}$ versus cylindrical radius $R$ (in units of gravitational radius $R_{g}$ on the bottom axis, or light-days on the top axis). Each panel shows a single disk example (value of BH mass $M_{\rm BH}$ and accretion rate $\dot{m}$), plotted from the ISCO ($R_{\rm ISCO}$) to the BHROI ($R_{\rm BHROI}$R), for the flux-frozen magnetized ($\beta \ll 1$) models here ({\em black solid line}). We contrast the prediction for the standard SS73 thermal-pressure-dominated $\alpha$-disk model ({\em pink dotted line}). 
	We compare different parameters: 
	${\rm M}_{\rm BH}=10^{7}\,{\rm M}_{\odot}$, $\dot{m}=1$ ({\em top left}); 
	${\rm M}_{\rm BH}=10^{9}\,{\rm M}_{\odot}$, $\dot{m}=1$ ({\em top right}); 
	${\rm M}_{\rm BH}=10^{7}\,{\rm M}_{\odot}$, $\dot{m}=0.01$ ({\em bottom left}); 
	${\rm M}_{\rm BH}=10^{7}\,{\rm M}_{\odot}$, $\dot{m}=100$ ({\em bottom right}). 
	Shaded bars along the horizontal axis represent the locations of different zones (colored as Figs.~\ref{fig:cartoon.definitions}-\ref{fig:cartoon.quadrants}) for each case, with their name ({\em bottom right}) and dividing radii ({\em bottom left}) labeled. Their $y$-axis values are not meaningful (they only highlight the range of radii for each zone). 
	Bottom row shows the midplane ($|z|\ll H$) zones: thermalized disk (6a), ionized (5) and neutral (4) optically-thin disks, and obscuring torus (2; with 2b the thermalized-dust subregion). The row above corresponds to the disk illuminated surface ($|z|\sim H$) zones: warm comptonizing skin (6b), the scattering/reprocessing (8) and optically-thin ionized illuminated disk (3) BLR-like region; and illuminated warm dust-reprocessing torus (2c). Above this we show the range of radii of the coronal gas ($|z| \gtrsim H$), zone (7). Galactic ISM (Zone (1)) resides to the right of the plot ($R > R_{\rm BHROI}$) and is not modeled here.
	At most radii, masses, and accretion rates, the proportionally much stronger Maxwell stresses in the flux-frozen disks translate to lower $\Sigma_{\rm gas}$ compared to a thermal-pressure-dominated disk. Note the declining central densities in SS73 at high $\dot{m}$ are a consequence of the inner disks becoming radiation-pressure dominated. This only influences the models here weakly, as there is a small change in slope in the $\dot{m}=100$ flux-frozen case at $x_{g} \equiv R/R_{g} \sim 1500$ interior to which the disk becomes radiatively inefficient (\S~\ref{sec:super.eddington}).
	\label{fig:surface.densities}}
\end{figure*} 

\begin{figure*}
	\centering\includegraphics[width=0.97\textwidth]{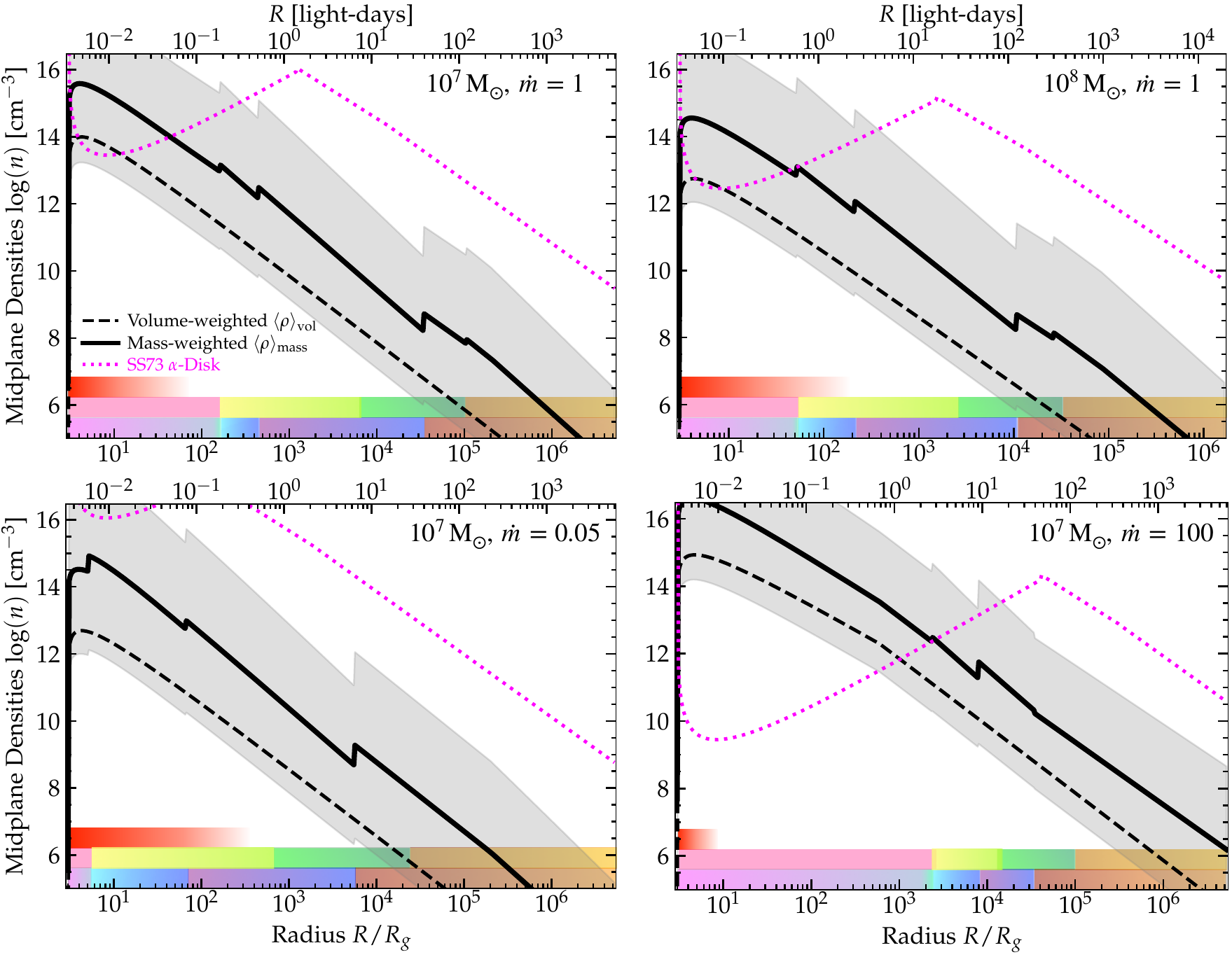} 
	\vspace{-0.2cm}
	\caption{Predicted midplane 3D gas densities $n\equiv \rho/m_{p}$, as Fig.~\ref{fig:surface.densities}. We compare the same BH mass and accretion rates (panels), range of radii (ISCO to BHROI), zone locations (shaded horizontal bars), and flux-frozen disk ({\em thick black}) versus thermal-pressure-dominated (SS73; {\em thin pink}) prediction. SS73-like models assume disks are weakly turbulent/laminar, so there is only one midplane density to plot. Flux-frozen disks are supersonically turbulent and multi-phase, so we plot our estimate of the volume-weighted mean midplane density $\langle \rho \rangle_{\rm vol} \sim \Sigma_{\rm gas}/(2\,H)$,; mass-weighted mean midplane density $\langle \rho \rangle_{\rm mass} \equiv  M^{-1} \int \rho dm \approx \langle \rho \rangle_{\rm vol}  C^{1/2}$ where $C$ is the ``clumping factor''; and mass-weighted $\pm 2\sigma$ range of densities ({\em grey shaded}). We assume a lognormal density distribution with the standard variance-Mach number relation $S \approx \ln[1 + (\mathcal{M}_{s}/3)^{2}]$ for supersonic turbulence to compute these \citep[see][\&\ \S~\ref{sec:model:general:clumping}]{konstantin:mach.compressive.relation}. The discontinuities in $\langle \rho \rangle_{\rm mass}$ appear at midplane zone transitions because we use the analytic approximations from \S~\ref{sec:zone1.ism}-\ref{sec:zone8.warm.scattering} for the temperature/phase structure and dominant opacities, hence sound speed and $\mathcal{M}_{s}$. The mean profile crudely follows $\rho \propto R^{-2}$, with a broad range of densities (i.e.\ ``clumpy''/inhomogeneous structure) at all radii. In general the densities are orders-of-magnitude lower than in an SS73 disk (owing both to lower $\Sigma_{\rm gas}$, Fig.~\ref{fig:surface.densities}, and thicker disks with $H/R \sim 0.1-1$), except for interior regions at large $\dot{m}$ where radiation pressure modifies the magnetized solutions much more weakly. Note the similarity of the densities for flux-frozen disks to observational inferences for the BLR (\S~\ref{sec:zone3.blr}) and dusty torus (\S~\ref{sec:zone2.torus}) at corresponding radii (where the SS73 model is many orders-of-magnitude more dense). 
	\label{fig:3d.densities}}
\end{figure*}

\begin{figure}
	\centering\includegraphics[width=0.95\columnwidth]{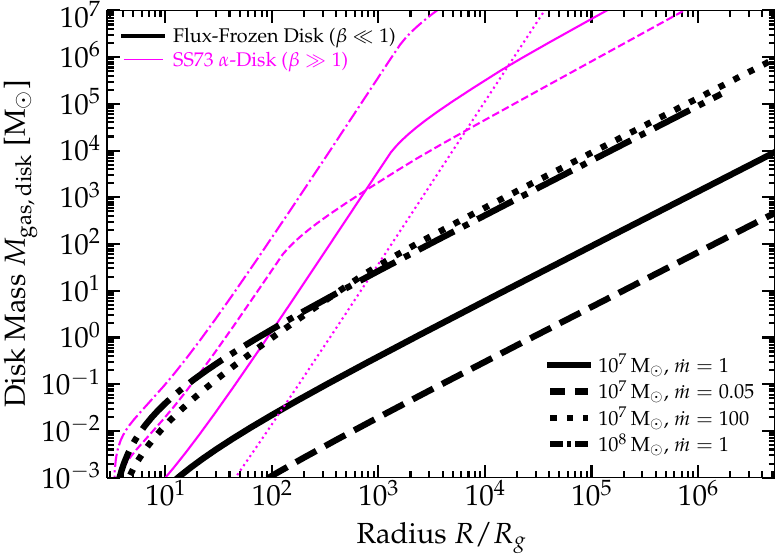} 
	\vspace{-0.2cm}
	\caption{Total enclosed disk gas mass $M_{\rm gas,\,disk}(<R) \equiv \int_{0}^{R} 2\pi\Sigma_{\rm gas}\,R\,dR$ versus radii $R$ for the flux-frozen models  ({\em thick black}) or thermal-pressure-dominated models ({\em thin pink}) with different BH masses and accretion rates (linestyles labeled). Per Fig.~\ref{fig:surface.densities}, the flux-frozen disks predict much lower-disk masses. The gas masses required for thermal pressure-dominated disks become enormous at large radii, which in turn raises many theoretical challenges \citep[e.g.][]{goodman:qso.disk.selfgrav}, and even comparable to the BH mass (so Keplerian approximations break down, the disk cannot stabilize, and orbits become strongly perturbed) at $\gtrsim 1000\,R_{g}$. In contrast the flux-frozen disks require far less disk mass for the same BH mass and accretion rate/luminosity.
	\label{fig:masses}}
\end{figure}

\section{Fundamental Model Assumptions \&\ Scalings}
\label{sec:model}

\subsection{Assumptions}
\label{sec:model:assumptions}

In \paperthree, we present a simple analytic model for the disk properties from \papertwo. This model makes two particularly important {\em ansatz}: 
\begin{enumerate}
\item{{\bf Magnetic Pressure Dominates} over Thermal Pressure ($\beta \sim P_{\rm thermal}/P_{\rm mag} \ll 1$) in the Disk, with in-plane (toroidal, radial, and turbulent) field amplification driven by flux-freezing. So $|{\bf B}|$ obeys a relation like $|{\bf B}|^{2} \propto \rho^{\gamma}$ with $\gamma\sim 1-5/3$ or $|{\bf B}|\propto H^{-1}$ or other similar scalings, depending on the assumed geometry (but the detailed choice has little effect on our results). This is the most important difference from classical $\alpha$-disk models which assume $\beta \gg 1$ ($P_{\rm mag} \ll P_{\rm thermal}$).}

\item{The turbulence is {\bf trans-\Alf{ic}} ($\mathcal{M}_{A} \equiv v_{\rm turb}/v_{A} \sim 1$), or more generally speaking the in-plane total stress driving angular momentum transport is comparable to the Maxwell stress. We note that this assumption (but not $\beta \ll 1$) is also made in many thermal-pressure dominated $\alpha$ disk models, and is implicit in e.g.\ any model where the MRI or the mean flux-frozen in-plane (toroidal-radial) field regulates the stresses (e.g.\ it follows naturally from physics like those in \citealt{balbus.hawley.review.1998}). The key difference is that, because $\beta \ll 1$ ($v_{A}\gg c_{s}$), this means the turbulence should be highly supersonic ($\mathcal{M}_{s} \equiv v_{\rm turb}/c_{s} \sim \mathcal{M}_{A} \beta^{-1/2} \gg 1$), while for classical $\alpha$-disk models it would imply sub-sonic turbulence.}
\end{enumerate} 

These two conditions appear to hold quite robustly at all radii (from near-horizon to BH radius of influence) in the simulations in \S~\ref{sec:intro}, and have now been seen in a wide variety of multi-physics simulations of BH accretion from larger (ISM) scales, despite widely different detailed parameters and physics treatments, numerical methods and resolution \citep[see][]{gaburov:2012.public.moving.mesh.code,hopkins:superzoom.disk,guo:2024.fluxfrozen.disks.lowmdot.ellipticals,shi:2024.imbh.growth.feedback.survey,shi:2024.seed.to.smbh.case.study.subcluster.merging.pairing.fluxfrozen.disk,most:2024.bh.circumbinary.acc.disk.decoupling.when.mad,luo:2024.magnetically.dominated.disk.like.our.zoomins.zoomin.on.first.supermassive.star.situation,kaaz:2024.hamr.forged.fire.zoom.to.grmhd.magnetized.disks}. And they appear independent of e.g.\ numerical resolution in these studies (references therein and Appendices of \citealt{hopkins:superzoom.agn.disks.to.isco.with.gizmo.rad.thermochemical.properties.nlte.multiphase.resolution.studies}). But given the recency of this work, the physical processes underlying all of this -- i.e.\ what actually produces conditions (1) and (2) -- remains very much the subject of active study \citep[e.g.][]{squire:2024.mri.shearing.box.strongly.magnetized.different.beta.states}. It is clear that (1), in broad strokes, depends on the ``outer boundary conditions'' of the disk -- i.e.\ accretion of sufficient flux from the ISM to ensure that flux is accreted through the disk faster than it is lost via turbulent reconnection or Parker instabilities (see discussion in \papertwo), though there could be important contributions to sustaining ${\bf B}$ through other processes like flux freezing or a local Parker dynamo \citep{johansen.levin:2008.high.mdot.magnetized.disks,squire:2024.mri.shearing.box.strongly.magnetized.different.beta.states}, turbulent/wind dynamos \citep{hopkins:superzoom.agn.disks.to.isco.with.gizmo.rad.thermochemical.properties.nlte.multiphase.resolution.studies}, or MRI-like instabilities \citep{begelman:2023.mri.saturation.estimates}. Meanwhile for (2), saturation at trans-\Alf{ic} speeds is generic in a low-$\beta$ disk, but there are many viable drivers of turbulence in realistic multi-physics simulations, including (but not limited to) Parker and global MRI-like instabilities, swing-amplified spiral modes, global disk eccentricity, multi-phase cooling dynamics, fountain flows/failed winds, advection/convection of extrinsic turbulence, convective cooling overstabilities, shear interface instabilities and vertical shear instabilities. We stress that for our purposes in this manuscript, we are agnostic to these driving physics, so long as any disk model satisfies the two {\em ansatz} outlined above.

\subsection{Basic Disk Properties}
\label{sec:model:disk}

Taking this together with the fact that the gravitational energy flux of accreting material is $F_{\rm grav} \approx (3/4\pi)\,\dot{M}\,\Omega^{2}$, \paperthree\ showed that this admits a simple similarity solution valid over all radii where the potential is approximately Keplerian (from outside the ISCO out to the BHROI). 
Taking the default ansatz {\bf (1)} and {\bf (2)} as therein, the resulting scaling for the turbulent velocity $v_{t}$, \Alf\ speed $v_{A}$, scale height $H$, and gas surface density $\Sigma_{\rm gas}$ becomes:
\begin{align}
\label{eqn:veff.model}\frac{v_{t}}{v_{\rm K}} &\sim \frac{v_{A}}{v_{\rm K}} \sim \frac{H}{R} \sim \left( \frac{R}{r_{\rm ff}}\right)^{1/6} \sim 0.07\,\left( \frac{m_{7}\,x_{g}}{r_{{\rm ff},5}} \right)^{1/6}, \\
\label{eqn:sigma.gas} \frac{\Sigma_{\rm gas}}{{\rm g\,cm^{-2}}} &\sim 0.014\, \frac{\dot{m}\,m_{7}^{1/2}}{r_{{\rm ff},5}^{1/2}}\,\left( \frac{R}{r_{\rm ff}}\right)^{-5/6} \sim 10^{4} \frac{\dot{m}\,r_{{\rm ff},5}^{1/3}}{m_{7}^{1/3}\,x_{g}^{5/6}}
\end{align}
in terms of the cylindrical radius $R$ (also defined in terms of the usual gravitational radius $x_{g} \equiv R/R_{g}$ with $R_{g}\equiv 2\,G\,M_{\rm BH}/c^{2}$), Keplerian speed $v_{\rm K}^{2}\equiv G\,M_{\rm BH}/R$, Eddington-scaled accretion rate $\dot{m} \equiv \dot{M}/\dot{M}_{\rm crit}$ (with $\dot{M}_{\rm crit} \equiv M_{\rm BH}/5\times10^{7}\,{\rm yr} = L_{\rm Edd}/(0.1\,c^{2})$ as defined here, for a reference radiative efficiency $\epsilon_{r,\,0.1}=0.1$), BH mass $m_{7} \equiv M_{\rm BH}/10^{7}\,{\rm M_{\odot}}$, and ``free-fall'' radius $r_{\rm ff} \equiv r_{\rm ff,\,5}\,5\,{\rm pc}$ defined as the radius where the solutions map onto free-fall into the BH potential from larger radii outside of the Keplerian potential. For reasonable models of accretion onto a SMBH from galactic scales via gravitational capture (e.g.\ Bondi-Hoyle accretion, or tidal disruption of ISM clouds or star clusters, or loss-cone capture, or gravitational torques from nested bars or mergers; \citealt{shlosman:bars.within.bars,jogee:review,hopkins:seyferts,hopkins:zoom.sims,daa:20.hyperrefinement.bh.growth}) $r_{\rm ff}$ will correspond to the BHROI, 
\begin{align}
r_{\rm ff} \sim R_{\rm BHROI} \sim G\,M_{\rm BH}/\sigma_{\rm galaxy}^{2} \approx 5\,{\rm pc}\,m_{7}^{1/2} 
\end{align}
in terms of the galactic nuclear velocity dispersion $\sigma_{\rm galaxy}$ (where the latter scaling inserts the observed $M_{\rm BH}$-$\sigma_{\rm galaxy}$ relation; \citealt{mcconnell:mbh.host.revisions}). This scale, the gravitational radius/ISCO, and other key scales, along with the (weak) scaling of $H/R$ with radius, are illustrated in Figs.~\ref{fig:cartoon.definitions}-\ref{fig:cartoon.quadrants}. The predicted surface density scalings are shown in Fig.~\ref{fig:surface.densities}, with 3D midplane densities in Fig.~\ref{fig:3d.densities} and disk mass in Fig.~\ref{fig:masses}.

From these, we can immediately derive other relevant scalings, e.g.\ the volume-averaged midplane density $\rho \approx \Sigma_{\rm gas}/2\,H$ or effective Toomre $Q$ (including magnetic+turbulent support):
\begin{align}
\label{eqn:rho} \frac{\rho}{\rm m_{p}\,cm^{-3}} &\sim 272\,\frac{m_{7}^{1/2}\,\dot{m}}{r_{\rm ff,\,5}^{3/2}}\,\left( \frac{R}{r_{\rm ff}}\right)^{-2} \sim 7\times10^{15} \frac{\dot{m}\,r_{\rm ff,\,5}^{1/2}}{m_{7}^{3/2}\,x_{g}^{2}} \\
\label{eqn:Qtot} Q_{\rm tot} &\sim 3000\,\frac{m_{7}^{1/2}}{r_{\rm ff,\,5}^{3/2}\,\dot{m}}\,\left( \frac{R}{r_{\rm ff}}\right)^{-1} \gg 1
\end{align}
These can be compared to the values in Fig.~\ref{fig:3d.densities} (where we also show the mass-weighted densities and range of densities, which depends on the turbulent Mach numbers and therefore indirectly on the thermal state, calculated below). Both $Q \gg 1$ as well as our assumption of an approximately Keplerian potential follow from these densities and $M_{\rm disk} \ll M_{\rm BH}$ as shown in Fig.~\ref{fig:masses}. 

Again, we note per \S~\ref{sec:model:assumptions} that these profiles and conclusions depend only on some viable physics providing the two conditions we assume therein, not on the precise details of the physical mechanisms seeding the magnetic fields or turbulence.

\subsection{General Considerations}
\label{sec:model:general}

Particularly convenient for us, precisely because these disks are magnetically-dominated, it means that {\em unlike} standard thermal-pressure dominated disk models, changing the disk thermal or opacity structure has no effect on the accretion rates or basic dynamical properties of the disk derived in \paperthree. So do not need to re-derive a self-consistent ``global'' disk model, but instead solve for the thermal structure given a fixed ``background'' disk dynamical structure. For example, the scale heights in Fig.~\ref{fig:cartoon.definitions}, or surface and 3D densities in Figs.~\ref{fig:surface.densities} \&\ \ref{fig:3d.densities}, are defined by Eqs.~\ref{eqn:veff.model}-\ref{eqn:sigma.gas} above, independent of the disk thermal/opacity  properties.

For calculating irradiation effects, we will define the bolometric accretion luminosity, dominated by the emission from the inner disk regions, as \begin{align}
L_{\rm bol} \equiv \epsilon_{r} \,\dot{M}\,c^{2} \approx 1.1\times10^{45}\,{\rm erg\,s^{-1}}\,m_{7}\,\dot{m}\,\epsilon_{r,\,0.1} 
\end{align} 
for some arbitrary radiative efficiency. 

\papertwo\ and \paperthree\ briefly considered upper limits to the disk temperature to show that these assumptions are self-consistent, i.e.\ the disk maintains $\beta \ll 1$ at all radii. This also means, with ansatz {\bf (2)}, that the turbulence is supersonic at all radii 
\begin{align}
\mathcal{M}_{s} \sim \frac{v_{\rm turb}}{c_{s}} \sim \frac{v_{A}}{c_{s}} \sim \beta^{-1/2} \gg 1,
\end{align}
 and that the ratio of cooling time $t_{\rm cool}$ to dynamical time $t_{\rm dyn}\sim 1/\Omega$ is $\ll1$ (for cooling flux balancing $F_{\rm grav}$, $t_{\rm cool}/t_{\rm dyn} \sim \beta \sim 1/\mathcal{M}_{s}^{2} \ll 1$). Other implicit assumptions we will make here, like that ideal MHD and a collisional fluid description is a reasonable approximation (the ionization fractions are high, and the atomic/molecular/ion/electron mean free paths are all very small compared to global length scales)\footnote{For example, the ion collision time in the ionized disk is approximately $\sim 10^{-9}\Omega^{-1} (T/T_{\rm eff})^{3/2}\,(n_{\rm midplane}/n)\,(\dot{m} x_{g})^{-5/8} \ll \Omega^{-1}$.} are easily validated and checked explicitly in \paperone. We confirm this ourselves in e.g.\ Fig.~\ref{fig:thermal.profile}. In \S~\ref{sec:super.eddington}, we also explicitly confirm that we can neglect radiative viscosity and damping effects. 

Implicit in this is that the gravitational flux/energy change (as gas flows in) appears in kinetic (turbulent) and magnetic energy, via simple infall+compression/flux-freezing and mixing via various instabilities and nonlinear processes. Since this is supersonic and trans-\Alf{ic} the driving scale is naturally $\mathcal{O}(H)$ and turnover time of the largest eddies (containing most of the power) is $\sim \Omega^{-1}$. The heat transfer to gas is therefore mediated by shocks and (turbulent) reconnection, before energy is radiated away.

\subsubsection{Clumping, Inhomogeneity, \&\ the Sonic Scale}
\label{sec:model:general:clumping}

In a SS73-like ($\beta \gg 1$) $\alpha$-disk, the turbulence is subsonic ($\mathcal{M}_{s} \ll 1$) and locally adiabatic ($t_{\rm cool} \gg t_{\rm dyn}$), so the midplane is weakly-compressible and near uniform density. But here, as noted above, the turbulence is highly supersonic ($\mathcal{M}_{s} \gg 1$) and radiative ($t_{\rm cool} \ll t_{\rm dyn}$). This means the turbulence and shocks generate large density inhomogeneities, and we would expect the gas density fluctuations to follow the usual log-normal distribution seen in simulations of radiative supersonic turbulence under a wide range of conditions \citep[see][]{ostriker:2001.gmc.column.dist,hopkins:2012.intermittent.turb.density.pdfs,konstantin:mach.compressive.relation,beattie:2021.turb.intermittency.mhd.subalfvenic}, with the standard variance-Mach number relation $S \approx \ln[1 + (\mathcal{M}_{s}/3)^{2}]$ (so the volumetric probability of local $s \equiv \ln{[\rho/\langle \rho\rangle_{V}]}$, with $\langle \rho\rangle_{V} \approx \Sigma_{\rm gas}/2\,H$ the volumetric mean midplane density, is $P[s] \equiv (2\pi S)^{-1/2} \exp{[-(s+S/2)^{2}/2\,S]}$). This also implies dense structures persist down to the sonic scale (equivalent, up to an order-unity constant, to the post-shock width), $R_{\rm sonic} \sim H/\mathcal{M}_{s}^{2}$ (for a driving scale $\mathcal{O}(H)$ as we have here), and that the maximum temperatures reached will be shock temperatures $T_{\rm shock} \sim (3/16)\,m_{p}\,v_{\rm turb}^{2}/k_{B} \sim \mathcal{M}_{s}^{2} T_{\rm midplane}$ (with the turbulence and rapid cooling maintaining clumping despite this shock-heating/dissipation). These approximate relations are validated directly in the simulations in \papertwo\ and \citet{hopkins:superzoom.agn.disks.to.isco.with.gizmo.rad.thermochemical.properties.nlte.multiphase.resolution.studies}, so we assume them to calculate the relevant density/clumping statistics.

\subsubsection{Temperatures in the Disk}
\label{sec:model:general:temperature}

To calculate the disk temperature structure, e.g.\ the mean midplane temperature $T_{\rm midplane}$ in Fig.~\ref{fig:thermal.profile}, we follow standard accretion disk models \citep[e.g.][]{abramowicz:accretion.theory.review} and (unless otherwise stated) assume each radial annulus can be approximated as a vertical plane-parallel atmosphere in local quasi-steady-state (heating balancing cooling), subject to accretion/shock/reconnection/turbulent dissipation heating (all of which scale together, given our assumption of equilibrium trans-\Alf{ic} turbulence) with flux $F_{\rm grav} \approx (3/4\pi)\,\dot{M}\,\Omega^{2}$ defined above plus external illumination $F_{\rm illum}$ (defined below, \S~\ref{sec:model:illumination}) incident at height $|z|\sim H$.
We account self-consistently for whether the cooling is optically thick or thin (and for effective vertical heat transport by turbulence), as described in the relevant sub-section for each zone.
The key difference from classic accretion disk calculations here is the different accretion disk structure, giving rise to different opacities and illumination, and therefore different temperatures.

Because the disks are not single-phase, there are several other salient temperatures shown in Fig.~\ref{fig:thermal.profile}. 
The effective temperature $T_{\rm eff}$ is defined by historical convention to be the temperature at which the blackbody cooling flux $2\,\sigma_{B}\,T_{\rm eff}^{4}$ would equal $F_{\rm grav}+F_{\rm illum}$. 
The virial temperature is defined by the potential $T_{\rm vir} \sim (1/2) (G\,M_{\rm BH}/r)\,(\mu\,m_{p}/k_{B})$. 
And the post-shock/reconnection temperature is set by the turbulence per \S~\ref{sec:model:general:clumping} as $T_{\rm shock} \sim (3/16)\,m_{p}\,v_{\rm turb}^{2}/k_{B} \sim (3/16)\,m_{p}\,v_{A}^{2}/k_{B}$. 
The temperature $T_{\rm illum,\,thin}^{\rm warm}$ or $T_{\rm skin}$ of optically-thin surface or ``skin'' layers at $|z| \approx H$ (illuminated by the flux from the disk underneath, and $F_{\rm illum}$) is calculated as described in \S~\ref{sec:warm.skin} by balancing their optically-thin cooling rates (including Compton cooling from irradiation) with heating by reconnection/shocks and irradiation. 
Similarly the ``hard'' coronal electron temperatures $T^{\rm cor}_{e}$ are defined (where coronal solutions are supported) in \S~\ref{sec:zone7.corona} by setting optically-thin cooling (primarily Compton cooling) equal to heating, accounting for two-temperature plasma effects (which define this regime), or (where the cooling time becomes longer than the dynamical time) to the virial temperature. 

Other temperatures are defined in the text where they are introduced, as relevant for specific comparisons and zones.

\subsection{Self-Illumination}
\label{sec:model:illumination}

We will throughout consider cases where the illumination of the surface of the outer disk by the central inner disk (which dominates the bolometric luminosity) may be important. Assuming $L_{\rm bol}$ comes from the inner disk, then the flux per unit cylindrical area for a flared disk scales as
\begin{align}
 F_{\rm illum} \sim \left( \frac{\partial \ln{H}}{\partial \ln{R}}-1\right) \left(\frac{H}{R}\right)\,F_{\rm incident} \approx \frac{1}{6}\frac{H}{R}\,F_{\rm incident}
 \end{align}
  (Eq.~\ref{eqn:veff.model}), where 
 \begin{align}
 F_{\rm incident} \equiv \frac{L_{\rm bol}\,f_{\theta}}{4\pi\,r^{2}}
 \end{align}
  depends on $f_{\theta}$ which defines the illumination pattern as a function of polar angle. For a flat, geometrically thin ($H/R \ll 1$), effectively optically-thick ($\tau^{\ast} \gg 1$), and aligned (coplanar with the outer disk) inner disk, 
  \begin{align} 
  f_{\theta} \approx 2\,\cos{\theta} = \frac{2\,(H/R)}{\sqrt{1+(H/R)^{2}}}. 
  \end{align}
  For an isotropic, or spherical/point-source-like, or optically thin central source, 
  \begin{align}
  f_{\theta} = 1\ .
  \end{align}

In practice, we show below that the disk models here should be ``in between'' these extremes: the inner disk is effectively optically-thick but not by a large margin, and the extended inner-disk scattering layer is not; the inner disk has $H/R < 1$ but again not by a large margin, and is flared; and the inner disk will often be mis-aligned with the outer disk (e.g.\ in the simulations, the disk has warps of $\sim 40^{\circ}$ at multiple radii owing to accretion of material with different angular momentum, and is typically mis-aligned with the BH spin which will realign the inner disk; see \papertwo). In these cases an exact solution requires detailed radiative transfer calculations, so we instead simply consider both extremes to bracket the range of possibilities analytically.

Figs.~\ref{fig:cartoon.definitions}-\ref{fig:cartoon.quadrants} show, as we discuss below, the importance of self-illumination. After calculating the relevant opacity effects in Fig.~\ref{fig:opacity.profile} (and pressure profiles in Fig.~\ref{fig:pressure.profile}), we illustrate this in more detail in Fig.~\ref{fig:cartoon.illumination}. 

We stress that future insights and more detailed predictions for self-illumination will depend on non-linear self-consistent radiation-hydrodynamic simulations, of the sort just beginning to be explored in e.g.\ \citet{kaaz:2024.hamr.forged.fire.zoom.to.grmhd.magnetized.disks,hopkins:superzoom.agn.disks.to.isco.with.gizmo.rad.thermochemical.properties.nlte.multiphase.resolution.studies}.

\subsection{Comparison to Thermal-Pressure Dominated (SS73) Disks}
\label{sec:model:ss73}

We will refer to SS73 disks throughout for comparison (e.g.\ Figs.~\ref{fig:surface.densities}, \ref{fig:3d.densities}, \ref{fig:masses},  \ref{fig:thermal.profile}, \ref{fig:opacity.profile}, \ref{fig:pressure.profile}, \ref{fig:turb.props}), as a reference model for the category of thermal-pressure dominated disks. We take their scalings directly (converting to our definitions of $\dot{m}$, etc.), assuming the Maxwell stress is comparable to the total stress (determined by the $\alpha$ parameter in their model which we will take by default to be $\alpha\sim0.1$). While details in some models differ, for a radiatively efficient, thermal-pressure-dominated disk, the scalings in SS73 are all qualitatively similar (for our purposes) to other thermal-pressure-dominated model variants in the literature (see \citealt{abramowicz:accretion.theory.review}), and we review below how the most general conclusions from this comparison are robust to other detailed assumptions once one assumes a midplane $\beta \gg 1$. 

In this sense, one should note that in the scalings we presented above, there are a couple of key differences in the structural properties of magnetically-dominated disks compared to thermal-pressure-dominated disks, which will appear repeatedly in our analysis. The magnetically-dominated disks are flared, and extremely geometrically thick (with $H/R \sim 1$ in the outer regions) compared to a thermal pressure-dominated disk (for which $H/R \ll 0.01$ in the outer regions). For the same accretion rate, magnetically-dominated disks have both much lower 3D density $\rho$ (by factors up to $\sim 10^{7}$; Fig.~\ref{fig:3d.densities}) {\em and} much lower surface density $\Sigma_{\rm gas}$ (by a factor $\sim 10^{5}$; Fig.~\ref{fig:surface.densities}) owing to the much stronger Maxwell stresses (for a detailed comparison of these basic structural properties, see \paperthree). As a result, we will show that the thermal properties of the different disk models are qualitatively distinct (e.g.\ Figs.~\ref{fig:thermal.profile}, \ref{fig:opacity.profile}, \ref{fig:pressure.profile}).

Again, we stress that all of our key conclusions and comparisons to SS73-like disks in this paper are agnostic to the detailed microphysics of the disk: they derive from the different central assumption (1) in \S~\ref{sec:model:assumptions}, that magnetic pressure is much larger than thermal pressure.

\section{Zone 1: The Galactic Nucleus/ISM/Starburst Environment}
\label{sec:zone1.ism}

We now consider different regions or ``zones'' around the BH, roughly proceeding ``outside-in'' with the accretion flow (Fig.~\ref{fig:cartoon.definitions}). We divide the zones by their qualitative thermochemical and radiative behaviors (Fig.~\ref{fig:cartoon.quadrants}), which are determined by the dominant gas phases and heating/cooling processes (Figs.~\ref{fig:surface.densities}, \ref{fig:3d.densities}, \ref{fig:thermal.profile}, \ref{fig:opacity.profile}, \ref{fig:pressure.profile}). Fig.~\ref{fig:cartoon.definitions} shows the definitions of these zones; Fig.~\ref{fig:cartoon.quadrants} illustrates the key properties of the dominant gas phases in each zone and their connections to different classic structures known around AGN.

Beginning from the largest radii, therefore, at $R \gtrsim r_{\rm ff} \sim R_{\rm BHROI}$, we are fully outside the regime where the BH dominates the gravitational potential, so cannot in any meaningful way talk about being in an ``accretion disk.'' Of course gas may lose angular momentum and flow into this general region from the outer galaxy or intergalactic medium (IGM), but this is the galactic ISM in the galaxy nucleus/bulge center, and such flows are governed by galactic dynamics. This should be multi-phase, dusty, clumpy, and can have a covering factor for optical obscuration of the AGN ranging from nil (in e.g.\ ``passive'' elliptical host galaxies) to $\sim 100\%$ (in e.g.\ starburst nuclei with isotropic, Compton-thick nuclear dust columns), as discussed in e.g.\ \citet{simcoe:1997.agn.host.alignment,hopkins:lifetimes.letter,hopkins:lifetimes.methods,hopkins:lifetimes.interp,hopkins:lifetimes.obscuration,hopkins:qso.all,ghosh:all.obscuration,trump:lowl.agn.dilution,trump:2011.host.agn.morph.discussion,hopkins:torus,gilli:2022.host.galaxy.obscuration,glikman:2024.accretion.obscuration.merger.luminous.red.quasar.evidence}. So certainly part of the clumpy torus and some re-emission/reprocessing can arise from these radii: e.g.\ dust directly exposed to an unobscured QSO sightline can reach dust temperatures 
\begin{align}
%T_{\rm dust} \sim 200\,{\rm K}\,(m_{7}\,\dot{m}\,\epsilon_{r,\,0.1}/r_{\rm ff,\,5}^{2})^{1/4}\,(r_{\rm ff}/R)^{1/2}, \\ 
T_{\rm dust} \sim 200\,{\rm K} \left( \frac{m_{7}\,\dot{m}\,\epsilon_{r,\,0.1}}{r_{\rm ff,\,5}^{2}}\right)^{1/4} \left(\frac{r_{\rm ff}}{R}\right)^{1/2}, 
\end{align}
modestly higher than the median $\sim 70-100\,$K expected from illumination just by stars in a starburst environment \citep{sanders88:warm.ulirgs,narayanan:2005.co32.lirgs,veilleux:ulirg.to.qso.sample.big.mdot.changes,younger:warm.ulirg.evol,pope:2008.pah.agn.dont.dominate.smgs,greve:2009.sb.molgas.props,younger:mm.obs.z2.ulirgs,narayanan:2011.xco,hayward:2011.smg.merger.rt}. And of course, the NLR will arise from gas on these scales \citep{starkcarlson:m82.nlr,bennert:nlr.structure,rice:nlr.kinematics,meena:2022.strong.nlr.rad.pressure.driven.winds}. But modeling these scales in more detail depends on the galactic environment, star formation properties, morphology, etc., and most of the gas properties will depend on the detailed physics of star formation and stellar feedback self-regulating the galactic ISM (see \paperone\ and \citealt{hopkins:zoom.sims,hopkins:qso.stellar.fb.together,daa:20.hyperrefinement.bh.growth,hopkins:2021.bhs.bulges.from.sigma.sfr,wellons:2022.smbh.growth,byrne:2023.fire.elliptical.galaxies.with.agn.feedback,byrne:2023.fb.lim.bh.growth.center.formation.critical,mercedes.feliz:2023.agn.feedback.positive.negative,cochrane:2023.agn.winds.galaxy.size.effects,mercedes.feliz:2023.qso.feedback.fire.induced.clumps.in.gas.and.stars}), and so is outside the scope of our focus here.

\section{Zone 2: The Cold, Dusty Outer Disk/Torus}
\label{sec:zone2.torus}

The first region where our model applies is just interior to $r_{\rm ff} \sim R_{\rm BHROI}$ (Figs.~\ref{fig:cartoon.definitions}-\ref{fig:cartoon.quadrants}). We expect, just like in the ISM, the medium to be dusty (given the dust temperature scaling above, we are well exterior to the sublimation radius $R_{\rm sub}$). For dust one can approximate the ratio of absorption to total opacities as $\kappa_{a}/(\kappa_{a}+\kappa_{s}) \sim 1-0.5\,(1 + 10^{5.72}\,(1+T_{\rm rad}^{2})^{-1})^{-1}$ \citep[compare][]{weingartner:2001.dust.size.distrib,semenov:2003.dust.opacities,draine:ism.book} for a given radiation temperature $T_{\rm rad}$, which is $\kappa_{a}/(\kappa_{a}+\kappa_{s}) \approx 0.5$ for $T_{\rm rad} \gtrsim 1000\,$K and $\kappa_{a}/(\kappa_{a}+\kappa_{s}) \approx 1$ for $T_{\rm rad} \lesssim 1000$\,K. So at the cold temperatures of interest in the outermost disk, the dust opacity will be primarily absorption (and in the warmest parts the scattering opacity is at most comparable to absorption). This means the ``effective'' optical depth from the midplane to infinity, $\tau^{\ast} \equiv \int_{0}^{\infty} \sqrt{\kappa_{a}\,(\kappa_{a} + \kappa_{s})}\,\rho(R,\,z)\,dz \approx \langle \sqrt{\kappa_{a}\,(\kappa_{a} + \kappa_{s})} \rangle\,\Sigma_{\rm gas}/2$ (where the surface densities follow the fiducial scalings for these disks from Eq.~\ref{eqn:sigma.gas}, illustrated in Fig.~\ref{fig:surface.densities}) and the ``total'' optical depth $\tau \equiv \langle \kappa_{a} + \kappa_{s} \rangle\,\Sigma_{\rm gas}/2$ are roughly equal to the absorption optical depth $\tau_{\rm abs} = \kappa_{a}\,\Sigma_{\rm gas}/2$. The absorption opacity $\kappa_{a}$ can be approximated (at the level of approximation here, ignoring detailed substructure within the disk and chemical inhomogenity, etc.) as ${\rm MIN}[5\, , 3\,(T_{\rm rad}/100\,{\rm K})^{\beta}]\,\tilde{Z}\,{\rm cm^{2}\,g^{-1}}$ with $\beta\approx 1.5$ and $\tilde{Z} \equiv Z/Z_{\odot}$,\footnote{Our scalings can be extrapolated to any $\tilde{Z}$, but motivated by modern numerical simulations and more detailed QSO full-spectrum line modeling \citep[e.g.][]{temple:2021.qso.spectra.solar.metallicity.not.supersolar.components.just.come.from.different.radii}, we will often refer to a reference $\tilde{Z} \sim 1$, as opposed to the highly super-Solar $\tilde{Z}\sim 10-50$ sometimes invoked in older quasar models (these highly super-solar values are based on more simplified older line models which assumed all observed lines come from a single density and ionization parameter and source distance).} until $T_{\rm dust} \gtrsim 1500-2000\,$K where the dust sublimates ($\tilde{Z}\rightarrow 0$) \citep{semenov:2003.dust.opacities,draine:ism.book}. These tabulations for the opacities are used for the calculation of the opacities in this zone shown in Fig.~\ref{fig:opacity.profile}. Given that the free electron fractions in this region are $\lesssim 0.01$ (\paperone), and metallicities are $\tilde{Z} \sim 1$, this will dominate over other sources of opacity. So we have:
\begin{align}
\tau^{\ast} & \approx {\rm MIN}[0.021 T_{100}^{2},\,0.035] \frac{m_{7}^{1/2}\,\dot{m}\,\tilde{Z}}{r_{\rm ff,\,5}^{1/2}} \left(\frac{R}{r_{\rm ff}} \right)^{-5/6}
\end{align}
where $T_{100} \equiv T_{\rm rad}/100\,{\rm K}$. Fig.~\ref{fig:cartoon.definitions} shows the range of scales (calculated below) over which this ``zone'' applies; Fig.~\ref{fig:cartoon.quadrants} illustrates the characteristic thermochemical properties qualitatively, and breaks the zone up into sub-zones which we describe below. The quantitative values of the temperature, opacity, and pressure properties, given the approximations in this section, are shown for various values of the BH mass and accretion rate in Figs.~\ref{fig:thermal.profile}, \ref{fig:opacity.profile}, \ref{fig:pressure.profile}, respectively.

Note the covering factor reaches $\sim 0.7$ for the very outer edge of this disk as $R\rightarrow r_{\rm ff}$, but only some of this is able to be illuminated by the central disk (illustrated in Fig.~\ref{fig:cartoon.illumination}). At the (order-unity) values of $H/R$ predicted for the radii of interest here, it makes a negligible difference for our conclusions whether we assume the central source is an optically-thick flat aligned disk or assume quasi-isotropic illumination (i.e.\ either form of the function $f_{\theta}$ in \S~\ref{sec:model:illumination} we assume gives nearly identical results).

\subsection{2a: The Outer, Optically-Thin Cold Dust Region}
\label{sec:zone2a.cold.torus}

The outer regions will therefore be optically-thin to the cooling radiation of the dust. We show its properties as Zone (2a) in Fig.~\ref{fig:cartoon.quadrants}, and quantitatively see the effects of optically-thin cooling in e.g.\ the temperature calculation in Fig.~\ref{fig:thermal.profile} and corresponding pressures (Fig.~\ref{fig:pressure.profile}). The optically-thin cooling flux $F_{\rm cool,\,thin} = 4\pi\,\kappa_{\rm em}\,\Sigma_{\rm gas}\,\sigma_{B}\,T_{\rm dust}^{4}$ exceeds $F_{\rm grav}$ so long as 
$T_{\rm dust} \gtrsim 12\,{\rm K}\,m_{7}^{3/8}\,r_{\rm ff,\,5}^{-5/8}\,(\kappa_{\rm em}/5\,{\rm cm^{2}\,g^{-1}})^{-1/4}\,(R/r_{\rm ff})^{-13/24}$, which is easily maintained. In practice the midplane regions of this zone, while shielded from the central source, will be {\em externally} heated by the massive star irradiation and cosmic ray production in Zone (1), and maintain typical nuclear dust and gas temperatures in equilibrium with one another at these densities (but not necessarily with $T_{\rm rad}$) at 
\begin{align}
T_{\rm dust} \sim T_{\rm gas} \sim 20-100\,K.
\end{align} 
The upper parts of the disk near $|z| \sim H$ will be directly illuminated by the central disk, and maintain warmer dust temperatures as discussed below: 
\begin{align}
%T_{\rm gas} \sim T_{\rm dust} \sim 200\,{\rm K}\,(m_{7}\,\dot{m}\,\epsilon_{r,\,0.1}/r_{\rm ff,\,5}^{2})^{1/4}\,(r_{\rm ff}/R)^{1/2},
T_{\rm gas} \sim T_{\rm dust} \sim 200\,{\rm K} \left( \frac{m_{7}\,\dot{m}\,\epsilon_{r,\,0.1}}{r_{\rm ff,\,5}^{2}}\right)^{1/4} \left(\frac{r_{\rm ff}}{R}\right)^{1/2}, 
\end{align} 
until sublimation (noted below).\footnote{If the irradiated layer can efficiently transfer heat to the central layers below it and reach equilibrium, then the optically-thin cooling balances heating for $T_{\rm dust} \sim 200\,{\rm K}\,m_{7}^{1/8}\,r_{\rm ff,\,5}^{-3/8}\,\epsilon_{r,\,0.1}^{1/4}\,(\kappa_{\rm em}/5\,{\rm cm^{2}\,g^{-1}})^{-1/4}\,(R/r_{\rm ff})^{-5/24}$, similar to the directly heated value but much more weakly dependent on radius and other parameters. It is not obvious how this would occur, but we note it for completeness.}

In most ways, this is akin to Zone (1) and ``normal'' GMCs at least qualitatively (through quantitatively it is ``more extreme'' in most respects): it is optically-thin to its own cooling radiation, the dust and gas are both ``cool'' and (given the low densities) may not be in exact equilibrium. The plasma $\beta \ll 1$ and the turbulence is super-sonic 
\begin{align}
\mathcal{M}_{s} \equiv \frac{v_{\rm turb}}{c_{s}} \sim \beta^{-1/2} \sim 100\, m_{7}^{1/2} r_{\rm ff,\,5}^{-1/2} T_{100}^{-1/2} \left(\frac{r_{\rm ff}}{R}\right)^{1/3}
\end{align}
 (quantitative examples of $\mathcal{M}_{s}$ are shown in Fig.~\ref{fig:turb.props}, for different BH masses and accretion rates), with cooling times much shorter than dynamical times $t_{\rm cool} \ll t_{\rm dyn} \sim 1/\Omega$. Just like in GMCs the gas should therefore be multiphase (with the phases still neutral but akin to a warm and cold atomic and cold molecular medium) with large density contrasts owing to the supersonic motions, with the largest density contrasts reaching $\sim \mathcal{M}_{s}^{2} \sim 10^{4}$ (for shocks along the field lines, as given the rapid cooling these are radiative/isothermal shocks; see Fig.~\ref{fig:turb.props}). The characteristic size of such extreme overdensities will reach as small as the sonic length or post-shock length in supersonic turbulence, 
\begin{align}
R_{\rm sonic} \sim \frac{H}{\mathcal{M}_{s}^{2}} \sim 10^{15}\,{\rm cm}\,r_{\rm ff,\,5}^{2}\,m_{7}^{-1}\,T_{100}\,(R/r_{\rm ff})^{11/6}.
\end{align}
 The variation of that size scale with radius (again for different accretion rates and BH masses) is shown in Fig.~\ref{fig:sobolev.size}. Note that (like in GMCs, but even moreso here) it is {\em not} correct to assume that different phases or clumps are in thermal pressure equilibrium, since thermal pressure is vastly subdominant to both magnetic and turbulent ram pressure. 

While some very small amount of star formation can (and does, in simulations; see \citealt{hopkins:superzoom.imf}) persist at these radii, it is strongly suppressed (compared to e.g.\ the expectations for classical $\alpha$-disks discussed in \citealt{goodman:qso.disk.selfgrav}), as the Toomre $Q \gg 1$ (Eq.~\ref{eqn:Qtot}) and similarly the magnetic critical mass is formally larger than the entire disk mass (so collapse can only occur in very special regions, e.g.\ local magnetic field line polarity switches with strong shocks along the field lines). But unlike in the ISM, young/massive stars are not needed to power either the turbulence or ``maintain'' the large scale heights here. The magnetic field (which can grow to these values purely via flux-freezing) and turbulence powered by the fields and gravitational flux, before being dissipated, supports the material vertically and prevents denser gas from ``sinking'' or sedimenting to the midplane (by definition in the models here, but easily verified from the scalings of Eq.~\ref{eqn:veff.model}-\ref{eqn:Qtot}).

The covering factors of this zone 
\begin{align}
f_{\rm cover} \approx \cos{\left(\tan^{-1}\frac{R}{H}\right)}=\frac{H/R}{\sqrt{1+(H/R)^{2}}} \sim 0.5-0.7
\end{align}
 extend from $f_{\rm cover} \sim 0.5\,(m_{7}/r_{\rm ff,\,5})^{1/10}\,(\dot{m}\,\tilde{Z})^{1/5}$ at its inner boundary to $f_{\rm cover} \sim 0.7$ at its outer boundary (see Fig.~\ref{fig:cartoon.illumination}). This is therefore important for re-radiation/reprocessing (discussed below) and obscuration, but the intrinsic gravitational cooling luminosity (i.e.\ distinct from the reprocessed radiation) coming from this region is small, $\sim 10^{41}\,{\rm erg\,s^{-1}}\,m_{7}^{7/5}\,r_{\rm ff,\,5}^{-2/5}\,\tilde{Z}^{-6/5}\,\dot{m}^{-1/5}$ (note this can {\em decrease} with $\dot{m}$ because of how the location of this region varies with $\dot{m}$), a tiny fraction of the bolometric luminosity for the parameter space of interest.

\subsection{2b: The Inner, Thermalized Warm Dust Region}
\label{sec:zone2b.thermalized.dust}

At radii interior to 
\begin{align}
R \lesssim r_{\rm dust,\,therm} \approx 0.1\,{\rm pc}\,m_{7}^{3/5}\,r_{\rm ff,\,5}^{2/5}\,(\dot{m}\,\tilde{Z})^{6/5}, 
\end{align}
$\tau^{\ast}$ will become $\gtrsim 1$, and the dust will thermalize and radiate as a blackbody (zone (2b) in Fig.~\ref{fig:cartoon.quadrants}). The gas densities here are 
\begin{align}
n_{\rm gas} \sim 10^{6}\,\dot{m}^{-7/5}\,m_{7}^{-7/10}\,r_{\rm ff,\,5}^{-3/10}\,\tilde{Z}^{-12/5}\,(R/r_{\rm dust,\,therm})^{-2}, 
\end{align}
so dust and gas are efficiently collisionally coupled, and now $T_{\rm gas} \approx T_{\rm dust} \approx T_{\rm rad}$. Equating the blackbody cooling flux $F_{\rm cool,\,thick} = 2\,\sigma_{B}\,T_{\rm eff}^{4}$ to $F_{\rm grav}$ gives an effective temperature $T_{\rm eff} \approx 200\,{\rm K}\,m_{7}^{1/20}\,\dot{m}^{-13/20}\,r_{\rm ff,\,5}^{-3/10}\,\tilde{Z}^{-9/10}\,(R/r_{\rm dust,\,therm})^{-3/4}$, so the dust is at hundreds of ${\rm K}$ (justifying the opacity used to calculate where $\tau^{\ast}>1$). But here, the re-radiation from the illuminated layer above should indeed be reprocessed within the disk, so it is more appropriate to equate to $F_{\rm illum}$ rather than $F_{\rm grav}$, giving 
\begin{align}
T_{\rm eff} \approx 1000\,(\epsilon_{r,\,0.1}/\dot{m}\,r_{\rm ff,\,5}\,\tilde{Z}^{2})^{1/4}\,(R/r_{\rm dust,\,therm})^{-5/12}
\end{align}
 (Fig.~\ref{fig:thermal.profile}). Under these conditions the radiation transport is diffusive ($\tau^{\ast} \sim \tau \gtrsim 1$, and the vertical turbulent transport speed $v_{\rm turb}$ as well as radial/inflow advection transport $v_{r} \sim v_{\rm turb}^{2}/v_{\rm K}$ are much smaller than the diffusive speed $v_{\rm diff} \sim c/\tau$ here), so the dust becomes somewhat more mono-phase at a given position and can set up a stratified plane-parallel radiation temperature gradient with midplane temperature $T_{\rm mid} \approx T_{\rm eff}\,\tau^{1/4} \approx T_{\rm eff}\,(R/r_{\rm dust,\,therm})^{-5/24}$, which is a modest enhancement to the midplane over effective temperature, over the range of interest. For this reason, in Fig.~\ref{fig:thermal.profile}, we show multiple temperatures corresponding to different regions or viable phases of the gas. This plus the calculated densities (Fig.~\ref{fig:3d.densities}) and radiation fluxes gives us the resulting pressures in Fig.~\ref{fig:pressure.profile}. The covering factor $f_{\rm cover}$ is
 \begin{align}
 f_{\rm cover} \sim 0.5\,(m_{7}/r_{\rm ff,\,5})^{1/10}\,(\dot{m}\,\tilde{Z})^{1/5}\,(R/r_{\rm dust,\,therm})^{1/6}
 \end{align}
  (Fig.~\ref{fig:cartoon.illumination}). But the turbulence is still highly supersonic $\mathcal{M}_{s} \gtrsim 100$ (Fig.~\ref{fig:turb.props}) and $t_{\rm cool} \ll t_{\rm dyn}$ so the medium can still be clumpy/inhomogeneous down to the sonic scale 
 \begin{align}
 \nonumber R_{\rm sonic} \sim& 7\times10^{12}\,{\rm cm}\,m_{7}^{1/10}\,r_{\rm ff,\,5}^{13/20}\,\dot{m}^{39/20}\,\tilde{Z}^{17/10}\,\epsilon_{r,\,0.1}^{1/4}\,\times
\\
 & \ \ \ \ (R/r_{\rm dust,\,therm})^{29/24} 
 \end{align}
  (Fig.~\ref{fig:sobolev.size}).

When $T_{\rm dust}$ at the midplane $T_{\rm mid}$ exceeds the sublimation temperature $T_{\rm sub} \sim 1500\,$K, it will begin to be destroyed: this will begin at $R_{\rm sub,\,T_{\rm mid}}/r_{\rm dust,\,therm} \sim 0.12\,m_{7}^{6/115}\,r_{{\rm ff},\,5}^{-36/115}\,\dot{m}^{-78/115}\,\tilde{Z}^{-108/115}\,T_{\rm sub,\,1500}^{-24/23}$ ignoring external illumination, or $\sim 0.5\,\epsilon_{r,\,0.1}^{2/5}\,\dot{m}^{-2/5}\,r_{\rm ff,\,5}^{-2/5}\,\tilde{Z}^{-4/5}\,T_{\rm sub,\,1500}^{-8/5}$ including it (the illuminated surface layer is discussed more below). But this will lower the midplane opacity, making the midplane cooler, and slowing sublimation. Sublimation will therefore not be complete until it occurs at the effective temperature $T_{\rm eff} \gtrsim T_{\rm sub}$, which occurs at $R_{\rm sub,\,T_{\rm eff}}/r_{\rm dust,\,therm} \sim 0.07\,m_{7}^{1/15}\,\dot{m}^{-13/15}\,r_{\rm ff,\,5}^{-2/5}\,\tilde{Z}^{-6/5}\,T_{\rm sub,\,1500}^{-4/3}$ or 
\begin{align}
R_{\rm sub,\,T_{\rm eff}} \sim 2\times10^{16}\,{\rm cm}\,m_{7}^{2/3}\,\dot{m}^{1/3}\,T_{\rm sub,\,1500}^{-4/3}
\end{align}
 ignoring external illumination, or 
 \begin{align}
 R_{\rm sub,\,T_{\rm eff}} \sim 10^{17}\,{\rm cm}\,\epsilon_{r,\,0.1}^{3/5}\,m_{7}^{3/5}\,\dot{m}^{3/5}\,r_{\rm ff,\,5}^{-1/5}\,T_{\rm sub,\,1500}^{-12/5}
 \end{align}
 including it. This inner boundary appears in Figs.~\ref{fig:cartoon.definitions}-\ref{fig:cartoon.quadrants}.

The covering factor at the inner radius $R_{\rm sub,\,T_{\rm eff}}$ of this zone (Fig.~\ref{fig:cartoon.illumination}) is $f_{\rm cover}^{\rm inner} \approx 0.33\,m_{7}^{1/9}\,\dot{m}^{1/18}\,r_{\rm ff,\,5}^{-1/6}\,T_{\rm sub,\,1500}^{-2/9}$ (ignoring illumination) or $f_{\rm cover}^{\rm inner}  \approx 0.43\,(m_{7}\,\dot{m}\,\epsilon_{r,\,0.1}/r_{\rm ff,\,5}^{2}\,T_{\rm sub,\,1500}^{4})^{1/10}$ (including it), remarkably independent of the BH accretion properties (Fig.~\ref{fig:cartoon.mdot} illustrates how the boundaries of these zones move in radius and $H/R$ with varying $\dot{m}$), and again $\mathcal{O}(1)$ so clearly important for re-radiation and obscuration, but the intrinsic (neglecting reprocessing) gravitational cooling emission luminosity is 
$\sim 3\times10^{41}\,{\rm erg\,s^{-1}}\,m_{7}^{7/5}\,\dot{m}^{2/5}\,T_{\rm sub,\,1500}^{12/5}\,\epsilon_{r,\,0.1}^{-3/5}$, 
still a small fraction ($\sim 0.1\%$) of the bolometric output. The maximum dust column density for a sightline through the midplane (integrating from the sublimation radius to infinity) is equivalent to a gas column of 
\begin{align}
N_{\rm H} \sim 3\times10^{23}\,{\rm cm^{-2}}\,\dot{m}^{1/5}\,m_{7}^{-1/10}\,r_{\rm ff,\,5}^{1/2}\,\tilde{Z}^{3/5}\,T_{\rm sub,\,1500}^{8/5}
\end{align} 
(the usual units quoted assuming a standard dust-to-gas ratio, or $A_{v} \sim 160\,\dot{m}^{1/5}\,m_{7}^{-1/10}\,r_{\rm ff,\,5}^{1/2}\,\tilde{Z}^{3/5}\,T_{\rm sub,\,1500}^{8/5}$). 
This ``midplane'' section of this zone will therefore not contribute much to the direct emission/SED, but will be important for obscuration of the central source, as viewed from angles intercepted by its covering factor. Notably these predicted covering factors, clumping factors, and column densities are very similar to the canonical ``obscuring torus'' properties inferred from observations \citep{lawrence:1982.torus.alignment,antonucci:1982.torus,krolik:clumpy.torii,mor:2009.torus.structure.from.fitting.obs,hatziminaoglou:2009.torus.properties.inferred.obs,hoenig:clumpy.torus.modeling,2011ApJ...736...82A,koshida:2014.agn.torii.sizes,hoenig:2019.ir.submm.torus.wind.review,cackett:2021.reverberation.mapping.multiwavelength.review}.

\subsection{2c: The Directly-Illuminated Dusty Region}
\label{sec:zone2c.illum.torus}

Because the disk is flared with a covering factor $\mathcal{O}(1)$ at these radii, above the scale height of the inner sublimation zone $\sim 0.33$ (below this is shielded by the entire disk with Compton-thick column densities), there will a region near the top of the disk ($|z| \sim H$) ``exposed'' to the direct radiation of the central source (zone (2c) in Figs.~\ref{fig:cartoon.definitions}, \ref{fig:cartoon.quadrants}, \ref{fig:cartoon.illumination}). This region will be optically-thin or only modestly optically-thick to electron scattering ($\tau_{\rm es} \sim 0.1-10$ even if fully-ionized, and much smaller in the neutral part of this region where free electron fractions range from $\sim 10^{-8}-10^{-2}$; Fig.~\ref{fig:opacity.profile}), so this can be neglected here. Given the non-linearly large $H/R$ and scattering regions (discussed below, see Fig.~\ref{fig:cartoon.illumination}) around the central source it makes little difference if we take the incident flux $F_{\rm incident} = f_{\theta}\, L_{\rm bol}/(4\pi\,r^{2})$ to be isotropic or follow $f_{\theta}=2\,\cos{\theta}$, along rays from the center (recall though this is distinct from the heating rate per unit area, which scales as $F_{\rm illum}$). In either case this will sublimate dust out to a radial distance 
\begin{align}
 r_{\rm sub,\,ext} &\sim (L_{\rm bol}\,\tilde{Q}/16\pi\,\sigma_{B}\,T_{\rm sub}^{4})^{1/2} \\
\nonumber & \sim 3\times10^{17}\,{\rm cm}\,(\epsilon_{r,\,0.1}\,m_{7}\,\dot{m}\,\tilde{Q}_{\rm abs})^{1/2}\,T_{\rm sub,\,1500}^{-2}.
\end{align}
 Here $\tilde{Q}=Q_{\rm abs}/Q_{\rm em}$ depends on the ratio of the absorption and emission coefficients which themselves depend on the incident spectral shape, dust composition, and temperature, but in general this is not a large correction. This is the more commonly-quoted ``sublimation radius.'' Generically this is $\gg R_{\rm sub,\,T_{\rm eff}}$ (where 
$H/R=(H/R)_{\rm in} \sim 0.43$ 
above) but $\ll r_{\rm ff,\,5}$ (generally comparable to $r_{\rm dust,\,therm}$), with $H/R$ at $r_{\rm sub,\,ext}$ giving a covering factor if we include the entire disk up to this height of $f_{\rm cover}^{\rm out,\,illum} \sim (H/R)_{\rm out} \sim 0.51\,(m_{7}\,\dot{m}\,\tilde{Q}\,\epsilon_{r,\,0.1}/r_{\rm ff,\,5}^{2}\,T_{\rm sub,\,1500})^{1/6}$. 

So the covering fraction specifically of directly illuminated ``hot'' dust, where $T_{\rm dust}$ approaches an appreciable fraction of the sublimation temperature is given by 
\begin{align}
f_{\rm cover}^{\rm illum} \sim f_{\rm cover}^{\rm out,\,illum} - f_{\rm cover}^{\rm inner}\sim 0.08
\end{align}
at $H/R \sim 0.4-0.5$ (Fig.~\ref{fig:cartoon.illumination}). So (again roughly independent of $f_{\theta}$) it will re-radiate a similar fraction of $L_{\rm bol}$ at hot dust temperatures, and a fraction $\sim 0.2-0.5$ at cooler dust temperatures (corresponding to the illuminated surface reaching $H/R \sim 0.7$ at the outer $r_{\rm ff,\,5}$, and as large as $\sim 1$ depending on if there is colder dust in zone 1 from the ISM). Again, this is consistent with what is typically inferred from observations of the mid-IR emission spectra of quasars (see references above and \citealt{alonso.herrero:2021.torus.emission.properties,lyu:2022.torus.variability.sed.models}) and/or more recent spatially-resolved imaging studies \citep{garcia.burillo:2019.alma.torus.imaging,garcia.burillo:2021.torus.imaging.alma,cackett:2021.reverberation.mapping.multiwavelength.review,gravity:2021.resolved.blr.disk.hot.dust.coronal.regions,izumi:2023.imaging.nuclear.gas.disk.circinus.accretion.rate} and associated with the ``warm/hot dusty torus.''

\begin{figure*}
	\centering\includegraphics[width=0.97\textwidth]{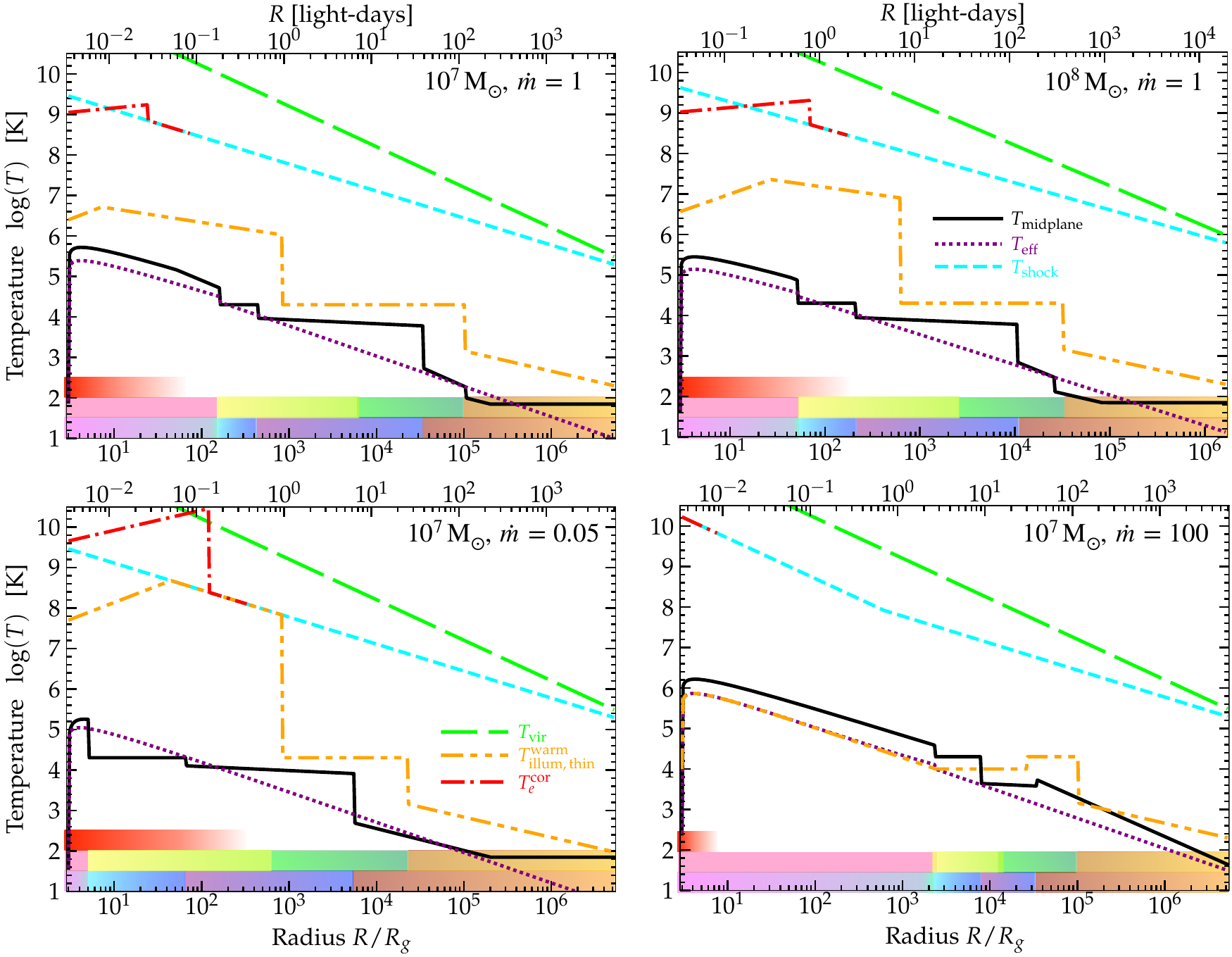} 
	\vspace{-0.2cm}
	\caption{Predicted disk temperature profiles for the flux-frozen magnetized ($\beta \ll 1$) disks versus radius as Figs.~\ref{fig:surface.densities} \&\ \ref{fig:3d.densities}.
	We compare different temperatures (\S~\ref{sec:model:general:temperature}) at each radius calculated in \S~\S~\ref{sec:zone1.ism}-\ref{sec:zone8.warm.scattering}: the mean midplane gas temperature $T_{\rm midplane}$; effective temperature including heating by gravitational dissipation and (approximate) self-illumination $T_{\rm eff}$; characteristic maximum post-shock/reconnection temperature for the disk internal turbulence $T_{\rm shock} \sim (3/16)\,m_{p}\,v_{\rm turb}^{2}/k_{B} \sim (3/16)\,m_{p}\,v_{A}^{2}/k_{B}$; virial temperature $T_{\rm vir} \sim (1/2) (G\,M_{\rm BH}/r)\,(\mu\,m_{p}/k_{B})$; temperature of optically-thin, directly illuminated surface layers or ``skin'' (at $|z|\approx H$; taking the ``warmer'' solution when multiple quasi-stable solutions exist), $T^{\rm warm}_{\rm illum,\,thin}$ or $T_{\rm skin}$; and the estimated ``hard'' electron coronal temperature $T^{\rm cor}_{e}$ where coronal solutions are supported (\S~\ref{sec:zone7.corona}). 
	We see clear transitions corresponding to the predicted zone/phase boundaries. 
	Flux-frozen disks have cooler $T_{\rm midplane}$ than SS73 owing to their lower optical depths, but more importantly exhibit obvious multi-phase structure with gas at a range of temperatures supported. The similarity between the predicted temperatures and those needed to explain e.g.\ the soft excess in the warm skin, hard coronal X-rays, BLR emission lines, torus reprocessed warm dust emission, and the thermal continuum/big-blue-bump are discussed in \S~\ref{sec:zone1.ism}-\ref{sec:zone8.warm.scattering}. Note that one cannot directly translate $T_{\rm eff}$ versus $R$ into a predicted emission region size or observed effective temperature because most of the radiation from the inner region will be absorbed or reprocessed by the large $H/R$, flared disk (Fig.~\ref{fig:cartoon.quadrants} and below).
	\label{fig:thermal.profile}}
\end{figure*}

\begin{figure*}
	\centering\includegraphics[width=0.97\textwidth]{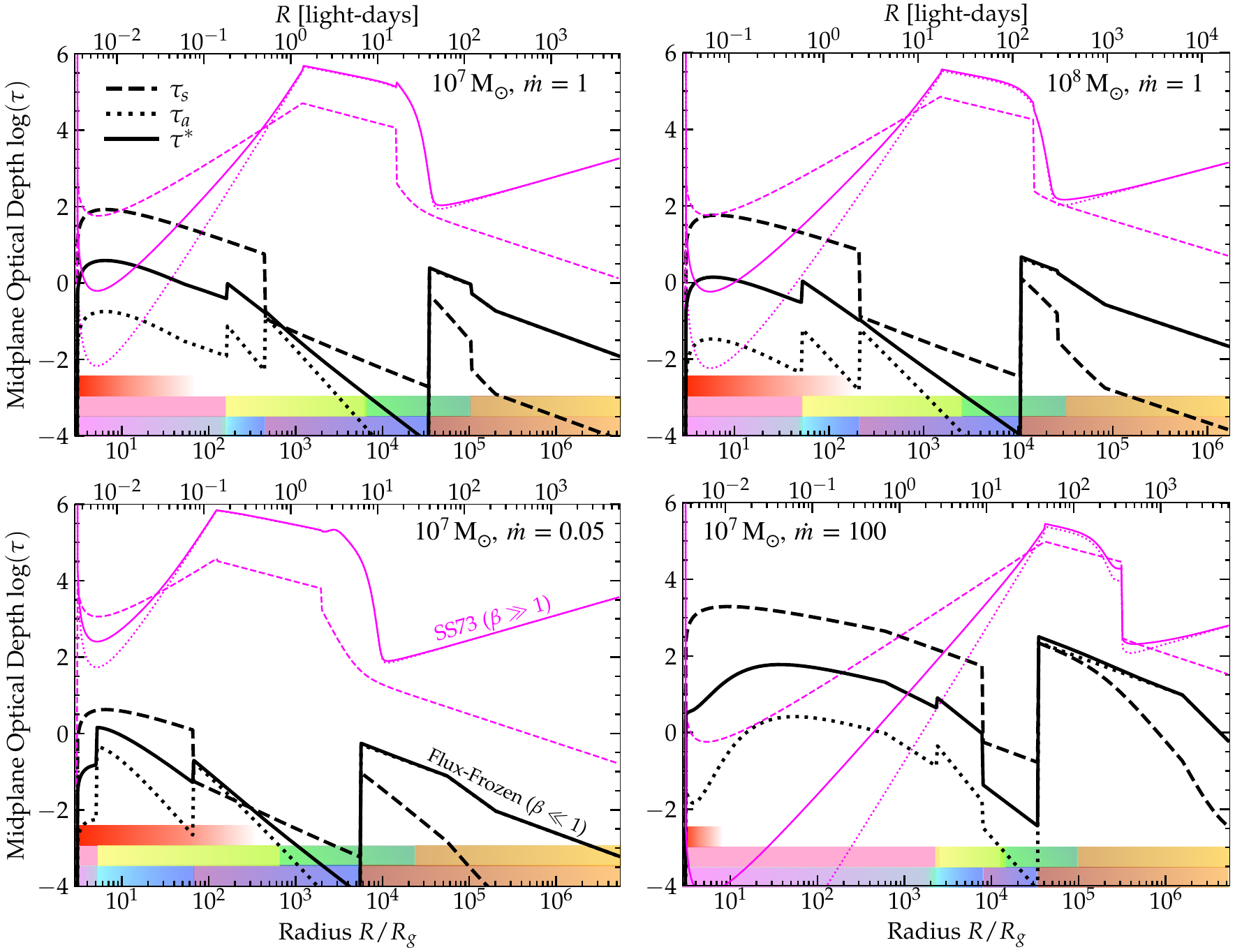} 
	\vspace{-0.2cm}
	\caption{Opacity structure of the disks, as Fig.~\ref{fig:surface.densities}. We plot the vertically-integrated scattering $\tau_{s}$, absorption $\tau_{a}$, and ``effective'' absorption $\tau^{\ast} \equiv \sqrt{\tau_{a}\,(\tau_{a} + \tau_{s})}$ optical depths to the midplane (linestyles labeled). We compare flux-frozen ($\beta\ll 1$; {\em thick black}) and thermal-pressure-dominated ($\beta\gg1$; {\em thin pink}) disks. For the sake of direct comparison note we have recalculated the SS73 models using our more detailed opacity models here (including e.g.\ dust, atomic, and iron absorption line opacities), but this does not change the qualitative results. The optical depths are generally much lower in the magnetized disks as expected from Fig.~\ref{fig:surface.densities}, but the different temperature and density structure leads to qualitatively different scalings of the opacities in many regions. Discontinuities owing to phase changes (e.g.\ dust sublimation) are pronounced here. The ratio of e.g.\ $\tau_{s} > \tau_{a}$ in the thermalized disk is important for Comptonization (\S~\ref{sec:zone6.disk}) while (given the large $H/R$ and reprocessing; Fig.~\ref{fig:cartoon.quadrants}) the effective temperature of e.g.\ the big blue bump is influenced by reprocessing at the radii around $R\sim 100\,R_{g}$ where the absorption optical depths first drop below unity, and low mean absorption $\tau_{a} \lesssim 1$ depths in the BLR-like and scattering zones (3 \&\ 8) are critical for observed line emission and scattering (\S~\ref{sec:zone3.blr} \&\ \ref{sec:zone8.warm.scattering}). Below $\dot{m} \lesssim 0.01$, optical depths become so low that some regions of the disk cannot cool efficiently.
	\label{fig:opacity.profile}}
\end{figure*}

\begin{figure*}
	\centering\includegraphics[width=0.95\textwidth]{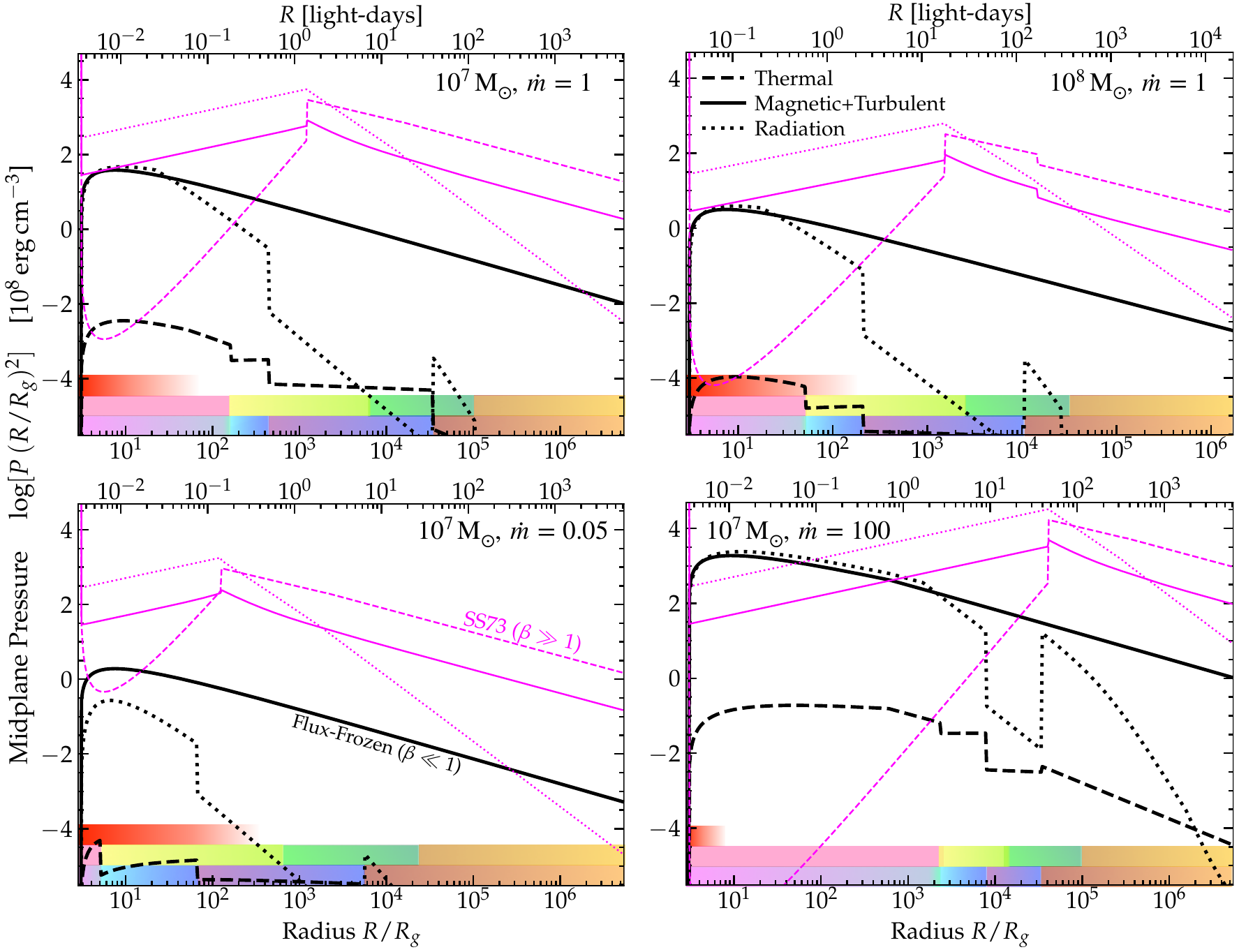} 
	\vspace{-0.2cm}
	\caption{Midplane pressures of the disks, as Fig.~\ref{fig:surface.densities}. We plot the compensated pressure, $P_{i} \times (R/R_{g})^{2}$ (for ease of visualization), at the midplane ($|z|\ll H$), for thermal pressure ($P_{\rm thermal} \equiv \rho_{\rm mid}\,c_{s,\,{\rm mid}}^{2} = n_{\rm mid} k_{B} T_{\rm midplane}$; Figs.~\ref{fig:3d.densities}, \ref{fig:thermal.profile}), magnetic+thermal pressure (which scale together as $P_{\rm mag} \sim |B_{\rm midplane}|^{2}/8\pi \sim (1/2) \rho_{\rm mid} v_{A,\,{\rm mid}}^{2} \sim P_{\rm turb} \sim (1/2) \rho_{\rm mid}\,v_{\rm turb}^{2}$  in both the fluz-frozen and SS73-like $\alpha$-disk models), and radiation pressure ($P_{\rm rad}$, accounting for finite optical depth effects by defining this as the radiation pressure/force per unit area {\em exerted on gas} in the midplane, $P_{\rm rad} \sim \langle \Sigma_{\rm gas} \kappa {F}_{{\rm rad,\,mid}} / c \rangle$ in terms of the midplane flux $F_{\rm rad,\,mid}$). Owing to the much lower densities (Fig.~\ref{fig:3d.densities}), the pressures -- even magnetic pressure/field strength $|{\bf B}|^{2}$ -- in the flux-frozen disks are almost everywhere much lower than in thermal-pressure-dominated models (see \paperthree). While SS73 models become radiation-pressure-dominated at small radii even at modest accretion rates $\dot{m} \sim 0.05-1$, the radiation pressure even at $\dot{m} \gg 1$ never becomes much larger in the flux-frozen models than the magnetic pressure (\S~\ref{sec:super.eddington}).
	\label{fig:pressure.profile}}
\end{figure*}

The non-sublimated but illuminated ``wedge'' at $R \gtrsim r_{\rm sub,\,ext}$ will otherwise have properties (besides dust temperature) similar to region (2a), per Figs.~\ref{fig:surface.densities}-\ref{fig:sobolev.size}. We discuss the sublimated, externally-illuminated wedge at $R\lesssim r_{\rm sub,\,ext}$ next. Its properties, along with those of potential line-driven outflows discussed in \S~\ref{sec:super.eddington:outflows}, are predicted to scale together with the inner boundary of the torus-like region (forming its outer boundary), both moving outwards (therefore increasing $H/R$ and their corresponding covering factors) at higher Eddington ratios $\dot{m}$, in a manner that appears quite similar to recent observational inferences in e.g.\ \citet{temple:2021.agn.outflows.assoc.hot.dust.covering,temple:2023.outflow.lines.wind.association.eddington.ratio}.

\section{Zone 3: The Directly-Illuminated, Ionized Upper Disk as the Broad Line Region}
\label{sec:zone3.blr}

As noted above and shown in Fig.~\ref{fig:cartoon.illumination}, rays from the central disk with $\cos{\theta} \gtrsim 0.7$ will not intersect the disk, but travel to Zone (1) (the ISM). Rays with $0.5\lesssim \cos{\theta}\lesssim 0.7$ will intersect between $r_{\rm sub,\,ext} \lesssim R \lesssim r_{\rm ff}$, where the dust will not be sublimated, i.e.\ Zone (2c), where the dust will strongly shield the gas against ionization by the central source. Rays with $\cos{\theta} \lesssim 0.5$ will intersect at $r \lesssim r_{\rm sub,\,ext}$, where the dust is efficiently sublimated. Over an intermediate range of angle $\theta$ or disk height $H/R$, this will create an ionized layer reaching out to $r_{\rm sub,\,ext}$. 

\subsection{3a: The ``Fully'' Ionized (Dust-Limited) Layers}
\label{sec:zone3a.ionized}

For the directly illuminated, dust-sublimated (effectively dust-free) disk, two effects will limit ionization. First, the ionization layer could be photon-bounded. We can estimate this via a simple Stromgren type argument. Consider all rays from the central source in some azimuthally-symmetric $\Delta \cos{\theta}$ opening angle, with some number of ionizing photons per unit bolometric luminosity $(Q/L)$  which depends on the spectrum of the inner disk, so $\dot{N}_{\rm ion}/dA \sim 2\pi\,\Delta\cos{\theta}\,f_{\theta}\,(Q/L)\,L_{\rm bol}/(4\pi\,r^{2})$. If this ionizes a ``wedge'', there must be sufficient number of photons per unit time per unit solid angle to offset recombination in the wedge assuming it is fully-ionized, given by $\dot{N}_{\rm rec} \approx \int \alpha_{\rm rec}\,n(r,\,\theta,\,\phi)^{2}\,r^{2}\,{\rm d}\cos{\theta}\,{\rm d}\phi\,{\rm d}r$ where $\alpha_{\rm rec} \approx 4\times10^{-13}\,{\rm cm^{3}}$ for the conditions of interest. Comparing the two (noting $n\propto R^{-2}$ in the disk model), we see that the ``fully ionized'' condition (requirement to ionize to $r\rightarrow r_{\rm sub,\,ext}$) is completely dominated by the ability to ionize the innermost radii where $n(R)$ is maximized (Eq.~\ref{eqn:rho} and Figs.~\ref{fig:surface.densities}, \ref{fig:3d.densities}, \ref{fig:opacity.profile}), and this can occur for rays intercepting the disk initially at $R \gtrsim R_{\rm min,\,strom}$ (so at $\cos{\theta} \gtrsim H/R|_{R=R_{\rm min,\,strom}}$) given by 
\begin{align}
R_{\rm min,\,strom} &\approx 1.6 \times10^{16}\,{\rm cm}\,r_{\rm ff,\,5}\,\dot{m}^{6/7}\,\tilde{Q}^{-6/7} \\ 
\nonumber &\sim 6\,{\rm ld}\,(L_{\rm bol}/10^{45}\,{\rm erg\,s^{-1}})^{1/2}\,\epsilon_{r,\,0.1}^{-1/2} \tilde{Q}^{-6/7} \dot{m}^{5/14}
\end{align}
($\tilde{Q} \equiv \epsilon_{r,\,0.1}\,(Q/L)/(0.1/13.6\,{\rm eV})\sim 1$ for a typical quasar spectrum; \citealt{shen:bolometric.qlf.update}), where 
\begin{align}
H/R|_{R=R_{\rm min,\,strom}} \approx 0.3\,(\dot{m}/\tilde{Q})^{1/7}.
\end{align}
 At heights larger than this, but less than $\sim 0.5$ where the ray intersects the disk outside $r_{\rm sub,\,ext}$ (Figs.~\ref{fig:cartoon.illumination} \&\ \ref{fig:cartoon.mdot}), the wedge will be ionized but only a fraction of the incident ionizing flux is ``needed'' and therefore some of the ionizing radiation will reach the sublimation region and be attenuated by dust, giving rise to intermediate zones with lower-excitation ions. 

For incident heights $\lesssim 0.3 \sim H/R|_{R=R_{\rm min,\,strom}}$, the ionizing photons will be fully-consumed in a surface layer of radial size $\ll R$, beyond which (noting that $R_{\rm therm,\,atomic} < R_{\rm min,\,strom} < R_{\rm sub,\,T_{\rm eff}}$) the layer will resemble the underlying atomic Zone (4) described below, until reaching the dusty Zone (2c). But note that the Stromgren approximation assumes negligible electron scattering or continuum (Kramers) absorption optical depth. The latter assumption is quite good over the specific range of radii here, but the former assumption becomes invalid at small $R$, as electron scattering can further limit ionization by scattering photons out of the narrow ``wedge'' and into polar angles where they can escape. Assuming a fully-ionized surface ``layer'' at some $R$ and $H$, the electron scattering optical depth to some depth $d\ell \ll R$ is $\sim 0.35\,\rho(R)\,d\ell$, if we solve for $\ell_{\rm es}$ where this exceeds a factor of $\tau_{\rm crit}\sim$\,a couple, we have $\ell_{\rm es}/R \sim 0.26\,\tau_{\rm crit}\,(R/10^{16}\,{\rm cm})\,m_{7}^{-1/2}\,r_{\rm ff,\,5}^{-1/3}\,\dot{m}^{-1}$. When $\ell_{\rm es} \ll R$ ($R \ll R_{\rm es,\,ion}$) where
\begin{align}
R_{\rm es,\,ion} \approx 3.8\times10^{16}\,{\rm cm}\,\dot{m}\,(m_{7}\,r_{\rm ff,\,5})^{1/2}\,\tau_{\rm crit}^{-1},
\end{align}
 this effectively means that the wedge further exterior is shielded by a relatively narrow Thompson-thick scattering layer. At $R < R_{\rm es,\,ion}$, the maximum possible reprocessed ionizing luminosity will scale $\propto R^{4/3}$ or steeper. Thus the reprocessed ionizing luminosity will come predominantly from the radii outside the maximum of either $R_{\rm es,\,ion}$ or $R_{\rm min,\,strom}$. But it is notable that these are very similar to one another, so basically $R_{\rm es,\,ion}$ serves to further ``cut off'' the fraction reprocessed at $R \lesssim R_{\rm min,\,strom}$ more rapidly. This defines the boundaries in Figs.~\ref{fig:cartoon.definitions}, \ref{fig:cartoon.quadrants}, \ref{fig:surface.densities} and others. 
 
 Note that under some conditions -- mostly notably when $\dot{m} \gg 1$ (especially if the accretion becomes radiatively inefficient as expected in this regime) -- $R_{\rm min,\,strom}$ and $R_{\rm es,\,ion}$ can expand outwards to become larger than the outer radius $r_{\rm sub,\,ext}$ of this zone. We stress that this does not mean the BLR-like region would vanish. Rather it simply means that, as above, the fractional contribution from each radius $R < r_{\rm sub,\,ext}$ will fall off as $\propto (R/r_{\rm sub,\,ext})^{4/3}$ or so as the gas able to freely absorb and emit such lines will be confined to a narrower surface layer. This simply means that the BLR-like emission will be more strongly dominated by the gas just inside the dust sublimation radius $r_{\rm sub,\,ext}$ when $\dot{m} \gg 1$. Qualitatively (as discussed in \S~\ref{sec:extreme.mdot}) as these move to larger radii and therefore larger $H/R$, this implies a larger covering/reprocessing factor for hot dust and the BLR, and stronger outflows, associated with weaker X-rays, at higher $\dot{m}$.

Depending on the photon energies of interest, there will be a region from $\sim R_{\rm min,\,strom}$ to $r_{\rm sub,\,ext}$, corresponding to a covering factor of 
\begin{align}
f_{\rm cover}^{\rm 3a} \sim f_{\rm cover}^{\rm r_{\rm sub,\,ext}} -  f_{\rm cover}^{R_{\rm min,\,strom}} \sim  0.2
\end{align}
 (from $H/R \sim 0.3-0.5$; Figs.~\ref{fig:cartoon.illumination} \&\ \ref{fig:cartoon.mdot}) which can reprocess a fraction up to $f_{\rm reprocessed} \sim f_{\theta} f_{\rm cover}^{\rm 3a} \sim 0.12-0.16$ of the bolometric luminosity from the central source (depending on how isotropic or not its emission is). The total mass or volume subtended by this will be dominated by the largest radii, $R_{\rm max,\,3a} \sim 100\,{\rm ld}\,(L_{\rm bol}/10^{45}\,{\rm erg\,s^{-1}})^{1/2}$ (ld = light-days), while the smaller radii where the efficiency of reprocessing peaks correspond to 
$R_{\rm min,\,3a} \sim 6\,{\rm ld}\,(L_{\rm bol}/10^{45}\,{\rm erg\,s^{-1}})^{1/2}$.
  Taking a typical linewidth to be twice the Keplerian velocity (for disky material), this range of radii corresponds to $\Delta v \sim 2\,v_{K}$ in the range 
  \begin{align} 
   1400\,m_{7}^{1/4}\,\dot{m}^{-1/4}\,T_{\rm sub,\,1500}  \lesssim \Delta v \lesssim 6200\, m_{7}^{1/4}\,\dot{m}^{-3/7}\,\tilde{Q}^{3/7}.
  \end{align} 

The temperatures in this zone will be set by photonionization equilibrium to $\sim 10^{4}\,$K (Fig.~\ref{fig:thermal.profile}), so the sonic Mach numbers will be $\sim 100$ (for more details, see Fig.~\ref{fig:turb.props}) and there should be inhomogeneous/clumpy/turbulent structure down to scales of order the sonic length $R_{\rm sonic} \sim H/\mathcal{M}_{s}^{2}$ as noted above (see Fig.~\ref{fig:3d.densities}), which also corresponds (by definition) roughly to the Sobolev length in a supersonically turbulent medium, relevant for the coherence length of resonant line emission/absorption even in the smoother or lower-density gas (Fig.~\ref{fig:sobolev.size}). Calculating this properly with our model scalings, 
\begin{align}
R_{\rm sonic}  \sim 0.5\times10^{12}\,{\rm cm}\,\dot{m}^{11/7}\,\tilde{Q}^{-11/7}\,m_{7}^{1/12}\,r_{\rm ff,\,5}^{1/6}
\end{align}
at the inner radii and $\sim 10^{14}\,{\rm cm}\,\dot{m}^{11/12}\,T_{\rm sub,\,1500}^{-11/3}$ at the outer radii of this zone. The volume-weighted mean disk densities in the zone range from $\sim 3\times10^{8}\,{\rm cm^{-3}}\,\tilde{Q}^{12/7}\,r_{\rm ff,\,5}^{1/2}\,m_{7}^{-1/2}\,\dot{m}^{-5/7}$ (near the inner region of the zone where most of the emission occurs) to $\sim 10^{6}\,{\rm cm^{-3}}\,(Q_{\rm em}/Q_{\rm abs})\,T_{\rm sub,\,1500}^{4}\,\epsilon_{r,0.1}^{-1}\,(m_{7}\,r_{\rm ff,\,5})^{-1/2}$ (at the outer zone boundary), but the high sonic Mach numbers mean that the {\em local} gas density fluctuations should follow something like the usual lognormal for these mach numbers (with $\mathcal{M}_{A} \sim 1$; see \citealt{ostriker:2001.gmc.column.dist,hopkins:2012.intermittent.turb.density.pdfs,beattie:2021.turb.intermittency.mhd.subalfvenic}), so $\sim 90\%$ of the gas mass will be between $\sim 0.2-6000\times\langle n_{\rm gas}(R)\rangle_{\rm vol}$. The gas-mass weighted mean or median will be around 
\begin{align}
\langle n\rangle_{\rm mass} \sim 50\,\langle n_{\rm gas}(R)\rangle_{\rm vol} \sim 10^{10}\,{\rm cm^{-3}}\,\tilde{Q}^{12/7}\,r_{\rm ff,\,5}^{1/2}\,m_{7}^{-1/2}\,\dot{m}^{-5/7}
\end{align}
 giving $\sim 10\%$ of the gas in densities more like 
 \begin{align}
 n_{10\%}^{\rm inner} \gtrsim 2\times10^{12}\,{\rm cm^{-3}}\,m_{7}^{1/4}\,\tilde{Q}^{16/7}\,\dot{m}^{-9/7}
 \end{align}
  (at the radii which dominate the total reprocessed emission of the zone) or 
  \begin{align}
   n_{10\%}^{\rm outer} \gtrsim 10^{9}\,{\rm cm^{-3}}\,m_{7}^{1/4}\,T_{\rm sub,\,1500}^{16/3}\,\dot{m}^{-1/3}
   \end{align}
    (at the outer zone boundary), as shown more explicitly in Fig.~\ref{fig:3d.densities}. The dimensionless version of this in terms of $\rho/\rho_{\rm midplane}$ or corresponding ``clumping factors'' of the gas is shown in Fig.~\ref{fig:turb.props} in more detail. Given this, the gas will be multi-phase for the same reasons and with the same character as in Zone (5) below.

Because of how the various disk properties scale, the volume-weighted mean ionization parameter for directly-illuminated gas depends quite weakly on radius or on any other parameter except $\tilde{Q}$ (the shape of the actual ionizing spectrum) and the local density relative to mean: $\xi  \equiv ({\rm d}Q/{\rm d}A)/(c\,n_{e}) \sim 3\,(f_{\theta}/0.5) \tilde{Q}\,(n_{\rm gas}/\langle n_{\rm gas}(R) \rangle_{\rm mass})^{-1}\, (m_{7}/r_{\rm ff,\,5})^{1/2} \sim \tilde{Q}\,(m_{7}\,R/10^{16}\,{\rm cm})^{1/6}\,(n_{\rm gas}/\langle n_{\rm gas}(R) \rangle_{\rm mass})^{-1}$ (where in the latter expression we take the expected $r_{\rm ff,\,5} \sim m_{7}^{1/2}$ and the geometric mean of the two limiting cases for $f_{\theta}$ given in \S~\ref{sec:model:illumination}, as the difference between the two is small at these radii of interest). So $\xi$ will vary primarily with these same density fluctuations, giving $\xi \sim 0.01-1\tilde{Q}$ for the denser gas in the $1-2\sigma$ overdensity clumps (i.e.\ the $\sim 10-30\%$ mass range, at densities $n \sim 10^{12}\,{\rm cm^{-3}}$ at the smallest BLR radii and $n \sim 10^{9}\,{\rm cm^{-3}}$ at the largest BLR radii) quoted above. We show this computed more explicitly, for directly ionized gas at the surface layer, in Fig.~\ref{fig:sobolev.size}. 

As discussed below, these scalings are largely robust (modulo order-unity coefficients) to whether this upper layer is in inflow with the disk or outflow (if the incident luminosity is high enough to drive e.g.\ a line-driven wind off the surface; discussed quantitatively below).

All of these properties are remarkably similar to the canonical observationally-inferred properties of the BLR. Specifically, this includes the emitting gas densities, temperatures, and ionization parameters (see typical observed values in \citealt{Peterson2006:BLR.review} and references therein);  strength of the density fluctuations or range of densities at a given radius \citep[compare][]{dexter:2011.qso.accretion.disks.strongly.inhomogneous}; observed effective ``emitter'' or ``cloud'' size which should be similar to the predicted Sobolev length in both the discrete-cloud scenario (per \citealt{krolik:1981.twophase.model.quasar.emission.lines,krolik:1999.agn.book}) and smooth emitting-structures case (per \citealt{arav:1998.blr.not.discrete.clouds.but.inhomogeneous.system,laor:2006.blr.could.be.smooth.disk.not.clumpy.but.must.be.turb}); linewidths \citep[compare][]{laor:1991.blr.disk.line.profiles}; BLR covering factors and/or scale-heights constrained by direct imaging and reverberation mapping in e.g.\ \citet{gravity:2018.sturm.blr.rotating.thick.disk}; luminosities/reprocessing fraction \citep[compare][]{kaspi:2005.blr.size.reverb.mapping,gravity:2024.blr.infrared.size.luminosity.relation.agn}; inner and outer characteristic BLR radii \citep{du:2015.smbh.reverb.map.supereddington.in.lowmass.bhs,gravity:2020.resolved.blr.size.disk.inside.dust.sub,gravity:2021.resolved.blr.disk.hot.dust.coronal.regions,woo:2023.reverb.map.updated.compilation.higher.lum}; emitting gas mass (compare e.g.\ the BLR emitting gas mass  of $\sim 40\,{\rm M_{\odot}}$ for NGC 3227 in \citealt{devereux:2021.blr.mass.low.40msun.or.less.and.interior.to.sublimation.and.transition.to.xray.interior.favors.blr.as.disk} to the predicted $\sim 30\,{\rm M}_{\odot}$ here from Fig.~\ref{fig:masses} for similar luminosity and BH mass); and thick-disk like geometry and global kinematics \citep[as observed in][]{gravity:2018.sturm.blr.rotating.thick.disk,gravity:2021.resolved.blr.disk.hot.dust.coronal.regions,gravity:2024.blr.infrared.size.luminosity.relation.agn}. This also trivially ensures that the BLR is ``virialized'' as it is fundamentally part of the accretion disk itself (no additional process or interactions are actually needed to ensure that the BLR clouds meet the traditional observational definition of a virialized BLR), and the relative contribution of turbulence versus rotation closely resembles those inferred observationally in the studies above.

All of these will be the subject of more detailed study in future papers in preparation, in which we will make more detailed quantitative comparisons with specific well-observed AGN. A rigorous comparison, however, requires detailed radiation transport calculations performed on the full numerical simulations, in order to properly compare observables, and so is outside the scope of our predictions here, as we discuss further in \S~\ref{sec:discussion}. But compared this to, for example, a thermal-pressure-dominated SS73-like $\alpha$-disk. In that case, at the radii of observed BLRs, every disk property is many orders of magnitude different from the properties observed in the BLR (the thermal-pressure-dominated disk prediction is that the disk would be vastly more dense, hotter, geometrically thinner, more massive, less turbulent, etc.). We note in \S~\ref{sec:discussion} that the same is true of e.g.\ marginally-self-gravitating disk models. So in this model classes, it is simply not possible that the BLR comes from the disk itself. Of course, in those models, as noted in \S~\ref{sec:intro}, one could posit that the BLR is something different entirely which sits ``above'' the disk, held up by different physics. Of course, if one invokes an arbitrary magnetic field or other physics to hold up BLR ``clouds'' and fits or adjusts the model to reproduce all the same observables as above, then there will be no obvious way to observationally discriminate between these models (though there may still be measurable differences in the kinematics owing to the effects of the thin disk in the midplane). The key difference between those models from \S~\ref{sec:intro} and the model here is that the BLR properties we discuss above are predictions of the disk model itself -- they arise inevitably from an accretion disk that obeys our two simple ansatz in \S~\ref{sec:model:disk}.

\subsection{3b: The Shielded, Mostly Neutral Layer}
\label{sec:zone3b.neutral}

The inner boundary of the fully ionized zone above corresponds roughly with the radius $R_{\rm sub,\,T_{\rm eff}}$ -- i.e.\ the radii where at the midplane, the gas transitions to mostly neutral but dust-free. At $|z| \sim H$ at these radii, the gas will be partially ionized but the electron-scattering effect above shields most of the layer which would naively be directly illuminated (Fig.~\ref{fig:opacity.profile}). Thus the neutral zone can extend through nearly to the disk surface, and this zone is basically identical to Zone (4) we calculate next, except for the shielding layer (Fig.~\ref{fig:cartoon.quadrants}).

\section{Zone 4: The Cool, Neutral Dust-Free Disk Midplane}
\label{sec:zone4.neutral}

Inside of the midplane (shielded/un-illuminated) sublimation radius, the disk is dust free but still relatively cool with temperatures $T\sim 10^{3}-10^{4}\,$K (Fig.~\ref{fig:thermal.profile}) so the gas should be mostly neutral, with free electron fractions $x_{e} \sim 0.01\,x_{e,\,0.01}$ typical of gas in this temperature range (see \paperone, \citealt{wolfire:1995.neutral.ism.phases,tielens:2005.book,draine:ism.book} for more detailed discussion). The optical depth to starlight or ionizing photons through the midplane from the central source (directly through the midplane of the disk) is enormous (Compton-thick) and so this remains shielded to those sources of irradiation. This zone will be less interesting observationally, though we calculate its properties for the sake of completeness and defining its boundaries.

Here the scattering opacity is now dominated by electron scattering with $\kappa_{\rm es} \sim 0.35\,x_{e}$, but the absorption opacity is largely dominated by H$^{-}$ and warm molecular opacities. In this temperature range we can adopt the usual approximations $\kappa_{H^{-}} \approx 1.1 \times10^{-25}\,{\rm cm^{2}\,g^{-1}}\,Z^{1/2}\,\rho^{1/2}\,T^{7.7} \sim 10^{-7}\,(n_{\rm gas}/10^{8}\,{\rm cm^{-3}})^{1/2}\,\tilde{Z}^{1/2}\,T_{3000}^{77/10}$, and $\kappa_{\rm mol} \sim 0.0014\,\tilde{Z}\,f_{\rm mol}$ where typical $f_{\rm mol} \sim 0.001\,f_{\rm mol,\,0.001}$ at the midpoint of the temperature range (\paperone\ and \citealt{hollenback.mckee:co.cooling,tielens:2005.book}). So we have $\kappa_{s} \gg \kappa_{a}$ and the effective opacity for thermalization is $\kappa^{\ast} = \sqrt{\kappa_{a}\,(\kappa_{s}+\kappa_{s})}\approx \sqrt{\kappa_{a}\,\kappa_{s}}$. This gives a vertically-integrated scattering optical depth 
\begin{align}
\tau_{s} \approx \tau \sim 0.0017\,x_{e,\,0.01}\,(\dot{m}\,r_{\rm ff,\,5}/\epsilon_{r,\,0.1})^{1/2}\,T_{\rm sub,\,1500}^{2}\,\hat{R}^{-5/6},
\end{align} 
scaling to the outer radius $R_{\rm sub,\,T_{\rm eff}}$ (defined including illumination, though this makes no important differences here) with $\hat{R} \equiv R/R_{\rm sub,\,T_{\rm eff}}$, and effective opacity 
\begin{align}
\nonumber \tau^{\ast} \approx & \, \sqrt{\tau_{s}\,\tau_{s}} \sim 3.3\times10^{-5}\,(x_{e,\,0.01}\,f_{\rm mol,\,0.001}\,\dot{m}\,r_{\rm ff,\,5}\,\tilde{Z})^{1/2}\\
& \times \epsilon_{r,\,0.1}^{-1/2}\, T_{\rm sub,\,1500}^{2}\,\hat{R}^{-5/6}
\end{align}
or (if H$^{-}$ dominates) 
\begin{align}
\nonumber \tau^{\ast} \sim & \, 6\times10^{-8}\,T_{1000}^{3.9}\,x_{e,\,0.01}^{0.5}\,\tilde{Z}^{0.25}\,\dot{m}^{0.45}\,m_{7}^{-0.18}\,r_{\rm ff,\,5}^{0.73}\\ 
& \times T_{\rm sub,\,1500}^{3.2}\,\epsilon_{r,\,0.1}^{-0.8}\,\hat{R}^{-4/3},
\end{align}
 all shown in Fig.~\ref{fig:opacity.profile}.
So the disk is optically-thin, so long as temperatures remain below $\sim 10^{4}$\,K (given the steep H$^{-}$ opacity dependence). 

Initially, given the $T_{\rm eff} \sim 1500\,$K at the sublimation radius, this would give the result that the absorption, and therefore continuum emission opacity is dominated by optically-thin molecular transitions. The optically-thin cooling rate from said transitions, 
$F_{\rm cool,\,thin} = 4\pi\,\kappa_{\rm em,\,mol}\,\Sigma_{\rm gas}\,\sigma_{B}\,T_{\rm mol}^{4} \sim 5\times10^{6}\,f_{\rm mol}\,\dot{m}^{1/2}\,r_{\rm ff,\,5}^{1/2}\,T_{\rm sub,\,1500}^{2}\,\epsilon_{r,\,0.1}^{-1/2}\,\tilde{Z}\,\hat{R}^{-5/6}\,(T_{\rm mol}/1500\,{\rm K})^{4}$ 
(in cgs) is much less than the gravitational heating flux 
$F_{\rm grav} \sim 4\times10^{6}\,m_{7}^{1/5}\,r_{\rm ff,\,5}^{3/5}\,\dot{m}^{-4/5}\,T_{\rm sub,\,1500}^{36/5}\,\epsilon_{r,\,0.1}^{-9/5}\,\hat{R}^{-3}$ at $R< R_{\rm sub,\,T_{\rm eff}}$, even for $f_{\rm mol} \rightarrow 1$. 
For more realistic lower $f_{\rm mol}$ the cooling luminosity would be dominated by CII at these temperatures,\footnote{We adopt the standard expressions for effectively optically-thin cooling: $\Lambda_{\rm CII} \sim 9.6\times 10^{-24}\,\tilde{Z}\,x_{e}\,T^{-1/2}\,e^{-92\,{\rm K}/T}$ and $\Lambda_{\rm Ly\alpha} \sim 1.8 \times10^{-11}\,x_{e}^{2}\,T^{-1/2}\,e^{-120000\,{\rm K}/T}$ in ${\rm erg\,s^{-1}\,cm^{3}}$ \citep[see e.g.][]{field:1969.ism.phases.heating.cooling,dalgarno:1972.heating.ionization.HII.regions}.} but that also gives too-low a cooling luminosity $F_{\rm cool,\,thin}^{\rm CII} \sim 7500\,x_{e,\,0.01}\,m_{7}^{-7/10}\,r_{\rm ff,\,5}^{7/5}\,\dot{m}^{-3/10}\,\tilde{Z}\,\epsilon_{r,\,0.1}^{-17/10}\,T_{\rm sub,\,1500}^{34/5}\,\hat{R}^{-17/6}\,T_{1500}^{-1/2}$, so the molecular and atomic gas would quickly heat to much larger equilibrium temperatures 
$\sim 2\times10^{4}\,{\rm K}\,m_{7}^{1/72}\,\dot{m}^{-13/72}\,r_{\rm ff,\,5}^{-1/12}\,\tilde{Z}^{-1/4}\,T_{\rm sub,\,1500}^{13/18}\,f_{\rm mol,\,0.001}^{-1/4}\,\hat{R}^{-13/24}$ 
to balance heating, but as the temperature rises the molecules dissociate further and the CII cooling rate actually decreases, while H$^{-}$ opacity further increases. The medium therefore becomes primarily atomic and the temperature quickly rises until it reaches the equilibrium where optically-thin H cooling balances the gravitational heat flux. 
The cooling at these temperatures will be dominated by a combination of H$^{-}$ and Ly-$\alpha$, with either or both in combination giving an extremely similar, nearly-constant temperature (owing to the extremely strong temperature dependence of both) around 
\begin{align}
T\sim 6000\,{\rm K}\,T_{\rm sub,\,1500}^{0.24}\,(\dot{m}\,\hat{R})^{-0.1}
\end{align}
 or 
 \begin{align}
 T\sim 6100\,{\rm K}\,(1-0.04\,\ln{[x_{e,\,0.01}^{2}\,\dot{m}]}),
 \end{align} respectively. 
Thus the temperature rapidly jumps within this zone to $\sim 6000\,$K, then rises more slowly with decreasing $R$ (Fig.~\ref{fig:thermal.profile}), with $f_{\rm ion} \sim 0.01-0.1$, akin to the warm neutral medium (WNM) in the ISM \citep{field:1969.ism.phases.heating.cooling,wolfire:1995.neutral.ism.phases,wolfire.2003:neutral.atomic.cooling}. Note that in the above, we have used the gravitational flux -- again since the region is extinction and effectively optically-thin, the upper layer of the disk should be directly illuminated (discussed below) but the midplane should see little reprocessed indirect illumination flux, and this will not be effective at heating it. Even if we did assume that $F_{\rm illum}$ were thermalized, the results for the temperature and chemical structure would be the nearly identical, given the strong temperature dependence of the cooling/opacity physics. Moreover at the inner regions of this zone, the effective illumination flux $F_{\rm illum}$ becomes comparable to and smaller than $F_{\rm grav}$, with $F_{\rm grav} \gtrsim F_{\rm illum}$ for $\hat{R} \lesssim 0.03\,\dot{m}^{-0.6}\,m_{7}^{0.15}\,r_{\rm ff,\,5}^{0.45}$. 

\begin{figure}
	\centering\includegraphics[width=0.95\columnwidth]{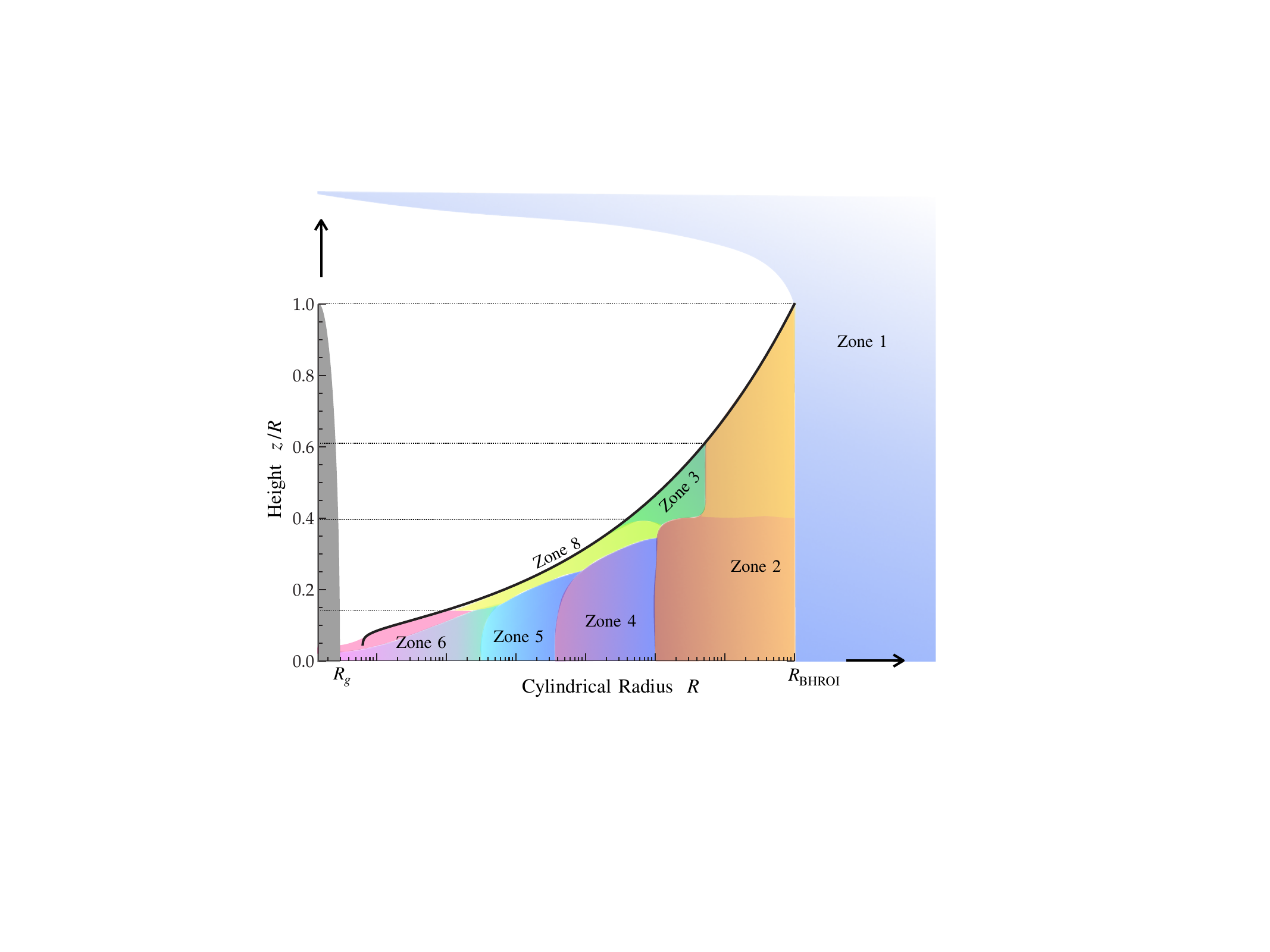} \\
	\centering\includegraphics[width=0.95\columnwidth]{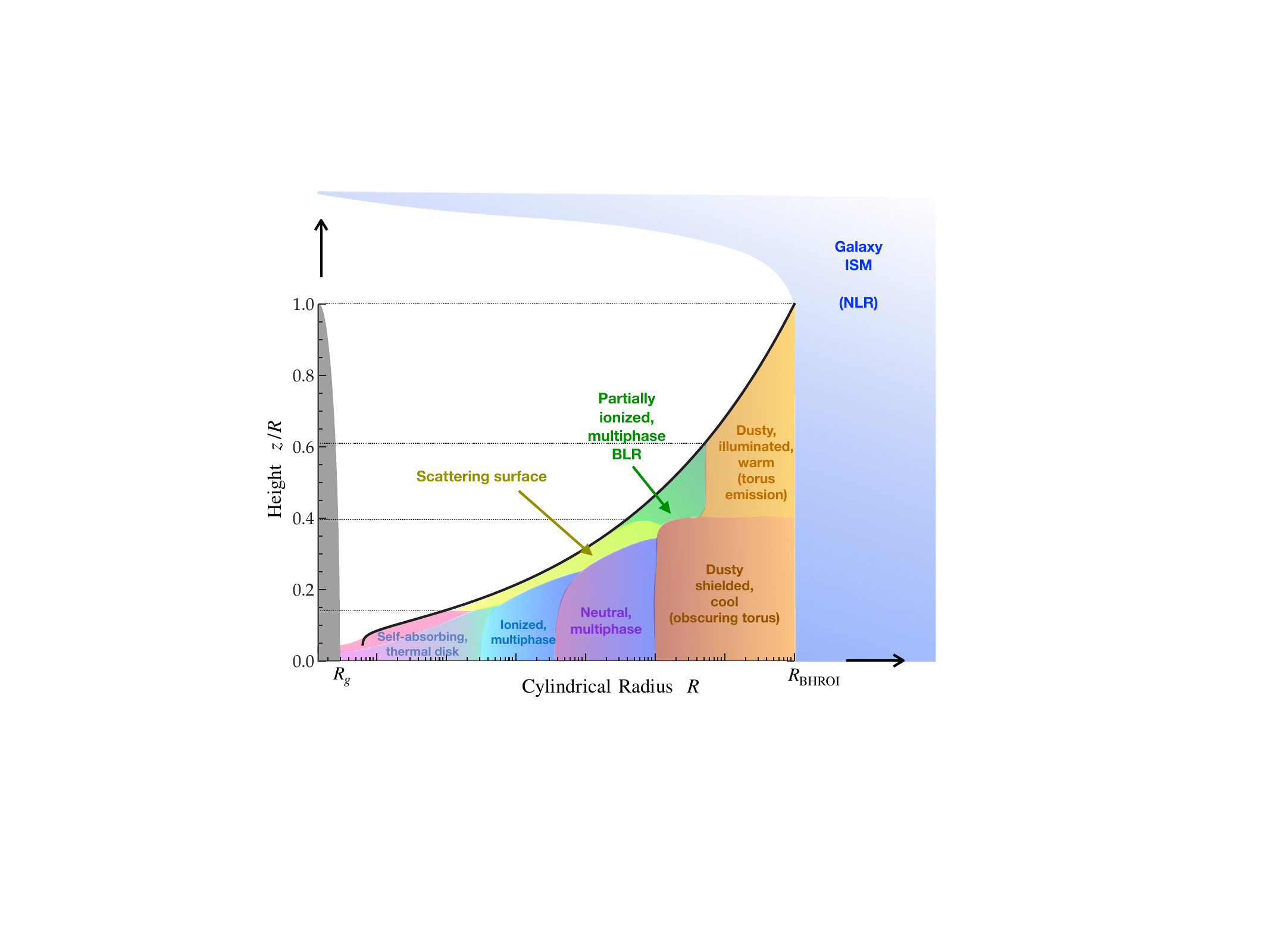} \\
	\centering\includegraphics[width=0.95\columnwidth]{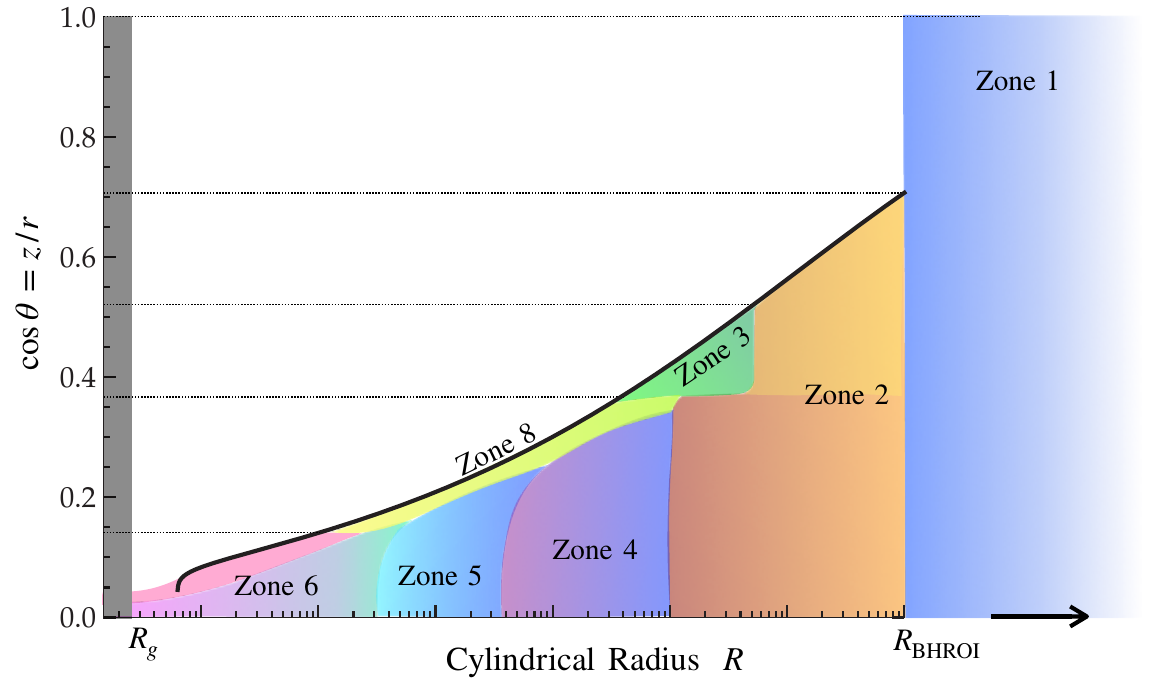} \\
	\caption{Illustration (as Figs.~\ref{fig:cartoon.definitions}-\ref{fig:cartoon.quadrants}) of covering factors for the different zones. 
	We plot the relative height $z/R$ ({\em top} and {\em middle}) and polar angle from the assumed disk axis $\cos{\theta} = z/r=z/\sqrt{R^{2}+z^{2}}$ ({\em bottom}) of the zones ($z/R=H/R$ as {\em solid black} line). Zones are labeled (interior to horizon in {\em grey shaded}), versus radius with their numbers ({\em top}/{\em bottom}) and observational descriptors ({\em middle}). 
	On the linear scale here, the predicted flaring of the disk and large scale-heights ($H/R \approx (R/R_{\rm BHROI})^{1/6}$) are more obvious. As discussed in \S~\ref{sec:zone8.warm.scattering} and seen here, this leads to large $\sim 10-20\%$ covering factors for specific regions like the BLR and directly illuminated hot torus, as well as an extended scattering zones (6-8). Together we see that {\em most} of the light from the innermost disk should be scattered or reprocessed in some form (\S~\ref{sec:zone8.warm.scattering}). 
	\label{fig:cartoon.illumination}}
\end{figure}

In any case, the warmer temperatures (which mean much larger H$^{-}$ opacities) plus rising surface densities as $R$ decreases means that $\tau^{\ast}$ increases, and crosses unity at $R \lesssim R_{\rm therm,\,atomic}$ where $R_{\rm therm,\,atomic} \sim  0.014\,R_{\rm sub,\,T_{\rm eff}}\,\tilde{Z}^{0.05}\,m_{7}^{0.003}\,\dot{m}^{0.03}\,r_{\rm ff,\,5}^{0.36}\,\epsilon_{r,\,0.1}^{-0.6}\,T_{\rm sub,\,1500}^{2.4}$ or 
\begin{align}
R_{\rm therm,\,atomic} \sim 1.3 \times 10^{15}\,{\rm cm}\,m_{7}^{0.6}\,\dot{m}^{0.63}\,r_{\rm ff,\,5}^{0.16}\,\tilde{Z}^{0.05}
\end{align}
 (Fig.~\ref{fig:opacity.profile}).
The disk will then thermalize and radiate with an effective temperature balancing the gravitational accretion flux, giving 
\begin{align}
\nonumber T_{\rm eff}(R<R_{\rm therm,\,atomic}) \sim & \, 11000\,{\rm K}\,m_{7}^{0.05}\,r_{\rm ff,\,5}^{-0.12}\,\dot{m}^{-0.22}\,\tilde{Z}^{-0.04}\\ 
& \times (R/R_{\rm therm,\,atomic})^{-3/4}
\end{align}
 (Figs.~\ref{fig:thermal.profile}, \ref{fig:pressure.profile}). 
The H$^{-}$ rapidly dissociates and the disk begins to efficiently ``self-ionize'' from its own thermal emission at these temperatures. Thus the midplane radial dynamic range of this zone in radius is modest but significant ($\sim 1\,$dex; Figs.~\ref{fig:cartoon.definitions}-\ref{fig:cartoon.quadrants}). At this self-ionization radius, the covering factor of the disk is 
$H/R \sim 0.21\,m_{7}^{0.1}\,r_{\rm ff,\,5}^{-0.14}\,\dot{m}^{0.1}\,\tilde{Z}^{0.008}$ (Fig.~\ref{fig:cartoon.illumination}). For the reasons above, these scalings are nearly identical with or without including external illumination effects.

The intrinsic gravitational luminosity from this region is $\sim 2\times10^{43}\,{\rm erg\,s^{-1}}\,m_{7}^{1.4}\,\dot{m}^{0.37}\,r_{\rm ff,\,5}^{-0.16}\,\tilde{Z}^{-0.05} \sim 0.017\,L_{\rm bol}\,m_{7}^{0.32}\,\dot{m}^{-0.63}$, so still a small fraction of $L_{\rm bol}$. Since most of this comes from the inner (and therefore warmer) regions, it will primarily emerge in a mix of Ly$\alpha$ and H$^{-}$ continuum. The former will be very strongly scattered in the external medium around this zone and is much smaller than the reprocessed Ly$\alpha$ luminosity of the directly-illuminated regions, while the latter will be much smaller than the thermal continuum at the same wavelengths from the warm disk at smaller radii, so both will be largely undetectable. Given the cooling rates and temperatures, a smaller fraction $\sim 6\times10^{-5}\,(x_{e}/0.1)\,\dot{m}^{1/2}\,\tilde{Z}\,m_{7}^{-0.16}$ of $L_{\rm bol}$ will emerge in CII lines. 

\section{Zone 5: The Warm Ionized Multi-Phase Disk}
\label{sec:zone5.multiphase.disk}

\begin{figure*}
	\centering\includegraphics[width=0.95\textwidth]{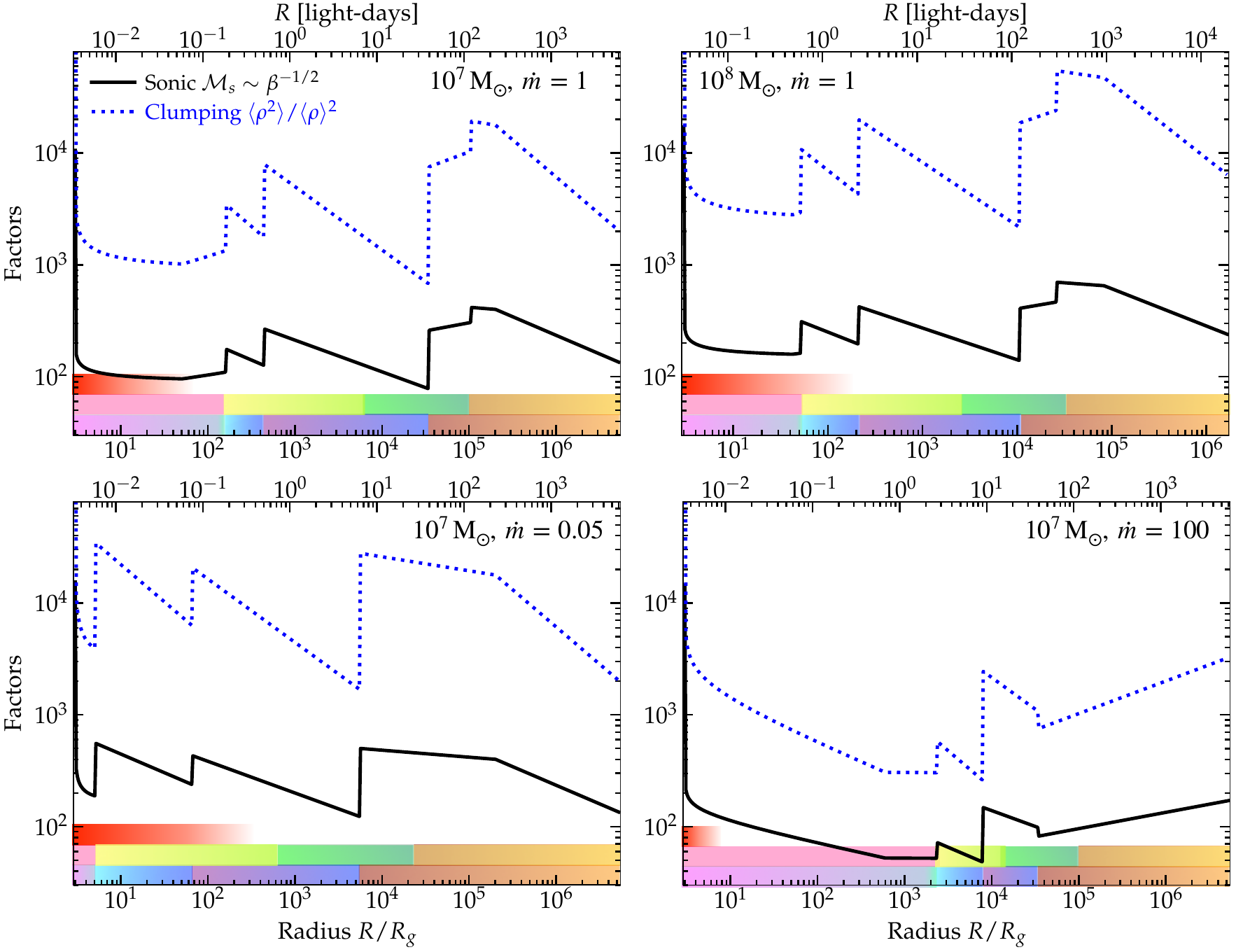} 
	\vspace{-0.2cm}
	\caption{Radial profile of the predicted midplane turbulent sonic Mach number $\mathcal{M}_{s} \equiv \langle v_{\rm turb} \rangle/\langle c_{s,\,{\rm mid}} \rangle$ (as Fig.~\ref{fig:surface.densities}) and corresponding expected gas clumping factor $\langle \rho^{2} \rangle \sim 1 + (b \mathcal{M}_{s})^{2}$ (where $b\approx 1/3$ reflects the compressive component expected; Fig.~\ref{fig:3d.densities}) given the predicted trans-\Alf{ic}, supersonic turbulence in the flux-frozen $\beta\ll 1$ disks.
	Despite various changes in temperature and structure, typical sonic Mach numbers for a range of conditions tend to be around $\sim 100$, corresponding to $\sim 5-10\%$ of the gas mass being in multi-phase denser structures with densities $\sim 10^{3}-10^{4}$ times the volume-averaged mean density within the disk at a given radius.
 	\label{fig:turb.props}}
\end{figure*}

\begin{figure}
	\centering\includegraphics[width=0.95\columnwidth]{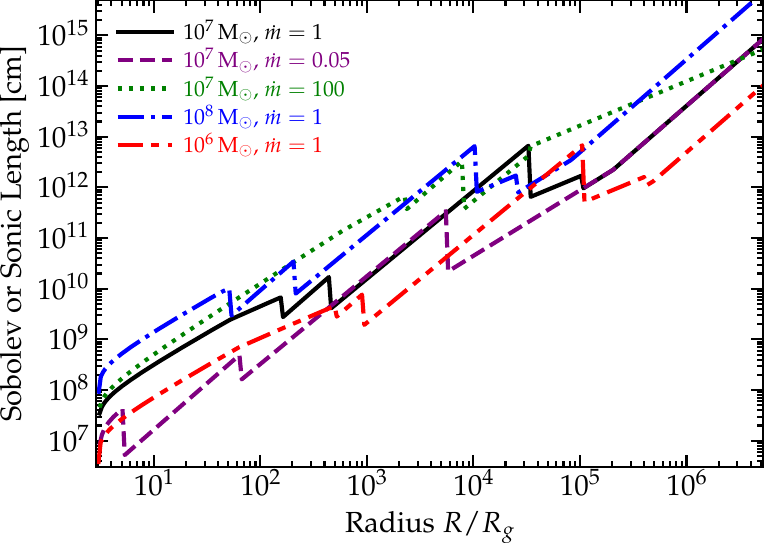} 
	\centering\includegraphics[width=0.95\columnwidth]{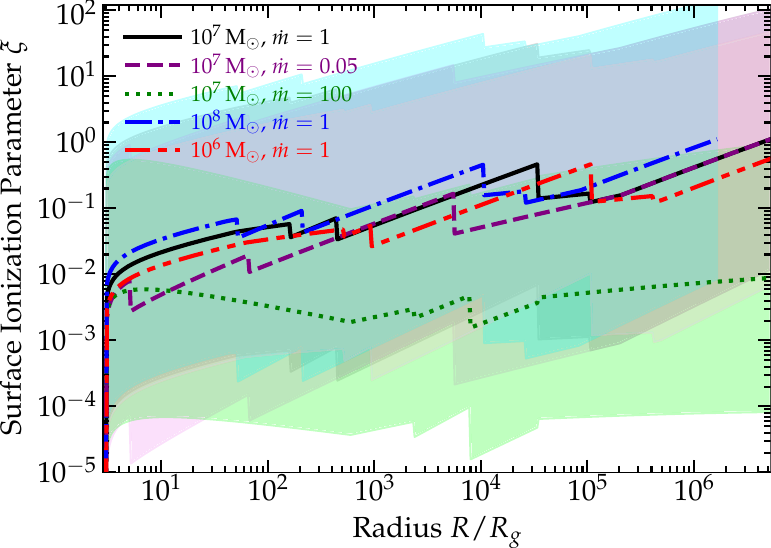} 
	\caption{{\em Top:} Radial profile of the Sobolev velocity coherence or sonic length (the two are similar in supersonic turbulence) of the disk, for different parameters (labeled). Note this is defined in the midplane, so is slightly different in the surface/illuminated layers, but the order-of-magnitude values are similar. At BLR-like radii there is a plateau around $\sim 10^{10}-10^{13}\,{\rm cm}$, while in the inner disk there is another extended region where this is around $\sim 10^{10}\,{\rm cm}$. These will correspond to the characteristic sizes, in optically-thin regions like the BLR, of individual velocity-and-density coherent gas emission structures (e.g.\ BLR ``clouds,'' loosely defined).
	{\em Bottom:} Same for the mass-weighted median surface ionization parameter $\xi \equiv \langle (Q/L)\,F_{\rm illum}/(n\,c) \rangle_{\rm mass}$, where $Q/L$ gives the ratio of ionizing photon production rate to total luminosity, $F_{\rm illum}$ is the mean directly-illuminated flux, and $n$ the gas density at $|z|\sim H$, all evaluated assuming illumination with a flat-disk pattern ($f_{\theta} \propto \cos{\theta}$) and ignoring scattering/reprocessing and optical depth effects interior to the disk. The shaded range shows the gas-mass-weighted $90\%$ inclusion interval owing to the range of densities predicted (Fig.~\ref{fig:3d.densities}). The range of $\xi$ is remarkably stable, only dropping weakly for $\dot{m} \gg 1$ because the disks become radiatively inefficient while being denser, while the scatter is large and similar to that observed in e.g.\ the BLR (\S~\ref{sec:zone3.blr}).
	\label{fig:sobolev.size}}
\end{figure}

Now consider the midplane zone interior to the thermalization radius of Zone (4), which occurs (as we noted before) very close to where the effective temperature exceeds $\sim 10^{4}$\,K (compare Figs.~\ref{fig:cartoon.quadrants} \&\ \ref{fig:thermal.profile}). Interior to this radius $R_{\rm therm,\,atomic}$, the increasing temperatures rapidly dissociate any atomic hydrogen. For convenience we will in this section scale the radii to the outer boundary of the zone: $\tilde{R} \equiv R/R_{\rm therm,\,atomic}$. Again this midplane zone is of less direct interest observationally, as we note below it does not much influence the observed spectrum.

The scattering opacity in this (now mostly-ionized) regime is dominated by electron scattering with $\kappa_{s} \approx \kappa_{\rm es} \approx 0.35$. Absorption opacity comes primarily from a combination of the usual Kramers continuum opacity $\kappa_{\rm K} \approx 7\times10^{25}\,(0.014\,\tilde{Z}+0.001)\,\rho\,T^{-7/2} \sim 0.0006\,m_{7}^{-0.71}\,r_{\rm ff,\,5}^{0.19}\,\dot{m}^{-0.26}\,\tilde{Z}^{0.90}\,T_{4}^{-7/2}\,\tilde{R}^{-2}$ (where $T_{4} \equiv T/10^{4}\,{\rm K}$; \citealt{mihalas:1984oup..book.....M,rybicki.lightman:1986.radiative.processes.book,krolik:1999.agn.book}), plus iron line opacities (which effectively cover the continuum of interest here), which over the temperature range $10^{4}\,{\rm K} \lesssim T \lesssim 10^{6}\,{\rm K}$ can be reasonably approximated by $\kappa_{\rm Fe} \sim 0.00036\,(\rho/10^{10}\,{m_{p}\,{\rm cm^{-3}}})\,T_{4}^{-2}\,\tilde{Z}$ \citep{jiang:2015.rhd.star.sims.metal.opacities.for.agn.disks.as.well} -- larger by a factor $\sim 2.2\,T_{4}^{3/2}$ compared to $\kappa_{\rm K}$. We are again in the regime with $\kappa_{s} \gg \kappa_{a}$, so the effective opacity is $\kappa^{\ast}\approx \sqrt{\kappa_{a}\,\kappa_{s}}$. This gives a scattering optical depth 
\begin{align}
\tau_{s} \sim 6\,\dot{m}^{0.47}\,r_{\rm ff,\,5}^{0.2}\,m_{7}^{-0.003}\,\tilde{Z}^{-0.04}\,\tilde{R}^{-5/6},
\end{align} 
and effective optical depth (assuming Fe dominates since we are near $10^{4}\,$K) 
\begin{align}
\tau^{\ast} \sim 0.36\,r_{\rm ff,\,5}^{0.3}\,m_{7}^{-0.36}\,\dot{m}^{0.34}\,\tilde{Z}^{0.41}\,T_{4}^{-1}\,\tilde{R}^{-11/6}
\end{align}
 (Fig.~\ref{fig:opacity.profile}).

From this, we see that there can be a relatively small continuation of the multi-phase structure from Zone (4), as the ionization causes a drop in the absorption opacity to reduce $\tau^{\ast} < 1$. In this region, the optically-thin line cooling around $T \sim 10^{4}-10^{5}\,$K ($\Lambda \sim 10^{-22}\,{\rm erg\,s^{-1}\,cm^{-3}}$) can easily offset the gravitational heating $F_{\rm grav}$,\footnote{Even if we ignored the dominant H and He recombination line cooling mechanisms and just used the effective iron continuum opacity, this would balance $F_{\rm grav}$ for a slightly higher temperature $\sim 2\times10^{4}\,$K.} and we noted above for Zone (4) that even if external illumination were possible it is now at these radii smaller than $F_{\rm grav}$ (but here we are shielded by the scattering layer since $\tau_{s} > 1$, and effectively optically-thin, so this radiation should not effectively heat this region anyways). The regime will therefore be akin to the usual optically-thin, supersonically turbulent ISM at these temperatures, and form multi-phase structure (the sonic scale can reach as small as 
\begin{align}
R_{\rm sonic} \sim H/\mathcal{M}_{s}^{2} \sim 10^{11}\,{\rm cm}\,m_{7}^{0.11}\,r_{\rm ff,\,5}^{0.45}\,T_{4}\,\dot{m}^{1.16}\,\tilde{Z}^{0.09}\,\tilde{R}^{11/6};
\end{align}
 Fig.~\ref{fig:turb.props}) with shocks forming hot gas at temperatures 
 \begin{align}
 T_{\rm shock} \sim \mathcal{M}_{s}^{2}\,10^{4}\,{\rm K}
 \end{align}
  (Fig.~\ref{fig:thermal.profile}) which can cool but slowly enough (given the decreasing cooling time for hotter gas) that a non-negligible fraction of such gas persists while the rest cools to $\sim 10^{4}\,$K below which temperature the cooling rate drops rapidly.\footnote{We can estimate the steady-state fraction $f_{\rm hot}^{\rm midplane}$ of still-hot post-shock gas by computing the ratio of optically-thin cooling time to inter-shock timescale $\Delta t_{\rm shock}$. Given the turbulence is supersonic with turnover scale $\mathcal{O}(H)$, the latter is necessarily about equal to the dynamical time $\Delta t_{\rm shock} \sim t_{\rm dyn} \sim 1/\Omega$ \citep{hopkins:frag.theory}. So computing $t_{\rm cool}^{\rm shock}/\Delta t_{\rm shock} = \Omega\,(3/2)\,k_{B}\,T_{\rm shock}/(4\pi\,\kappa_{\rm K}\,\sigma_{B}\,T_{\rm shock}^{4}\,m_{p})$ for typical post-shock gas, assuming an initial strong shock $T_{\rm shock} \approx (3/16)\,(m_{p}/k_{B})\,v_{\rm turb}^{2}$, $\rho_{\rm shock} \approx 4\,\rho_{\rm gas}$ (before cooling), gives $f_{\rm hot}^{\rm midplane} \sim t_{\rm cool}^{\rm shock}/t_{\rm dyn}\sim 0.005\,\tilde{R}^{1/6}\,m_{7}^{0.6}\,r_{\rm ff,\,5}^{-0.64}\,\dot{m}^{-0.89}\,\tilde{Z}^{-1}$.} We discuss this hotter gas below, in particular in the discussion of the extended warm skin and patchy coronal gas, but emphasize that the majority of the gas and emission at these radii is coming from gas at temperatures below $T_{\rm shock}$, closer to $T_{\rm eff}$ (itself the same $T_{\rm eff}$, by definition, that would appear in e.g.\ an SS73-like disk). 

Given the $\tau_{\ast}$ calculated above, as we move inwards in radius $R$, the disk will (for high accretion rates at least) soon exceed $\tau^{\ast}\sim1$ and thermalize (with $T \rightarrow T_{\rm eff}$) at 
$\tilde{R} \lesssim 0.36\,\dot{m}^{0.52}\,m_{7}^{-0.37}\,r_{\rm ff,\,5}^{0.38}\,\tilde{Z}^{0.41}$, or $R \lesssim R_{\rm therm,\,ion}$ where
\begin{align}
\nonumber R_{\rm therm,\,atomic} \sim & 4.8\times10^{14}\,{\rm cm}\,m_{7}^{3/13}\,\dot{m}^{15/13}\,r_{\rm ff,\,5}^{7/13}\,\tilde{Z}^{6/13} \\ 
& \sim 160\,R_{g}\,\dot{m}^{15/13}\,r_{\rm ff,\,5}^{7/13}\,m_{7}^{-10/13}\,\tilde{Z}^{6/13}
\end{align}
 (Fig.~\ref{fig:cartoon.quadrants}). 
Note that at sufficiently large $\dot{m}$ or $m_{7}$ this expression will over-estimate $R_{\rm therm,\,ion}$ because the scalings we used will extrapolate beyond the radii where $T_{\rm eff}$ drops too low to maintain the disk as fully-ionized and we need to change our opacity expressions. In that limit we would have instead 
\begin{align}
\nonumber R_{\rm therm,\,ion} \sim & 7.7 \times 10^{14}\,{\rm cm}\,m_{7}^{9/22}\,r_{\rm ff,\,5}^{7/22}\,\dot{m}^{9/11}\,\tilde{Z}^{3/11}\\ 
& \sim 260\,R_{g}\,m_{7}^{-13/22}\,r_{\rm ff,\,5}^{7/22}\,\dot{m}^{9/11}\,\tilde{Z}^{3/11}
\end{align}
 (Fig.~\ref{fig:cartoon.mdot}). These expressions are not strongly modified by illumination: for the most extreme ``total'' illumination case anticipated at very large $\dot{m}$ (e.g.\ physically interior to an extended scattering layer), where we neglect gravitational flux but assume all of the interior light is reprocessed, we obtain $R_{\rm therm,\,ion}^{\rm illum} \rightarrow 230 R_{g} m_{7}^{-5/8} r_{\rm ff,\,5}^{7/16} \dot{m}^{15/16} \tilde{Z}^{3/8} \epsilon_{r,\,0.1}^{-3/16}$, very similar for most practical purposes. So in any case zone (5a) is small at high $\dot{m}$. 

At this thermalization radius -- the transition radius between zones (5) and (6) -- the effective temperature will be 
\begin{align}
T_{\rm eff}(R\sim R_{\rm therm,\,ion}) \sim 2\times10^{4}\,{\rm K}\,m_{7}^{0.33}\,r_{\rm ff,\,5}^{-0.4}\,\dot{m}^{-0.62}\,\tilde{Z}^{-0.35}
\end{align} 
(Fig.~\ref{fig:thermal.profile}), and the gravitational cooling luminosity is $\sim 3\times10^{43}\,m_{7}^{23/13}\,r_{\rm ff,\,5}^{-7/13}\,\dot{m}^{-2/13}\,\tilde{Z}^{-6/13} \sim 0.03\,L_{\rm bol}\,m_{7}^{10/13}\,\dot{m}^{-15/13}\,r_{\rm ff,\,5}^{-7/13}\,\tilde{Z}^{-6/13}\,\epsilon_{r,\,0.1}^{-1}$. So the direct emission is small but not completely negligible. However, most of the line emission will be strongly scattered and sub-dominant to the line emission from the directly illuminated region, and potentially Comptonized, as we discuss below. But the continuum radiation can contribute non-negligibly to the total NIR emission, potentially. For these reasons it is also worth noting that at the inner boundary of this zone the scale height will be 
\begin{align}
H/R|_{R=R_{\rm therm,\,ion}} \sim 0.18\,m_{7}^{1/26}\,r_{\rm ff,\,5}^{-1/13}\,\dot{m}^{5/26}\,\tilde{Z}^{1/13}
\end{align}
(Fig.~\ref{fig:cartoon.illumination}).

\section{Zone 6: The Central Disk}
\label{sec:zone6.disk}

\subsection{6a: The Thermalized, Modified Black-Body, UV-Emitting Disk} 
\label{sec:zone6a.blackbody}

Interior to $R \lesssim R_{\rm therm,\,ion}$, in the midplane, the disk is effectively optically-thick (Fig.~\ref{fig:opacity.profile}) and ionized (Fig.~\ref{fig:thermal.profile}). At these radii we have $F_{\rm grav} \gg F_{\rm illum}$ (Figs.~\ref{fig:cartoon.quadrants} \&\ \ref{fig:cartoon.illumination}), so direct illumination owing purely to the disk flaring is negligible as a heating mechanism for the disk at a given radius (though reprocessing in general, from scattering by the gas at $|z| > H$, may not be negligible, as we discuss below). 

Now scaling by $x_{g} \equiv R/R_{g}$ for these smaller radii, it is useful to note that the scattering opacity is still primarily from Thompson $\kappa_{s} \approx \kappa_{\rm es} \approx 0.35$, with optical depth 
\begin{align}
\tau_{s} \approx \tau \approx 1000\,\dot{m}\,m_{7}^{-1/3}\,r_{\rm ff,\,5}^{1/3}\,x_{g}^{-5/6}
\end{align}
($\gg 1$ at all radii of interest out to $R_{\rm therm,\,ion}$ for large $\dot{m}$, but not always so if $\dot{m}$ is very small, as we discuss below; see Fig.~\ref{fig:opacity.profile}). The effective temperature {\em assuming} (for now) local balance of cooling with $F_{\rm grav}$ and no illumination (we will free this assumption below) is 
\begin{align}
T_{\rm eff}^{\rm local,\,grav-only} \sim 10^{6}\,{\rm K}\,m_{7}^{-1/4}\,\dot{m}^{1/4}\,x_{g}^{-3/4}.
\end{align}
 So except for fairly small BHs (especially if we include reprocessing effects as noted below), we should be in the regime where we can reasonably approximate the opacity at $T_{\rm eff}$ from the iron opacities plus Kramers, giving $\kappa_{a} \sim \kappa_{\rm Fe} \sim 0.02\,(\dot{m}\,r_{\rm ff,\,5}/x_{g})^{1/2}\,\tilde{Z}\,m_{7}^{-1}\,(T/T_{\rm eff}^{\rm local})^{-2}$ which is smaller than $\kappa_{s}$.\footnote{Note if we do have small BHs where this would naively appear to exceed $\kappa_{s}$, this means the central $T_{\rm eff}^{\rm local}$ becomes large, and at sufficiently small $x_{g}$ we should use the usual Kramers opacity expression, giving $\kappa_{a} \rightarrow 10^{-5}\,r_{\rm ff,\,5}^{1/2}\,\dot{m}^{1/8}\,m_{7}^{-5/8}\,x_{g}^{5/8}\,(T/T_{\rm eff}^{\rm local})^{7/2}$, which changes some of our expressions below but not the qualitative behaviors of interest.} So the effective opacity is $\kappa^{\ast}\approx \sqrt{\kappa_{a}\,\kappa_{s}}$. We therefore have for the massive BH regime 
\begin{align}
\tau^{\ast} \approx 250\,\dot{m}^{5/4}\,m_{7}^{-5/6}\,r_{\rm ff,\,5}^{7/12}\,\tilde{Z}^{1/2}\,(T_{\rm eff}^{\rm local}/T)\,x_{g}^{-13/12}
\end{align}
(Fig.~\ref{fig:opacity.profile}). 

The disk is geometrically thinner than at much larger radii but still flaring and quite geometrically-thick compared to e.g.\ a thermal-pressure supported disk, with 
\begin{align}
H/R \sim 0.1\,(m_{7}/r_{\rm ff,\,5})^{1/6}\,(x_{g}/3)^{1/6}
\end{align}
 (Fig.~\ref{fig:cartoon.illumination}; assuming no additional radiation pressure support, which may be important in the supercritical limit discussed below). 

If the disk is radiatively efficient (which should apply at intermediate $\dot{m}$; see discussion in \S~\ref{sec:extreme.mdot}), then (like any $\alpha$-disk model) we predict a spin-dependent $\epsilon_{r} \sim 0.1$ as the total luminosity increases integrating from large radii towards the ISCO, with the majority of the {\em intrinsic} luminosity emitted coming {\em initially} from gas within $R \lesssim R_{\rm therm,\,ion}$. The emission from this zone will give rise to the usual NIR-optical-UV continuum and big blue bump, with an intrinsic spectrum in good agreement with that observed in quasars (see e.g.\ \citealt{kishimoto:qso.spectrum.ir.bluer}) given the standard $T_{\rm eff} \propto R^{-3/4}$ relation. However, just like in observed quasars, a majority of this light is reprocessed and/or scattered by material at larger radii \citep{kishimoto:qso.spectrum.ir.bluer,lawrence:2012.big.blue.bump.agn.theory.problems.needs.reprocessing.and.scattering,chelouche:2019.qso.reprocessing.inner.disk.light.further.out.obs}, which we discuss further below. 
Given the outermost radii, we expect the thermalized continuum to transition to the Rayleigh-Jeans tail (of the temperature at $R_{\rm therm,\,ion}$) at wavelengths 
\begin{align}
\lambda\gtrsim \lambda^{\rm char,\,R_{\rm therm,\,ion}} \sim {\rm \mu m}\,m_{7}^{-0.1}\dot{m}^{0.6} \tilde{Z}^{0.3},
\end{align} quite weakly dependent on AGN properties, except that the spectra are predicted to be slightly flatter/cooler at higher $\dot{m}$. However at these wavelengths the emission will be filled in by re-emission from the illuminated torus (zone (2c), in particular), also as observed \citep{kishimoto:qso.spectrum.ir.bluer}. In very low accretion rate systems $\dot{m} \lesssim 0.01$, this would predict the thermal optical/NIR continuum begins to vanish, as the thermalized disk radius moves inwards; this involves a fundamental state change in the accretion disk which we discuss below (\S~\ref{sec:extreme.mdot}).

\subsubsection{Illumination/Reprocessing and the Disk Effective Temperature}
\label{sec:illumination.disk.teff}

At high $\dot{m}$, as noted above, we expect the thermalized disk to be highly self-illuminated and interior to an extended scattering atmosphere/corona. This therefore represents a thermalized ``re-processing zone,'' with size $\sim R_{\rm therm,\,ion}^{\rm illum}$ (i.e.\ size given by the radius interior to which the scattering optical depth above the disk and effective absorption optical depth of the disk are both $>1$), as illustrated quantitatively in Fig.~\ref{fig:cartoon.illumination}. Thus the {\em emergent} thermal emission from this region will have an effective temperature given not by $T_{\rm eff}$ at each interior radius $R$, but $T_{\rm eff}(R_{\rm phot})$ at the effective photospheric radius of this region $R\rightarrow R_{\rm phot} \sim R_{\rm therm,\,ion}^{\rm illum}$. Calculating this for the illuminated limit gives: 
\begin{align}
T_{\rm eff}(R_{\rm therm,\,ion}^{\rm illum}) \sim 4\times10^{4}\,{\rm K}\,\epsilon_{r,\,0.1}^{11/32}\,m_{7}^{-3/64}\,\dot{m}^{-7/32}\,\tilde{Z}^{-3/16}
\end{align} 
(Fig.~\ref{fig:thermal.profile}) -- very weakly dependent on any of the properties of the disk, and much cooler than classic SS73-like $\alpha$-disk prediction which is $T_{\rm eff}^{\rm SS73} \sim 4\times10^{5}\,{\rm K}\,(\dot{m}\,\epsilon_{r,\,0.1}/m_{7})^{1/4}$ (which we also compare in Fig.~\ref{fig:thermal.profile}). Crucially, in the latter, the extremely geometrically-thin scale-heights (which we also show in Fig.~\ref{fig:cartoon.ss73}) mean that there is negligible reprocessing at larger radii (the radiation flux from the innermost disk directly escapes to infinity). Note that even if we neglected the details of illumination as calculated above in shifting this boundary, and simply assumed most of the emission came from the largest thermalized radii within the disk, we would obtain similar expressions: $T_{\rm eff} \sim 4\times10^{4}\,{\rm K}\,\epsilon_{r,\,0.1}^{1/4}\,m_{7}^{-3/88}\,\dot{m}^{-7/44}\,\tilde{Z}^{-3/22}$ or $T_{\rm eff}^{\rm alt} \sim 5\times10^{4}\,{\rm K}\,\epsilon_{r,\,0.1}^{1/4}\,\dot{m}^{-17/52}\,\tilde{Z}^{-3/13}$, given the scalings in \S~\ref{sec:zone5.multiphase.disk}.

This predicted effective temperature agrees quite well with the canonical values for bright quasars \citep{peterson:1997.agn.book,vandenberk01:composite.qso.seds,richards:seds,shen:bolometric.qlf.update,cai:2023.qso.sed.universal.w.flat.intrinsic.spectrum.into.uv.larger.than.expected.need.reprocessing.in.blr} especially after correcting for dust reddening \citep{hopkins:dust,gaskell07:qso.reddening.curves}. But it is not just the value of the effective temperature itself, but the scaling which is important. It is well-established that the simple prediction of an SS73-like razor-thin disk model (in which any such reprocessing is impossible) extending to the ISCO, namely $T_{\rm eff} \propto L^{1/4}\,M_{\rm BH}^{-1/2}$ (as above) is not observed in AGN, with most exhibiting a much weaker, and {\em oppositely-signed} mass dependence, see e.g.\ \citealt{bonning:2007.accretion.disk.temps.opposite.scaling.observed.thindisk.models,bonning:2013.qso.temps.emission.lines.too.low.and.wrong.scaling.with.mass.vs.ss73,davis:2007.continuum.qsos.wrong.shape.vs.simplest.models,hall:2018.non.blackbody.models.agn.try.explain.sizes.spectral.shapes,antonucci:2023.galaxies.agn.review,cai:2023.qso.sed.universal.w.flat.intrinsic.spectrum.into.uv.larger.than.expected.need.reprocessing.in.blr}.\footnote{Or, conversely, fitting AGN to model SS73-like $\alpha$-disks which assume $T_{\rm eff} \propto L^{1/4}\,M_{\rm BH}^{-1/2}$, the fit appears to prefer systematically higher $M_{\rm BH}$ and spin values than inferred from any other methods, see \citealt{capellupo:2015.fitting.agn.spectra.forced.to.higher.mbh.and.spin.because.temps.too.low.in.obs}.} 
For the flux-frozen disks, because of the reprocessing effects above, we predict (depending weakly on exactly which of the above scalings we adopt, and how $\epsilon_{r}$ and $\tilde{Z}$ scale with mass or accretion rate) $T_{\rm eff} \propto L^{-0.2}\,M_{\rm BH}^{0.1-0.2}$, a much weaker and oppositely-signed scaling (much closer to the behavior observed).

\subsubsection{Vertical Structure within the Thermalized Zone}
\label{sec:zone6a.vertical}

For a laminar ($\mathcal{M}_{s} \ll 1$), stable, hydrostatic, optically-thick, geometrically thin disk, we would naively expect the vertical heat flux to be governed by radiative diffusion with diffusive speed 
\begin{align} 
v_{\rm diff} \approx c/\tau, 
\end{align} 
which establishes a radiation energy density gradient with midplane $T_{\rm mid} \sim T_{\rm eff}\,\tau^{1/4}$. 

However, as discussed in \papertwo\ and \paperthree, turbulence also transports radiation advectively in the vertical direction with speed $v_{\rm adv,\,z} \approx v_{\rm turb}$, and by definition the supersonic, trans-\Alf{ic} turbulence means the eddies reach heights $\approx H$ on turnover times $\sim 1/\Omega$, which means they carry {\em past} the effectively optically-thick photosphere of the disk (i.e.\ they can carry trapped radiation to heights where $\tau^{\ast} \ll 1$), to heights where it can become effectively optically-thin radiation transport in the outer disk. The criterion for advection to be faster than diffusion is
\begin{align}
1 < v_{\rm turb}/v_{\rm diff} \sim 52\,\dot{m}\,(r_{\rm ff,\,5}/m_{7})^{1/6}\,x_{g}^{-7/6},
\end{align}
 i.e.\ 
 \begin{align}
 x_{g} \lesssim x_{g}^{\rm turb,z} \approx 30\,(r_{\rm ff,\,5}/m_{7})^{1/7}\,\dot{m}^{6/7}
 \end{align}
  (noting $r_{\rm ff,\,5} \approx m_{7}^{1/2}$, this is basically $\sim 30\,\dot{m}^{6/7}$). Interior to these radii, the system will become vertically well-mixed by turbulence. If the medium is sufficiently clumpy or inhomogeneous, or the turbulent transport behaves advectively, turbulent mixing will basically set $T_{\rm mid} \sim T_{\rm eff}$ throughout the disk. If we more conservatively allow for some vertical stratification and assume the disk remains smooth on average with vertical turbulent transport being quasi-diffusive with ``turbulent diffusivity'' $\sim \ell_{\rm turb}\,v_{\rm turb}$, then solving the vertical steady-state flux equations gives a local maximum possible midplane temperature relative to effective temperature of $T_{\rm mid}^{\rm max}/T_{\rm eff} \approx {\rm MIN}[(c/v_{\rm turb}),\,\tau]^{1/4} \sim {\rm MIN}[2\,(r_{\rm ff,\,5}/m_{7})^{1/24}\,x_{g}^{1/12},\,\tau^{1/4}]$ with this ratio reaching its global maximum at $x_{g} \approx x_{g}^{\rm turb,\,z}$ of 
  \begin{align}
  T_{\rm mid}^{\rm max}/T_{\rm eff} \lesssim 2.7\,\dot{m}^{1/14}\,(r_{\rm ff,\,5}/m_{7})^{3/56}.
  \end{align} 
  We adopt the latter for calculating the midplane temperatures in this limit in Fig.~\ref{fig:thermal.profile}. 

As a result, basically independent of accretion and BH properties, $T_{\rm mid}$ never exceeds a couple times $T_{\rm eff}$ at the surface (Fig.~\ref{fig:thermal.profile}). As discussed below, these physics mean that even in the highly super-critical limit, radiation pressure should never inflate the disk by more than a factor $\sim 1.7$ or so in $H$ (shown quantitatively in the dependence of $H/R$ on scale and $\dot{m}$ in Figs.~\ref{fig:cartoon.mdot}-\ref{fig:cartoon.mdot.smallest} as well as directly in pressure in Fig.~\ref{fig:pressure.profile}), which does not fundamentally alter any of our conclusions, but could (since the inflation would be highest at small radii) make the radiation from the central disk somewhat more isotropic ($f_{\theta} \sim 1$) instead of vertically collimated ($f_{\theta} \sim 2\,\cos{\theta}$). It also means that $\beta$ remains $\ll 1$ and the turbulence remains highly supersonic and compressible (sonic Mach numbers $\sim 50-90$, weakly dependent on disk parameters; Fig.~\ref{fig:turb.props}) and likewise the sonic scale $\sim H/\mathcal{M}_{s}^{2}$ remains very small (Fig.~\ref{fig:sobolev.size}).

\subsection{6b: The Warm, Comptonizing Skin}
\label{sec:warm.skin}

At radii interior to $R \lesssim {\rm a\ few} \times R_{\rm es,\,ion}$ and $R_{\rm therm,\,ion}$, the disk necessarily will have an electron-scattering optically-thick but absorption-optically-thin layer or ``skin'' on top of the thermalized effectively-optically thick disk (Fig.~\ref{fig:cartoon.quadrants}). The depth from the surface of the disk to thermalization, between which this layer exists, is approximately (for a constant density disk midplane as we have assumed) $\sim H/\tau^{\ast}$, which means that the scattering optical depth through the layer is (Fig.~\ref{fig:opacity.profile})
\begin{align}
\tau_{s}^{\rm skin} & \approx \kappa_{s}\,\rho_{\rm gas}\,H/\tau^{\ast} = \sqrt{\kappa_{s}/\kappa_{a}}  \\ 
\nonumber & \approx 8\,(x_{g}/\dot{m}\,r_{\rm ff,\,5})^{1/4}\,(m_{7}/\tilde{Z})^{1/2}\,(T_{\rm mid}/2\,T_{\rm eff}^{\rm local}) \sim 1-20
\end{align}
 for a wide range of parameters in this zone (recalling from above that within this zone of the disk $T_{\rm mid} \sim 2\,T_{\rm eff}^{\rm local}$ is a typical value;  Fig.~\ref{fig:thermal.profile} \&\ \S~\ref{sec:zone6a.vertical}). 

The gas in this skin is, by definition, effectively optically-thin, so should cool according to the optically-thin rates. Those cooling rates still give $t_{\rm cool} \ll t_{\rm dyn}$, however, so the gas temperature should quickly reach some equilibrium even if (again as noted above) it is being advected up from the midplane. The heating rate per particle, whether mediated by viscosity or turbulent dissipation or shocks or magnetic reconnection, is the necessarily the same in the disk models here up to an $\mathcal{O}(1)$ constant, $\sim v_{\rm turb}^{2}/t_{\rm eddy} \sim v_{\rm turb}^{2}\,\Omega \sim F_{\rm grav}\,m_{p}/\Sigma_{\rm gas} \sim ((B^{2}/8\pi)/\rho) / t_{\rm reconnect,\,turb}$, and recall from above that these eddies transport said energy fully into $|z|\gtrsim H$. The gas can cool according to either the optically-thin Kramers+Fe rates, or via Compton cooling\footnote{Note that despite the relative importance of magnetic pressure, we can largely neglect synchrotron/cyclotron cooling. As calculated in detail in Appendix~A of \citet{pariev:2003.mag.dominated.disk.models}, the electrons are highly non-relativistic here (except perhaps in the hottest regions of the ``hard'' Corona) and so cooling via cyclotron emission is both inefficient and strongly self-absorbed at the resonant frequencies where most of the radiation would be emitted. But even if we assumed ultra-relativistic electrons so that synchroton and Compton cooling scaled identically with the magnetic and radiation energy densities, in this particular environment, close to the surface of the disk, if $u_{\rm rad} \sim a\,T_{\rm eff}^{4}$ (a plausible lower limit) and $u_{\rm B}\sim u_{\rm B}^{\rm mid} \sim B_{\rm mid}^{2}/8\pi$ (its midplane value) is uniform up to the upper disk layer, than synchrotron cooling (already much larger than cyclotron) could only be more efficient than Compton cooling by at most a factor of a few, which translates to an equilibrium skin temperature difference of a factor $\lesssim 2$, insufficient to change our conclusions.} from the photons in the disk below. Technically if we include all of these and simply equate heating and cooling rates, there are two solution branches, the first branch being the one relevant here\footnote{The second solution branch involves much lower temperatures $\sim 3\times10^{4}\,{\rm K}$ dominated by optically-thin Fe line and H recombination cooling, which is relevant for cooler gas much further out in the disk or raining back into the disk near the photosphere, but not of interest here since we are interested in the warm gas skin heated by shocks or reconnection, where the characteristic post-shock or reconnection temperature is $\sim (3/16)\,m_{p}\,v_{\rm turb}^{2}/k_{B}\sim (3/16)\,m_{p}\,v_{A}^{2}/k_{B} \sim 6\times10^{9}\,{\rm K}\,(m_{7}/r_{\rm ff,\,5})^{1/3}\,x_{g}^{2/3}$, so the skin will reach the hotter solution first and remain there in a quasi-stable state.} is dominated by free-free cooling at larger radii ($x_{g} \gtrsim 8\,(m_{7}/r_{\rm ff,\,5})^{8/7}\,\dot{m}^{-6/7}\,\tilde{Z}^{-12/7}\,(u_{\rm rad}^{\rm skin}/a\,T_{\rm eff}^{4})$) and Compton cooling at smaller radii. This gives 
\begin{align}
T_{\rm skin} = T_{\rm illum,\,thin}^{\rm warm} \approx 10^{7}\,{\rm K}\,(m_{7}/r_{\rm ff,\,5})^{5/3}\,\dot{m}^{-2}\,\tilde{Z}^{-2}\,x_{g}^{-1/3}
\end{align}
 and 
 \begin{align}
 T_{\rm skin} \approx 10^{6}\,{\rm K}\,(m_{7}/r_{\rm ff,\,5})^{1/3}\,\dot{m}^{-1}\,x_{g}^{5/6}\,(u_{\rm rad}^{\rm skin}/a\,T_{\rm eff}^{4}),
 \end{align}
  respectively, with a maximum temperature of 
  \begin{align}
    T_{\rm skin}^{\rm max} \sim 5\times10^{6}\,{\rm K}\,\left( \frac{m_{7}}{r_{\rm ff,\,5}}\right)^{9/7} \tilde{Z}^{-10/7} \dot{m}^{-12/7} \left(\frac{a\,T_{\rm eff}^{4}}{u_{\rm rad}^{\rm skin}}\right)^{2/7}
  \end{align} 
  and minimum (at the outermost radii of the thermalized zone) of 
  \begin{align}
  \nonumber T_{\rm skin}^{\rm min}\sim {\rm MIN}[ & 2\times10^{6}\,{\rm K}\,(m_{7}/r_{\rm ff,\,5})^{1.9}\,\tilde{Z}^{-2.2}\,\dot{m}^{-2.4} \ , \\ 
  &  \ \  \ 7\times10^{7}\,{\rm K}\,m_{7}^{-1/4}\,\tilde{Z}^{5/13} ]
  \end{align}
   (shown in Fig.~\ref{fig:thermal.profile}). 

Given this characteristic temperature, we can estimate the Compton $y$ parameter of the skin as $y=(k_{B}\,T_{e}^{\rm skin}/m_{e}\,c^{2})\,(\tau_{s}^{\rm skin})^{2}$ giving 
\begin{align}
\nonumber y \sim & \, \frac{0.21}{\tilde{Z}} \,\left( \frac{T_{e,\,{\rm skin}}}{2\times10^{6}\,{\rm K}}\right) \left(\frac{T_{\rm mid}}{2\,T_{\rm eff}^{\rm local}}\right)^{2} \left( \frac{x_{g}}{100\, \dot{m}\, r_{\rm ff,\,5}} \right)^{1/2} 
\end{align}
At small radii approaching the ISCO, the skin will not Comptonize the gas, but at most radii covering the thermal gas disk out to $\sim 200\,R_{g}$, this is appreciable. 

Comparing to the electron temperatures ($k_{B}\,T_{e,\,{\rm skin}}$; \citealt{hubeny:2001.acc.disk.spectra.w.comptonization.temps.and.metal.lines.detailed.calcs,czerny:2003.soft.excess.and.uv.profile.in.agn.from.reprocessing.warm.skin.warm.absorber}); optical depths $\tau_{s}^{\rm skin}$ and Compton $y$ parameter \citep{czerny:2003.soft.excess.and.uv.profile.in.agn.from.reprocessing.warm.skin.warm.absorber,kubota:2018.soft.excess.comptonizing.layers,petrucci:2018.agn.comptonizing.layer.properties}; covering factors and geometry \citep{wilkins:2015.patchy.corona.for.comptonizing.and.reprocessing,petrucci:2018.agn.comptonizing.layer.properties} observationally inferred from AGN spectra -- most notably required to explain the soft excess -- the predicted properties all appear consistent. Note that while the radial extent of the Comptonizing region is predicted to expand with $L$ or $\dot{m}$ as does the disk itself (Fig.~\ref{fig:cartoon.mdot}), as implied by observations \citep{palit:2024.warm.covering.agn.corona.expands.with.luminosity.reprocesses.most.emission}, the temperatures above become lower and therefore the excess becomes softer at higher $\dot{m}$, again consistent with observed behaviors \citep{done:2012.soft.excess.comptonization.xr.vs.lum,krawczyk:2013.mean.qso.seds.softx.vs.lbol,stevans:2014.soft.excess.uv.slopes.corr.luminosity.agn,boissay:2016.softexcess.from.comptonization.intensity.corr.edd,ballantyne:2024.warm.corona.soft.excess.coronal.props.weaker.higheddrat.compilation}. 
Thus a Comptonized soft excess similar to that observed seems inevitable in these flux-frozen disks.

\subsubsection{6c: Towards the ISCO and Near-Horizon Region}
\label{sec:near.horizon}

At sufficiently small radii $x_{g}$ approaching and interior to the ISCO (Zone (6c) in Fig.~\ref{fig:cartoon.definitions}), the solutions above will need to be modified to map to the different orbit structure and ``plunge'' or rapid infall, as well as including the appropriate non-Newtonian terms in the potential and angular momentum. However, at such radii, the detailed results will depend on properties like the BH spin, and almost certainly on components like a jet and the detailed behavior of the magnetic fields in the near-horizon regime which we are not modeling here (though we discuss jets briefly in \S~\ref{sec:jets.winds}). Moreover as shown in \paperone, the accretion disks are often misaligned with the BH spin: this can introduce more radical changes like strong precession, warps, or even disk ``tearing,'' none of which we model here \citep[see e.g.][]{kaaz:2022.grmhd.sims.misaligned.acc.disks.spin}. Clearly, detailed modeling of this regime requires GRMHD simulations to extend the Newtonian simulations in \paperone\ and \papertwo.

However, to gain some insight, we can attempt to simply estimate how our solutions should be modified assuming an inner free-fall boundary at the ISCO for a simple quasi-Newtonian approximation around a non-rotating BH (assuming there is no warp or precession in the disk), as done in most $\alpha$-disk models. There is no conceptual difficulty here and it does not contradict any of our important assumptions from \paperthree. For example, retaining the foundational two assumptions from \S~\ref{sec:model:assumptions}, we can replace the potential with Paczynski-Wiita (PW) which multiplies the orbital frequency $\Omega$ by $\tilde{\Omega} \equiv {x_{g}/(x_{g}-1)}$, and multiply the angular momentum term by $\tilde{g}(x)\equiv 1-(x_{0}/x_{g})^{1/2}$ (where $x_{0} = 3$ for the ISCO of a non-rotating BH, as in \citealt{shakurasunyaev73}). In principle, we can represent any quasi-Newtonian potential (e.g.\ the galactic potential with contribution from stars and dark matter in Zone (1)) and zero angular momentum disk ``cutoff'' via appropriate $\tilde{\Omega}$ and $\tilde{g}$, respectively, with such an approach (though of course this cannot capture various relativistic effects). In fact, we use this in all of our more detailed calculations in Figs.~\ref{fig:surface.densities}-\ref{fig:sobolev.size}. 

Within such a simple approximation, this gives us the exact same global disk solutions from \S~\ref{sec:model} except we modify:
\begin{align}
v_{t} & \sim v_{A} \rightarrow v_{A}^{\rm Keplerian}\,\tilde{\Omega}^{2/9}\,\tilde{g}^{1/9}, \\ 
\Sigma_{\rm gas} & \rightarrow \Sigma_{\rm gas}^{\rm Keplerian}\,\tilde{\Omega}^{5/9}\,\tilde{g}^{7/9}.
\end{align} 
Or, for the modified (advection-dominated) solutions in \S~\ref{sec:super.eddington:rad.eff}, $v_{t}\sim v_{A} \propto v_{\rm c} \propto \tilde{\Omega}$ and $\Sigma_{\rm gas} \propto \tilde{\Omega}^{-1}\,\tilde{g}$.
At large $x_{g}$ these reduce to the expressions we have used throughout. 
At small $x_{g} \lesssim 10-20$, these introduce modest quantitative changes to the solutions, most notably, 
\begin{align}
T_{\rm eff}^{\rm local} \rightarrow T_{\rm eff}^{\rm local,\,Keplerian}\,\tilde{\Omega}^{1/2}\,\tilde{g}^{1/4}.
\end{align}  
(again, included in Fig.~\ref{fig:thermal.profile} and those like it for both our models and SS73 comparisons), and the optical depths change by a small amount. None of this changes our conclusions above about the disk being optically-thick, having $\kappa_{s} > \kappa_{a}$, being strongly turbulent, etc.\footnote{If anything they tend to slightly strengthen some of the above conclusions. For example, the upper limit -- assuming optically-thick diffusive transport only -- of the ratio of midplane radiation pressure to magnetic+turbulent pressure scales as $\propto \tilde{\Omega}^{7/9}\,\tilde{g}^{8/9} < 1$ (a factor of $\sim 0.5$ at $x_{g} \approx 10$, decreasing to $0$ as $x_{g}\rightarrow 3$). Similarly, the disk flaring and vertical structure is nearly identical, except that it becomes slightly more strongly flared ($H/R$ decreases with smaller $R$ more rapidly) in the very narrow range $3\lesssim x_{g} \lesssim 5$.} Of course, {\em very} close to $x_{g} \rightarrow 3$, these correction terms strongly modify the solution (formally extrapolating to a vacuum; Figs.~\ref{fig:surface.densities}-\ref{fig:3d.densities}), but this occurs so close to $x_{g} \approx 3$ that it is, for most purposes, akin to simply truncating the disk. This is precisely the regime where our SS73-like assumption of zero angular momentum at the ISCO is suspect and requires more detailed fully-GRMHD models to calibrate (see discussion in \citealt{abramowicz:accretion.theory.review}). 

In practice, the most important consequence of these correction terms for our results here, like with a standard thermal-pressure dominated $\alpha$-disk, is that they significantly suppress both the relative amount of direct radiation flux coming from small $x_{g}$ and its effective temperatures, so the thermal emitted continuum will be softer than if we ignored the term $\tilde{g}(x)$.

\section{Zone 7: The ``Hard'' Corona}
\label{sec:zone7.corona}

\begin{figure}
	\centering\includegraphics[width=0.95\columnwidth]{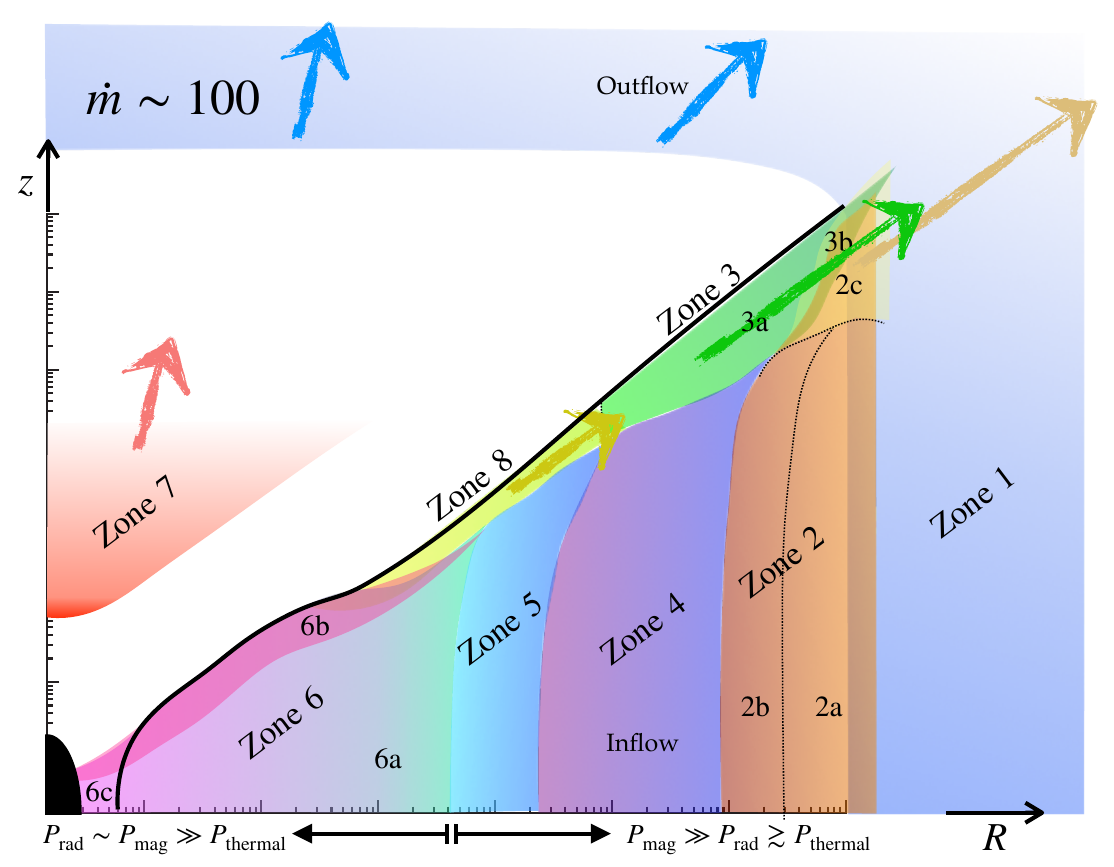} \\
	\centering\includegraphics[width=0.95\columnwidth]{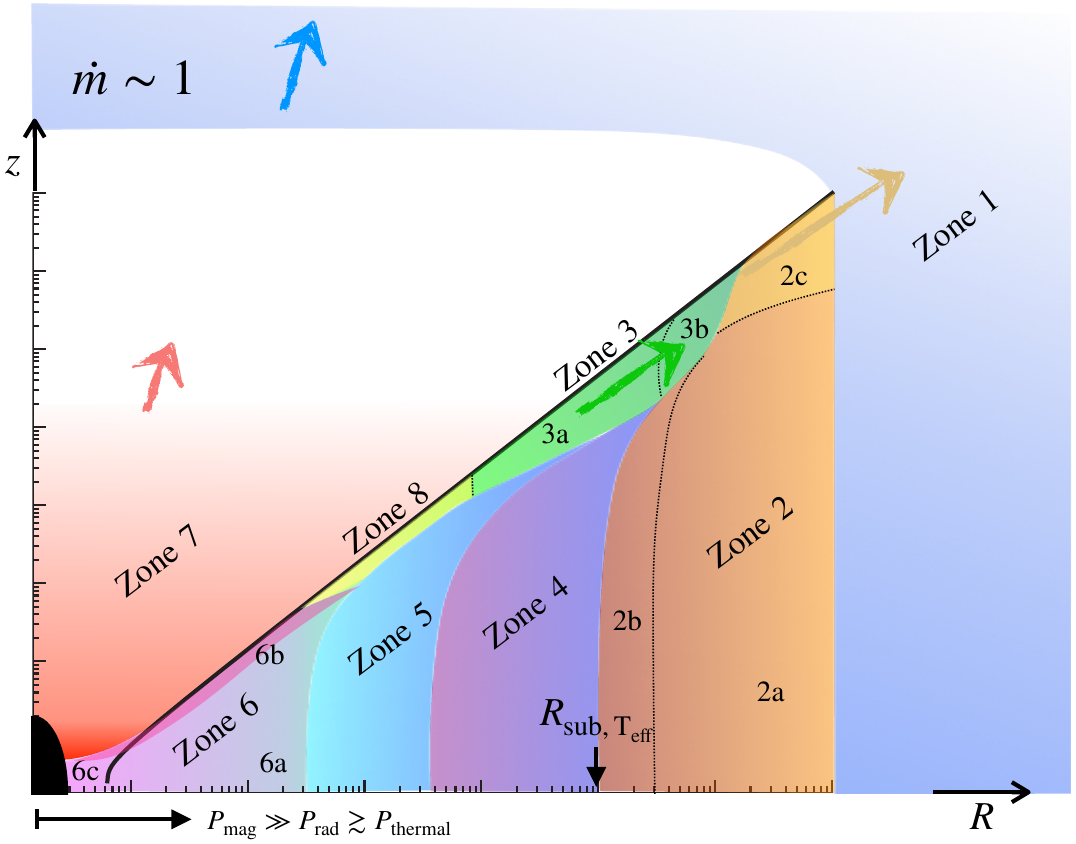} \\
	\centering\includegraphics[width=0.95\columnwidth]{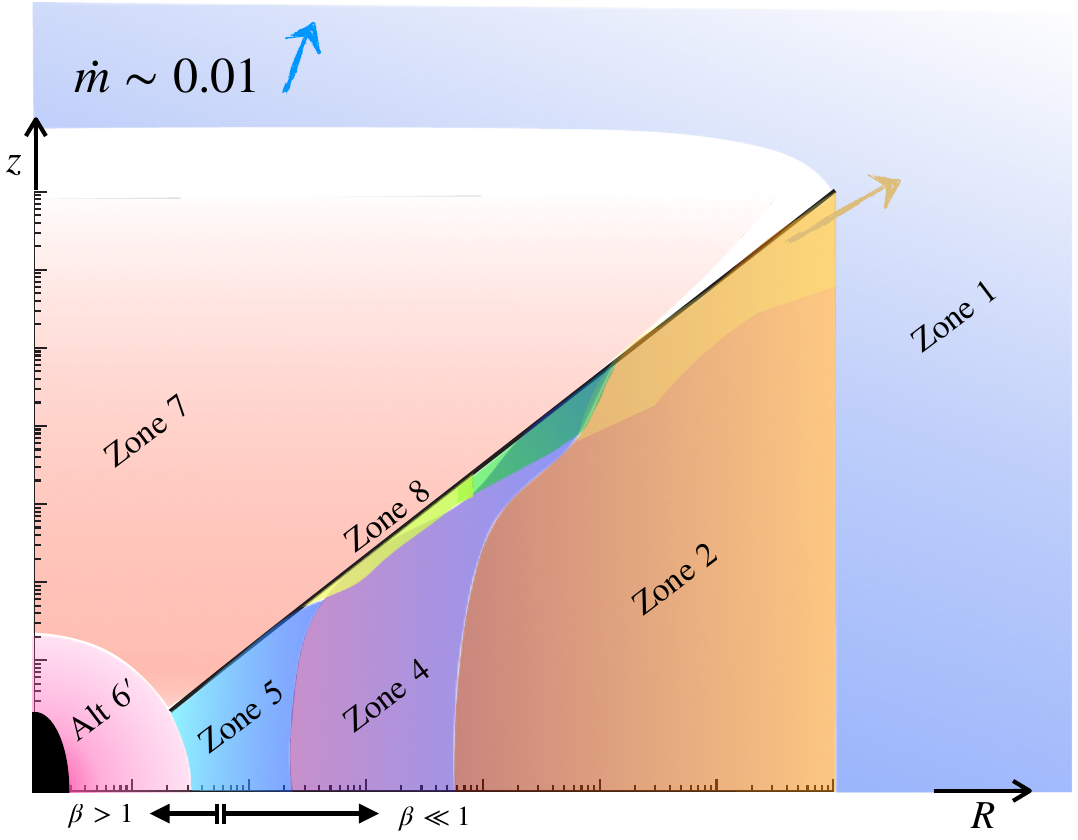} \\
	\caption{Zones as Figs.~\ref{fig:cartoon.definitions}-\ref{fig:cartoon.quadrants}, but now illustrating the effects of changing the Eddington-scaled accretion rate (\S~\ref{sec:extreme.mdot}). We compare $\dot{m}\sim100$ ({\em top}), $\sim 1$ ({\em middle}), $\sim 0.01$ ({\em bottom}), calculating the locations of different zones and how they move (black line at $z=H$). At $\dot{m} \gg 1$, most of the zones move outwards, the corona (if it still exists) must be much further above the disk in a more tenuous atmosphere, the inner zone 6/thermal disk is geometrically thickened by an order-unity factor owing to radiation pressure comparable to magnetic, and several of the directly illuminated zones are likely driven into outflows (arrows). We also label the radii interior to which radiation pressure becomes comparable to magnetic. At $\dot{m} \ll 1$ zones move inwards, and the innermost zone cannot effectively cool and becomes virialized and quasi-spherical in an inner ADAF-type flow, at which point it develops thermal pressure larger than magnetic, as shown. Outflows shut down completely or are suppressed (smaller arrows) in the NLR and edge of the dusty torus (where the opacity jump may still allow weak winds; \S~\ref{sec:super.eddington:outflows}). 
	\label{fig:cartoon.mdot}}
\end{figure}

\begin{figure}
	\centering\includegraphics[width=0.95\columnwidth]{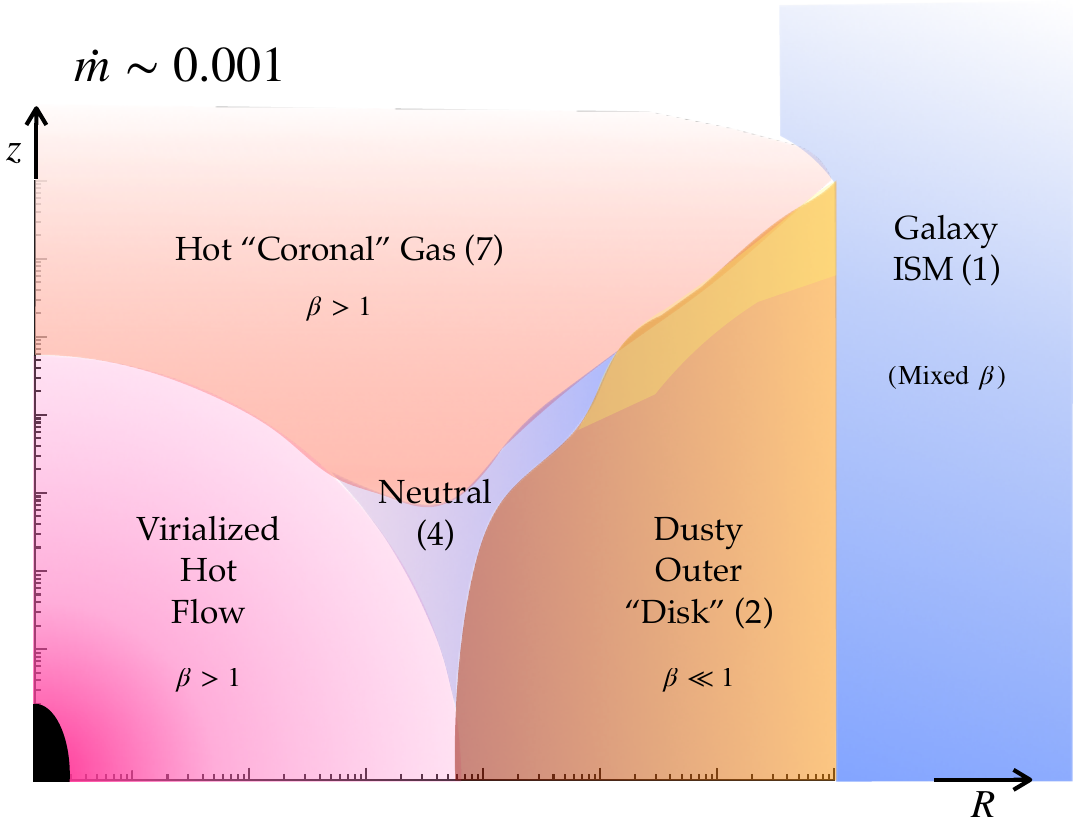} \\
	\caption{Illustration of the effects of Eddington ratio as Fig.~\ref{fig:cartoon.mdot} for an extremely sub-critical flow. The virialized hot flow expands as does coverage of coronal gas, while the dust sublimation radius moves inwards, until there is essentially no thermal-emitting ``disk'' in the inner regions (\S~\ref{sec:subcritical}). The models here can still apply to the outer disk/torus and neutral atomic/molecular region at large radii, until below the sublimation radius where these would join directly to the virialized hot flow. 
	\label{fig:cartoon.mdot.smallest}}
\end{figure}

The Comptonizing skin (zone (6b); \S~\ref{sec:warm.skin}) is made up of gas within one disk scale height $H\,(1-1/\tau^{\ast}) < |z| < H$ above the midplane of the inner, thermalized disk (see Figs.~\ref{fig:cartoon.definitions}, \ref{fig:cartoon.quadrants}, \ref{fig:cartoon.illumination}). Now consider the gas at somewhat larger $|z|\gtrsim H$, at similar and smaller radii (Figs.~\ref{fig:cartoon.mdot} \&\ \ref{fig:cartoon.mdot.smallest}). The vertical profiles of these disks at $|z|>H$ are discussed in detail in \paperone: the gas density (and rms vertical turbulent velocity) follows an approximate $\rho \propto {\rm sech}^{2}(z/H)$ or Gaussian ``core'' profile (at $|z|\lesssim H$) with an extended power-law falloff at larger radii, while the magnetic field strength $|{\bf B}|$ falls of less rapidly than $\rho$ (especially at $|z| \gtrsim H$), and the temperatures rise in the more rarified optically-thin coronal gas above the disk. The behavior of the vertical profiles -- most notably, that $|{\bf B}|$ falls off slowly compared to $\rho$ -- is expected from basic vertical equilibrium considerations given our {\em ansatz}. Consider: equating the (primarily toroidal) field support for the gas at $|z|\gtrsim H$, $(\nabla \times {\bf B})\times {\bf B}/4\pi \sim B_{\phi}^{2}/z$, and gravity, $-\rho\nabla \Phi_{\rm grav} = \rho \Omega^{2} z$, gives $B_{\phi}(r,z) \propto \rho(r,z)^{1/2} z$ at a given $r$. Provided this qualitative behavior holds, we find the details of the vertical profile are not so important, so consider a generalized $\rho(z) = \tilde{\rho}(|z|\gtrsim H)\,\rho_{\rm mid}$, $|{\bf B}(z)| = \tilde{B}(|z|\gtrsim H)\,|{\bf B}|_{\rm mid}$ (with $\tilde{B} \sim \tilde{\rho}^{1/2}\,(|z|/H)$). Assume (since we are not very far above the disk) the gas is illuminated primarily by the flux from the disk surface, and define its electron temperature $T_{e}^{\rm cor}$ and ion temperature $T_{i}^{\rm cor}$ (shown in Fig.~\ref{fig:thermal.profile}). In steady-state, this gas should be heated by magnetic reconnection (or shocks or turbulent dissipation, which give a similar rate since $v_{\rm turb} \sim v_{\rm shock} \sim v_{A} > c_{s}$) with time-averaged heating flux $F_{\rm heat}^{\rm cor} \sim |{\bf B}|^{2}\,v_{A}(z)/(8\pi)$ and characteristic post-shock/reconnection temperatures 
\begin{align} 
T_{\rm shock}^{\rm max} \sim T_{\rm recon}^{\rm max} \sim \frac{3 m_{p}}{16 k_{B}} v_{\rm shock}^{2} \sim  \frac{3 m_{p}}{16 k_{B}} (2\,\alpha_{2}\,v_{A})^{2}, 
\end{align} 
of which a fraction $\delta \approx 0.5$ and $(1-\delta)$ go into the electrons and ions respectively \citep{bisnovatyi:1997.bfield.heating.accretion.flows,bisnovatyi:2000.bfield.heating.expectations.acc.disks.corona.adafs}. Gas at these temperatures is cooled by a combination of Compton, cyclotron/synchrotron (if the electrons become highly relativistic), and Kramers (Bremsstrahlung, at the temperatures of interest) cooling of the electrons $F_{\rm cool}^{\rm cor} \sim F_{\rm Compton}^{\rm cor} + F_{\rm synch}^{\rm cor} + F_{\rm Brems}^{\rm cor}$. In the inner regions, since the radiation flux is strong and falls off less-rapidly than other quantities above with height, Compton cooling will be most important.

As we move above the disk, since $\tilde{\rho}$ decreases more rapidly than $\tilde{B}$, ``hard'' X-ray coronal solutions can appear at small enough $R$ and $H \lesssim |z| \lesssim r$. Specifically, at a given sufficiently low $\tilde{\rho}$, the timescale for full ion-electron temperature equilibration via Coulomb interactions $t_{\rm ie}^{\rm full}$ \citep{spitzer:1962.ionized.gases.book,narayan:bh.review.1998,cao:2009.coronal.model,cafg:2012.egy.cons.bal.winds} at the characteristic post-shock/reconnection temperature $T_{\rm shock}^{\rm max}$ becomes longer than the characteristic shock crossing/reconnection timescale $t_{\rm cross} \sim |z|/v_{\rm char}(z) \sim |z|/v_{A}(z) \sim \Omega^{-1} \sim t_{\rm dyn}$. If that occurs, the ions will be heated faster than they can cool and will therefore virialize at 
\begin{align}
T_{i}^{\rm cor} \sim T_{\rm vir} \sim \frac{1}{2} \frac{G M_{\rm BH}}{r}\,\frac{\mu\,m_{p}}{k_{B}},
\end{align}
 at which point the scale-height of the hard corona is $\sim r$, with $|{\bf B}|$ and $T_{e}$ sub-virial (so the gas remains bound with $c_{s}$ and $v_{A}$ modestly sub-virial). If the gas density in this region is initially more tenuous, it will build up rapidly with inflow (since this virialized component is not rapidly accreting) until reaching a quasi-steady-state density where continued heating is balanced by cooling for electrons and ions (at $T_{\rm mid} \ll T_{e}^{\rm cor} \le T_{i}^{\rm cor}$).

We can then calculate the approximate coronal properties following \citet{cao:2009.coronal.model} and equating the Coulomb, Compton, and heating rates (expressions given therein) for ions and electrons in steady-state, assuming a virialized ion system with $H/R \sim 1$ and electron-scattering optical depths in the corona of order unity. In the inner regions at sufficiently low $\dot{m}$, self-consistent coronal solutions can extend to $\tilde{B} \sim 1$, i.e.\ to the scale-height $|z|\sim H$, with ion 
\begin{align}
T_{i}^{\rm cor} \sim T_{\rm vir} \sim 10^{12}\,{\rm K}/x_{g},
\end{align} 
electron 
\begin{align}
T_{e}^{\rm cor} \sim 10^{9}\,{\rm K}\,(r_{\rm ff,\,5}/m_{7})^{1/19}\,(x_{g}/\dot{m})^{4/19} \sim 50-100\,{\rm keV},
\end{align} 
saturated density 
\begin{align}
\rho_{\rm cor}^{\rm sat}/\rho_{\rm mid} \sim 0.001\,(m_{7}/r_{\rm ff,\,5})^{7/19}\,(x_{g}/\dot{m})^{10/19},
\end{align} 
and approximate Thompson optical depth 
\begin{align}
\tau_{\rm es}^{\rm cor} \sim 13\,(r_{\rm ff,\,5}/m_{7})^{5/38}\,(\dot{m}/x_{g})^{9/19} \sim 4\,(x_{g}/10\dot{m})^{-0.5}
\end{align}
and Compton 
\begin{align}
y\sim 10\,m_{7}^{-0.2}\,(x_{g}/10\dot{m})^{-0.8}.
\end{align} 
This regime is valid approximately between radii 
$10\,\dot{m} \lesssim x_{g} \lesssim x_{g}^{\rm cor,\,H}$. 
At smaller radii $x_{g} \lesssim 10\,\dot{m}$ (only relevant at higher $\dot{m}$), radiation pressure and optical depth effects become important as expected (see discussion below, \S~\ref{sec:extreme.mdot}). At larger radii and/or lower heights above the midplane at large $x_{g}$ and $\dot{m}$, more specifically where 
\begin{align}
x_{g} \gtrsim x_{g}^{\rm cor,\,H} \sim 24\,(m_{7}/r_{\rm ff,\,5})^{30/33}\,\alpha_{2}^{130/33}\,\dot{m}^{-6/11}\,(|z|/H)^{20/33},
\end{align} 
coronal solutions cannot extend to the disk scale height $|z|\sim H$, or equivalently must have $\tilde{B} \ll 1$ (or else e.g.\ the implied \Alf\ speeds become super-virial and no self-consistent solution is possible). In this regime the coronal gas will reside at larger vertical heights $|z|$ above the disk. In this regime, from $
x_{g}^{\rm cor,\,H} \lesssim x_{g} \lesssim x_{g}^{\rm cor,\,z}$, 
using the virial limit for $\tilde{B}$ and $\tilde{\rho}$, we obtain the modified relations for the corona at sufficiently large height: 
\begin{align}
T_{e}^{\rm cor} & \sim 0.4\times10^{9}\,{\rm K}\,(x_{g}/10)^{1/2}\,\dot{m}^{-1}, \\  
\tilde{\rho} & \sim 0.0002\,(m_{7}/r_{\rm ff,\,5})^{1/2}\,(x_{g}/10)^{5/4}\,\dot{m}^{-5/2}, \\
\tau_{\rm es}^{\rm cor} & \sim 0.3\,(x_{g}/10)^{1/4}\,\dot{m}^{-3/2}, \\
y &= 4\,k_{B}\,T_{e}\,\tau_{\rm es}/(m_{e}\,c^{2}) \sim 0.07\,(x_{g}/10)^{3/4}\,\dot{m}^{-5/2}. 
\end{align}
In Fig.~\ref{fig:thermal.profile} we compare these coronal electron and ion temperature profiles to the midplane temperatures at different radii, BH mass, and accretion rates. Note that both the electron temperature (hardness of the corona) and Compton $y$ depend inversely on $\dot{m}$.

These solutions are able to persist at some $H < |z|_{\rm min} < z \lesssim R$ out to radii 
\begin{align}
x_{\rm cor,\,z} \sim 75\,(m_{7}/r_{\rm ff,\,5})^{80/109}\,\dot{m}^{-54/109}\,\alpha_{2}^{390/109}
\end{align}
 (see Fig.~\ref{fig:cartoon.mdot}). At larger radii, the Coulomb equilibration time and cooling time for new post-shock/reconnection gas at $T^{\rm max}_{\rm recon}$ trying to reach these densities will be shorter than $t_{\rm cross}$, and so the system cannot reach this state. 
Note that at this level of approximation, the details of the disk vertical profile factor out of the steady-state coronal properties because of their self-regulation via heating/cooling from basic physical processes. However, the outer radii of the corona do depend significantly on the $\alpha_{2}$ parameter here, indicating that detailed predictions are sensitive to the tails of the dissipation-rate PDF. A dependence on the dissipation rate is not unexpected, since we are discussing tenuous gas with low cooling rates being heated by processes like reconnection which tend to be fundamentally intermittent. However, this does emphasize that numerical simulations of this regime are needed for more concrete predictions. 

Observationally, many attempts have been made to infer these properties of the hard corona. The predicted electron temperatures $k_{B} T_{e}^{\rm cor}$, optical depths $\tau_{\rm es}^{\rm cor}$ and Compton $y$ parameters above all appear to lie reasonably within the range inferred from direct hard X-ray observations 
\citep{marinucci:2018.agn.coronae.review,kamraj:2022.hard.xray.agn.corona.properties,tortosa:2022.hard.coronal.constraints.in.hyper.eddington.qsos}, suggesting the predictions are at least order-of-magnitude plausible. And the qualitative trends of hardness decreasing with Eddington ratio, as well as overall coronal covering factor scalings (below) also appear similar to what is observed in AGN as a function of accretion rate \citep[see][]{petrucci:2018.agn.comptonizing.layer.properties,kubota:2018.soft.excess.comptonizing.layers,palit:2024.warm.covering.agn.corona.expands.with.luminosity.reprocesses.most.emission,ballantyne:2024.warm.corona.soft.excess.coronal.props.weaker.higheddrat.compilation}. 
But more detailed comparison is known to be sensitive to details of the non-linear scattering physics and geometry, and requires explicit X-ray radiation transfer calculations even in idealized geometries \citep[e.g.][]{george:1991.agn.xray.corona,magdziarz.zdziarski.95:compton.reflection.model}, which will be the subject of future work (Yun et al., in preparation).

The ``covering size'' of the hard corona to the thermal disk is therefore $\mathcal{O}[(x_{g}^{{\rm cor}})^{2}] \propto \dot{m}^{-1}$. Note this is defined in terms of the fraction of the thermal disk which is covered vertically, here, rather than an angular covering factor defined by the outer disk to the inner disk. A ``lamp-post-like'' vertical covering geometry, in a loose sense, automatically follows from the vertical scalings above, though of course since we are at non-negligible $z/R$ and invoking scattering our quasi-cylindrical approximation is not ideal, and as noted above we are sensitive to the tails of the dissipation rate PDF. So we caution again that numerical simulations are needed, but this can give some order-of-magnitude guidance to said simulations. 

In this sense, we can estimate that the ``steady-state'' virialized corona will collapse for sufficiently high $\dot{m} \gg 10$, as outside the ISCO it cannot stably persist in steady-state with a large covering factor (the volume-filling gas outside the ISCO will have lower $T_{e}$, $\tau_{\rm es}$, and $y$). Even in this regime, or at larger radii compared to $x_{g}^{\rm cor}$ above, hot coronal gas heated by reconnection will still exist, but it will be transient, as part of outflowing hot gas or occasional shocks or large-scale reconnection events above the disk, with an electron and ion temperature $\sim T_{\rm shock} \sim 10^{9}\,{\rm K}\,(m_{7}/r_{\rm ff,\,5})^{1/3}\,(x_{g}/10)^{-2/3}$ (as opposed to the ions being able to ``build up'' thermal energy and virialize, creating the steady-state large vertical scale-length and supporting the coronal pressure). The mass fraction of such hot gas will be suppressed by approximately a factor of $\sim t_{\rm eqm}/t_{\rm dyn}$ (as the gas cools back to the lower disk temperatures rapidly). Thus there would naturally be some ``patchy'' outer layers of the corona marking this transition zone, rather than a hard ``edge'' or sharp cutoff at a given $\dot{m}$ (illustrated in Fig.~\ref{fig:cartoon.mdot}). 

For more typical accretion rates $\dot{m} \lesssim 1$, comparing $x_{g}^{\rm cor}$ for the virialized estimate above to the size of the thermalized disk, and noting the intrinsically flared structure of the disk, the reflection fraction (discussed further below) will be comparable to this factor times the scattering fraction, giving typical fractions scaling like $\sim20\%$ for $m_{7}=\dot{m}=\alpha_{2}=1$, but scaling strongly with some of these properties (as e.g.\ $m_{7}^{2}\,\dot{m}^{-3}\,\alpha_{2}^{7}$).

\section{Zone 8: Extended Scattering Surfaces}
\label{sec:zone8.warm.scattering}

\subsection{The Disk As An Extended Scattering Source}
\label{sec:scattering}

As discussed above, the disk is strongly-flared and has large $H/R$, so a large fraction of the ``direct'' radiation from the central disk is intercepted. In the context of the ionized, directly illuminated layers and warm skin, at all radii interior to the dust sublimation radius $r_{\rm sub,\,ext}$, the directly-illuminated layer has a surface scattering opacity much larger than absorption opacity at most wavelengths of interest, and at all radii interior to $R_{\rm es,\,ion}$ that scattering layer is Thompson-thick and despite being geometrically ``thin'' (layer geometric thickness $\lambda \ll H$). Thus it acts as an effective scattering surface. Accounting for finite scattering probability even at $R>R_{\rm es}$, where one can estimate the scattering probability by the surface layer scattering optical depth $\tau_{s}<1$, can boost the covering factor of the scattering layer by an additional tens of percent. We can therefore summarize the behaviors of the extended scattering surface of the outer disk {\em alone} to the central source as follows, also illustrated in Fig.~\ref{fig:cartoon.illumination} (with a similar comparison for an SS73 disk in Fig.~\ref{fig:cartoon.ss73}).

\subsubsection{Summary of Disk Scattering/Reprocessing By Angle}
\label{sec:scattering:by.angle}

Rays with $\cos{\theta} \gtrsim 0.71$ will not intersect the disk on direct angles, but travel to the ISM (Zone (1)), where they could be reprocessed into NLR or ionization cones or ISM dust absorption (warm/cool disk) or escape entirely. Of course even these rays can be scattered by non-zero scattering opacities above the disk, so this angle $\theta$ should be considered to be the angle after emerging from the region where scattering opacities are non-negligible. The flux emitted into this range of $\theta$ amounts to a fraction of $\sim 30-50\%$ of the central region emission, depending on its geometry (lower if it is more isotropic/point-source like owing to scattering or flaring/reflection, higher if it is ``flat-disk-like''). 

Rays with $0.52 \lesssim \cos{\theta} \lesssim 0.71$ will intersect the disk between $r_{\rm sub,\,ext} \lesssim r \lesssim r_{\rm ff}$, where the dust will not be sublimated. These will therefore interact with the ``dusty torus-like'' regions (Zone (2c)), i.e.\ be reprocessed by warm/hot dust. Regardless of how we approximate the central emission geometry, this amounts to a fraction $\sim 20-25\%$ of the central emission. Note a larger fraction $\sim 0.5-0.7$ of the total central source can be obscured by the extended disk/torus-like structure, and there will be an additional contribution of one to a few percent emission from the intrinsic emission of the disk, but we are interested here in the reprocessed component.

Rays with $\cos{\theta} \lesssim 0.52$ will intersect at $r \lesssim r_{\rm sub,\,ext}$ and the dust will be sublimated. Rays with $0.38 \lesssim \cos{\theta} \lesssim 0.5$ will see sublimated dust but low enough gas densities to ionize their way ``through'' to $r_{\rm sub,\,ext}$ in a Stromgren sense. These will be reprocessed or interact with the ionized, multi-phase ``BLR-like'' zone[s] of the disk (Zones (3a) and (3b)). Again nearly independent of the central emission geometry this gives a fraction $\sim 10-15\%$ of the intrinsic emission being reprocessed (with a potential fraction $\sim 1-5\%$ from intrinsic emission). 

Rays at lower inclination $\cos{\theta} \lesssim 0.38$ will intercept at $r \lesssim R_{\rm es,\,ion} \sim 4\times 10^{16}\,{\rm cm}\,m_{7}^{3/5}\,\dot{m}\,r_{\rm ff,\,5}^{2/5}$, where they will see an ionized ``skin'' with $\lambda \lesssim H$ which is Thompson-thick to electron scattering and has $\kappa_{s} \gg \kappa_{a}$, so will be scattered off of the surface (but not re-processed). Since most of the effective area and covering factor and electron number lies at the outermost radii of this region, the effective size of the scattering region should be close to $R_{\rm es,\,ion}$. This amounts to a fraction $\sim 20-40\%$ of the central emission being scattered around these larger radii, here again more strongly dependent on the emission geometry (smaller if it is flat-disk-like and larger if more isotropic). Thus if there were no scattering anywhere outside of the disk (at $|z|>H$ anywhere), the ratio of scattered-to-direct luminosity along polar (non-obscured) sightlines could vary from $\sim 0.43-1.5$ (being slightly more careful about the integration of scattering, rather than simply treating the scattering as ceasing with some ``edge'' at a given $\cos{\theta}$, per \S~\ref{sec:scattering:warm.absorber}), depending on the emission geometry, but any additional scattering above the disk will increase this ratio. This scattering surface will both influence the inferred microlensing and reverberation mapping time delays, with the electron scattering explaining their observed achromatic results and the predicted surface size (e.g.\ $R_{\rm es\,ion}$) values broadly consistent with those observed \citep{dai:2010.agn.microlensing.xray.optical.larger.than.expected,blackburne:2011.sizes.qso.acc.disks.microlensing.too.big,jimenez:2014.qso.disk.temp.profile.size.from.microlensing.large.flat,cornachione:2020.accretion.disk.microlensing.profiles.shallow,cai:2023.qso.sed.universal.w.flat.intrinsic.spectrum.into.uv.larger.than.expected.need.reprocessing.in.blr,ren:2024.reverb.mapping.sizes.larger.than.expected} and its interior regions will give rise to roughly this reprocessing fraction of the hard coronal photons in an X-ray reflection component, again in agreement with observations \citep{czerny:2003.soft.excess.and.uv.profile.in.agn.from.reprocessing.warm.skin.warm.absorber,dai:2010.agn.microlensing.xray.optical.larger.than.expected,kamraj:2022.hard.xray.agn.corona.properties,tortosa:2022.hard.coronal.constraints.in.hyper.eddington.qsos}. 

In this sense, the large reprocessing arising from the flared disk with $H/R\sim 0.1-1$ is akin to what has often been invoked phenomenologically for scattering to explain microlensing and X-ray reflection observations as well as reverberation mapping, namely scattering from a broad-opening angle biconical wind centered on the accretion disk midplane (i.e.\ with an ``open'' or evacuated polar region and $z/R \sim H/R \sim 0.1-1$). They key difference is this structure and the specific geometry needed is predicted in these models, instead of invoked {\em ad hoc}, and this provides a natural explanation of its ubiquity even in systems that do not appear to exhibit strong outflows \citep[compare discussion in e.g.][]{chelouche:2019.qso.reprocessing.inner.disk.light.further.out.obs}.

\subsection{On the ``Warm Absorbers''}
\label{sec:scattering:warm.absorber}

The ``warm absorber'' \citep{halpern:1984.warm.absorber.discovery}, while occasionally discussed in the literature as a single structure, is really an umbrella term referring to any partially-ionized metals in gas detected in the UV through X-ray spectra of AGN \citep{laor:warm.absorber}. As such, observed gas associated with ``warm absorbers'' has been identified with more than four orders of magnitude spread in ionization parameter $\xi \propto L_{\rm ion}/n\,R^{2}$, radii from just above/outside the thermalized disk and corona ($\lesssim 100\,R_{g}$) to well into the galactic ISM/NLR/Zone (1) at $\gtrsim 100\,$pc, and velocities (often, but not always, in outflow) ranging from $\sim 100-30,000\,{\rm km\,s^{-1}}$ \citep{kinkhabwala:2002.warm.absorber.1068.requires.wide.range.of.densities.at.given.r.also.large.r.fewhundredpc.to.smaller,gofford:2013.ufos.to.warm.absorbers.continuous.family.absorber.properties.extended,tombesi:2013.ufo.warm.absorbers.may.be.connected.through.outflow}. The broadest observational definitions of the warm absorber would even include metal-line absorption systems observed in the circum-galactic medium (at $R \sim 100-300\,$kpc) around AGN \citep{tumlinson:2017.cgm.review,cafg:2023.cgm.review}. Therefore, no single ``structure'' or region/zone can uniquely be associated with the population of warm absorbers. And the majority of the observed ``warm absorbers'' are believed to come from radii well outside of the accretion disk regions we are explicitly modeling here -- e.g.\ from somewhere in the galactic Zone (1) or beyond, around or outside the BHROI (or at least the ``outer'' regions of the dusty/molecular torus) at $\gg$\,pc \citep{reynolds:1997.xray.agn.warm.absorbers.launched.from.dusty.torus,kaspi:2001.warm.absorber.3783.originates.outside.blr.and.torus,netzer:2003.detailed.modeling.warm.absorber.parsec.scale.outflow.weak,blustin:2005.warm.absorbers.accel.in.radii.outside.torus.bhroi,crenshaw:2012.warm.absorbers.strong.feedback.but.well.beyond.accretion.disk.nlr.accel,tombesi:2013.ufo.warm.absorbers.may.be.connected.through.outflow}. 

That said, given the broad opening angle of the geometrically-thick, flared disk and extensive reprocessing we have discussed above, many of the illuminated components we discussed above could potentially contain gas which would be classified in some sense as the rare ``compact'' (sub-pc-scale) warm absorbers \citep[e.g.\ akin to those observationally described in][]{krongold:2007.4051.example.compact.warm.absorber.from.inside.torus,tombesi:2013.ufo.warm.absorbers.may.be.connected.through.outflow} and/or intermediate/larger-linewidth absorbers (all the way to ``ultra-fast'' outflows) which appear to be more robustly associated with sub-parcsec radii \citep{gofford:2013.ufos.to.warm.absorbers.continuous.family.absorber.properties.extended,tombesi:2013.ufo.warm.absorbers.may.be.connected.through.outflow}. This includes the warm Comptonizing skin (Zone (6b)), extended scattering surfaces (Zone (8)), BLR (Zone (3)), illuminated torus (Zone (4)), and even some gas within the hard corona\footnote{Recall, in \S~\ref{sec:zone7.corona} we noted there can be two solutions for the equilibrium temperature of gas at larger heights $|z| \gtrsim H$ in the inner disk, one corresponding to strong shock/reconnection heating which produces the hard coronal gas properties, the other corresponding to more weakly-heated gas by e.g.\ weak shocks which should also be present at a smaller level and equilibrates at ``warm'' temperatures $\sim 10^{4}-10^{6}$\,K.} (Zone (7)), in addition to the NLR/galactic ISM/extragalactic gas (Zone (1)), or outer illuminated dusty torus. Many of these regions, but especially the illuminated dusty torus and illuminated galaxy/NLR, could also be in outflow or partially in outflow, as discussed below (\S~\ref{sec:extreme.mdot}). 
In addition to these, there is one more ``region'' which could potentially contain warm absorbers and scattering/reprocessing gas, namely gas ``above'' the disk scale-height ($|z| \gtrsim H$) but {\em outside} the radii of the hard corona ($x_{g} \gtrsim x_{g}^{\rm cor}$; \S~\ref{sec:zone7.corona}). As discussed in \S~\ref{sec:zone7.corona}, because the disk scale heights are large (especially at larger radii) with $H/R \sim 0.1-1$, even at $|z| \sim 1-3\,R$, the gas density is not vanishingly small but will still be a non-negligible fraction of the midplane density. Independent of the disk effective beaming pattern $f_{\theta}$ (\S~\ref{sec:model:illumination}), this will all be illuminated, with ``warm'' equilibrium temperatures ranging from $\sim 10^{4}-10^{6}$\,K (see \S~\ref{sec:warm.skin}) out to the directly-illuminated dust sublimation radius. 
Together, these span the range of radii and characteristic velocities of warm absorbers, with the ``compact'' warm absorbers being those that originate in the inner zones (see references above). The predicted structure also supports the proposed link between at least some of the warm absorbers and reprocessing/scattering of light from the AGN, as discussed widely in the observational and modeling literature \citep[see e.g.][]{laor:warm.absorber,krolik:2001.warm.absorbers.multiphase.winds,czerny:2003.soft.excess.and.uv.profile.in.agn.from.reprocessing.warm.skin.warm.absorber}, as we have explicitly associated most of these structures with scattering/reprocessing above.

\section{Behavior at Very Low and Very High Accretion Rates}
\label{sec:extreme.mdot}

Throughout, our discussion was largely focused on ``typical'' parameters for accretion, applicable to SMBHs with masses from $\sim 10^{4}-10^{10}\,{\rm M}_{\odot}$ with reasonable $r_{\rm ff} \sim R_{\rm BHROI}$ expected in galaxies, and dimensionless accretion rates $0.01 \lesssim \dot{m} \lesssim 100$. At very low and high accretion rates the behaviors above can be strongly modified, and some of our core assumptions can break down. We illustrate some of the salient quantitative changes in Figs.~\ref{fig:surface.densities}, \ref{fig:thermal.profile}, \ref{fig:opacity.profile}, \ref{fig:turb.props}, \ref{fig:sobolev.size}, but also summarize the qualitative changes and shifting of the positions of different zones in Figs.~\ref{fig:cartoon.mdot} \&\ \ref{fig:cartoon.mdot.smallest}. 

\subsection{On the Super-Critical Limit}
\label{sec:super.eddington}

In the super-critical limit ($\dot{m} \gg 1$), the various characteristic radii of different regions we describe above all tend to move outwards (e.g.\ Fig.~\ref{fig:cartoon.mdot}), but this does not alone change any of our qualitative conclusions about their nature and role. Some of the outermost radii, e.g.\ the dust thermalization radius $r_{\rm dust,\,therm}$, can even move beyond $r_{\rm ff}$, but again this does not actually mean anything meaningfully different for our results, except that our quantitative predictions would be slightly revised because we would need to account for the background potential of the stars in the host galaxy modifying $\Omega$ and therefore terms like the energy flux $\dot{M}\,\Omega^{2}$. A larger effect from the stellar potential at these radii is its ability to induce strong gravitational torques via non-axisymmetric structure in the collisionless component acting on the gas disk, which is universally seen to dominate the stresses (relative to classical disk Maxwell/Reynolds stresses) for the strong inflows outside $\gtrsim r_{\rm ff}$ in galaxy-scale simulations \citep{garcia.burillo:torques.in.agn.nuclei.obs.maps.no.inflow,haan:nuga.gas.dynamics.maps,hopkins:zoom.sims,hopkins:m31.disk,hopkins:slow.modes,hopkins:cusp.slopes,hopkins:inflow.analytics,prieto:2016.zoomin.sims.to.fewpc.hydro.cosmo.highz,prieto:2017.zoomin.sims.agn.fueling.sne.fb,angles.alcazar:grav.torque.accretion.cosmo.sim.implications,querejeta:grav.torque.obs.m51,williamson:2022.gizmo.rhd.psph.sims.binary.smbh.torii.radiation.reduces.grav.torques}.

At large radii, it is also always the case that the disk is highly sub-Eddington in a {\em local} sense, meaning that the radiation pressure forces in the midplane are much smaller than those from magnetic+turbulent pressure, let alone gravity. But two regions need to be reconsidered at sufficiently high $\dot{m}$, owing to the effects of radiation pressure: first, the innermost thermalized disk, where this may no longer hold for $\dot{m} > 1$, and second, the directly-illuminated outer disk regions, where we need to consider the direct radiation flux from the central disk (potentially larger than the local cooling flux). We will show, however, that everywhere in the super-critical limit, the solutions here remain distinct in some important ways from the simplest ``slim disk'' analytic solutions which still assume $\beta \gg 1$ (negligible magnetic pressure everywhere; \citep{paczynsky.wiita:1980.slim.disk,abramowicz:1988.slim.disks}). 

\subsubsection{Thermal-Viscous Stability}
\label{sec:thermal.viscous.stability}

Briefly, recall that for standard thermal+radiation $\alpha$ disks with $\beta \gg 1$, the disk becomes unstable to the runaway (and potentially catastrophic) thermal-viscous instability if the radiation pressure becomes comparable to the thermal pressure (interior to $x \lesssim 1600\,\dot{m}^{16/21}$ in the SS73 model; \citealt{piran:1978.thermal.viscous.accretion.disk.instability}). For the magnetized disks here (or any disk where magnetic pressure is much larger than thermal), this instability vanishes, because the dissipation rate per unit area from the stress, $\mathcal{F}^{+} \sim r H \mathbb{\Pi}_{r\phi} d\Omega/dr \propto \Omega\,H\,B^{2}$ is independent of or only weakly-dependent on the disk temperature structure \citep[see e.g.][for discussion]{begelman.pringle:2007.acc.disks.strong.toroidal.fields}.

\subsubsection{Gravitational Stability}
\label{sec:super.eddington:stability}

Also briefly, we recall that the Toomre $Q$ parameter even at the largest radii (where it is minimized), is $\sim 3000/(m_{7}^{1/4}\,\dot{m})$ -- thus, as discussed in detail in \paperone-\paperthree, these disks are gravitationally stable at all radii from $r_{g}$ out to $r_{\rm ff} \sim R_{\rm BHROI} \sim 5\,m_{7}^{1/2}\,{\rm pc}$, up to accretion rates as large as $\dot{m} \sim 3000\,m_{7}^{1/4}$. This is vastly larger than any thermal-pressure dominated disk -- in contrast a standard thermal-pressure dominated $\alpha$ disk becomes gravitationally unstable outside $x \gtrsim 200\,(10^{8}\,M_{\odot}/M_{\rm BH})\,\dot{m}^{-(0.3-0.5)}$ (depending on opacity scalings). So gravitational instability does not meaningfully limit whether the disks can be super-critical, though it may still limit the maximum accretion rates to these very large values.

\subsubsection{Does the Inner Disk Midplane Ever Become Radiation-Pressure Dominated?}
\label{sec:super.eddington:rad.pressure}

For a laminar, optically-thick disk, in which the radiation flux from the midplane is dominated by diffusive transport, we would have midplane radiation pressure $P_{\rm rad} = (4\sigma_{B}/3c) T_{\rm mid}^{4}$ with $T_{\rm mid}=f_{\tau}^{1/4} T_{\rm eff}^{4}$ and $f_{\tau} \approx \tau \approx \tau_{\rm es}$ (the electron-scattering optical depth). Using this, we can calculate ratio $P_{\rm rad}/(P_{\rm mag}+P_{\rm turb}) \approx 2\,(x/x_{r})^{-7/6}$ where $x_{r} \equiv 30\,\dot{m}^{6/7}\,(r_{\rm ff,\,5}/m_{7})^{1/7}$. At $x \lesssim 2\,x_{r} \sim 60\,\dot{m}$, therefore, this suggestions radiation pressure should not be neglected. While for small $\dot{m} \lesssim 0.04\,m_{7}^{1/12}$ this moves inside the ISCO, larger accretion rates will always have some radii where this could be important. Of course, at {\em very} large $\dot{m}$, this outer radius $\sim 2\,x_{r}$ is bounded by the radius where the disk is absorption optically-thick, i.e.\ it cannot be larger than $\sim R_{\rm therm,\,ion}$, beyond which the radiation will not be effectively trapped to build up.

But as noted in \S~\ref{sec:zone6a.vertical} (and \paperthree), that is not consistent at small radii and high accretion rates. Noting that the turbulent velocity scales as $v_{\rm turb} \sim (H/R)\,v_{\rm K} \sim 1.6\times10^{4}\,{\rm km\,s^{-1}}\,m_{7}^{1/6}\,r_{{\rm ff},\,5}^{-1/6}\,x_{g}^{-1/3}$, while the diffusive radiation speed (in the inner disk) is $v_{\rm diff} \sim c/\tau_{\rm es} \sim 3.1\times10^{2}\,{\rm km\,s^{-1}}\,m_{7}^{1/3}\,r_{{\rm ff},\,5}^{-1/3}\,\dot{m}^{-1}\,x_{g}^{-5/6}$, for $x_{g} \lesssim x_{g}^{\rm turb,\,z} \equiv x_{r}$, we have $v_{\rm turb} > v_{\rm diff}$. Thus vertical mixing of radiation by turbulence becomes faster than radiative diffusion, which means that the midplane temperature either remains at $\sim T_{\rm eff}$ (if leakage were very efficient) -- which one can easily verify (and we noted in \paperthree) gives radiation pressure much smaller than magnetic/turbulent pressure everywhere -- or, at most, assuming a standard ``turbulent diffusion'' rate and equating the midplane flux to the surface flux, limits to $T_{\rm mid}/T_{\rm eff} \approx (c/v_{\rm turb})^{1/4} \sim 2\,(r_{\rm ff,\,5}/m_{7})^{1/24}\,x_{g}^{1/12}$ (i.e.\ $f_{\rm tau} = c/v_{\rm turb}$). Since this is an increasing function of $x_{g}$ while the expectation for diffusive transport is a decreasing function of $x_{g}$, one can calculate the maximum ratio $T_{\rm mid}/T_{\rm eff} \sim 2.7$ (appearing around $x_{g} \sim x_{r}$). 

This, in turn, means that the radiation pressure in the disk midplane\footnote{Note that (as required for consistency) we obtain the identical result using the midplane radiation flux and defining the {\em coupled} radiation pressure (a more useful proxy in optically-thin regions) as $\int dz \rho \kappa F_{\rm rad}/c$, shown in Fig.~\ref{fig:pressure.profile}.} $P_{\rm rad} \sim (1/3)\,(4\sigma_{B}/c_{L})\,T_{\rm mid}^{4}$ can never exceed a factor of $\approx 2$ times the magnetic/turbulent pressure, i.e.\ 
\begin{align}
P_{\rm rad} \lesssim 2\,(P_{\rm mag} + P_{\rm turb}) \ , 
\end{align}
which would inflate the disk by at most a factor of $\sim 1.7$ in height. But since $H/R \sim 0.07\,(m_{7}\,x_{g}/r_{\rm ff,\,5})^{1/6}$ at these radii is predicted in the absence of radiation pressure, this is a modest $\mathcal{O}(1)$ correction at most and the {\em local} optically-thick Eddington limit is never exceeded (the vertical and radial radiation pressure forces are weaker than gravity, equivalent to the equilibrium $H/R < 1$) at any $\dot{m}$.\footnote{This also allows us to immediately and simply verify that radiative viscosity and/or damping effects can be safely neglected in this models for any parameters. In the optically-thick limit ($\tau \gg 1$), the radiative damping acts as a viscosity with $\nu_{\rm rad} \sim (U_{\rm rad}/\rho c)\,(\rho \kappa)^{-1}$ \citep{loeb.laor:1992.radiative.viscosities.in.accretion.disks}. If we take the upper limit to $P_{\rm rad}\sim U_{\rm rad}/3$ provided by the expressions here, and compare $\nu_{\rm rad}$ to the standard Maxwell/Reynolds effective viscosities in the disk models here, we immediately obtain $\nu_{\rm rad}/\nu_{\rm Maxwell/Reynolds} \lesssim (3/2) v_{\rm turb}/(c \tau) \ll 1$, i.e.\ this is always extremely small. In the optically-thin ($\tau \gg 1$) limit the radiative damping can act as fast as $\Gamma_{\rm rad} \sim (U_{\rm rad}/\rho\,c)\,(\rho\,\kappa)$. Comparing this to $\Omega$ and noting $U_{\rm rad} \sim F_{\rm rad}/c$ (since $\tau \ll 1$) and $\kappa F_{\rm rad}/c$ is the radiative acceleration, and noting the vertical gravitational acceleration $a_{\rm grav} \sim \Omega^{2} H \sim \Omega v_{\rm turb}$, we obtain $\Gamma_{\rm rad}/\Omega \sim (v_{\rm turb}/c)\,(a_{\rm rad}/a_{\rm grav})$. But since $v_{\rm turb} \ll c$ and we showed $a_{\rm rad} \lesssim a_{\rm grav}$ everywhere (moreso in the outer, optically-thin regions), this is also small.} We see this illustrated directly in Fig.~\ref{fig:pressure.profile} where the pressures are calculated. For the innermost radii at $\dot{m} \sim 1$, and out to $x_{g} \gtrsim 1000$ for $\dot{m} \sim 100$, the disks do approach this limit, but never exceed it.

So the prediction here is that the inner regions with $\dot{m} \gg 1$ should saturate at $P_{\rm rad} \sim P_{\rm turb} \sim P_{\rm mag,\,turb}$, where the latter represents the turbulent/tangled magnetic component (which can easily be smaller than the smooth/laminar mean toroidal field by a factor of a few in the simulations; see \papertwo). That saturation, in turn, means our fiducial {\em ansatz} for the disk structure and predictions should still apply at least at the order of magnitude level.

\subsubsection{Do Super-Critical Disks Become Radiatively Inefficient?} 
\label{sec:super.eddington:rad.eff}

Meanwhile radial advective transport/accretion occurs with a speed between $v_{R} \sim v_{\rm turb}^{2}/v_{\rm K}$ (for a purely turbulent/fluctuating trans-\Alf{ic} Maxwell/Reynolds stress) to $v_{R} \sim v_{\rm turb}$ (for e.g.\ accretion dominated by a mean Maxwell stress in a toroidal-dominated field so $\langle B_{\phi} B_{R} \rangle \propto v_{\rm turb} \propto v_{R}$). The inflow speed will generally be smaller than or at most comparable to $v_{\rm turb}$ (because $v_{\rm turb} \lesssim v_{\rm K}$ everywhere), but can also easily exceed the naive $v_{\rm diff}$. Specifically taking the more conservative (slower) $v_{R} \sim v_{\rm turb}^{2}/v_{\rm K}$ assumption and inserting our predictions for $v_{\rm turb}$, we obtain $v_{\rm R}/v_{\rm diff}\approx 2.6\,\dot{m}/x_{g}$ so $v_{\rm R} > v_{\rm diff}$ at $x_{g} < 2.6\,\dot{m}$. Adopting $v_{\rm R} \sim v_{\rm turb}$, $v_{R} > v_{\rm diff}$ at all $x_{g} \lesssim x_{g}^{\rm turb,\,z}$ (defined in \S~\ref{sec:super.eddington:rad.pressure}).

If the radiation can never move faster than $v_{\rm diff}\equiv c/\tau$, this would naively imply something like the slim disk regime in the super-critical ($\dot{m} > 1$) limit outside of the ISCO, i.e.\ that the radiation is advected and swallowed by the BH, rather than escaping. However by definition in this regime $v_{\rm turb} \gtrsim v_{R}$ also governs the vertical transport as noted above. If we instead say the radiation is being bulk transported vertically by the ``turbulent diffusion'' or advection as above, then some of it could reach the ``top'' of the disk ($|z| \sim H$) where $\tau$ becomes smaller and therefore the radiation might escape before being radially accreted. In this sense, the disk would be more akin to a system with e.g.\ strong photon-bubble instabilities, rather than a classic ``slim'' disk, in the supercritical regime, and could remain radiatively efficient down to smaller radii. 

However, that would only be true if the disk were analogous to a star and had a very ``sharp'' upper boundary, or were geometrically very thin. Here, the disks are geometrically quite thick with $H/R\sim 0.1-1$, and so the gas density only decreases by factors of $\sim 10-100$ even from the midplane to $|z| \sim 3-10\,R$ (e.g.\ \S~\ref{sec:zone7.corona} \&\ \papertwo). Per the calculation in \S~\ref{sec:illumination.disk.teff}, because of the large scale heights, gas at $|z| \gtrsim H$, coronal/warm components, etc.\ described above, we expect the surface where $\tau$ falls to $\ll 1$ to be at much larger distances $|z| \gg H$. Thus radiation cannot, in practice, immediately escape upon reaching $|z|\sim H$ (the coherence length of the turbulence, by definition in this supersonic, trans-\Alf{ic} regime). So it follows that the disks should become radiatively inefficient in more or less the standard fashion of supercritical slim disks \citep{abramowicz:1988.slim.disks}, with the maximum flux emerging from each logarithmic interval in radius interior to the radii where $\sim 4\pi\,R^{2}\,F_{\rm grav}$ would naively exceed the Eddington limit saturating at of order the Eddington limit. Of course, because the illumination/heating is even more strongly non-local in this limit, this regime needs to be studied in more detail in numerical simulations. 

One important difference between this and the most common slim-disk models would be that the advected energy is necessarily primarily in the form of magnetic, rather than thermal, energy, at least at the outer radii. Per \paperthree, consider the usual vertically-integrated accretion disk equations in a Keplerian potential, making the same {\em ansatz} that $\beta \ll 1$ from a strong toroidal field with trans-\Alf{ic} turbulence, but now consider (as \citealt{abramowicz:1988.slim.disks,narayan.yi.95:adaf.self.similarity.outflows,narayan:bh.review.1998}) advection-dominated similarity solutions (i.e.\ taking the radiation flux to be negligible or $f_{\rm advection}=1$ in the energy equation but retaining the advected magnetic, instead of thermal energy, in \citealt{narayan:bh.review.1998} Eq.~2.4). It is straightforward to see that the alternative solution discussed in \paperthree\ (namely their Eqs.~10-12, with $\gamma=5/3$ as parameterized therein) is the similarity solution in this limit. If the flow becomes advection-dominated (radiatively inefficient) interior to some radius $R < R_{\rm adv} \equiv \tilde{x}_{\rm adv} 2.6\dot{m}$, we therefore predict our disk model from \S~\ref{sec:model} (with $v_{t}/v_{\rm K} \sim v_{A}/v_{\rm K} \sim H/R \sim (R/r_{\rm ff})^{1/6}$ and $\Sigma_{\rm gas} \propto R^{-5/6}$) to transition to:
\begin{align}
\label{eqn:vt.highmdot}\frac{v_{t}}{v_{\rm K}} \sim \frac{v_{A}}{v_{\rm K}} \sim \frac{H}{R} \rightarrow \psi \ , \\ 
\label{eqn:S.highmdot}\Sigma_{\rm gas} \rightarrow \frac{\dot{M}}{2\pi\,\psi^{2} \sqrt{G M_{\rm BH} R}}  \ .
\end{align}
Other properties like $Q \sim 2^{3/2}\psi^{3} \Omega M_{\rm BH} /\dot{M}$, $\langle \rho\rangle_{\rm vol} \sim \Sigma_{\rm gas}/2H \propto R^{-3/2}$, $|{\bf B}| \sim |B_{\phi}| \sim v_{A} \sqrt{4\pi \rho} \propto R^{-5/4}$, $v_{R} \sim \psi^{2} v_{\rm K}$ follow immediately in the same way they do from our default model (\S~\ref{sec:model}). These solutions are characterized by the dimensionless similarity parameter $\psi$, which (if we assume a rapid transition between regimes at $R \approx R_{\rm adv}$) is determined by continuity to be:
\begin{align}
\psi \rightarrow \left(\frac{R_{\rm adv}}{r_{\rm ff}}\right)^{1/6} \sim 0.1\left(  \frac{\tilde{x}_{\rm adv} m_{7} \dot{m} }{r_{\rm ff,\,5}} \right)^{1/6}
\end{align} 
or $\psi \sim 0.2\,m_{7}^{1/12}(\dot{m}/100)^{1/6}$ using the expected $r_{\rm ff,\,5}$, $\tilde{x}_{\rm adv}$. We adopt this generalization of the solution (with corresponding change in the radiative efficiency, $\epsilon_{r} \propto 1/\dot{m}$ for $\dot{m} \gg 1$) for our more detailed calculations throughout (e.g.\ Figs.~\ref{fig:surface.densities}-\ref{fig:cartoon.mdot}). In Fig.~\ref{fig:surface.densities}, for example, this produces the subtle change in slope of $\Sigma_{\rm gas}(R)$ around a few hundred $R_{g}$. Indeed, as discussed in \papertwo\ and \paperthree, for a simulation with $\dot{m} \sim 50-200$, this solution (Eqs.~\ref{eqn:vt.highmdot}-\ref{eqn:S.highmdot}) does potentially better fit the simulations inside a few hundred $R_{g}$, where predicted here, while our default \S~\ref{sec:model} solution provides a more accurate fit at larger radii. But importantly, as our Figures demonstrate, this is a quite weak effect -- the difference between this and simply extrapolating our default model for the structural disk properties is small. The most important effects are that $H/R$ goes from very weakly decreasing with $R$ to roughly constant, while $\rho$ and $\Sigma_{\rm gas}$ continue to increase but more weakly at smaller $R$, weakly modifying the predicted opacity properties of the inner disk.

\subsubsection{Outflows from Externally-Illuminated Regions}
\label{sec:super.eddington:outflows}

{\em General Considerations: } The situation is more complex for regions which are externally-illuminated (where the incident flux from the central disk is much larger than the local ``self'' flux from cooling). Consider rays incident to the disk surface at some angle $\cos{\theta}$ with a corresponding radiation flux $F_{\rm illum}$ defined above: the total radiation force per unit area on the radial ``column'' seen by the rays is 
\begin{align}
F_{\rm rad,\,\hat{r}} = F_{\rm illum}\,\mathcal{G}(\tau_{\theta})/c
\end{align} 
where $\mathcal{G}(\tau_{\theta})$ is a function (discussed below) of the optical depth of the column 
\begin{align}
\tau_{\theta} = \int (\kappa_{s}+\kappa_{a})\,\rho\,d\ell
\end{align}
 along $\theta$. Using the total opacity and integrating through the disk profile here (accounting for its flaring and radial structure, so most of this integral comes from the inner illuminated surface at some $R$ where the ray is first incident on the disk) this is 
 \begin{align}
 \tau_{\theta} \approx \langle \kappa \rangle\,\rho_{\rm gas}\,R \approx \kappa(R)\,\Sigma_{\rm gas}(R)/(H/R)_{R}.
 \end{align} 
 It is convenient to also rewrite $ \mathcal{G}(\tau_{\theta})$ via 
 \begin{align}
\mathcal{F}(\tau_{\theta}) \equiv \mathcal{G}(\tau_{\theta})/\tau_{\theta,\rm es}^{R} = \mathcal{G}(\tau_{\theta})/[\kappa_{\rm es}\,\Sigma_{\rm gas}/(H/R)].
 \end{align}
  Comparing this to the radial gravitational force per unit area in the radial direction $F_{\rm grav,\,\hat{r}} =\int g(r)\,\rho\,d\ell \approx (G\,M_{\rm BH}/R^{2})\,\Sigma_{\rm gas}(R)$, we obtain the ratio of (radial, not vertical here) forces or accelerations: $F_{\rm rad,\,\hat{r}}/F_{\rm grav,\,\hat{r}} = \langle a_{\rm rad} \rangle/\langle a_{\rm grav} \rangle \sim 0.26\,\dot{m}\,\epsilon_{r,\,0.1}\,\mathcal{F}(\tau_{\theta})\,\tilde{r}^{1/3}$. where $\tilde{r} \equiv R/r_{\rm ff}$. So this will exceed unity and launch outflows along $\theta$ for $\dot{m} \gtrsim \dot{m}^{\rm w} \equiv 4/(\mathcal{F}(\tau_{\theta})\,\epsilon_{r,\,0.1}\,\tilde{r}^{1/3}) \sim 700/[\mathcal{F}(\tau_{\theta})\,\epsilon_{r,\,0.1}\,(m_{7}\,x_{g}/r_{\rm ff,\,5})^{1/3}]$, or (noting $L = \dot{m} \epsilon_{r,\,0.1}\,L_{\rm Edd}$):
 \begin{align}
 \frac{L}{L_{\rm Edd}} > \lambda_{\rm Edd,\,w} \equiv \frac{4}{\mathcal{F}(\tau_{\theta})\,\tilde{r}^{1/3}} \sim \frac{700}{\mathcal{F}(\tau_{\theta}) (m_{7} x_{g}/r_{\rm ff,\,5})^{1/3}}
 \end{align} 

Of course, it is obvious from the above that for any $\mathcal{F}(\tau_{\theta})$, one can allow large $\dot{m}$ by suppressing the radiative efficiency, which is expected. But even for a fixed, large radiative efficiency $\epsilon_{r,\,0.1} \sim 1$, flaring and differential illumination and the non-trivial behavior of $\mathcal{F}(\tau_{\theta})$ introduce interesting behavior here as a function of the incident angle $\cos{\theta} \sim H/R$ and corresponding incident radius $R$. The consequences of these behaviors for which zones could and could not support outflows under such conditions are illustrated in Fig.~\ref{fig:cartoon.mdot}.

{\em Optically-Thin (Outer) Illuminated Regions:} First consider sufficiently large incident radii (high inclination $\theta$) where $\tau_{\theta} = \kappa\,\Sigma_{\rm gas}/(H/R) = 0.005\,(\kappa/\kappa_{\rm es})\,(m_{7}/r_{\rm ff,\,5})^{1/2}\,\dot{m}\,\tilde{r}^{-1}$ is $<1$ ($R \gtrsim r_{\rm thin}^{\rm eff} \sim 0.005\,(\kappa/\kappa_{\rm es})\,(m_{7}/r_{\rm ff,\,5})^{1/2}\,\dot{m}\,r_{\rm ff}$). In the optically-thin ($\tau_{\theta} \ll 1$) limit, $\mathcal{G}(\tau_{\theta}) \approx \tau_{\theta}$, so $\mathcal{F}(\tau_{\theta}) \sim \kappa/\kappa_{\rm es}$. For $\dot{m} \lesssim \dot{m}_{\rm min} \sim 4\,(\kappa_{\rm es}/\kappa)\,\epsilon_{r,\,0.1}^{-1}$ (or $L \lesssim 4\,(\kappa_{\rm es}/\kappa)\,L_{\rm Edd}$), no outflows will be driven inside $r_{\rm ff}$; for $\dot{m}_{\rm min} \lesssim \dot{m}_{\rm min} \lesssim \dot{m}_{1}$ with $\dot{m}_{1} \approx 11\,(\kappa_{\rm es}/\kappa)\,(m_{7}/r_{\rm ff,\,5})^{-1/8}\,\epsilon_{r,\,0.1}^{-3/4}$, or 
\begin{align}
\lambda^{\rm Edd,\,w}_{\rm min} = \frac{4\kappa_{\rm es}}{\kappa} \lesssim \frac{L}{L_{\rm Edd}} \lesssim \frac{11 \kappa_{\rm es}}{\kappa}\,\frac{r_{\rm ff,\,5}^{1/8} \epsilon_{r,\,0.1}^{1/4}}{m_{7}^{1/8}} = \lambda_{1}^{\rm Edd,\,w}
\end{align}
outflows will be driven from the outer optically-thin illuminated surface at 
\begin{align}
R \gtrsim R^{\rm disk}_{\rm min,\,w} \equiv r_{\rm ff}\,(\dot{m}/\dot{m}_{\rm min})^{-3} = r_{\rm ff}\,(\lambda_{\rm Edd}/\lambda_{\rm Edd,\,w}^{\rm min})^{-3}; 
\end{align}
and for $\dot{m} \gtrsim \dot{m}_{1}$ outflows could be driven throughout the entire optically-thin illuminated region down to $R \gtrsim r_{\rm thin}^{\rm eff}$. 

Note that in the uppermost incident regions where the incident layer is still dusty ($\cos{\theta} \gtrsim 0.5$), $\mathcal{F}(\tau_{\theta})$ can be larger than unity by a factor $\sim (\kappa_{\rm dust}/\kappa_{\rm es})$, which leads to the well-known result that even for $\dot{m}\,\epsilon_{r,\,0.1} <1$ (i.e.\ sub-Eddington {\em luminosities}), the directly-illuminated dusty region, aka Zone (2c) can be locally super-Eddington and driven into outflow. This is why, in e.g.\ Fig.~\ref{fig:cartoon.mdot}, we still show some outflow potentially emerging from both Zone (1) (the NLR) and Zone (2c) (the illuminated dusty torus) even at $\dot{m} \ll 1$, as both can have large effective opacities while still falling into this regime when radiation from the AGN along a relatively un-obscured sightline first encounters such an opacity jump \citep[and this appears consistent with observations like those in][]{leighly:2024.agn.winds.stronger.line.driven.highedd.dust.driven.alledd}. In the sublimated upper zones ($\cos{\theta} \lesssim 0.5$), as discussed above, $\kappa \approx \kappa_{\rm es}$, so this ``boost'' no longer appears given Thompson scattering alone. 

{\em Optically-Thick (Inner) Illuminated Regions:} In the inner or lower-inclination optically-thick regions, the scaling of $\mathcal{G}$ (or $\mathcal{F}$) is more ambiguous. For a homogeneous, laminar, spherical or plane-parallel atmosphere with strictly grey opacities, $\mathcal{G}\approx \tau_{\theta}$ (or $\mathcal{F}\sim 1$ where electron scattering dominates). However many have argued that inhomogeneity and instabilities and outflows introduced by the radiation enforce a maximum $\mathcal{G}(\tau_{\theta})\sim$\,a few, and even without these effects the geometry here (being distinct from spherical or plane-parallel) may enforce this as photons will diffuse out through the vertical direction or reflect and escape. But even if we were to allow for the upper-limit of multiple scattering, $\mathcal{G}(\tau_{\theta}) \approx \tau_{\theta}$, we would still obtain $F_{\rm rad,\hat{r}} < F_{\rm grav,\,\hat{r}}$, i.e.\ $\lambda_{\rm Edd} < \lambda_{\rm Edd,\,w}$, for disk surface layers at sufficiently small radii $R \lesssim R^{\rm disk}_{\rm min,\,w}$ (as derived above because of the same scaling with $\tau_{\theta}$). In more convenient units $R^{\rm disk}_{\rm min,\,w}/R_{g} \sim 3\times10^{8}\,(m_{7}/r_{\rm ff,\,5})^{-1}\,(\epsilon_{r,\,0.1}\,\dot{m})^{-3} \sim 300\,(m_{7}/r_{\rm ff,\,5})^{-1}\,\epsilon_{r,\,0.1}^{-3}\,(\dot{m}/100)^{-3}$.

Interior to this includes the illuminated layers of the inner disk and scattering zone (the electron-scattering thick zones which define e.g.\ the scattering layer and ``warm skin''). So these disk layers will be robust against outflow even for $\dot{m} \sim 100$. Much more tenuous gas well {\em above} the disk, in e.g.\ the hard corona, could still be accelerated however, for two reasons. First because the polar angle $\cos{\theta}$ is larger, the direct flux is also larger, and second because the gas is more tenuous, the flux-to-mass ratio is larger still, and the local sightline can fall into the optically-thin regime above, where an outflow would be expected at sufficiently high $\lambda_{\rm Edd}$ (Fig.~\ref{fig:cartoon.mdot}). 

Among the optically-thick disk layers, the ionized, non-shielded incident zones (e.g.\ Zone (3a)) would be driven into outflows if $\mathcal{G}(\tau_{\theta}) \sim \tau_{\theta} > 1$, the maximum upper-scattering limit. But it is not obvious how large this factor can, in practice, reach.  
Consider a more limited $\mathcal{G}(\tau_{\theta})\sim \mathcal{G} \sim $\,constant. Then $F_{\rm rad,\hat{r}} \gtrsim F_{\rm grav,\hat{r}}$ for $R \gtrsim R^{\rm multiple}_{\rm min,\,w}$ 
where 
\begin{align}
\nonumber R_{\rm min,\,w}^{\rm multiple}& \sim 0.05\,r_{\rm ff}\,(m_{7}/r_{\rm ff,\,5})^{3/8}\,(\mathcal{G}\,\epsilon_{r,\,0.1})^{-3/4} \\
\nonumber &\sim 3\times10^{5}\,R_{g}\,(m_{7}/r_{\rm ff,\,5})^{-5/8}\,(\mathcal{G}\,\epsilon_{r,\,0.1})^{-3/4} \\
&\sim 10^{18}\,{\rm cm}\,m_{7}^{3/8}\,r_{\rm ff,\,5}^{5/8}/(\mathcal{G}\,\epsilon_{r,\,0.1})^{-3/4},
\end{align}
 {\em independent} of $\dot{m}$ (because the dependence of ``weight'' and radiation pressure forces cancel, in this limit). Thus the ambiguous zone is precisely zone (3a), the fully-ionized (dust-limited) layer or ``BLR-like'' region. In this zone the dominant opacities will come from line opacities; if there is no multiple-scattering ($\mathcal{G}\le 1$), then even in the highly supercritical regime, this will not be efficiently accelerated into a wind (there may be some acceleration but the winds would be relatively weak or ``failed'' or ``fountain-like''), but if there is significant multiple-scattering the entire zone could be efficiently accelerated into outflow for $\dot{m} \gtrsim 14\,(m_{7}/r_{\rm ff,\,5})^{-1/8}\,\epsilon_{r,\,0.1}^{-3/4}\,(\kappa/\kappa_{\rm es})^{-3/4}$, plausibly similar to the near-Eddington thresholds observationally inferred in \citet{leighly:2004.agn.winds,temple:2023.outflow.lines.wind.association.eddington.ratio,leighly:2024.agn.winds.stronger.line.driven.highedd.dust.driven.alledd} for reasonably line opacities $\kappa$.  Clearly, detailed radiation-hydrodynamics calculations with line transfer are needed to address this in more detail \citep[see e.g.][]{higginbottom:2014.line.driven.wind.sims.rad.transfer,higginbottom:2024.sims.rad.transfer.line.driven.winds.weaker.than.thought,nomura:2020.line.driven.winds.agn.modest.massloss.rates,nomura:2021.agn.wind.massloss.modest.strongly.sensitive.to.metallicity,dyda:2023.agn.disc.wind.simulations.w.xray.rad.only.modest.massloss.rates}, informed by the initial/boundary conditions here.

{\em Consequences:} Of course, our models do not self-consistently include the non-linear effects of such outflows. The mass fraction of the disk contained in these layers is modest but non-negligible (tens of percent), so depending on the detailed interplay of the timescales to accelerate and blow out these surface layers and their ``replenishment'' time from material pushed up from the inner disk, it is plausible to expect outflow rates of order the accretion rates, which would modify $\dot{M}$ at interior radii. Moreover they could reduce the height of the outer surface layers if ejection is too rapid/efficient (e.g.\ giving rise to a ``receding torus''). Properly modeling these regimes is an important subject for future work. That said, even if we included a wind in the standard semi-analytic fashion modifying $\dot{M}$, the conclusions for these regions would not be too strongly modified so long as the wind mass loss was not much larger than accretion rates, because along radial trajectories the density profile of the disk is already the standard wind $\rho \propto r^{-2}$ and its thermal properties are determined by external illumination (so the predicted emission properties and boundaries of e.g.\ the BLR or torus do not change much if the illuminated surface regions of these were driven into outflow). And of course details of the thermal state and surface layers do not fundamentally change the global accretion dynamics from \S~\ref{sec:model} unless mass loss depletes the disk strongly \citep[e.g.][]{laor:2014.disk.winds.low.temp}. But those extremes generally involve much higher mass-loss rates than more recent simulations of even supercritical accretion disks give \citep[see e.g.][]{nomura:2020.line.driven.winds.agn.modest.massloss.rates,dyda:2023.agn.disc.wind.simulations.w.xray.rad.only.modest.massloss.rates}.

\begin{figure*}
	\centering\includegraphics[width=0.97\textwidth]{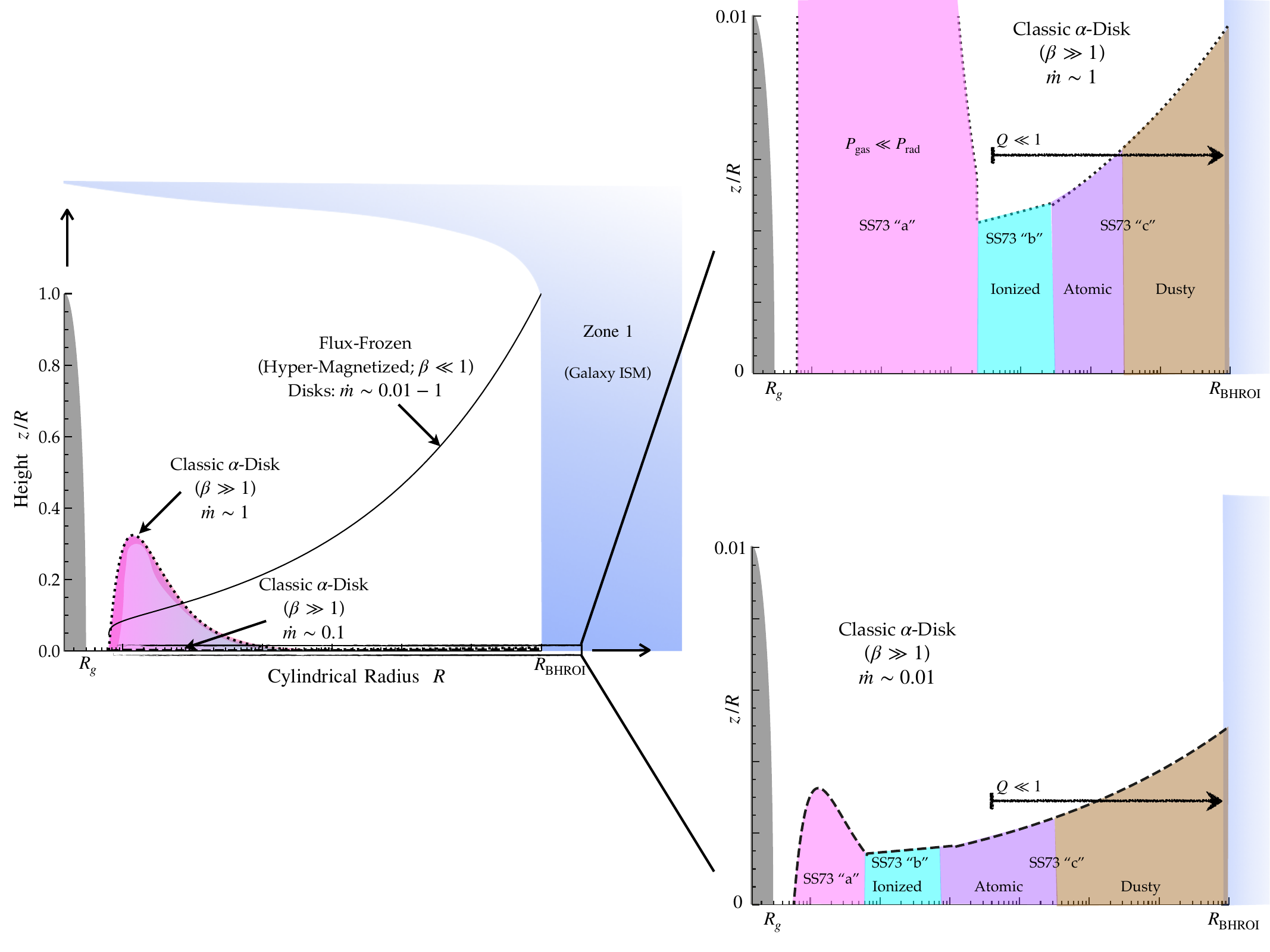} \\
	\caption{Illustration of the flared covering factor $z/R$ of disks as Fig.~\ref{fig:cartoon.illumination}, but now comparing the flux-frozen disks here ($H/R$ shown nearly independent of $\dot{m}$ in {\em left} panel) to a classic SS73 $\alpha$-disk model with $\dot{m}=1$, $0.1$, and $0.01$.  
	The ``zoom in'' panels ({\em right}) shrink the range of $z/R$ for the $\alpha$-disk models, so that the outer disk can be seen. 
	In a thermal-pressure dominated disk, the predicted scale heights are generally tiny, $H/R \lesssim 0.01$, with almost no flaring, so the outer disk covers a vanishingly small fraction of the central source emission. Note the ``bump'' in $H/R$ of the $\alpha$ disks at small radii (especially prominent for $\dot{m} \sim 1$) appears because the SS73 disk becomes strongly radiation-pressure dominated in this region -- with only thermal pressure, these models would predict $H/R \ll 0.01$ at the same radii.
	Together with the differences in density and corresponding opacity (Figs.~\ref{fig:3d.densities}, \ref{fig:opacity.profile}) and turbulence  (Fig.~\ref{fig:turb.props}), this dramatic difference in self-illumination/covering factor/scale-heights $H/R$ explains many of the most important differences between the flux-frozen ($\beta \ll 1$) models here and SS73-like $\alpha$-disk or thermal-pressure-dominated ($\beta \gg 1$) models for disk emission and scattering properties (\S~\ref{sec:discussion}).
	We label the ionized, atomic, and dusty portions of the SS73 outer $\alpha$-disks, and where they would have Toomre $Q \ll 1$ (almost all radii at $\gtrsim 100\,R_{g}$). The SS73 $\alpha$-disk models are unstable to global runaway thermal-viscous instabilities interior to where $P_{\rm gas} \ll P_{\rm rad}$ ({\em pink} region ``a'') and gravitationally unstable to catastrophic fragmentation and star formation exterior to $Q \ll 1$, so are only actually quasi-stable in a quite narrow range of radii. In contrast the flux-frozen models are stable against these effects from the ISCO to the BHROI (\paperthree).
	\label{fig:cartoon.ss73}}
\end{figure*}

\subsection{Highly Subcritical Disks}
\label{sec:subcritical}

At $\dot{m} \lesssim 0.01$, several qualitative changes to the disk begin to occur, illustrated in Fig.~\ref{fig:cartoon.mdot}.

(1) The disk never becomes effectively optically-thick and thermalized: the radius of this transition predicted above is $x_{g}^{\rm therm} \sim 160\,\dot{m}^{15/13}\,r_{\rm ff,\,5}^{7/13}\,\tilde{Z}^{6/13}\,m_{7}^{-10/13}$, so (using $r_{\rm ff,\,5}\sim m_{7}^{1/2}$) this cannot occur outside the ISCO for 
\begin{align}
\dot{m} \lesssim 0.03\,m_{7}^{0.43}\,\tilde{Z}^{-0.4}.
\end{align} 

(2) Even if thermalization does occur, it cannot cool efficiently. The upper, effectively optically-thin layer of the disk (which by definition must contain an $\mathcal{O}(1)$ fraction of the disk mass at radii just interior to the thermalization radius) cools by some combination of Compton+Kramers+line opacities. Sustaining the disk in steady-state, particularly $\beta \ll 1$ (and related aspects like supersonic turbulence) requires $t_{\rm cool} \ll t_{\rm dyn}$. And in particular, it is not just the cooling time $t_{\rm cool}$ at the effective or mean midplane temperature which must be considered, but rather in a disk with supersonic and/or $\beta \ll 1$ Maxwell stresses, where the energy transfer is mediated by shocks and reconnection, we need to consider the cooling rate at the post-shock/reconnection temperature. Assuming a maximal shock velocity $v_{\rm shock} \sim 2\,\psi_{s,\,2}\,v_{\rm turb} \sim (6\,\psi_{s,\,2}^{2}\,v_{\rm acc,\,r}\,v_{\rm K})^{1/2}$, this gives $T_{\rm shock}^{\rm max} \sim 2\times10^{10}\,{\rm K}\,(m_{7}/r_{\rm ff,\,5})^{1/3}\,x_{g}^{-2/3}\,\psi_{s,\,2}^{2}$. At the temperatures of interest the line cooling rates are not relevant, so this is governed by the faster of Compton or free-free cooling rates, which can lead to a hysteresis: there might persist in some regimes a radiatively-efficient, rapidly-cooling solution dominated by very strong Compton cooling (though this would be strongly modified by effect [3] below), but there is also a radiatively-inefficient, slowly-cooling self-consistent solution dominated by free-free cooling, where the ratio of cooling to dynamical time 
\begin{align}
t_{\rm cool}\,\Omega \sim 0.016\,(m_{7}/r_{\rm ff,\,5})^{2/3}\,x_{g}^{1/6}\,\psi_{s,\,2}\,\tilde{Z}^{-1}\,\dot{m}^{-1},
\end{align}
 which will be $\gtrsim 1$ at a wide range of radii for $\dot{m} \lesssim 0.01$, or more specifically at 
 \begin{align}
 \dot{m} \lesssim 0.017\,m_{7}^{2/3}\,r_{\rm ff,\,5}^{-5/7}\,\tilde{Z}^{-8/7}\,\psi_{s,\,2}^{26/21}, 
 \end{align} 
 at the nominal outer radius of the thermalized disk.

(3) Two temperature-effects will become important. The post-shock timescale for electron and proton temperatures to equilibrate to $T_{e} \approx T_{p}$ via Coulomb collisions (at the temperatures of interest; \S~\ref{sec:zone7.corona}) is given by $t_{\rm Coulomb} \sim 3\,m_{e}\,m_{p}\,(k_{B}\,T_{e}/m_{e} + k_{B}\,T_{p}/m_{p})^{3/2}/(8\,\sqrt{2\pi}\,n_{p}\,e_{\rm el}^{4}\,\ln{\Lambda})$ with $\ln{\Lambda} \approx 39 + \ln{(T/10^{10}\,{\rm K})} - 0.5\,\ln{(n_{p}/{\rm cm^{-3}})}$. Inserting relevant values, assuming $T_{e} \approx T_{p}$, and defining $n_{p} \equiv \tilde{n}\,n_{\rm gas}$ relative to the midplane density, we have 
\begin{align}
t_{\rm Coulomb}\,\Omega \sim 0.042\,(m_{7}/r_{\rm ff,\,5})\,\psi_{s,\,2}^{3}\,\tilde{n}^{-1}\,\dot{m}^{-1}\,x_{g}^{-1/2},
\end{align} 
so this will exceed unity at 
\begin{align}
x_{g} \lesssim 18\,(m_{7}/r_{\rm ff,\,5})^{2}\,\psi_{s,\,2}^{6}\,\tilde{n}^{-2}\,(\dot{m}/0.01)^{-2}.
\end{align}

So, around $\dot{m} \lesssim 0.01$, the thermalized, effectively optically-thick disk shrinks to below the ISCO size and the medium becomes more optically-thin. Cooling becomes less and less efficient, with $t_{\rm cool} \gtrsim t_{\rm dyn}$ near the inner radii, which cannot sustain highly super-sonic turbulence or $\beta \ll 1$. Two-temperature effects begin to appear from the innermost radii, making cooling even less efficient. This means the gas will heat up, lowering $\mathcal{M}_{s}$ and increasing $\beta$, further lowering the opacity and cooling rates and making two-temperature effects even more important, until the gas is effectively virialized with $\mathcal{M}_{s} \lesssim 1$, $\beta \gtrsim 1$, $t_{\rm cool}\,\Omega \gg 1$. 

This is just the classic optically-thin, geometrically-thick ADAF/RIAF/ADIOS-type disk solution. So these magnetically-dominated disks will undergo a qualitatively similar state change, at broadly similar $\dot{m}$ and inner radii, as the classic geometrically-thin and optically-thick thermal-pressure-dominated disks. This is illustrated in Fig.~\ref{fig:cartoon.mdot}, showing a partial transition to such a state at $\dot{m} \sim 0.01$, and Fig.~\ref{fig:cartoon.mdot.smallest}, showing a more ``complete'' transition to the ADAF-type regime at $\dot{m} \sim 0.001$. The main differences in the predicted transition (compared if we ``started from'' an SS73-like thermal-pressure-dominated $\alpha$-disk) appear to be that this model would naturally explain the relatively strong, but still $\beta \gtrsim 1$ (i.e.\ non-negligible $\beta\sim 1-10$ or so) magnetic field strengths often invoked in ADAF disk models \citep{yuan:2014.hot.accretion.flows.review} appearing immediately as a consequence of the same magnetic flux from the ISM. Basically, the transition occurs with roughly constant $|{\bf B}|$, and only modest ``puffing up'' of the disk, since it is transitioning from $H/R \sim 0.15 \rightarrow 1$ in the innermost regions, as opposed to $H/R \sim 10^{-3} \rightarrow 1$ as predicted for a thermal-pressure-dominated geometrically thin disk. And unlike SS73-type disks where the ``transition'' $\dot{m}$ is a rather sensitive function of the free parameter $\alpha$ (going as $\alpha^{2}$; \citealt{narayan.yi.95:adaf.lowmass.bhs}), here it is more robustly predicted to be around $\sim 0.01-0.05$. Finally, and perhaps most importantly, we note that the transition here is ``direct'' from the hyper-magnetized, flux-frozen disk to the ADAF state: the system never ``passes through'' some intermediate thermal-pressure-dominated geometrically-thin disk (SS73-like) state. 

Finally as shown in Fig.~\ref{fig:cartoon.mdot.smallest}, at lower $\dot{m}$ the scalings here predict an expansion of the virialized ADAF-like hot flow replacing the thermalized and multi-phase disk (zones (6) and (5)), and eating into the neutral disk (zone (4)), with the corona (zone (7)) expanding and connecting to the disk at the expense of the scattering layers (zone (8)) and BLR-like illuminated zone (zone (3)). But even at $\dot{m} \sim 0.001$, the changes listed above only operate at modest distance from the SMBH -- the flow is still single-temperature and able to cool efficiently at much larger radii approaching the torus and BHROI (let alone galactic ISM) scales. So zones (1) and (2) should persist in some form, as of course observed in low-luminosity AGN \citep{zhang:2009.agn.vs.hubble.type,rowanrobinson:xr.ir.comparison.of.torii,ramosalmeida:pc.scale.torus.emission,hatziminaoglou:2009.torus.properties.inferred.obs}. Of course, at sufficiently low accretion rates from galactic scales, as one might expect from Bondi accretion of only very tenuous, hot gas in extremely low-luminosity AGN (as observed in e.g.\ M87 or Sgr A$^{\ast}$), many orders-of-magnitude lower than the cases we focus on here, there may be no dust or efficient cooling or star formation of the gas at any radii owing to its high temperatures \citep[as assumed in e.g.][]{guo:2022.superzoom.riaf.in.m87.nosf.nomhd.etc,guo:2024.fluxfrozen.disks.lowmdot.ellipticals,cho:2023.multiscale.accretion.sims.bondi.inflow.model,cho:2024.multizone.grmhd.sims.bondi.flow.lowaccrate}. But at that point, the structure of these outer zones is more a function of the galactic boundary conditions, than it is of accretion disk physics.

\section{A Note on Jets \&\ Outflows}
\label{sec:jets.winds}

Briefly, it is worth mentioning various types of outflows (see also  \S~\ref{sec:scattering:warm.absorber}). These are interesting in their own right, as agents of AGN ``feedback,'' as potential scattering sources and ways to elevate gas ``above the disk,'' and in that they can modify the disk scalings (\S~\ref{sec:model}). We neglect outflows in our baseline accretion disk model for simplicity and predictive power, but we are not saying they cannot occur, and future work could generalize the models in \paperthree\ for some outflow scalings motivated by simulations. 

Indeed in  \S~\ref{sec:super.eddington:outflows}, we calculate and discuss conditions where we predict illuminated various zones to have radiation-pressure-driven outflows. 

We also expect some form of jets may be ubiquitous given the strong magnetization of the disks, but these will be launched in the near-horizon regions and be sensitive to properties like spin and how magnetic fields approach the horizon where our model is not really applicable (\S~\ref{sec:near.horizon}), and therefore should be studied in numerical GRMHD simulations like those in \citet{tchekhovskoy:2011.mad.disk.jets,davis.tchekhovskoy:2020.accretion.disk.sims.review,kaaz:2022.grmhd.sims.misaligned.acc.disks.spin,guo:2022.superzoom.riaf.in.m87.nosf.nomhd.etc,cho:2023.multiscale.accretion.sims.bondi.inflow.model,cho:2024.multizone.grmhd.sims.bondi.flow.lowaccrate}. We can briefly speculate that at sufficiently low accretion rates $\dot{m} \lesssim 0.01$, where we predict the usual transition to a low-hard state where the thermalized, blackbody-like (UV/optical-emitting) accretion disk should be replaced with an ADAF-like virialized flow with magnetization $\beta \sim 1$ (a hospitable environment for jets), the suppressed optical and higher radio emission will produce a more radio-loud low-luminosity AGN population. On the other hand, at very high $\dot{m}$, the extended gas envelopes/photospheres/coronal gas above the disk (e.g.\ \S~\ref{sec:illumination.disk.teff}, \ref{sec:zone7.corona}, \ref{sec:super.eddington:rad.eff}) may pose a challenge for escape of jets from the near-horizon regime. But again, more detailed study is needed.

There might also be some magnetocentrifugal (e.g.\ \citealt{blandfordpayne:mhd.jets}-type) outflows, but the expectation here is more ambiguous, owing to the facts that (1) even though the disks are strongly magnetized, the field is primarily toroidal, not vertical (as usually assumed in analytic models); (2) the disks are thick and turbulent, not thin; and closely-related (3) the density above the disk falls off slowly (ram pressure could stall/inhibit such flows; see \papertwo). Theoretically, although the minimal requirements for outflow existence might be met \citep[see][]{seifried:2012.mhd.outflow.disk.criterion}, if one wanted to involve appreciable mass-loss, i.e.\ launch the outflow from the disk ($|z|\lesssim H$, as opposed to more tenuous gas at $|z| \gg H$), then it would be in the ``highly-mass-loaded'' regime \citep{spruit:1996.mhd.disk.winds.review,ouyed:1999.numerical.sims.mhd.winds.vs.massloading,anderson:2005.mass.loading.mhd.winds}. Specifically, to have a meaningful ``outflow'' (even a failed/fountain flow) from the body of the disk, the poloidal flow velocity $v_{w,\,z}$ would need to exceed the turbulent velocity ($v_{\rm turb} \sim v_{A} \sim v_{A,\,\phi} \sim |\langle B_{\phi}\rangle|/\sqrt{4\pi\rho}$), so the dimensionless mass-loading-parameter defined by \citet{spruit:1996.mhd.disk.winds.review} $\mu_{\rm w} \equiv 4\pi\rho v_{\rm w,z} R \Omega / |\langle B_{z} \rangle|^{2}$ (in terms of density, poloidal flow velocity, and mean poloidal field $|\langle B_{z} \rangle|$ at the launch point) must be $\mu_{\rm w}\gtrsim 4\pi \rho v_{\rm turb} R \Omega/|\langle B_{z} \rangle|^{2} \sim v_{A,\,\phi} v_{\rm K}/ \langle v_{A,z}\rangle^{2} \sim (v_{\rm K}/v_{A,\,\phi})\,(v_{A,\,\phi}/v_{A,\,z})^{2} \sim (R/H)\,(|\langle B_{\phi} \rangle|/|\langle B_{z} \rangle|)^{2} \gtrsim 100-1000$. This is well above the limit where \citet{ouyed:1999.numerical.sims.mhd.winds.vs.massloading,anderson:2005.mass.loading.mhd.winds} found that no cold, steady wind solution exists, and idealized winds transition to chaotic/turbulent flows. Notably, the simulations of flux-frozen disks in \paperone\ \&\ \papertwo\ and those around smaller BHs in \citet{shi:2024.seed.to.smbh.case.study.subcluster.merging.pairing.fluxfrozen.disk}, all at high $\dot{m} >1$, do not see much evidence for coherent magnetocentrifugal outflows from the disk, and even the lower $\dot{m} \ll 1$ flux-frozen disk in \citet{guo:2024.fluxfrozen.disks.lowmdot.ellipticals} only exhibits a very weak outflow in the hot coronal/atmospheric diffuse gas at $|z| \gtrsim R$ (see e.g.\ their Fig.~8) at the radii exterior to where their accretion flow transitions to a hot/ADAF-like flow (akin to the scenario in Fig.~\ref{fig:cartoon.mdot.smallest}). So magnetocentrifugal outflows would probably need to either be launched from more tenuous, coronal gas ($|z|\gg H$, where they would not strongly modify the disk scalings themselves), or on near-horizon scales (akin to the jets).

\section{Discussion: The Importance of Flux-Frozen, Hyper-Magnetized Disks}
\label{sec:discussion}

\subsection{Key Differences from ``Classic'' Disk Models}
\label{sec:discussion:differences}

We have shown that the thermal properties and structure of these hyper-magnetized, flux-frozen disks differ {\em dramatically} from classic $\alpha$-disk models which assume thermal pressure dominates over magnetic pressure. But recall, the model here is defined by just two simple ansatz (\S~\ref{sec:model:assumptions}): (1) that the disks have midplane $\beta \ll 1$ (magnetic pressure much larger than thermal) and (2) trans-\Alf{ic} turbulence (turbulent velocities comparable to \Alf\ speeds, i.e.\ broadly similar Maxwell and Reynolds stresses). 
This leads to three critical differences, compared to thermal-pressure-dominated (SS73-like) $\alpha$-disks: 

\begin{enumerate}

\item{Flux-frozen disks (disks with $\beta \ll 1$ wherein the mean toroidal field is amplified primarily by flux-freezing and advection of magnetic flux from larger radii, rather than some strictly-local dynamo amplifying fields {\em in situ} from trace values) are geometrically much thicker, and strongly-flared (compare Fig.~\ref{fig:cartoon.ss73}). Magnetic fields with $\beta \ll 1$ support the gas vertically with height $H\sim R$ in the outer disk. This means the substantial covering factors of the BLR, torus, X-ray reflection components, etc.\ emerge naturally and need nothing special to ``hold them up.'' This does not preclude that there are strong winds in these regions or even that most of the BLR emission comes from a wind: instead, it ensures such a wind has the correct covering factor if launched by e.g.\ radiation pressure from the central disk acting on the surface of the outer disk (while if $H/R$ were small, as predicted for any thermal-pressure-dominated disk at these radii, such a wind could not have the correct BLR/torus geometry).}

\item{Flux-frozen disks have much lower surface densities (and 3D densities), owing to magnetic pressure support and much stronger (relative) Maxwell stresses sustaining accretion, given their stronger fields. Thus their opacity structure is completely different. This allows for the existence of regions like the dusty torus, multi-phase BLR, warm comptonizing skin, and extended scattering and/or reflection layers {\em as a part of} the disk. If we only assumed thermal pressure support, the vastly higher densities and surface densities means both the effective and total optical depths are much larger and such phases could not exist self-consistently ``within'' the disk. This also means that even though the absolute value of $|{\bf B}|$ is actually {\em smaller} at all radii in hyper-magnetized disks than in the midplane of an SS73-like disk (see \papertwo-\paperthree), the \Alf\ speed and relative importance of reconnection heating, etc.\ is much larger.}

\item{Flux-frozen disks have much stronger turbulence (as measured by the sonic Mach number $\mathcal{M}_{s}$ or absolute turbulent velocities). As assumed in most thermal-pressure dominated disks, Maxwell and Reynolds stresses are generally order-of-magnitude comparable. Given this, it follows that the turbulence here is highly super-sonic and compressible, with driving scales of order the (large) disk scale height, and cooling/dissipation times much shorter than dynamical times, whereas in a thermal-pressure-dominated disk the opposite is true (the turbulence is sub-sonic, approximately incompressible, with driving/coherence scales smaller than the already-small $H$, and cooling times much longer than dynamical times). This promotes inhomogeneous density structure, co-existence of multiple phases of gas, and creation of coronal gas via strong shocks and reconnection; strongly modifies the vertical thermal structure of the disk; and can prevents the disks from becoming strongly radiation-pressure-dominated in the supercritical limit.}

\end{enumerate}

These differences emphasize that the predictions for flux-frozen disks are highly non-trivial: i.e.\ it is not enough to simply ``put gas at the right distance with the right accretion rate,'' in which case models like SS73 or any other disk models in the literature would make similar predictions for the different structures in the disk. To illustrate just how important these differences are, let us compare for example the predictions for the model here versus SS73 for gas at the distances of the BLR ($\sim 1-100\,$ld). In the SS73 model, the prediction is that gas at these radii ``in the disk'' (1) has an effective covering factor (reprocessed light fraction) of $\sim (1/20)\,(H/R)^{2}_{\rm SS73} \sim 10^{-7}$ (Fig.~\ref{fig:cartoon.ss73}); (2) has densities about 5-8 orders of magnitude larger ($\gtrsim 10^{17}\,{\rm cm^{-3}}$; see \paperthree\ and Figs.~\ref{fig:surface.densities}-\ref{fig:3d.densities}); (3) is violently thermal-viscously unstable at small radii and gravitationally unstable ($Q \ll 1$) at large radii (see Figs.~\ref{fig:masses} \&\ \ref{fig:cartoon.ss73}); (4) is extremely optically-thick to both its self-illumination and its own cooling radiation, with absorption optical depth ($\tau^{\ast} \gtrsim 10^{4}$) much greater than unity (and greater than the scattering depth) so cannot emit lines observed (Fig.~\ref{fig:opacity.profile}); and (5) is thermalized with a warm temperature $\gtrsim 10^{5}$\,K that over-ionizes most of the relevant lines. An SS73-like disk therefore cannot possibly represent the observed BLR. One can easily show the same for the dusty torus, the extended scattering structures, warm Comptonizing skin, coronae, and X-ray reflection components -- these are all qualitatively distinct from the structures predicted by the SS73 disk to exist at the same radii. 

Of course, we cannot immediately rule out the possibility that such structures are qualitatively separate from the disk and ``sit above it'' with completely different physics controlling their existence and giving rise to properties akin to those predicted here. Those are the models which have generally been discussed in the literature (see references in \S~\ref{sec:intro}). But not only does that require the (often ad-hoc) introduction of new physical components and fitting/tuning parameters to reproduce the same observations, it clearly cannot be considered a true prediction. Moreover, in future work (in preparation), we show that some observations can indeed already rule out the idea that such components (like the BLR or torus) simply sit ``on top'' of a thermal-pressure dominated $\alpha$-disk, {\em purely} from the constraints on the un-avoidable gravitational influence of such a disk on the orbital structure of the emitting gas observed. And as we discuss below, it is straightforward to rule out at least some forms of pressure besides magnetic providing the dominant support for the gas in some of these regions.

\subsection{Can Some Other Pressure Replace Magnetic Fields?}
\label{sec:discussion:other}

Given the discussion in \S~\ref{sec:discussion:differences}, it is natural to ask whether one might imagine some {\em other} form of pressure, besides magnetic fields, supporting the disk at these radii and therefore giving rise to the same key structural properties distinct from classic $\alpha$ disks. To our knowledge, no self-consistent disk models exist which predict similar total pressures as the hyper-magnetized disks but arising from thermal or radiation or some other form of pressure entirely (these automatically predict very different pressure profiles, as we showed above). But in this section, we will ignore these sorts of consistency arguments, and simply assume some arbitrary pressure profile and effective stress profile in order to reproduce qualitatively the same disk structure (scale heights, densities, etc) predicted by the flux-frozen disks, and ask whether the salient pressure/stress could (in principle) be provided alternative (non-magnetic) pressure sources.

\subsubsection{Stronger Thermal Pressure}
\label{sec:discussion:other:thermal}

First, consider thermal pressure. Again, we showed above the classic thermal-pressure dominated $\alpha$ disks cannot possibly provide such pressure, but let us for now simply assume some arbitrary thermal pressure and arbitrary stress (i.e.\ arbitrary $\alpha(R)$, which can be $\gg 1$, and temperature independent of heating/cooling rate calculations) in order to support the desired $H/R$, $\Sigma_{\rm gas}$, etc, at a given $\dot{m}$. Recalling that for a thermal pressure-dominated disk, $H \sim c_{s}/\Omega$ depends (for a given BH mass and distance) just on the temperature, then if we wish to obtain the same $H/R$ (Eq.~\ref{eqn:veff.model}) as in the magnetically-dominated disks from thermal support alone we immediately obtain the required temperature 
\begin{align} 
\nonumber T^{\rm therm\,only}_{\beta\gg 1} &\sim 3\times10^{11}\,{\rm K} \,x_{g}^{-2/3}\,(m_{7}/r_{\rm ff,\,5})^{1/3} \\ 
& \sim 3\times10^{8}\,{\rm K}\,(R/{\rm ld})^{-2/3}\,m_{7}^{5/6}.
\end{align}
This immediately poses several problems. 
(1) Where does the thermal energy come from? The disk temperatures in the outer regions would have to be near-virial to thermally support the large $H/R \sim 0.1-1$ required for e.g.\ the torus and BLR and scattering structures. This is vastly larger than can be maintained by direct illumination from the central disk (for any radiative efficiency $<1$) or the heating from gravitational energy release (effective viscosity), as we showed above (assuming just those, one arrives at SS73-like models, with $H/R \ll 0.01$ at these radii). Invoking e.g.\ stellar sources to provide the heat would require an enormous density of young/massive stars (e.g.\ to maintain the BLR as ionized, the stellar-to-gas mass density ratio would need to be $\gg 1000$, implying SFRs $\gtrsim 10^{5}\,{\rm M_{\odot}\,yr^{-1}}$ in steady state just within $<100\,$light-days), and moreover this is not dynamically stable at any radii interior to the BHROI (see e.g.\ \citealt{torrey.2016:fire.galactic.nuclei.star.formation.instability}, who show the entire disk would be destroyed as soon as the first stars explode). 
(2) Any denser, cooler gas ``clumps'' in the torus and BLR would be buoyantly unstable and sink to a much smaller scale-height midplane layer on of order a dynamical time, and the ``diffuse'' gas would have to thermally pressure-confine said clumps making it too hot (super-virial). 
(3) The turbulence would be by definition subsonic so could not generate the desired density fluctuations for said structure, nor prevent the inner disk from becoming radiation-pressure dominated. 
(4) The phases of the BLR and dusty torus could not exist -- the gas would be far too hot (e.g.\ $\sim 10^{8}$\,K at $10$\,ld or $\sim10^{7}$\,K at the sublimation radius or $\sim 0.1$\,pc) to allow the existence of any of the observed phases/lines/emission mechanisms (like partially-ionized atomic gas, molecular maser emission, etc.). 
(5) The inner disk, optically-thick disk would be extremely hot ($T_{\rm disk} \gg 10^{10}\,{\rm K}$), so the ``big blue bump'' would vanish and the emission would be entirely X-ray dominated. 
(6) The observed luminosity from large radii would be vastly too-large. At these temperatures, the cooling luminosity from the disk would be at least $\gtrsim 10^{4}$ times larger than the Eddington luminosity, independent of $\dot{m}$ (i.e.\ they would be this high for every observed BH).

Thus we can strongly rule out a thermal pressure-dominated disk with anything like the pressure profile and $H/R$ of the magnetically-dominated disks. 

\subsubsection{Radiation Pressure}
\label{sec:discussion:other:rad}

Radiation pressure-dominated disks are well-known to be geometrically thicker, and indeed, in the innermost disk regions ($\lesssim 30-300\,R_{g}$) have been shown to reproduce at least some of the phenomenology discussed above at accretion rates $\dot{m} \gg 1$, being geometrically thick with $P_{\rm rad} \sim P_{\rm mag} \gg P_{\rm thermal}$ (so they are similar to the models here by definition), strongly turbulent, and potentially self-generating a Comptonizing skin and larger electron scattering layers \citep[see e.g.][]{jiang:2019.superedd.sims.smbh.prad.pmag.modest.outflows}. And as discussed in \S~\ref{sec:super.eddington}, it is plausible that at supercritical accretion rates $\dot{m} \gg 1$, the magnetically-dominated disks here could develop radiation pressure comparable to their magnetic pressure, a case which should be explored in more detail in future work and simulations.

However, what we are asking here is whether one can support the entire disk (out to and including the BLR and torus) with radiation pressure with a comparable $H/R \sim 0.1-1$ to the magnetically-dominated disk prediction. This requires a vertical radiation flux $F_{\rm rad}(R)$ at each radius coming from the midplane to support the disk versus gravity, $\kappa F_{\rm rad}/c \approx g \approx \Omega^{2} H \sim \Omega^{2} R$. Again in the innermost radii at $\dot{m} \gtrsim 1$, that is plausible (albeit with $P_{\rm rad} \sim P_{\rm mag}$, not $P_{\rm rad} \gg P_{\rm mag}$, per \S~\ref{sec:super.eddington}), but in the outer disk or at lower $\dot{m}$, this immediately poses several problems. 
(1) Where does the radiation come from? The required flux $F_{\rm rad}$ in the outer disk at $R \sim 100\,R_{g} - 10\,$pc is many orders-of-magnitude larger than the local accretion luminosity $\sim \dot{M}\,\Omega^{2}$ or cooling luminosity. Attempting to provide it via starlight produces the same problem as noted (\S~\ref{sec:discussion:other:thermal}) for heating via stars (indeed the implied SFR is even higher), except far outside the BHROI \citep{thompson:rad.pressure}, and is again not dynamically stable.
(2) Radiation pressure-dominated accretion disks with $P_{\rm rad} \gg P_{\rm mag},\ P_{\rm thermal}$ are thermally and viscously unstable in a manner not observed \citep{abramowicz:accretion.theory.review}. 
(3) The temperature structure of the optically-thick regions would be incorrect. 
(4) Most important, the required emergent luminosity from each radius would be enormous: 
\begin{align}
\nonumber L(R)^{\rm rad\ only}_{\beta \gg 1} &\sim 2\pi R^{2} F_{\rm rad} \sim 10^{45}\,{\rm erg\,s^{-1}}\,m_{7}\,(r/r_{\rm ff})^{1/6}\,(\kappa/\kappa_{\rm es})^{-1}  \\ 
&\sim L_{\rm Edd}\,(r/r_{\rm ff})^{1/6}\,(\kappa/\kappa_{\rm es})^{-1} \ . 
\end{align}
In other words, to support the disk with $H/R \sim 0.1-1$ via radiation pressure, {\em every} radial annulus would have to be emitting at roughly the Eddington limit, {\em independent} of the actual accretion rate $\dot{M}$ (i.e.\ no sub-Eddington luminosities are possible), and the total luminosity would be dominated by the emission from the furthest radii from the BH.

Thus we can easily rule out a radiation pressure-dominated disk for these components (BLR, torus, etc.) at large radii.

\subsubsection{Cosmic Ray Pressure}
\label{sec:discussion:other:cr}

In the extremely diffuse gas of the circum and inter-galactic medium (CGM/IGM) and perhaps most diffuse phases of the ISM (densities $n \ll 1\,{\rm cm^{-3}}$), cosmic ray (CR) pressure may dominate \citep{hopkins:cr.mhd.fire2,hopkins:2020.cr.outflows.to.mpc.scales,ji:fire.cr.cgm,butsky:2022.cr.kappa.lower.limits.cgm}, but again attempting to replace magnetic with CR pressure in the disks here immediately produces several problems. 
(1) Where do the CRs come from? If CRs are efficiently diffusive/streaming with anything like observed diffusion coefficients, they cannot be ``carried into'' the nucleus with ISM magnetic fields \citep{derome:usine.cr.transport.fits,hopkins:cr.multibin.mw.comparison,delaTorre:2021.dragon2.methods.new.model.comparison,dimauro:2023.cr.diff.constraints.updated.galprop.very.similar.our.models.but.lots.of.interp.re.selfconfinement.that.doesnt.mathematically.work}, and the confinement is actually predicted to be much weaker in neutral gas \citep{farber:decoupled.crs.in.neutral.gas}. But even if they were strongly confined and carried ``adiabatically'' (what is needed to reach anything like the requisite pressures), the CR loss rate at these densities would be much faster than their advection rate \citep{bustard:2020.crs.multiphase.ism.accel.confinement,hopkins:cr.transport.constraints.from.galaxies,hopkins:2021.sc.et.models.incompatible.obs,krumholz:2023.cosmic.ray.ionization.gamma.ray.loss.budgets}. Local shocks cannot source them in the outer disk owing to much too-low shock velocities, and the required energy supply to offset losses would be much larger than the gravitational/turbulent dissipation rates. Invoking stars (SNe) as in the ISM would again imply enormous SFRs ($\gtrsim 10^{6}\,{\rm M_{\odot}\,yr^{-1}}$ SFR needed from {\em interior} to $\sim 1$\,ld) and be unstable as above. Invoking CRs from the central source (e.g.\ inner disk or base of the jet) would give the incorrect pressure profile and require unphysical acceleration efficiencies ($>100\%$ of $\dot{M}\,c^{2}$). 
(2) Related to the above, the CR loss rates would be huge in such environments: even on much larger, lower-density scales of galactic nuclei (the central $\sim$\,kpc of galaxy starbursts, with gas densities $\sim 10-1000\,{\rm cm^{-3}}$), observations indicate that almost all CR energy is lost and they do little work on the ISM \citep{lacki:2011.cosmic.ray.sub.calorimetric,chan:2018.cosmicray.fire.gammaray,zhang:2019.new.cosmic.ray.compilation.vs.calorimetry.sub.calor}. 
(3) A CR-pressure supported geometrically-thin or slim disk is dynamically unstable for most CR transport models \citep{chan:2021.cosmic.ray.vertical.balance,kempski:2021.reconciling.sc.et.models.obs}, even neglecting losses: if the effective diffusion/streaming is too fast, CRs simply escape without doing any work on the medium, while if it is too slow, the gas behaves adiabatically and virializes. 
(4) The CR ionization rate would be enormous ($\sim 10^{17}$ times larger than in the Solar neighborhood for e.g.\ a $\sim 10^{8}\,{\rm M_{\odot}}$ BH near-Eddington at $\sim$\,a few light-days), strongly over-ionizing all species in at BLR radii and suppressing any lines. 
(5) Simply assuming such a large CR pressure existed, the implied $\gamma$-ray luminosities (given by CR collisions with the disk H, He producing pion to $\gamma$-ray emission) would also be enormous. Assuming a broadly Voyager-like CR spectral slope, the $\gamma$-ray flux from a given radial annulus is $\sim 3\times10^{8}\,{\rm erg\,s^{-1}\,cm^{-2}}\,P_{\rm cr}\,\Sigma_{\rm gas}$ \citep{guo.oh:cosmic.rays,chan:2018.cosmicray.fire.gammaray,hopkins:cr.transport.constraints.from.galaxies}; integrating down to the ISCO with $P_{\rm cr}$ equal by the midplane pressure in our default model would produce a total $\gamma$-ray luminosity $\sim 200\,\dot{m}/\epsilon_{r,\,0.1}$ times larger than the ``normal'' accretion luminosity ($\sim 0.1\,\epsilon_{r,\,0.1}\,\dot{M} c^{2}$), i.e.\ 
\begin{align}
L_{\gamma}( \gtrsim {\rm GeV})^{\rm CR\ only}_{\beta \gg 1} \sim 200\,\dot{m}\,\epsilon_{r,\,0.1}^{-1}\, L_{\rm NIR+Optical+UV}
\end{align} 
Technically this is a lower limit as in the inner disk CR-photon interactions could also produce $\gamma$-rays. This is obviously vastly larger than observed in any AGN.

Thus we can strongly rule out a CR pressure-dominated disk with these pressures.

\subsubsection{Turbulent Pressure ``Alone''}
\label{sec:discussion:other:turb}

If we imagine a disk with {\em only} turbulent pressure support, i.e.\ $P_{\rm turb} \sim \rho\,v_{\rm turb}^{2}$ much larger than magnetic, thermal, radiation, and cosmic ray pressures, and arbitrarily set the turbulent velocities to be both isotropic and have the same magnitude as in our hyper-magnetized disk model, then in fact most of the predicted disk properties are the same. This of course is because we assumed trans-\Alf{ic} turbulence, so even in our default hyper-magnetized model $P_{\rm turb} \sim P_{\rm mag} \sim P_{\rm tot}$, the only difference would be much weaker magnetic fields at otherwise similar pressure (i.e.\ the turbulence goes from trans-\Alf{ic} to highly super-\Alf{ic}).\footnote{Note that as discussed above, the magnetic field only weakly enters the heating and cooling rates for the thermodynamics, at fixed stress/accretion rate/scale-height/density. Magnetic reconnection may be important for heating the coronal/Comptonizing gas but as noted above in the hyper-magnetized disks this is by definition comparable to the turbulent/shock heating rates, and cyclotron cooling is generally a weak effect.}

So this model does not immediately present a contradiction nor predict obviously unphysical or wildly different-from-observed behaviors. The difficulty with this model comes from attempting to imagine how such a situation could self-consistently arise. 

(1) Even if such turbulence existed, we would naively expect the turbulent dynamo to amplify ${\bf B}$ until it was closer to trans-\Alf{ic}, more similar to our hyper-magnetized model. 

(2) Given this model necessarily predicts the turbulence is highly super-sonic and super-\Alf{ic} with $t_{\rm cool} \ll \Omega^{-1}$, the turbulence must be driven, and with (by definition) $P_{\rm turb} \gg P_{\rm therm}+P_{\rm mag}+P_{\rm rad}+P_{\rm CR}$, magnetic/thermal/radiation/CR energies cannot be the ultimate driving energy source. Unlike in the ISM where (in some phases) super-sonic and super-\Alf{ic} turbulence can be driven by stellar feedback, this (like the scenarios above) would require an unphysically large population of massive stars (equivalent to SFRs $\gg 10^{5}\,{\rm M_{\odot}\,yr^{-1}}$ coming from {\em inside} the BHROI). 
It would also (given the relative scaling of radiation pressure to momentum flux from young stellar populations) necessarily also push the system to $P_{\rm rad} \sim P_{\rm turb}$ under optically-thick conditions \citep{thompson:rad.pressure,hopkins:rad.pressure.sf.fb}. So the only theoretical source for the turbulence driving in such a model is gravitational, akin to gravito-turbulent models. But assuming this, in turn, introduces other problems. 

(3) It is well established that for the large value of $1/(\Omega\,t_{\rm cool}) \gg 1$, without a strong mean magnetic field, such gravito-turbulence not only cannot prevent but will in fact net promote runaway fragmentation and star formation \citep{gammie:2001.cooling.in.keplerian.disks,rice:2005.disk.frag.firstlook,paardekooper:2012.stochastic.disk.frag,federrath:2014.low.sfe,riols:2016.mhd.ppd.gravitoturb,forgan:2017.mhd.gravitoturb.sims,hopkins:frag.theory,hopkins:2013.turb.planet.direct.collapse,hopkins:superzoom.disk}, giving a SFR at each radius of $\sim 2\pi\,R^{2}\,\Sigma_{\rm gas}\,\Omega$ much larger than the accretion rate (by factors $\sim (v_{\rm K}/v_{\rm t})^{2} \gtrsim 100$). 

(4) It is also the case that these gravitational modes, in simulations absent strong magnetic fields or strong stellar feedback, produce strong but highly-anisotropic turbulence almost exclusively in the midplane $R\phi$ directions, so the disk collapses to be razor-thin (geometrically) even if highly turbulent \citep{bournaud:2010.grav.turbulence.lmc,hopkins:fb.ism.prop,hopkins:superzoom.disk,ceverino:2012.clump.rotation,orr:2020.resolved.dispersions.sfrs.correlations,bending:2022.supernova.feedback.spiral.arms.clouds}. In other words, gravito-turbulence alone with $\Omega\,t_{\rm cool} \ll 1$ can produce large $v_{\rm t}$ in-plane, but {\em cannot} produce the actual desired result of $H/R \sim 0.1-1$.

(5) The generic, extremely robust prediction for gravity-driven turbulence in the absence of a stronger magnetic/thermal/other pressure is to self-regulate with a turbulent $Q^{\rm turb} \sim 1$ \citep[see references above and][]{kim.ostriker:2001.gravitoturb.galactic.disks.mhd.conditions,deng:gravito.turb.frag.convergence.gizmo.methods,riols:2018.mhd.sims.ppd.gravitoturb.fx.on.mri,zier:2022.gravitoturb.sims.validating.stochastic.frag}, but the models here require $Q = Q^{\rm mag}  \sim 10^{10}/(\dot{m}\,x_{g})$ at near-horizon scales to $\sim 3000/(\dot{m}\,r/r_{\rm ff})$ at outer scales. Equivalently, even if we assumed the turbulence were isotropic, the required $H/R$ implies $Q \gg 1$ by a huge factor, which should shut down the gravito-turbulent driving mechanism completely. Thus there is no plausible way for this driving mechanism to actually produce the desired scalings.

We note that issues (2)-(5) are directly demonstrated in the simulations of \paperone-\papertwo, where a test simulation without magnetic fields was presented, which catastrophically fragments, produces SFRs vastly larger than accretion rates, gives rise to a geometrically razor-thin disk which actually spends most of the time as a decretion disk, with $Q\sim1$ and vastly larger densities and negligible covering factors. 

Note that if we were to instead assume self-regulation to $Q = Q^{\rm turb} \sim 1$ and isotropic turbulence, as in most ``marginally self-gravitating'' or ``self-regulating'' disk models \citep[e.g.][]{paczynski:1978.selfgrav.disk,sirko:qso.seds.from.selfgrav.disks,thompson:rad.pressure,hopkins:rad.pressure.sf.fb,forbes:2011.thick.disk.torque.evol,ostriker.shetty:2011.turb.disk.selfreg.ks}, then in addition to problems (1)-(4) above (which this model still does not solve), this immediately does lead to conflict with observations, as the midplane density would necessarily be $\sim Q^{\rm mag}/Q^{\rm turb}$ times larger, i.e.\ this gives a predicted midplane density 
\begin{align}
\frac{\langle n \rangle^{Q=1,\,{\rm turb}}_{\beta\gg1}}{\rm cm^{-3}} &\sim 10^{6}\,m_{7}^{-1/2}\,(R/r_{\rm ff})^{-3} \\
\nonumber & \sim 10^{18}\,(M_{\rm BH}/10^{8}\,{\rm M_{\odot}})\,(R /{\rm ld})^{-3} \sim 10^{26}\,m_{7}^{-2}\,x_{g}^{-3}.
\end{align}
 So the mean predicted $\alpha$-disk densities at BLR radii are vastly larger than those of observed BLR emitting gas, the disk is optically-thick and line emission/reflection cannot occur in either the BLR or dusty torus regions (like with SS73), and the ratio of absorption to scattering optical depths in the inner regions changes completely preventing both the extended scattering regions and Comptonizing skin from existing. Again even assuming the (implausible) case where the turbulence would be isotropic under such driving, at all radii the predicted disk scale height would collapse to 
\begin{align}
\nonumber (H/R)^{Q=1,\,{\rm turb}}_{\beta \gg 1} &\rightarrow \left( \frac{G\dot{M} Q}{3 c^{3}}\right)^{1/3}\,x_{g}^{1/2} \sim 10^{-5}\,(m_{7}\,\dot{m})^{1/3}\,x_{g}^{1/2}\\ 
& \sim 6\times10^{-4}\,\dot{m}^{1/3}\,m_{7}^{-1/6}\,(R/{\rm ld})^{1/2}, 
\end{align}
which means the effective covering/reflection factor of the BLR drops to $f_{\rm cover,\,BLR}^{\rm turb\ only}\sim 10^{-7}\,\dot{m}^{2/3}$, akin to the SS73 model, and even the covering of the outermost-possible, cold dusty torus regions at $R \sim R_{\rm BHROI}$ drops to $f_{\rm cover,\,torus}^{\rm turb\ only}\sim 10^{-3}\,\dot{m}^{2/3}$. So this model is not able to explain how these regions are ``held up'' in any physical sense.

Thus, attempting to construct a ``pure'' turbulence-dominated model (with weak magnetic fields) with these pressures, while not {\em strictly} ruled out by the basic spectral properties of AGN, can only be viable if one somehow constructed a gravito-turbulent model which differed by orders of magnitude and in important qualitative behaviors (e.g.\ self-regulating with out producing star formation, with something driving vertical motions, and $Q$ larger by a radial-dependent factor increasing from $\sim 3000$ to $\sim 10^{10}$) from almost all predictions of numerical gravito-turbulence simulations to date.

\subsection{Relation to Previous Work}
\label{sec:discussion:previous}

As we have repeatedly discussed, we are far from the first to note that the various components discussed above (torus, BLR, corona) cannot be ``part of'' a geometrically-thin thermal-pressure-dominated classical $\alpha$-disk. Indeed as mentioned in \S~\ref{sec:intro}, it has been known for decades that if one invokes these components phenomenologically, residing ``above'' the midplane with some thick-disk-like (large covering factor/opening angle) geometry, one can reproduce a wide range of observational phenomena. The same is true for extended scattering and reprocessing surfaces in the AGN ecosystem. The ideas of geometrically-thick, low-density structures which reprocess the inner thermalized disk emission in the form of Comptonizing skin/layers, warm absorbers, and/or scattering layers to explain microlensing and reverberation mapping, and hard X-ray reflecting structures have been discussed extensively in the literature with phenomenologically-inferred properties similar to those predicted here, as we note throughout. 

We are also far from the first to propose that some of these components and structures could be ``held up'' by magnetic fields. For example, the idea goes back decades that the BLR in particular could be supported  by magnetic pressure  either in  some quasi-static magnetized atmosphere/coronal gas \citep{rees:1987.mag.confined.blr,begelman:2017.mag.elevated.disks.potential.blr.levitation} or in a dynamic magnetocentrifugal outflow \citep{emmering:1992.blr.in.mag.winds,koniglkartje:disk.winds,elitzur:torus.wind}, and that this would solve various problems posed in e.g.\ \citet{krolik:1981.twophase.model.quasar.emission.lines}. The key difference is that those models all required introducing some {\em ad-hoc} additional physical structures and parameters ``on top of'' the accretion disk, usually assumed to be SS73-like, with some parameters fitted/assumed to explain the observational phenomena. The novel contribution here is that we show all of these properties are {\em predicted}, with no new introduced components or parameters, from incredibly simple disk similarity model. While more rigorous comparisons to AGN spectra depend on more detailed radiation-transport calculations (in preparation), the simple fact that these phenomenological structures and components all appear to fall out naturally from the two extremely simple {\em ansatz} in \S~\ref{sec:model} is remarkable.

\section{Summary \&\ Conclusions}
\label{sec:conclusions}

Recent simulations and analytic models have argued for a novel type of accretion disk: hyper-magnetized, flux-frozen disks, where magnetic flux from the ISM is sufficient to support the disk and provide strong stresses (\paperone-\paperthree). In this paper, we adopt a simple analytic model for the structure of such disks from near-horizon scales to ISM scales (outside the BHROI), and calculate in more detail the thermo-chemical properties such disks should have as a function of position in the disk, BH mass, and accretion rate. We show that these are orders-of-magnitude distinct from the predictions of thermal-pressure dominated disks (e.g.\ classic $\alpha$-disk models like SS73), and that they appear to naturally explain a wide variety of the classic components of the AGN ``ecosystem.'' 

Specifically, the large scale-heights, flaring, and much lower densities of these disks make it is inevitable that almost all of the inner disk emission observed is ``reprocessed'' at some level: thermal emission from the innermost disk is trapped/reprocessed by scattering layers in the more extended thermal disk (Zone 6), lowering the effective temperatures and making them nearly independent of BH mass; it passes through a warm Comptonizing layer producing a soft excess (Zone 6b), and a hard corona (Zone 7) producing hard X-rays. An order-unity fraction of this radiation scatters off the extended scattering surface of the high-covering flared ionized disk (Zone 8) making the optical/UV emission ``effective'' size nearly wavelength-independent and extending to $\gtrsim 10^{16}\,{\rm cm}$, and naturally producing an X-ray reflection component. Some $\sim 10-20\%$ of the light is reprocessed by a multi-phase, geometrically-thick disk (potentially outflow at high $\dot{m} \gtrsim 1$), partially ionized, multi-phase clumpy medium having all the properties of the BLR (Zone 3), while an order-unity fraction intercepts the outermost portions of the flared disk outside the dust sublimation radius (Zone 2) where the radiation will be absorbed by dust and re-emitted in the infrared. Finally, this light must pass through the actual host galaxy ISM (Zone 1), where (depending on the galaxy properties, not modeled here) a fraction ranging from nil to unity of the light could be re-processed by galactic ISM dust and gas (e.g.\ cold dust in a more isotropic configuration in e.g.\ starburst galaxies, or the NLR). 

What is remarkable is that properties like those of the BLR, including its covering/reprocessed light fraction, characteristic line-emitting structure sizes $\sim 10^{11}-10^{13}\,{\rm cm}$, location at $\sim 1-100\,$ld, densities $\sim 10^{8}-10^{12}\,{\rm cm^{-3}}$, geometrically thick-disk dynamics/kinematics, temperatures, and ionization parameters $\sim 0.001-1$, all appear naturally as predictions of this model for the gas in the disk. The same appears to be true for other components like the dusty torus or Comptonizing and scattering and reflection structures. We stress that we have not introduced various free or tunable parameters or ``fudge factors'' or ``additional'' physical components to our model. These are simply the emergent thermochemical properties of the disk model which follows our basic {\em ansatz}: that magnetic pressure dominates with fields supplied by flux-freezing and trans-\Alf{ic} turbulence. We also stress that as we show above, it is fundamentally {\em not possible} for a thermal-pressure dominated accretion disk -- regardless of the details of what supplies the thermal pressure or determines the effective stress -- to contain or reproduce such structures {\em within the disk}, and the same is true for radiation pressure or cosmic ray pressure-dominated disks. That does not mean a BLR could not, in principle, exist {\em alongside} a thermal-pressure-dominated accretion disk, but it would have to be distinct physically. Indeed that is how these components have traditionally been modeled, as some additional, distinct physical system which happens to ``sit above'' the disk at the appropriate radii. But even in those models -- where many additional ``components'' are separately attached in a phenomenological manner to the AGN ecosystem -- it is not obvious how their characteristic properties would actually arise. For example, if one assumes the BLR is some wind launched by a classic SS73-like disk, then it is quite difficult, in practice, to explain the necessary covering/reprocessing factors, and there is no unique prediction for the characteristic gas densities \citep[e.g.][]{luo:2015.thick.disk.agn.shielding.wind.unification,naddaf:2022.frado.failed.wind.model,zhu:2022.idealized.wind.driving.sims.start.thick.configuration}. 

The key differences from classic thermal-pressure-dominated models like traditional $\alpha$-disks which enable these components to not just exist but be naturally predicted in hyper-magnetized flux-frozen disks here are identified, and at their most essential come down to (1) the disk is vastly geometrically thicker and strongly flared; (2) the strong stresses make the disk much lower mass and lower density and so radically alter the opacities; and (3) these properties mean that the turbulence is highly super-sonic and cooling times short compared to dynamical times, generating strong clumping and multi-phase structure.

Because the disk is {\em not} extremely geometrically-thin like SS73, and the reprocessed fractions in each zone is significant, photons can in principle have many different interactions between these zones, non-linearly modifying the details of the disk thermal structure in the process. And of course, to make analytic predictions here, we have adopted a simple analytic similarity model for the disk structure, and convenient analytic approximations (highly simplified fitting functions) for the opacities and heating/cooling rates of the gas (often taking various limits to obtain analytic scalings, leading to somewhat artificial ``breaks'' in the predicted disk profiles). Therefore the scalings we present for temperatures, opacities, ionization states, etc., and especially the ``boundaries'' between zones, should be considered order-of-magnitude estimates. It is clear that more detailed predictions, especially for e.g.\ line properties or the precise fraction of light attributable to a given ``zone,'' etc., necessarily require explicit, detailed radiation-transfer calculations (coupled to the thermo-chemistry to correctly calculate opacities and ionization states) encompassing the {\em global} disk structure and extremely multi-wavelength, multi-scale range of the system. This will be an important subject for future work, using the simulations described in \S~\ref{sec:intro} to make quantitative comparisons to observations of X-ray, optical/UV, and IR spectra of observed AGN. Since we have assumed steady-state throughout, it will also be important to study the predicted variability properties of such disks, both in analytic theoretical models like those here and in numerical simulations, to compare to detailed observations (potentially resolving various open puzzles, see e.g.\ \citealt{dexter.begelman:2019.mag.dom.disk.variability.explanations}).

\begin{acknowledgements}
We thank Norm Murray, Gordon Richards, Todd Thompson, Eliot Quataert, Dalya Baron, Joanna Piotrowska, and Ryan Hickox for many helpful and insightful conversations. We also thank the anonymous referee for a number of helpful suggestions. Support for PFH was provided by NSF Research Grants 1911233, 20009234, 2108318, NSF CAREER grant 1455342, NASA grants 80NSSC18K0562, HST-AR-15800. 
\end{acknowledgements}

\bibliographystyle{mn2e}
%\bibliography{/Users/phopkins/Dropbox/Public/ms}
\bibliography{ms_extracted}

\begin{thebibliography}{}
\makeatletter
\relax
\def\mn@urlcharsother{\let\do\@makeother \do\$\do\&\do\#\do\^\do\_\do\%\do\~}
\def\mn@doi{\begingroup\mn@urlcharsother \@ifnextchar [ {\mn@doi@}
  {\mn@doi@[]}}
\def\mn@doi@[#1]#2{\def\@tempa{#1}\ifx\@tempa\@empty \href
  {http://dx.doi.org/#2} {doi:#2}\else \href {http://dx.doi.org/#2} {#1}\fi
  \endgroup}
\def\mn@eprint#1#2{\mn@eprint@#1:#2::\@nil}
\def\mn@eprint@arXiv#1{\href {http://arxiv.org/abs/#1} {{\tt arXiv:#1}}}
\def\mn@eprint@dblp#1{\href {http://dblp.uni-trier.de/rec/bibtex/#1.xml}
  {dblp:#1}}
\def\mn@eprint@#1:#2:#3:#4\@nil{\def\@tempa {#1}\def\@tempb {#2}\def\@tempc
  {#3}\ifx \@tempc \@empty \let \@tempc \@tempb \let \@tempb \@tempa \fi \ifx
  \@tempb \@empty \def\@tempb {arXiv}\fi \@ifundefined
  {mn@eprint@\@tempb}{\@tempb:\@tempc}{\expandafter \expandafter \csname
  mn@eprint@\@tempb\endcsname \expandafter{\@tempc}}}

\bibitem[\protect\citeauthoryear{{Abramowicz} \& {Fragile}}{{Abramowicz} \&
  {Fragile}}{2013}]{abramowicz:accretion.theory.review}
{Abramowicz} M.~A.,  {Fragile} P.~C.,  2013, \mn@doi [Living Reviews in
  Relativity] {10.12942/lrr-2013-1}, \href
  {https://ui.adsabs.harvard.edu/abs/2013LRR....16....1A} {16, 1}

\bibitem[\protect\citeauthoryear{{Abramowicz}, {Czerny}, {Lasota}  \&
  {Szuszkiewicz}}{{Abramowicz} et~al.}{1988}]{abramowicz:1988.slim.disks}
{Abramowicz} M.~A.,  {Czerny} B.,  {Lasota} J.~P.,   {Szuszkiewicz} E.,  1988,
  \mn@doi [\apj] {10.1086/166683}, \href
  {https://ui.adsabs.harvard.edu/abs/1988ApJ...332..646A} {332, 646}

\bibitem[\protect\citeauthoryear{{Alonso-Herrero} et~al.,}{{Alonso-Herrero}
  et~al.}{2011}]{2011ApJ...736...82A}
{Alonso-Herrero} A.,  et~al., 2011, \mn@doi [\apj]
  {10.1088/0004-637X/736/2/82}, \href
  {https://ui.adsabs.harvard.edu/abs/2011ApJ...736...82A} {736, 82}

\bibitem[\protect\citeauthoryear{{Alonso-Herrero} et~al.,}{{Alonso-Herrero}
  et~al.}{2021}]{alonso.herrero:2021.torus.emission.properties}
{Alonso-Herrero} A.,  et~al., 2021, \mn@doi [\aap]
  {10.1051/0004-6361/202141219}, \href
  {https://ui.adsabs.harvard.edu/abs/2021A&A...652A..99A} {652, A99}

\bibitem[\protect\citeauthoryear{{Anderson}, {Li}, {Krasnopolsky}  \&
  {Blandford}}{{Anderson} et~al.}{2005}]{anderson:2005.mass.loading.mhd.winds}
{Anderson} J.~M.,  {Li} Z.-Y.,  {Krasnopolsky} R.,   {Blandford} R.~D.,  2005,
  \mn@doi [\apj] {10.1086/432040}, \href
  {https://ui.adsabs.harvard.edu/abs/2005ApJ...630..945A} {630, 945}

\bibitem[\protect\citeauthoryear{{Angl{\'e}s-Alc{\'a}zar}, {Dav{\'e}},
  {Faucher-Gigu{\`e}re}, {{\"O}zel}  \& {Hopkins}}{{Angl{\'e}s-Alc{\'a}zar}
  et~al.}{2017}]{angles.alcazar:grav.torque.accretion.cosmo.sim.implications}
{Angl{\'e}s-Alc{\'a}zar} D.,  {Dav{\'e}} R.,  {Faucher-Gigu{\`e}re} C.-A.,
  {{\"O}zel} F.,   {Hopkins} P.~F.,  2017, \mn@doi [\mnras]
  {10.1093/mnras/stw2565}, \href
  {http://adsabs.harvard.edu/abs/2017MNRAS.464.2840A} {464, 2840}

\bibitem[\protect\citeauthoryear{{Angl{\'e}s-Alc{\'a}zar}
  et~al.,}{{Angl{\'e}s-Alc{\'a}zar}
  et~al.}{2021}]{daa:20.hyperrefinement.bh.growth}
{Angl{\'e}s-Alc{\'a}zar} D.,  et~al., 2021, \mn@doi [\apj]
  {10.3847/1538-4357/ac09e8}, \href
  {https://ui.adsabs.harvard.edu/abs/2021ApJ...917...53A} {917, 53}

\bibitem[\protect\citeauthoryear{{Antonucci}}{{Antonucci}}{1982}]{antonucci:1982.torus}
{Antonucci} R.~R.~J.,  1982, \mn@doi [\nat] {10.1038/299605a0}, \href
  {http://adsabs.harvard.edu/abs/1982Natur.299..605A} {299, 605}

\bibitem[\protect\citeauthoryear{{Antonucci}}{{Antonucci}}{1993}]{antonucci:agn.unification.review}
{Antonucci} R.,  1993, \mn@doi [\araa] {10.1146/annurev.aa.31.090193.002353},
  \href {http://adsabs.harvard.edu/abs/1993ARA%26A..31..473A} {31, 473}

\bibitem[\protect\citeauthoryear{{Antonucci}}{{Antonucci}}{2023}]{antonucci:2023.galaxies.agn.review}
{Antonucci} R. R.~J.,  2023, \mn@doi [Galaxies] {10.3390/galaxies11050102},
  \href {https://ui.adsabs.harvard.edu/abs/2023Galax..11..102A} {11, 102}

\bibitem[\protect\citeauthoryear{{Arav}, {Barlow}, {Laor}, {Sargent}  \&
  {Blandford}}{{Arav}
  et~al.}{1998}]{arav:1998.blr.not.discrete.clouds.but.inhomogeneous.system}
{Arav} N.,  {Barlow} T.~A.,  {Laor} A.,  {Sargent} W. L.~W.,   {Blandford}
  R.~D.,  1998, \mn@doi [\mnras] {10.1046/j.1365-8711.1998.297004990.x}, \href
  {https://ui.adsabs.harvard.edu/abs/1998MNRAS.297..990A} {297, 990}

\bibitem[\protect\citeauthoryear{{Balbus} \& {Hawley}}{{Balbus} \&
  {Hawley}}{1998}]{balbus.hawley.review.1998}
{Balbus} S.~A.,  {Hawley} J.~F.,  1998, \mn@doi [Reviews of Modern Physics]
  {10.1103/RevModPhys.70.1}, \href
  {http://adsabs.harvard.edu/abs/1998RvMP...70....1B} {70, 1}

\bibitem[\protect\citeauthoryear{{Ballantyne} et~al.,}{{Ballantyne}
  et~al.}{2024}]{ballantyne:2024.warm.corona.soft.excess.coronal.props.weaker.higheddrat.compilation}
{Ballantyne} D.~R.,  et~al., 2024, \mn@doi [\mnras] {10.1093/mnras/stae944},
  \href {https://ui.adsabs.harvard.edu/abs/2024MNRAS.530.1603B} {530, 1603}

\bibitem[\protect\citeauthoryear{{Beattie}, {Mocz}, {Federrath}  \&
  {Klessen}}{{Beattie}
  et~al.}{2021}]{beattie:2021.turb.intermittency.mhd.subalfvenic}
{Beattie} J.~R.,  {Mocz} P.,  {Federrath} C.,   {Klessen} R.~S.,  2021, arXiv
  e-prints, \href {https://ui.adsabs.harvard.edu/abs/2021arXiv210910470B} {p.
  arXiv:2109.10470}

\bibitem[\protect\citeauthoryear{{Begelman} \& {Armitage}}{{Begelman} \&
  {Armitage}}{2023}]{begelman:2023.mri.saturation.estimates}
{Begelman} M.~C.,  {Armitage} P.~J.,  2023, \mn@doi [\mnras]
  {10.1093/mnras/stad914}, \href
  {https://ui.adsabs.harvard.edu/abs/2023MNRAS.521.5952B} {521, 5952}

\bibitem[\protect\citeauthoryear{{Begelman} \& {Pringle}}{{Begelman} \&
  {Pringle}}{2007}]{begelman.pringle:2007.acc.disks.strong.toroidal.fields}
{Begelman} M.~C.,  {Pringle} J.~E.,  2007, \mn@doi [\mnras]
  {10.1111/j.1365-2966.2006.11372.x}, \href
  {https://ui.adsabs.harvard.edu/abs/2007MNRAS.375.1070B} {375, 1070}

\bibitem[\protect\citeauthoryear{{Begelman} \& {Silk}}{{Begelman} \&
  {Silk}}{2017}]{begelman:2017.mag.elevated.disks.potential.blr.levitation}
{Begelman} M.~C.,  {Silk} J.,  2017, \mn@doi [\mnras] {10.1093/mnras/stw2533},
  \href {https://ui.adsabs.harvard.edu/abs/2017MNRAS.464.2311B} {464, 2311}

\bibitem[\protect\citeauthoryear{{Bending}, {Dobbs}  \& {Bate}}{{Bending}
  et~al.}{2022}]{bending:2022.supernova.feedback.spiral.arms.clouds}
{Bending} T. J.~R.,  {Dobbs} C.~L.,   {Bate} M.~R.,  2022, \mn@doi [\mnras]
  {10.1093/mnras/stac965}, \href
  {https://ui.adsabs.harvard.edu/abs/2022MNRAS.513.2088B} {513, 2088}

\bibitem[\protect\citeauthoryear{{Bennert}, {Falcke}, {Schulz}, {Wilson}  \&
  {Wills}}{{Bennert} et~al.}{2002}]{bennert:nlr.structure}
{Bennert} N.,  {Falcke} H.,  {Schulz} H.,  {Wilson} A.~S.,   {Wills} B.~J.,
  2002, \mn@doi [\apjl] {10.1086/342420}, \href
  {http://adsabs.harvard.edu/abs/2002ApJ...574L.105B} {574, L105}

\bibitem[\protect\citeauthoryear{{Bisnovatyi-Kogan} \&
  {Lovelace}}{{Bisnovatyi-Kogan} \&
  {Lovelace}}{1997}]{bisnovatyi:1997.bfield.heating.accretion.flows}
{Bisnovatyi-Kogan} G.~S.,  {Lovelace} R.~V.~E.,  1997, \mn@doi [\apjl]
  {10.1086/310826}, \href
  {https://ui.adsabs.harvard.edu/abs/1997ApJ...486L..43B} {486, L43}

\bibitem[\protect\citeauthoryear{{Bisnovatyi-Kogan} \&
  {Lovelace}}{{Bisnovatyi-Kogan} \&
  {Lovelace}}{2000}]{bisnovatyi:2000.bfield.heating.expectations.acc.disks.corona.adafs}
{Bisnovatyi-Kogan} G.~S.,  {Lovelace} R.~V.~E.,  2000, \mn@doi [\apj]
  {10.1086/308288}, \href
  {https://ui.adsabs.harvard.edu/abs/2000ApJ...529..978B} {529, 978}

\bibitem[\protect\citeauthoryear{{Blackburne}, {Pooley}, {Rappaport}  \&
  {Schechter}}{{Blackburne}
  et~al.}{2011}]{blackburne:2011.sizes.qso.acc.disks.microlensing.too.big}
{Blackburne} J.~A.,  {Pooley} D.,  {Rappaport} S.,   {Schechter} P.~L.,  2011,
  \mn@doi [\apj] {10.1088/0004-637X/729/1/34}, \href
  {https://ui.adsabs.harvard.edu/abs/2011ApJ...729...34B} {729, 34}

\bibitem[\protect\citeauthoryear{{Blandford} \& {Payne}}{{Blandford} \&
  {Payne}}{1982}]{blandfordpayne:mhd.jets}
{Blandford} R.~D.,  {Payne} D.~G.,  1982, \mnras, \href
  {http://adsabs.harvard.edu/abs/1982MNRAS.199..883B} {199, 883}

\bibitem[\protect\citeauthoryear{{Blustin}, {Page}, {Fuerst},
  {Branduardi-Raymont}  \& {Ashton}}{{Blustin}
  et~al.}{2005}]{blustin:2005.warm.absorbers.accel.in.radii.outside.torus.bhroi}
{Blustin} A.~J.,  {Page} M.~J.,  {Fuerst} S.~V.,  {Branduardi-Raymont} G.,
  {Ashton} C.~E.,  2005, \mn@doi [\aap] {10.1051/0004-6361:20041775}, \href
  {https://ui.adsabs.harvard.edu/abs/2005A&A...431..111B} {431, 111}

\bibitem[\protect\citeauthoryear{{Boissay}, {Ricci}  \& {Paltani}}{{Boissay}
  et~al.}{2016}]{boissay:2016.softexcess.from.comptonization.intensity.corr.edd}
{Boissay} R.,  {Ricci} C.,   {Paltani} S.,  2016, \mn@doi [\aap]
  {10.1051/0004-6361/201526982}, \href
  {https://ui.adsabs.harvard.edu/abs/2016A&A...588A..70B} {588, A70}

\bibitem[\protect\citeauthoryear{{Bonning}, {Cheng}, {Shields}, {Salviander}
  \& {Gebhardt}}{{Bonning}
  et~al.}{2007}]{bonning:2007.accretion.disk.temps.opposite.scaling.observed.thindisk.models}
{Bonning} E.~W.,  {Cheng} L.,  {Shields} G.~A.,  {Salviander} S.,   {Gebhardt}
  K.,  2007, \mn@doi [\apj] {10.1086/510712}, \href
  {https://ui.adsabs.harvard.edu/abs/2007ApJ...659..211B} {659, 211}

\bibitem[\protect\citeauthoryear{{Bonning}, {Shields}, {Stevens}  \&
  {Salviander}}{{Bonning}
  et~al.}{2013}]{bonning:2013.qso.temps.emission.lines.too.low.and.wrong.scaling.with.mass.vs.ss73}
{Bonning} E.~W.,  {Shields} G.~A.,  {Stevens} A.~C.,   {Salviander} S.,  2013,
  \mn@doi [\apj] {10.1088/0004-637X/770/1/30}, \href
  {https://ui.adsabs.harvard.edu/abs/2013ApJ...770...30B} {770, 30}

\bibitem[\protect\citeauthoryear{{Bournaud}, {Elmegreen}, {Teyssier}, {Block}
  \& {Puerari}}{{Bournaud} et~al.}{2010}]{bournaud:2010.grav.turbulence.lmc}
{Bournaud} F.,  {Elmegreen} B.~G.,  {Teyssier} R.,  {Block} D.~L.,   {Puerari}
  I.,  2010, \mn@doi [\mnras] {10.1111/j.1365-2966.2010.17370.x}, \href
  {http://adsabs.harvard.edu/abs/2010MNRAS.409.1088B} {409, 1088}

\bibitem[\protect\citeauthoryear{{Burtscher} et~al.,}{{Burtscher}
  et~al.}{2013}]{burtscher:2013.agn.torii.compilation}
{Burtscher} L.,  et~al., 2013, \mn@doi [\aap] {10.1051/0004-6361/201321890},
  \href {https://ui.adsabs.harvard.edu/abs/2013A&A...558A.149B} {558, A149}

\bibitem[\protect\citeauthoryear{{Bustard} \& {Zweibel}}{{Bustard} \&
  {Zweibel}}{2021}]{bustard:2020.crs.multiphase.ism.accel.confinement}
{Bustard} C.,  {Zweibel} E.~G.,  2021, \mn@doi [\apj]
  {10.3847/1538-4357/abf64c}, \href
  {https://ui.adsabs.harvard.edu/abs/2021ApJ...913..106B} {913, 106}

\bibitem[\protect\citeauthoryear{{Butsky}, {Nakum}, {Ponnada}, {Hummels}, {Ji}
  \& {Hopkins}}{{Butsky} et~al.}{2023}]{butsky:2022.cr.kappa.lower.limits.cgm}
{Butsky} I.~S.,  {Nakum} S.,  {Ponnada} S.~B.,  {Hummels} C.~B.,  {Ji} S.,
  {Hopkins} P.~F.,  2023, \mn@doi [\mnras] {10.1093/mnras/stad671}, \href
  {https://ui.adsabs.harvard.edu/abs/2023MNRAS.521.2477B} {521, 2477}

\bibitem[\protect\citeauthoryear{{Byrne} et~al.,}{{Byrne}
  et~al.}{2023a}]{byrne:2023.fire.elliptical.galaxies.with.agn.feedback}
{Byrne} L.,  et~al., 2023a, \mn@doi [arXiv e-prints]
  {10.48550/arXiv.2310.16086}, \href
  {https://ui.adsabs.harvard.edu/abs/2023arXiv231016086B} {p. arXiv:2310.16086}

\bibitem[\protect\citeauthoryear{{Byrne}, {Faucher-Gigu{\`e}re}, {Stern},
  {Angl{\'e}s-Alc{\'a}zar}, {Wellons}, {Gurvich}  \& {Hopkins}}{{Byrne}
  et~al.}{2023b}]{byrne:2023.fb.lim.bh.growth.center.formation.critical}
{Byrne} L.,  {Faucher-Gigu{\`e}re} C.-A.,  {Stern} J.,
  {Angl{\'e}s-Alc{\'a}zar} D.,  {Wellons} S.,  {Gurvich} A.~B.,   {Hopkins}
  P.~F.,  2023b, \mn@doi [\mnras] {10.1093/mnras/stad171}, \href
  {https://ui.adsabs.harvard.edu/abs/2023MNRAS.520..722B} {520, 722}

\bibitem[\protect\citeauthoryear{{Cackett}, {Bentz}  \& {Kara}}{{Cackett}
  et~al.}{2021}]{cackett:2021.reverberation.mapping.multiwavelength.review}
{Cackett} E.~M.,  {Bentz} M.~C.,   {Kara} E.,  2021, \mn@doi [iScience]
  {10.1016/j.isci.2021.102557}, \href
  {https://ui.adsabs.harvard.edu/abs/2021iSci...24j2557C} {24, 102557}

\bibitem[\protect\citeauthoryear{{Cai} \& {Wang}}{{Cai} \&
  {Wang}}{2023}]{cai:2023.qso.sed.universal.w.flat.intrinsic.spectrum.into.uv.larger.than.expected.need.reprocessing.in.blr}
{Cai} Z.-Y.,  {Wang} J.-X.,  2023, \mn@doi [Nature Astronomy]
  {10.1038/s41550-023-02088-5}, \href
  {https://ui.adsabs.harvard.edu/abs/2023NatAs...7.1506C} {7, 1506}

\bibitem[\protect\citeauthoryear{{Cao}}{{Cao}}{2009}]{cao:2009.coronal.model}
{Cao} X.,  2009, \mn@doi [\mnras] {10.1111/j.1365-2966.2008.14347.x}, \href
  {https://ui.adsabs.harvard.edu/abs/2009MNRAS.394..207C} {394, 207}

\bibitem[\protect\citeauthoryear{{Capellupo}, {Netzer}, {Lira}, {Trakhtenbrot}
  \& {Mej{\'\i}a-Restrepo}}{{Capellupo}
  et~al.}{2015}]{capellupo:2015.fitting.agn.spectra.forced.to.higher.mbh.and.spin.because.temps.too.low.in.obs}
{Capellupo} D.~M.,  {Netzer} H.,  {Lira} P.,  {Trakhtenbrot} B.,
  {Mej{\'\i}a-Restrepo} J.,  2015, \mn@doi [\mnras] {10.1093/mnras/stu2266},
  \href {https://ui.adsabs.harvard.edu/abs/2015MNRAS.446.3427C} {446, 3427}

\bibitem[\protect\citeauthoryear{{Ceverino}, {Dekel}, {Mandelker}, {Bournaud},
  {Burkert}, {Genzel}  \& {Primack}}{{Ceverino}
  et~al.}{2012}]{ceverino:2012.clump.rotation}
{Ceverino} D.,  {Dekel} A.,  {Mandelker} N.,  {Bournaud} F.,  {Burkert} A.,
  {Genzel} R.,   {Primack} J.,  2012, \mn@doi [\mnras]
  {10.1111/j.1365-2966.2011.20296.x}, \href
  {http://adsabs.harvard.edu/abs/2012MNRAS.420.3490C} {420, 3490}

\bibitem[\protect\citeauthoryear{{Chan}, {Kere{\v{s}}}, {Hopkins}, {Quataert},
  {Su}, {Hayward}  \& {Faucher-Gigu{\`e}re}}{{Chan}
  et~al.}{2019}]{chan:2018.cosmicray.fire.gammaray}
{Chan} T.~K.,  {Kere{\v{s}}} D.,  {Hopkins} P.~F.,  {Quataert} E.,  {Su} K.~Y.,
   {Hayward} C.~C.,   {Faucher-Gigu{\`e}re} C.~A.,  2019, \mn@doi [\mnras]
  {10.1093/mnras/stz1895}, \href
  {https://ui.adsabs.harvard.edu/abs/2019MNRAS.488.3716C} {488, 3716}

\bibitem[\protect\citeauthoryear{{Chan}, {Kere{\v{s}}}, {Gurvich}, {Hopkins},
  {Trapp}, {Ji}  \& {Faucher-Gigu{\`e}re}}{{Chan}
  et~al.}{2022}]{chan:2021.cosmic.ray.vertical.balance}
{Chan} T.~K.,  {Kere{\v{s}}} D.,  {Gurvich} A.~B.,  {Hopkins} P.~F.,  {Trapp}
  C.,  {Ji} S.,   {Faucher-Gigu{\`e}re} C.-A.,  2022, \mn@doi [\mnras]
  {10.1093/mnras/stac2236}, \href
  {https://ui.adsabs.harvard.edu/abs/2022MNRAS.517..597C} {517, 597}

\bibitem[\protect\citeauthoryear{{Chelouche}, {Pozo Nu{\~n}ez}  \&
  {Kaspi}}{{Chelouche}
  et~al.}{2019}]{chelouche:2019.qso.reprocessing.inner.disk.light.further.out.obs}
{Chelouche} D.,  {Pozo Nu{\~n}ez} F.,   {Kaspi} S.,  2019, \mn@doi [Nature
  Astronomy] {10.1038/s41550-018-0659-x}, \href
  {https://ui.adsabs.harvard.edu/abs/2019NatAs...3..251C} {3, 251}

\bibitem[\protect\citeauthoryear{{Cho}, {Prather}, {Narayan}, {Natarajan},
  {Su}, {Ricarte}  \& {Chatterjee}}{{Cho}
  et~al.}{2023}]{cho:2023.multiscale.accretion.sims.bondi.inflow.model}
{Cho} H.,  {Prather} B.~S.,  {Narayan} R.,  {Natarajan} P.,  {Su} K.-Y.,
  {Ricarte} A.,   {Chatterjee} K.,  2023, \mn@doi [\apjl]
  {10.3847/2041-8213/ad1048}, \href
  {https://ui.adsabs.harvard.edu/abs/2023ApJ...959L..22C} {959, L22}

\bibitem[\protect\citeauthoryear{{Cho}, {Prather}, {Su}, {Narayan}  \&
  {Natarajan}}{{Cho}
  et~al.}{2024}]{cho:2024.multizone.grmhd.sims.bondi.flow.lowaccrate}
{Cho} H.,  {Prather} B.~S.,  {Su} K.-Y.,  {Narayan} R.,   {Natarajan} P.,
  2024, \mn@doi [arXiv e-prints] {10.48550/arXiv.2405.13887}, \href
  {https://ui.adsabs.harvard.edu/abs/2024arXiv240513887C} {p. arXiv:2405.13887}

\bibitem[\protect\citeauthoryear{{Cochrane} et~al.,}{{Cochrane}
  et~al.}{2023}]{cochrane:2023.agn.winds.galaxy.size.effects}
{Cochrane} R.~K.,  et~al., 2023, \mn@doi [\mnras] {10.1093/mnras/stad1528},
  \href {https://ui.adsabs.harvard.edu/abs/2023MNRAS.523.2409C} {523, 2409}

\bibitem[\protect\citeauthoryear{{Cornachione} \& {Morgan}}{{Cornachione} \&
  {Morgan}}{2020}]{cornachione:2020.accretion.disk.microlensing.profiles.shallow}
{Cornachione} M.~A.,  {Morgan} C.~W.,  2020, \mn@doi [\apj]
  {10.3847/1538-4357/ab8aed}, \href
  {https://ui.adsabs.harvard.edu/abs/2020ApJ...895...93C} {895, 93}

\bibitem[\protect\citeauthoryear{{Crenshaw} \& {Kraemer}}{{Crenshaw} \&
  {Kraemer}}{2012}]{crenshaw:2012.warm.absorbers.strong.feedback.but.well.beyond.accretion.disk.nlr.accel}
{Crenshaw} D.~M.,  {Kraemer} S.~B.,  2012, \mn@doi [\apj]
  {10.1088/0004-637X/753/1/75}, \href
  {https://ui.adsabs.harvard.edu/abs/2012ApJ...753...75C} {753, 75}

\bibitem[\protect\citeauthoryear{{Czerny}, {Niko{\l}ajuk},
  {R{\'o}{\.z}a{\'n}ska}, {Dumont}, {Loska}  \& {Zycki}}{{Czerny}
  et~al.}{2003}]{czerny:2003.soft.excess.and.uv.profile.in.agn.from.reprocessing.warm.skin.warm.absorber}
{Czerny} B.,  {Niko{\l}ajuk} M.,  {R{\'o}{\.z}a{\'n}ska} A.,  {Dumont} A.~M.,
  {Loska} Z.,   {Zycki} P.~T.,  2003, \mn@doi [\aap]
  {10.1051/0004-6361:20031441}, \href
  {https://ui.adsabs.harvard.edu/abs/2003A&A...412..317C} {412, 317}

\bibitem[\protect\citeauthoryear{{Dai}, {Kochanek}, {Chartas}, {Koz{\l}owski},
  {Morgan}, {Garmire}  \& {Agol}}{{Dai}
  et~al.}{2010}]{dai:2010.agn.microlensing.xray.optical.larger.than.expected}
{Dai} X.,  {Kochanek} C.~S.,  {Chartas} G.,  {Koz{\l}owski} S.,  {Morgan}
  C.~W.,  {Garmire} G.,   {Agol} E.,  2010, \mn@doi [\apj]
  {10.1088/0004-637X/709/1/278}, \href
  {https://ui.adsabs.harvard.edu/abs/2010ApJ...709..278D} {709, 278}

\bibitem[\protect\citeauthoryear{{Dalgarno} \& {McCray}}{{Dalgarno} \&
  {McCray}}{1972}]{dalgarno:1972.heating.ionization.HII.regions}
{Dalgarno} A.,  {McCray} R.~A.,  1972, \mn@doi [\araa]
  {10.1146/annurev.aa.10.090172.002111}, \href
  {https://ui.adsabs.harvard.edu/abs/1972ARA&A..10..375D} {10, 375}

\bibitem[\protect\citeauthoryear{{Davidson} \& {Netzer}}{{Davidson} \&
  {Netzer}}{1979}]{davidson.netzer:1979.qso.emission.lines.reviews}
{Davidson} K.,  {Netzer} H.,  1979, \mn@doi [Reviews of Modern Physics]
  {10.1103/RevModPhys.51.715}, \href
  {https://ui.adsabs.harvard.edu/abs/1979RvMP...51..715D} {51, 715}

\bibitem[\protect\citeauthoryear{{Davis} \& {Tchekhovskoy}}{{Davis} \&
  {Tchekhovskoy}}{2020}]{davis.tchekhovskoy:2020.accretion.disk.sims.review}
{Davis} S.~W.,  {Tchekhovskoy} A.,  2020, \mn@doi [\araa]
  {10.1146/annurev-astro-081817-051905}, \href
  {https://ui.adsabs.harvard.edu/abs/2020ARA&A..58..407D} {58, 407}

\bibitem[\protect\citeauthoryear{{Davis}, {Woo}  \& {Blaes}}{{Davis}
  et~al.}{2007}]{davis:2007.continuum.qsos.wrong.shape.vs.simplest.models}
{Davis} S.~W.,  {Woo} J.-H.,   {Blaes} O.~M.,  2007, \mn@doi [\apj]
  {10.1086/521393}, \href
  {https://ui.adsabs.harvard.edu/abs/2007ApJ...668..682D} {668, 682}

\bibitem[\protect\citeauthoryear{{De La Torre Luque}, {Mazziotta}, {Loparco},
  {Gargano}  \& {Serini}}{{De La Torre Luque}
  et~al.}{2021}]{delaTorre:2021.dragon2.methods.new.model.comparison}
{De La Torre Luque} P.,  {Mazziotta} M.~N.,  {Loparco} F.,  {Gargano} F.,
  {Serini} D.,  2021, \mn@doi [\jcap] {10.1088/1475-7516/2021/03/099}, \href
  {https://ui.adsabs.harvard.edu/abs/2021JCAP...03..099D} {2021, 099}

\bibitem[\protect\citeauthoryear{{Deng}, {Mayer}  \& {Meru}}{{Deng}
  et~al.}{2017}]{deng:gravito.turb.frag.convergence.gizmo.methods}
{Deng} H.,  {Mayer} L.,   {Meru} F.,  2017, \mn@doi [\apj]
  {10.3847/1538-4357/aa872b}, \href
  {http://adsabs.harvard.edu/abs/2017ApJ...847...43D} {847, 43}

\bibitem[\protect\citeauthoryear{{Derome}, {Maurin}, {Salati}, {Boudaud},
  {G{\'e}nolini}  \& {Kunz{\'e}}}{{Derome}
  et~al.}{2019}]{derome:usine.cr.transport.fits}
{Derome} L.,  {Maurin} D.,  {Salati} P.,  {Boudaud} M.,  {G{\'e}nolini} Y.,
  {Kunz{\'e}} P.,  2019, \mn@doi [\aap] {10.1051/0004-6361/201935717}, \href
  {https://ui.adsabs.harvard.edu/abs/2019A&A...627A.158D} {627, A158}

\bibitem[\protect\citeauthoryear{{Devereux}}{{Devereux}}{2021}]{devereux:2021.blr.mass.low.40msun.or.less.and.interior.to.sublimation.and.transition.to.xray.interior.favors.blr.as.disk}
{Devereux} N.,  2021, \mn@doi [\mnras] {10.1093/mnras/staa3005}, \href
  {https://ui.adsabs.harvard.edu/abs/2021MNRAS.500..786D} {500, 786}

\bibitem[\protect\citeauthoryear{{Dexter} \& {Agol}}{{Dexter} \&
  {Agol}}{2011}]{dexter:2011.qso.accretion.disks.strongly.inhomogneous}
{Dexter} J.,  {Agol} E.,  2011, \mn@doi [\apjl] {10.1088/2041-8205/727/1/L24},
  \href {https://ui.adsabs.harvard.edu/abs/2011ApJ...727L..24D} {727, L24}

\bibitem[\protect\citeauthoryear{{Dexter} \& {Begelman}}{{Dexter} \&
  {Begelman}}{2019}]{dexter.begelman:2019.mag.dom.disk.variability.explanations}
{Dexter} J.,  {Begelman} M.~C.,  2019, \mn@doi [\mnras]
  {10.1093/mnrasl/sly213}, \href
  {https://ui.adsabs.harvard.edu/abs/2019MNRAS.483L..17D} {483, L17}

\bibitem[\protect\citeauthoryear{{Di Mauro}, {Korsmeier}  \& {Cuoco}}{{Di
  Mauro}
  et~al.}{2023}]{dimauro:2023.cr.diff.constraints.updated.galprop.very.similar.our.models.but.lots.of.interp.re.selfconfinement.that.doesnt.mathematically.work}
{Di Mauro} M.,  {Korsmeier} M.,   {Cuoco} A.,  2023, \mn@doi [arXiv e-prints]
  {10.48550/arXiv.2311.17150}, \href
  {https://ui.adsabs.harvard.edu/abs/2023arXiv231117150D} {p. arXiv:2311.17150}

\bibitem[\protect\citeauthoryear{{Done}, {Davis}, {Jin}, {Blaes}  \&
  {Ward}}{{Done} et~al.}{2012}]{done:2012.soft.excess.comptonization.xr.vs.lum}
{Done} C.,  {Davis} S.~W.,  {Jin} C.,  {Blaes} O.,   {Ward} M.,  2012, \mn@doi
  [\mnras] {10.1111/j.1365-2966.2011.19779.x}, \href
  {https://ui.adsabs.harvard.edu/abs/2012MNRAS.420.1848D} {420, 1848}

\bibitem[\protect\citeauthoryear{{Draine}}{{Draine}}{2011}]{draine:ism.book}
{Draine} B.~T.,  2011, {Physics of the Interstellar and Intergalactic Medium}.
Princeton University Press, Princeton, NJ, USA

\bibitem[\protect\citeauthoryear{{Du} et~al.,}{{Du}
  et~al.}{2015}]{du:2015.smbh.reverb.map.supereddington.in.lowmass.bhs}
{Du} P.,  et~al., 2015, \mn@doi [\apj] {10.1088/0004-637X/806/1/22}, \href
  {https://ui.adsabs.harvard.edu/abs/2015ApJ...806...22D} {806, 22}

\bibitem[\protect\citeauthoryear{{Dyda}, {Davis}  \& {Proga}}{{Dyda}
  et~al.}{2023}]{dyda:2023.agn.disc.wind.simulations.w.xray.rad.only.modest.massloss.rates}
{Dyda} S.,  {Davis} S.~W.,   {Proga} D.,  2023, \mn@doi [arXiv e-prints]
  {10.48550/arXiv.2310.18557}, \href
  {https://ui.adsabs.harvard.edu/abs/2023arXiv231018557D} {p. arXiv:2310.18557}

\bibitem[\protect\citeauthoryear{{Elitzur} \& {Shlosman}}{{Elitzur} \&
  {Shlosman}}{2006}]{elitzur:torus.wind}
{Elitzur} M.,  {Shlosman} I.,  2006, \mn@doi [\apjl] {10.1086/508158}, \href
  {http://adsabs.harvard.edu/abs/2006ApJ...648L.101E} {648, L101}

\bibitem[\protect\citeauthoryear{{Emmering}, {Blandford}  \&
  {Shlosman}}{{Emmering} et~al.}{1992}]{emmering:1992.blr.in.mag.winds}
{Emmering} R.~T.,  {Blandford} R.~D.,   {Shlosman} I.,  1992, \mn@doi [\apj]
  {10.1086/170955}, \href
  {https://ui.adsabs.harvard.edu/abs/1992ApJ...385..460E} {385, 460}

\bibitem[\protect\citeauthoryear{{Farber}, {Ruszkowski}, {Yang}  \&
  {Zweibel}}{{Farber} et~al.}{2018}]{farber:decoupled.crs.in.neutral.gas}
{Farber} R.,  {Ruszkowski} M.,  {Yang} H.-Y.~K.,   {Zweibel} E.~G.,  2018,
  \mn@doi [\apj] {10.3847/1538-4357/aab26d}, \href
  {http://adsabs.harvard.edu/abs/2018ApJ...856..112F} {856, 112}

\bibitem[\protect\citeauthoryear{{Faucher-Gigu{\`e}re} \&
  {Oh}}{{Faucher-Gigu{\`e}re} \& {Oh}}{2023}]{cafg:2023.cgm.review}
{Faucher-Gigu{\`e}re} C.-A.,  {Oh} S.~P.,  2023, \mn@doi [\araa]
  {10.1146/annurev-astro-052920-125203}, \href
  {https://ui.adsabs.harvard.edu/abs/2023ARA&A..61..131F} {61, 131}

\bibitem[\protect\citeauthoryear{{Faucher-Gigu{\`e}re} \&
  {Quataert}}{{Faucher-Gigu{\`e}re} \&
  {Quataert}}{2012}]{cafg:2012.egy.cons.bal.winds}
{Faucher-Gigu{\`e}re} C.-A.,  {Quataert} E.,  2012, \mn@doi [\mnras]
  {10.1111/j.1365-2966.2012.21512.x}, \href
  {http://adsabs.harvard.edu/abs/2012MNRAS.425..605F} {425, 605}

\bibitem[\protect\citeauthoryear{{Federrath}}{{Federrath}}{2015}]{federrath:2014.low.sfe}
{Federrath} C.,  2015, \mn@doi [\mnras] {10.1093/mnras/stv941}, \href
  {http://adsabs.harvard.edu/abs/2015MNRAS.450.4035F} {450, 4035}

\bibitem[\protect\citeauthoryear{{Field}, {Goldsmith}  \& {Habing}}{{Field}
  et~al.}{1969}]{field:1969.ism.phases.heating.cooling}
{Field} G.~B.,  {Goldsmith} D.~W.,   {Habing} H.~J.,  1969, \mn@doi [\apjl]
  {10.1086/180324}, \href
  {https://ui.adsabs.harvard.edu/abs/1969ApJ...155L.149F} {155, L149}

\bibitem[\protect\citeauthoryear{{Forbes}, {Krumholz}  \& {Burkert}}{{Forbes}
  et~al.}{2012}]{forbes:2011.thick.disk.torque.evol}
{Forbes} J.,  {Krumholz} M.~R.,   {Burkert} A.,  2012, \mn@doi [\apj]
  {10.1088/0004-637X/754/1/48}, \href
  {http://adsabs.harvard.edu/abs/2011arXiv1112.1410F} {754, 48}

\bibitem[\protect\citeauthoryear{{Forgan}, {Price}  \& {Bonnell}}{{Forgan}
  et~al.}{2017}]{forgan:2017.mhd.gravitoturb.sims}
{Forgan} D.,  {Price} D.~J.,   {Bonnell} I.,  2017, \mn@doi [\mnras]
  {10.1093/mnras/stw3314}, \href
  {https://ui.adsabs.harvard.edu/abs/2017MNRAS.466.3406F} {466, 3406}

\bibitem[\protect\citeauthoryear{{Frank}, {King}  \& {Raine}}{{Frank}
  et~al.}{2002}]{frank:2002.accretion.book}
{Frank} J.,  {King} A.,   {Raine} D.~J.,  2002, {Accretion Power in
  Astrophysics: Third Edition}, isbn 0521620538 edn.
Cambridge, UK: Cambridge University Press, Cambridge, UK

\bibitem[\protect\citeauthoryear{{GRAVITY Collaboration} et~al.,}{{GRAVITY
  Collaboration}
  et~al.}{2020}]{gravity:2020.resolved.blr.size.disk.inside.dust.sub}
{GRAVITY Collaboration} et~al., 2020, \mn@doi [\aap]
  {10.1051/0004-6361/202039067}, \href
  {https://ui.adsabs.harvard.edu/abs/2020A&A...643A.154G} {643, A154}

\bibitem[\protect\citeauthoryear{{GRAVITY Collaboration} et~al.,}{{GRAVITY
  Collaboration}
  et~al.}{2021}]{gravity:2021.resolved.blr.disk.hot.dust.coronal.regions}
{GRAVITY Collaboration} et~al., 2021, \mn@doi [\aap]
  {10.1051/0004-6361/202040061}, \href
  {https://ui.adsabs.harvard.edu/abs/2021A&A...648A.117G} {648, A117}

\bibitem[\protect\citeauthoryear{{GRAVITY Collaboration} et~al.,}{{GRAVITY
  Collaboration}
  et~al.}{2024}]{gravity:2024.blr.infrared.size.luminosity.relation.agn}
{GRAVITY Collaboration} et~al., 2024, \mn@doi [arXiv e-prints]
  {10.48550/arXiv.2401.07676}, \href
  {https://ui.adsabs.harvard.edu/abs/2024arXiv240107676G} {p. arXiv:2401.07676}

\bibitem[\protect\citeauthoryear{{Gaburov}, {Johansen}  \& {Levin}}{{Gaburov}
  et~al.}{2012}]{gaburov:2012.public.moving.mesh.code}
{Gaburov} E.,  {Johansen} A.,   {Levin} Y.,  2012, \mn@doi [\apj]
  {10.1088/0004-637X/758/2/103}, \href
  {http://adsabs.harvard.edu/abs/2012ApJ...758..103G} {758, 103}

\bibitem[\protect\citeauthoryear{{Gammie}}{{Gammie}}{2001}]{gammie:2001.cooling.in.keplerian.disks}
{Gammie} C.~F.,  2001, \mn@doi [\apj] {10.1086/320631}, \href
  {http://adsabs.harvard.edu/abs/2001ApJ...553..174G} {553, 174}

\bibitem[\protect\citeauthoryear{Garc{\'\i}a-Burillo, Combes, Schinnerer, Boone
   \& Hunt}{Garc{\'\i}a-Burillo
  et~al.}{2005}]{garcia.burillo:torques.in.agn.nuclei.obs.maps.no.inflow}
Garc{\'\i}a-Burillo S.,  Combes F.,  Schinnerer E.,  Boone F.,   Hunt L.~K.,
  2005, \mn@doi [Astronomy and Astrophysics] {10.1051/0004-6361:20052900}, 441,
  1011

\bibitem[\protect\citeauthoryear{{Garc{\'\i}a-Burillo}
  et~al.,}{{Garc{\'\i}a-Burillo}
  et~al.}{2019}]{garcia.burillo:2019.alma.torus.imaging}
{Garc{\'\i}a-Burillo} S.,  et~al., 2019, \mn@doi [\aap]
  {10.1051/0004-6361/201936606}, \href
  {https://ui.adsabs.harvard.edu/abs/2019A&A...632A..61G} {632, A61}

\bibitem[\protect\citeauthoryear{{Garc{\'\i}a-Burillo}
  et~al.,}{{Garc{\'\i}a-Burillo}
  et~al.}{2021}]{garcia.burillo:2021.torus.imaging.alma}
{Garc{\'\i}a-Burillo} S.,  et~al., 2021, \mn@doi [\aap]
  {10.1051/0004-6361/202141075}, \href
  {https://ui.adsabs.harvard.edu/abs/2021A&A...652A..98G} {652, A98}

\bibitem[\protect\citeauthoryear{{Gaskell} \& {Benker}}{{Gaskell} \&
  {Benker}}{2007}]{gaskell07:qso.reddening.curves}
{Gaskell} C.~M.,  {Benker} A.~J.,  2007, \apj, in press, arXiv:0711.1013
  [astro-ph], \href {http://adsabs.harvard.edu/abs/2007arXiv0711.1013G} {}

\bibitem[\protect\citeauthoryear{{George} \& {Fabian}}{{George} \&
  {Fabian}}{1991}]{george:1991.agn.xray.corona}
{George} I.~M.,  {Fabian} A.~C.,  1991, \mn@doi [\mnras]
  {10.1093/mnras/249.2.352}, \href
  {https://ui.adsabs.harvard.edu/abs/1991MNRAS.249..352G} {249, 352}

\bibitem[\protect\citeauthoryear{{Ghosh}, {Pogge}, {Mathur}, {Martini}  \&
  {Shields}}{{Ghosh} et~al.}{2007}]{ghosh:all.obscuration}
{Ghosh} H.,  {Pogge} R.~W.,  {Mathur} S.,  {Martini} P.,   {Shields} J.~C.,
  2007, \mn@doi [\apj] {10.1086/510421}, \href
  {http://adsabs.harvard.edu/abs/2007ApJ...656..105G} {656, 105}

\bibitem[\protect\citeauthoryear{{Gilli} et~al.,}{{Gilli}
  et~al.}{2022}]{gilli:2022.host.galaxy.obscuration}
{Gilli} R.,  et~al., 2022, \mn@doi [\aap] {10.1051/0004-6361/202243708}, \href
  {https://ui.adsabs.harvard.edu/abs/2022A&A...666A..17G} {666, A17}

\bibitem[\protect\citeauthoryear{{Giustini} \& {Proga}}{{Giustini} \&
  {Proga}}{2019}]{giustini.proga:2019.summary.acc.states.winds.qual.phenomenology}
{Giustini} M.,  {Proga} D.,  2019, \mn@doi [\aap]
  {10.1051/0004-6361/201833810}, \href
  {https://ui.adsabs.harvard.edu/abs/2019A&A...630A..94G} {630, A94}

\bibitem[\protect\citeauthoryear{{Glikman}, {LaMassa}, {Piconcelli},
  {Zappacosta}  \& {Lacy}}{{Glikman}
  et~al.}{2024}]{glikman:2024.accretion.obscuration.merger.luminous.red.quasar.evidence}
{Glikman} E.,  {LaMassa} S.,  {Piconcelli} E.,  {Zappacosta} L.,   {Lacy} M.,
  2024, \mn@doi [\mnras] {10.1093/mnras/stae042}, \href
  {https://ui.adsabs.harvard.edu/abs/2024MNRAS.528..711G} {528, 711}

\bibitem[\protect\citeauthoryear{{Gofford}, {Reeves}, {Tombesi}, {Braito},
  {Turner}, {Miller}  \& {Cappi}}{{Gofford}
  et~al.}{2013}]{gofford:2013.ufos.to.warm.absorbers.continuous.family.absorber.properties.extended}
{Gofford} J.,  {Reeves} J.~N.,  {Tombesi} F.,  {Braito} V.,  {Turner} T.~J.,
  {Miller} L.,   {Cappi} M.,  2013, \mn@doi [\mnras] {10.1093/mnras/sts481},
  \href {https://ui.adsabs.harvard.edu/abs/2013MNRAS.430...60G} {430, 60}

\bibitem[\protect\citeauthoryear{{Goodman}}{{Goodman}}{2003}]{goodman:qso.disk.selfgrav}
{Goodman} J.,  2003, \mn@doi [\mnras] {10.1046/j.1365-8711.2003.06241.x}, \href
  {http://adsabs.harvard.edu/abs/2003MNRAS.339..937G} {339, 937}

\bibitem[\protect\citeauthoryear{{Gravity Collaboration} et~al.,}{{Gravity
  Collaboration} et~al.}{2018}]{gravity:2018.sturm.blr.rotating.thick.disk}
{Gravity Collaboration} et~al., 2018, \mn@doi [\nat]
  {10.1038/s41586-018-0731-9}, \href
  {https://ui.adsabs.harvard.edu/abs/2018Natur.563..657G} {563, 657}

\bibitem[\protect\citeauthoryear{{Greve}, {Papadopoulos}, {Gao}  \&
  {Radford}}{{Greve} et~al.}{2009}]{greve:2009.sb.molgas.props}
{Greve} T.~R.,  {Papadopoulos} P.~P.,  {Gao} Y.,   {Radford} S.~J.~E.,  2009,
  \mn@doi [\apj] {10.1088/0004-637X/692/2/1432}, \href
  {http://adsabs.harvard.edu/abs/2009ApJ...692.1432G} {692, 1432}

\bibitem[\protect\citeauthoryear{{Guo} \& {Oh}}{{Guo} \&
  {Oh}}{2008}]{guo.oh:cosmic.rays}
{Guo} F.,  {Oh} S.~P.,  2008, \mn@doi [\mnras]
  {10.1111/j.1365-2966.2007.12692.x}, \href
  {http://adsabs.harvard.edu/abs/2008MNRAS.384..251G} {384, 251}

\bibitem[\protect\citeauthoryear{{Guo}, {Stone}, {Kim}  \& {Quataert}}{{Guo}
  et~al.}{2022}]{guo:2022.superzoom.riaf.in.m87.nosf.nomhd.etc}
{Guo} M.,  {Stone} J.~M.,  {Kim} C.-G.,   {Quataert} E.,  2022, arXiv e-prints,
  \href {https://ui.adsabs.harvard.edu/abs/2022arXiv221105131G} {p.
  arXiv:2211.05131}

\bibitem[\protect\citeauthoryear{{Guo}, {Stone}, {Quataert}  \& {Kim}}{{Guo}
  et~al.}{2024}]{guo:2024.fluxfrozen.disks.lowmdot.ellipticals}
{Guo} M.,  {Stone} J.~M.,  {Quataert} E.,   {Kim} C.-G.,  2024, \mn@doi [arXiv
  e-prints] {10.48550/arXiv.2405.11711}, \href
  {https://ui.adsabs.harvard.edu/abs/2024arXiv240511711G} {p. arXiv:2405.11711}

\bibitem[\protect\citeauthoryear{Haan, Schinnerer, Emsellem,
  Garc{\'\i}a-Burillo, Combes, Mundell  \& Rix}{Haan
  et~al.}{2009}]{haan:nuga.gas.dynamics.maps}
Haan S.,  Schinnerer E.,  Emsellem E.,  Garc{\'\i}a-Burillo S.,  Combes F.,
  Mundell C.~G.,   Rix H.-W.,  2009, \mn@doi [The Astrophysical Journal]
  {10.1088/0004-637X/692/2/1623}, 692, 1623

\bibitem[\protect\citeauthoryear{{Haardt} \& {Maraschi}}{{Haardt} \&
  {Maraschi}}{1991}]{haardt:1991.coronal.heating.model}
{Haardt} F.,  {Maraschi} L.,  1991, \mn@doi [\apjl] {10.1086/186171}, \href
  {https://ui.adsabs.harvard.edu/abs/1991ApJ...380L..51H} {380, L51}

\bibitem[\protect\citeauthoryear{{Hall}, {Sarrouh}  \& {Horne}}{{Hall}
  et~al.}{2018}]{hall:2018.non.blackbody.models.agn.try.explain.sizes.spectral.shapes}
{Hall} P.~B.,  {Sarrouh} G.~T.,   {Horne} K.,  2018, \mn@doi [\apj]
  {10.3847/1538-4357/aaa768}, \href
  {https://ui.adsabs.harvard.edu/abs/2018ApJ...854...93H} {854, 93}

\bibitem[\protect\citeauthoryear{{Halpern}}{{Halpern}}{1984}]{halpern:1984.warm.absorber.discovery}
{Halpern} J.~P.,  1984, \mn@doi [\apj] {10.1086/162077}, \href
  {https://ui.adsabs.harvard.edu/abs/1984ApJ...281...90H} {281, 90}

\bibitem[\protect\citeauthoryear{{Hatziminaoglou}, {Fritz}  \&
  {Jarrett}}{{Hatziminaoglou}
  et~al.}{2009}]{hatziminaoglou:2009.torus.properties.inferred.obs}
{Hatziminaoglou} E.,  {Fritz} J.,   {Jarrett} T.,  2009, \mn@doi [\mnras]
  {10.1111/j.1365-2966.2009.15390.x}, \href
  {http://adsabs.harvard.edu/abs/2009arXiv0907.2389H} {399, 1206}

\bibitem[\protect\citeauthoryear{{Hayward}, {Kere{\v s}}, {Jonsson},
  {Narayanan}, {Cox}  \& {Hernquist}}{{Hayward}
  et~al.}{2011}]{hayward:2011.smg.merger.rt}
{Hayward} C.~C.,  {Kere{\v s}} D.,  {Jonsson} P.,  {Narayanan} D.,  {Cox}
  T.~J.,   {Hernquist} L.,  2011, \mn@doi [\apj] {10.1088/0004-637X/743/2/159},
  \href {http://adsabs.harvard.edu/abs/2011arXiv1101.0002H} {743, 159}

\bibitem[\protect\citeauthoryear{{Higginbottom}, {Proga}, {Knigge}, {Long},
  {Matthews}  \& {Sim}}{{Higginbottom}
  et~al.}{2014}]{higginbottom:2014.line.driven.wind.sims.rad.transfer}
{Higginbottom} N.,  {Proga} D.,  {Knigge} C.,  {Long} K.~S.,  {Matthews} J.~H.,
    {Sim} S.~A.,  2014, \mn@doi [\apj] {10.1088/0004-637X/789/1/19}, \href
  {https://ui.adsabs.harvard.edu/abs/2014ApJ...789...19H} {789, 19}

\bibitem[\protect\citeauthoryear{{Higginbottom}, {Scepi}, {Knigge}, {Long},
  {Matthews}  \& {Sim}}{{Higginbottom}
  et~al.}{2024}]{higginbottom:2024.sims.rad.transfer.line.driven.winds.weaker.than.thought}
{Higginbottom} N.,  {Scepi} N.,  {Knigge} C.,  {Long} K.~S.,  {Matthews} J.~H.,
    {Sim} S.~A.,  2024, \mn@doi [\mnras] {10.1093/mnras/stad3830}, \href
  {https://ui.adsabs.harvard.edu/abs/2024MNRAS.527.9236H} {527, 9236}

\bibitem[\protect\citeauthoryear{{Hollenbach} \& {McKee}}{{Hollenbach} \&
  {McKee}}{1979}]{hollenback.mckee:co.cooling}
{Hollenbach} D.,  {McKee} C.~F.,  1979, \mn@doi [\apjs] {10.1086/190631}, \href
  {https://ui.adsabs.harvard.edu/abs/1979ApJS...41..555H} {41, 555}

\bibitem[\protect\citeauthoryear{{H{\"o}nig}}{{H{\"o}nig}}{2019}]{hoenig:2019.ir.submm.torus.wind.review}
{H{\"o}nig} S.~F.,  2019, \mn@doi [\apj] {10.3847/1538-4357/ab4591}, \href
  {https://ui.adsabs.harvard.edu/abs/2019ApJ...884..171H} {884, 171}

\bibitem[\protect\citeauthoryear{{H{\"o}nig} \& {Kishimoto}}{{H{\"o}nig} \&
  {Kishimoto}}{2010}]{hoenig:clumpy.torus.modeling}
{H{\"o}nig} S.~F.,  {Kishimoto} M.,  2010, \mn@doi [\aap]
  {10.1051/0004-6361/200912676}, \href
  {http://adsabs.harvard.edu/abs/2009arXiv0909.4539H} {523, A27}

\bibitem[\protect\citeauthoryear{{Hopkins}}{{Hopkins}}{2010}]{hopkins:slow.modes}
{Hopkins} P.~F.,  2010, arXiv e-prints, arXiv:1009.4702 [astro-ph], \href
  {http://adsabs.harvard.edu/abs/2010arXiv1009.4702H} {}

\bibitem[\protect\citeauthoryear{{Hopkins}}{{Hopkins}}{2013a}]{hopkins:frag.theory}
{Hopkins} P.~F.,  2013a, \mn@doi [\mnras] {10.1093/mnras/sts704}, \href
  {http://adsabs.harvard.edu/abs/2013MNRAS.430.1653H} {430, 1653}

\bibitem[\protect\citeauthoryear{{Hopkins}}{{Hopkins}}{2013b}]{hopkins:2012.intermittent.turb.density.pdfs}
{Hopkins} P.~F.,  2013b, \mn@doi [\mnras] {10.1093/mnras/stt010}, \href
  {http://adsabs.harvard.edu/abs/2013MNRAS.430.1880H} {430, 1880}

\bibitem[\protect\citeauthoryear{{Hopkins} \& {Christiansen}}{{Hopkins} \&
  {Christiansen}}{2013}]{hopkins:2013.turb.planet.direct.collapse}
{Hopkins} P.~F.,  {Christiansen} J.~L.,  2013, \mn@doi [\apj]
  {10.1088/0004-637X/776/1/48}, \href
  {http://adsabs.harvard.edu/abs/2013ApJ...776...48H} {776, 48}

\bibitem[\protect\citeauthoryear{{Hopkins} \& {Hernquist}}{{Hopkins} \&
  {Hernquist}}{2006}]{hopkins:seyferts}
{Hopkins} P.~F.,  {Hernquist} L.,  2006, \mn@doi [\apjs] {10.1086/505753},
  \href
  {http://adsabs.harvard.edu/cgi-bin/nph-bib_query?bibcode=2006ApJS..166....1H&db_key=AST}
  {166, 1}

\bibitem[\protect\citeauthoryear{{Hopkins} \& {Quataert}}{{Hopkins} \&
  {Quataert}}{2010a}]{hopkins:m31.disk}
{Hopkins} P.~F.,  {Quataert} E.,  2010a, \mn@doi [\mnras]
  {10.1111/j.1745-3933.2010.00855.x}, \href
  {http://adsabs.harvard.edu/abs/2010arXiv1002.1079H} {405, L41}

\bibitem[\protect\citeauthoryear{{Hopkins} \& {Quataert}}{{Hopkins} \&
  {Quataert}}{2010b}]{hopkins:zoom.sims}
{Hopkins} P.~F.,  {Quataert} E.,  2010b, \mn@doi [\mnras]
  {10.1111/j.1365-2966.2010.17064.x}, \href
  {http://adsabs.harvard.edu/abs/2009arXiv0912.3257H} {407, 1529}

\bibitem[\protect\citeauthoryear{{Hopkins} \& {Quataert}}{{Hopkins} \&
  {Quataert}}{2011a}]{hopkins:cusp.slopes}
{Hopkins} P.~F.,  {Quataert} E.,  2011a, \mn@doi [\mnras]
  {10.1111/j.1745-3933.2010.00995.x}, \href
  {http://adsabs.harvard.edu/abs/2011MNRAS.411L..61H} {411, L61}

\bibitem[\protect\citeauthoryear{{Hopkins} \& {Quataert}}{{Hopkins} \&
  {Quataert}}{2011b}]{hopkins:inflow.analytics}
{Hopkins} P.~F.,  {Quataert} E.,  2011b, \mn@doi [\mnras]
  {10.1111/j.1365-2966.2011.18542.x}, \href
  {http://adsabs.harvard.edu/abs/2010arXiv1007.2647H} {415, 1027}

\bibitem[\protect\citeauthoryear{{Hopkins} et~al.,}{{Hopkins}
  et~al.}{2004}]{hopkins:dust}
{Hopkins} P.~F.,  et~al., 2004, \mn@doi [\aj] {10.1086/423291}, \href
  {http://adsabs.harvard.edu/abs/2004AJ....128.1112H} {128, 1112}

\bibitem[\protect\citeauthoryear{{Hopkins}, {Hernquist}, {Martini}, {Cox},
  {Robertson}, {Di Matteo}  \& {Springel}}{{Hopkins}
  et~al.}{2005a}]{hopkins:lifetimes.letter}
{Hopkins} P.~F.,  {Hernquist} L.,  {Martini} P.,  {Cox} T.~J.,  {Robertson} B.,
   {Di Matteo} T.,   {Springel} V.,  2005a, \mn@doi [\apjl] {10.1086/431146},
  \href
  {http://adsabs.harvard.edu/cgi-bin/nph-bib_query?bibcode=2005ApJ...625L..71H&db_key=AST}
  {625, L71}

\bibitem[\protect\citeauthoryear{{Hopkins}, {Hernquist}, {Cox}, {Di Matteo},
  {Martini}, {Robertson}  \& {Springel}}{{Hopkins}
  et~al.}{2005b}]{hopkins:lifetimes.methods}
{Hopkins} P.~F.,  {Hernquist} L.,  {Cox} T.~J.,  {Di Matteo} T.,  {Martini} P.,
   {Robertson} B.,   {Springel} V.,  2005b, \mn@doi [\apj] {10.1086/432438},
  \href
  {http://adsabs.harvard.edu/cgi-bin/nph-bib_query?bibcode=2005ApJ...630..705H&db_key=AST}
  {630, 705}

\bibitem[\protect\citeauthoryear{{Hopkins}, {Hernquist}, {Cox}, {Di Matteo},
  {Robertson}  \& {Springel}}{{Hopkins}
  et~al.}{2005c}]{hopkins:lifetimes.interp}
{Hopkins} P.~F.,  {Hernquist} L.,  {Cox} T.~J.,  {Di Matteo} T.,  {Robertson}
  B.,   {Springel} V.,  2005c, \mn@doi [\apj] {10.1086/432463}, \href
  {http://adsabs.harvard.edu/cgi-bin/nph-bib_query?bibcode=2005ApJ...630..716H&db_key=AST}
  {630, 716}

\bibitem[\protect\citeauthoryear{{Hopkins}, {Hernquist}, {Cox}, {Di Matteo},
  {Robertson}  \& {Springel}}{{Hopkins}
  et~al.}{2005d}]{hopkins:lifetimes.obscuration}
{Hopkins} P.~F.,  {Hernquist} L.,  {Cox} T.~J.,  {Di Matteo} T.,  {Robertson}
  B.,   {Springel} V.,  2005d, \mn@doi [\apj] {10.1086/432755}, \href
  {http://adsabs.harvard.edu/cgi-bin/nph-bib_query?bibcode=2005ApJ...632...81H&db_key=AST}
  {632, 81}

\bibitem[\protect\citeauthoryear{{Hopkins}, {Hernquist}, {Cox}, {Di Matteo},
  {Robertson}  \& {Springel}}{{Hopkins} et~al.}{2006}]{hopkins:qso.all}
{Hopkins} P.~F.,  {Hernquist} L.,  {Cox} T.~J.,  {Di Matteo} T.,  {Robertson}
  B.,   {Springel} V.,  2006, \mn@doi [\apjs] {10.1086/499298}, \href
  {http://adsabs.harvard.edu/cgi-bin/nph-bib_query?bibcode=2006ApJS..163....1H&db_key=AST}
  {163, 1}

\bibitem[\protect\citeauthoryear{{Hopkins}, {Quataert}  \& {Murray}}{{Hopkins}
  et~al.}{2011}]{hopkins:rad.pressure.sf.fb}
{Hopkins} P.~F.,  {Quataert} E.,   {Murray} N.,  2011, \mn@doi [\mnras]
  {10.1111/j.1365-2966.2011.19306.x}, \href
  {http://adsabs.harvard.edu/abs/2011arXiv1101.4940H} {417, 950}

\bibitem[\protect\citeauthoryear{{Hopkins}, {Hayward}, {Narayanan}  \&
  {Hernquist}}{{Hopkins} et~al.}{2012a}]{hopkins:torus}
{Hopkins} P.~F.,  {Hayward} C.~C.,  {Narayanan} D.,   {Hernquist} L.,  2012a,
  \mn@doi [\mnras] {10.1111/j.1365-2966.2011.20035.x}, \href
  {http://adsabs.harvard.edu/abs/2011arXiv1108.3086H} {420, 320}

\bibitem[\protect\citeauthoryear{{Hopkins}, {Quataert}  \& {Murray}}{{Hopkins}
  et~al.}{2012b}]{hopkins:fb.ism.prop}
{Hopkins} P.~F.,  {Quataert} E.,   {Murray} N.,  2012b, \mn@doi [\mnras]
  {10.1111/j.1365-2966.2012.20578.x}, \href
  {http://adsabs.harvard.edu/abs/2012MNRAS.421.3488H} {421, 3488}

\bibitem[\protect\citeauthoryear{{Hopkins}, {Torrey}, {Faucher-Gigu{\`e}re},
  {Quataert}  \& {Murray}}{{Hopkins}
  et~al.}{2016}]{hopkins:qso.stellar.fb.together}
{Hopkins} P.~F.,  {Torrey} P.,  {Faucher-Gigu{\`e}re} C.-A.,  {Quataert} E.,
  {Murray} N.,  2016, \mn@doi [\mnras] {10.1093/mnras/stw289}, \href
  {http://adsabs.harvard.edu/abs/2016MNRAS.458..816H} {458, 816}

\bibitem[\protect\citeauthoryear{{Hopkins} et~al.,}{{Hopkins}
  et~al.}{2020}]{hopkins:cr.mhd.fire2}
{Hopkins} P.~F.,  et~al., 2020, \mn@doi [\mnras] {10.1093/mnras/stz3321}, \href
  {https://ui.adsabs.harvard.edu/abs/2020MNRAS.492.3465H} {492, 3465}

\bibitem[\protect\citeauthoryear{{Hopkins}, {Chan}, {Ji}, {Hummels},
  {Kere{\v{s}}}, {Quataert}  \& {Faucher-Gigu{\`e}re}}{{Hopkins}
  et~al.}{2021a}]{hopkins:2020.cr.outflows.to.mpc.scales}
{Hopkins} P.~F.,  {Chan} T.~K.,  {Ji} S.,  {Hummels} C.~B.,  {Kere{\v{s}}} D.,
  {Quataert} E.,   {Faucher-Gigu{\`e}re} C.-A.,  2021a, \mn@doi [\mnras]
  {10.1093/mnras/staa3690}, \href
  {https://ui.adsabs.harvard.edu/abs/2021MNRAS.501.3640H} {501, 3640}

\bibitem[\protect\citeauthoryear{{Hopkins}, {Squire}, {Chan}, {Quataert}, {Ji},
  {Kere{\v{s}}}  \& {Faucher-Gigu{\`e}re}}{{Hopkins}
  et~al.}{2021b}]{hopkins:cr.transport.constraints.from.galaxies}
{Hopkins} P.~F.,  {Squire} J.,  {Chan} T.~K.,  {Quataert} E.,  {Ji} S.,
  {Kere{\v{s}}} D.,   {Faucher-Gigu{\`e}re} C.-A.,  2021b, \mn@doi [\mnras]
  {10.1093/mnras/staa3691}, \href
  {https://ui.adsabs.harvard.edu/abs/2021MNRAS.501.4184H} {501, 4184}

\bibitem[\protect\citeauthoryear{{Hopkins}, {Wellons},
  {Angl{\'e}s-Alc{\'a}zar}, {Faucher-Gigu{\`e}re}  \& {Grudi{\'c}}}{{Hopkins}
  et~al.}{2022a}]{hopkins:2021.bhs.bulges.from.sigma.sfr}
{Hopkins} P.~F.,  {Wellons} S.,  {Angl{\'e}s-Alc{\'a}zar} D.,
  {Faucher-Gigu{\`e}re} C.-A.,   {Grudi{\'c}} M.~Y.,  2022a, \mn@doi [\mnras]
  {10.1093/mnras/stab3458}, \href
  {https://ui.adsabs.harvard.edu/abs/2022MNRAS.510..630H} {510, 630}

\bibitem[\protect\citeauthoryear{{Hopkins}, {Butsky}, {Panopoulou}, {Ji},
  {Quataert}, {Faucher-Gigu{\`e}re}  \& {Kere{\v{s}}}}{{Hopkins}
  et~al.}{2022b}]{hopkins:cr.multibin.mw.comparison}
{Hopkins} P.~F.,  {Butsky} I.~S.,  {Panopoulou} G.~V.,  {Ji} S.,  {Quataert}
  E.,  {Faucher-Gigu{\`e}re} C.-A.,   {Kere{\v{s}}} D.,  2022b, \mn@doi
  [\mnras] {10.1093/mnras/stac1791}, \href
  {https://ui.adsabs.harvard.edu/abs/2022MNRAS.516.3470H} {516, 3470}

\bibitem[\protect\citeauthoryear{{Hopkins}, {Squire}, {Butsky}  \&
  {Ji}}{{Hopkins} et~al.}{2022c}]{hopkins:2021.sc.et.models.incompatible.obs}
{Hopkins} P.~F.,  {Squire} J.,  {Butsky} I.~S.,   {Ji} S.,  2022c, \mn@doi
  [\mnras] {10.1093/mnras/stac2909}, \href
  {https://ui.adsabs.harvard.edu/abs/2022MNRAS.517.5413H} {517, 5413}

\bibitem[\protect\citeauthoryear{{Hopkins}, {Grudic}, {Kremer}, {Offner},
  {Guszejnov}  \& {Rosen}}{{Hopkins} et~al.}{2024a}]{hopkins:superzoom.imf}
{Hopkins} P.~F.,  {Grudic} M.~Y.,  {Kremer} K.,  {Offner} S. S.~R.,
  {Guszejnov} D.,   {Rosen} A.~L.,  2024a, \mn@doi [arXiv e-prints]
  {10.48550/arXiv.2404.08046}, \href
  {https://ui.adsabs.harvard.edu/abs/2024arXiv240408046H} {p. arXiv:2404.08046}

\bibitem[\protect\citeauthoryear{{Hopkins} et~al.,}{{Hopkins}
  et~al.}{2024b}]{hopkins:superzoom.overview}
{Hopkins} P.~F.,  et~al., 2024b, \mn@doi [The Open Journal of Astrophysics]
  {10.21105/astro.2309.13115}, \href
  {https://ui.adsabs.harvard.edu/abs/2024OJAp....7E..18H} {7, 18}

\bibitem[\protect\citeauthoryear{{Hopkins} et~al.,}{{Hopkins}
  et~al.}{2024c}]{hopkins:superzoom.disk}
{Hopkins} P.~F.,  et~al., 2024c, \mn@doi [The Open Journal of Astrophysics]
  {10.21105/astro.2310.04506}, \href
  {https://ui.adsabs.harvard.edu/abs/2024OJAp....7E..19H} {7, 19}

\bibitem[\protect\citeauthoryear{{Hopkins} et~al.,}{{Hopkins}
  et~al.}{2024d}]{hopkins:superzoom.analytic}
{Hopkins} P.~F.,  et~al., 2024d, \mn@doi [The Open Journal of Astrophysics]
  {10.21105/astro.2310.04507}, \href
  {https://ui.adsabs.harvard.edu/abs/2024OJAp....7E..20H} {7, 20}

\bibitem[\protect\citeauthoryear{{Hopkins} et~al.,}{{Hopkins}
  et~al.}{2025}]{hopkins:superzoom.agn.disks.to.isco.with.gizmo.rad.thermochemical.properties.nlte.multiphase.resolution.studies}
{Hopkins} P.~F.,  et~al., 2025, arXiv e-prints, \href
  {https://ui.adsabs.harvard.edu/abs/2025arXiv250205268H} {p. arXiv:2502.05268}

\bibitem[\protect\citeauthoryear{{Hubeny}, {Blaes}, {Krolik}  \&
  {Agol}}{{Hubeny}
  et~al.}{2001}]{hubeny:2001.acc.disk.spectra.w.comptonization.temps.and.metal.lines.detailed.calcs}
{Hubeny} I.,  {Blaes} O.,  {Krolik} J.~H.,   {Agol} E.,  2001, \mn@doi [\apj]
  {10.1086/322344}, \href
  {https://ui.adsabs.harvard.edu/abs/2001ApJ...559..680H} {559, 680}

\bibitem[\protect\citeauthoryear{{Izumi} et~al.,}{{Izumi}
  et~al.}{2023}]{izumi:2023.imaging.nuclear.gas.disk.circinus.accretion.rate}
{Izumi} T.,  et~al., 2023, \mn@doi [Science] {10.1126/science.adf0569}, \href
  {https://ui.adsabs.harvard.edu/abs/2023Sci...382..554I} {382, 554}

\bibitem[\protect\citeauthoryear{{Ji} et~al.,}{{Ji}
  et~al.}{2020}]{ji:fire.cr.cgm}
{Ji} S.,  et~al., 2020, \mn@doi [\mnras] {10.1093/mnras/staa1849}, \href
  {https://ui.adsabs.harvard.edu/abs/2020MNRAS.496.4221J} {496, 4221}

\bibitem[\protect\citeauthoryear{{Jiang}, {Cantiello}, {Bildsten}, {Quataert}
  \& {Blaes}}{{Jiang}
  et~al.}{2015}]{jiang:2015.rhd.star.sims.metal.opacities.for.agn.disks.as.well}
{Jiang} Y.-F.,  {Cantiello} M.,  {Bildsten} L.,  {Quataert} E.,   {Blaes} O.,
  2015, \mn@doi [\apj] {10.1088/0004-637X/813/1/74}, \href
  {https://ui.adsabs.harvard.edu/abs/2015ApJ...813...74J} {813, 74}

\bibitem[\protect\citeauthoryear{{Jiang}, {Stone}  \& {Davis}}{{Jiang}
  et~al.}{2019}]{jiang:2019.superedd.sims.smbh.prad.pmag.modest.outflows}
{Jiang} Y.-F.,  {Stone} J.~M.,   {Davis} S.~W.,  2019, \mn@doi [\apj]
  {10.3847/1538-4357/ab29ff}, \href
  {https://ui.adsabs.harvard.edu/abs/2019ApJ...880...67J} {880, 67}

\bibitem[\protect\citeauthoryear{{Jim{\'e}nez-Vicente}, {Mediavilla},
  {Kochanek}, {Mu{\~n}oz}, {Motta}, {Falco}  \&
  {Mosquera}}{{Jim{\'e}nez-Vicente}
  et~al.}{2014}]{jimenez:2014.qso.disk.temp.profile.size.from.microlensing.large.flat}
{Jim{\'e}nez-Vicente} J.,  {Mediavilla} E.,  {Kochanek} C.~S.,  {Mu{\~n}oz}
  J.~A.,  {Motta} V.,  {Falco} E.,   {Mosquera} A.~M.,  2014, \mn@doi [\apj]
  {10.1088/0004-637X/783/1/47}, \href
  {https://ui.adsabs.harvard.edu/abs/2014ApJ...783...47J} {783, 47}

\bibitem[\protect\citeauthoryear{{Jogee}}{{Jogee}}{2006}]{jogee:review}
{Jogee} S.,  2006, in {Alloin} D.,  ed.,  Lecture Notes in Physics, Berlin
  Springer Verlag Vol. 693, Physics of Active Galactic Nuclei at all Scales. pp
  143--+

\bibitem[\protect\citeauthoryear{{Johansen} \& {Levin}}{{Johansen} \&
  {Levin}}{2008}]{johansen.levin:2008.high.mdot.magnetized.disks}
{Johansen} A.,  {Levin} Y.,  2008, \mn@doi [\aap]
  {10.1051/0004-6361:200810385}, \href
  {https://ui.adsabs.harvard.edu/abs/2008A&A...490..501J} {490, 501}

\bibitem[\protect\citeauthoryear{{Kaaz}, {Liska}, {Jacquemin-Ide}, {Andalman},
  {Musoke}, {Tchekhovskoy}  \& {Porth}}{{Kaaz}
  et~al.}{2022}]{kaaz:2022.grmhd.sims.misaligned.acc.disks.spin}
{Kaaz} N.,  {Liska} M. T.~P.,  {Jacquemin-Ide} J.,  {Andalman} Z.~L.,  {Musoke}
  G.,  {Tchekhovskoy} A.,   {Porth} O.,  2022, \mn@doi [arXiv e-prints]
  {10.48550/arXiv.2210.10053}, \href
  {https://ui.adsabs.harvard.edu/abs/2022arXiv221010053K} {p. arXiv:2210.10053}

\bibitem[\protect\citeauthoryear{Kaaz, Liska, Tchekhovskoy, Hopkins  \&
  Jacquemin-Ide}{Kaaz
  et~al.}{2024}]{kaaz:2024.hamr.forged.fire.zoom.to.grmhd.magnetized.disks}
Kaaz N.,  Liska M.,  Tchekhovskoy A.,  Hopkins P.~F.,   Jacquemin-Ide J.,
  2024, H-AMR FORGE'd in FIRE I: Magnetic state transitions, jet launching and
  radiative emission in super-Eddington, highly magnetized quasar disks formed
  from cosmological initial conditions (\mn@eprint {arXiv} {2410.01877}), \url
  {https://arxiv.org/abs/2410.01877}

\bibitem[\protect\citeauthoryear{{Kamraj} et~al.,}{{Kamraj}
  et~al.}{2022}]{kamraj:2022.hard.xray.agn.corona.properties}
{Kamraj} N.,  et~al., 2022, \mn@doi [\apj] {10.3847/1538-4357/ac45f6}, \href
  {https://ui.adsabs.harvard.edu/abs/2022ApJ...927...42K} {927, 42}

\bibitem[\protect\citeauthoryear{{Kaspi} et~al.,}{{Kaspi}
  et~al.}{2001}]{kaspi:2001.warm.absorber.3783.originates.outside.blr.and.torus}
{Kaspi} S.,  et~al., 2001, \mn@doi [\apj] {10.1086/321333}, \href
  {https://ui.adsabs.harvard.edu/abs/2001ApJ...554..216K} {554, 216}

\bibitem[\protect\citeauthoryear{{Kaspi}, {Maoz}, {Netzer}, {Peterson},
  {Vestergaard}  \& {Jannuzi}}{{Kaspi}
  et~al.}{2005}]{kaspi:2005.blr.size.reverb.mapping}
{Kaspi} S.,  {Maoz} D.,  {Netzer} H.,  {Peterson} B.~M.,  {Vestergaard} M.,
  {Jannuzi} B.~T.,  2005, \mn@doi [\apj] {10.1086/431275}, \href
  {http://adsabs.harvard.edu/abs/2005ApJ...629...61K} {629, 61}

\bibitem[\protect\citeauthoryear{{Kempski} \& {Quataert}}{{Kempski} \&
  {Quataert}}{2022}]{kempski:2021.reconciling.sc.et.models.obs}
{Kempski} P.,  {Quataert} E.,  2022, \mn@doi [\mnras] {10.1093/mnras/stac1240},
  \href {https://ui.adsabs.harvard.edu/abs/2022MNRAS.514..657K} {514, 657}

\bibitem[\protect\citeauthoryear{{Kim} \& {Ostriker}}{{Kim} \&
  {Ostriker}}{2001}]{kim.ostriker:2001.gravitoturb.galactic.disks.mhd.conditions}
{Kim} W.-T.,  {Ostriker} E.~C.,  2001, \mn@doi [\apj] {10.1086/322330}, \href
  {https://ui.adsabs.harvard.edu/abs/2001ApJ...559...70K} {559, 70}

\bibitem[\protect\citeauthoryear{{Kinkhabwala} et~al.,}{{Kinkhabwala}
  et~al.}{2002}]{kinkhabwala:2002.warm.absorber.1068.requires.wide.range.of.densities.at.given.r.also.large.r.fewhundredpc.to.smaller}
{Kinkhabwala} A.,  et~al., 2002, \mn@doi [\apj] {10.1086/341482}, \href
  {https://ui.adsabs.harvard.edu/abs/2002ApJ...575..732K} {575, 732}

\bibitem[\protect\citeauthoryear{{Kishimoto}, {Antonucci}, {Blaes}, {Lawrence},
  {Boisson}, {Albrecht}  \& {Leipski}}{{Kishimoto}
  et~al.}{2008}]{kishimoto:qso.spectrum.ir.bluer}
{Kishimoto} M.,  {Antonucci} R.,  {Blaes} O.,  {Lawrence} A.,  {Boisson} C.,
  {Albrecht} M.,   {Leipski} C.,  2008, \mn@doi [\nat] {10.1038/nature07114},
  \href {https://ui.adsabs.harvard.edu/abs/2008Natur.454..492K} {454, 492}

\bibitem[\protect\citeauthoryear{{Konigl} \& {Kartje}}{{Konigl} \&
  {Kartje}}{1994}]{koniglkartje:disk.winds}
{Konigl} A.,  {Kartje} J.~F.,  1994, \mn@doi [\apj] {10.1086/174746}, \href
  {http://adsabs.harvard.edu/abs/1994ApJ...434..446K} {434, 446}

\bibitem[\protect\citeauthoryear{{Konstandin}, {Girichidis}, {Federrath}  \&
  {Klessen}}{{Konstandin} et~al.}{2012}]{konstantin:mach.compressive.relation}
{Konstandin} L.,  {Girichidis} P.,  {Federrath} C.,   {Klessen} R.~S.,  2012,
  \mn@doi [\apj] {10.1088/0004-637X/761/2/149}, \href
  {http://adsabs.harvard.edu/abs/2012arXiv1206.4524K} {761, 149}

\bibitem[\protect\citeauthoryear{{Koshida} et~al.,}{{Koshida}
  et~al.}{2014}]{koshida:2014.agn.torii.sizes}
{Koshida} S.,  et~al., 2014, \mn@doi [\apj] {10.1088/0004-637X/788/2/159},
  \href {https://ui.adsabs.harvard.edu/abs/2014ApJ...788..159K} {788, 159}

\bibitem[\protect\citeauthoryear{{Krawczyk}, {Richards}, {Mehta}, {Vogeley},
  {Gallagher}, {Leighly}, {Ross}  \& {Schneider}}{{Krawczyk}
  et~al.}{2013}]{krawczyk:2013.mean.qso.seds.softx.vs.lbol}
{Krawczyk} C.~M.,  {Richards} G.~T.,  {Mehta} S.~S.,  {Vogeley} M.~S.,
  {Gallagher} S.~C.,  {Leighly} K.~M.,  {Ross} N.~P.,   {Schneider} D.~P.,
  2013, \mn@doi [\apjs] {10.1088/0067-0049/206/1/4}, \href
  {https://ui.adsabs.harvard.edu/abs/2013ApJS..206....4K} {206, 4}

\bibitem[\protect\citeauthoryear{{Krolik}}{{Krolik}}{1999}]{krolik:1999.agn.book}
{Krolik} J.~H.,  1999, {Active galactic nuclei : from the central black hole to
  the galactic environment}.
Princeton, N. J. : Princeton University Press,

\bibitem[\protect\citeauthoryear{{Krolik} \& {Begelman}}{{Krolik} \&
  {Begelman}}{1988}]{krolik:clumpy.torii}
{Krolik} J.~H.,  {Begelman} M.~C.,  1988, \mn@doi [\apj] {10.1086/166414},
  \href {http://adsabs.harvard.edu/abs/1988ApJ...329..702K} {329, 702}

\bibitem[\protect\citeauthoryear{{Krolik} \& {Kriss}}{{Krolik} \&
  {Kriss}}{2001}]{krolik:2001.warm.absorbers.multiphase.winds}
{Krolik} J.~H.,  {Kriss} G.~A.,  2001, \mn@doi [\apj] {10.1086/323442}, \href
  {https://ui.adsabs.harvard.edu/abs/2001ApJ...561..684K} {561, 684}

\bibitem[\protect\citeauthoryear{{Krolik}, {McKee}  \& {Tarter}}{{Krolik}
  et~al.}{1981}]{krolik:1981.twophase.model.quasar.emission.lines}
{Krolik} J.~H.,  {McKee} C.~F.,   {Tarter} C.~B.,  1981, \mn@doi [\apj]
  {10.1086/159303}, \href
  {https://ui.adsabs.harvard.edu/abs/1981ApJ...249..422K} {249, 422}

\bibitem[\protect\citeauthoryear{{Krongold}, {Nicastro}, {Elvis}, {Brickhouse},
  {Binette}, {Mathur}  \& {Jim{\'e}nez-Bail{\'o}n}}{{Krongold}
  et~al.}{2007}]{krongold:2007.4051.example.compact.warm.absorber.from.inside.torus}
{Krongold} Y.,  {Nicastro} F.,  {Elvis} M.,  {Brickhouse} N.,  {Binette} L.,
  {Mathur} S.,   {Jim{\'e}nez-Bail{\'o}n} E.,  2007, \mn@doi [\apj]
  {10.1086/512476}, \href
  {https://ui.adsabs.harvard.edu/abs/2007ApJ...659.1022K} {659, 1022}

\bibitem[\protect\citeauthoryear{{Krumholz}, {Crocker}  \& {Offner}}{{Krumholz}
  et~al.}{2023}]{krumholz:2023.cosmic.ray.ionization.gamma.ray.loss.budgets}
{Krumholz} M.~R.,  {Crocker} R.~M.,   {Offner} S. S.~R.,  2023, \mn@doi
  [\mnras] {10.1093/mnras/stad459}, \href
  {https://ui.adsabs.harvard.edu/abs/2023MNRAS.520.5126K} {520, 5126}

\bibitem[\protect\citeauthoryear{{Kubota} \& {Done}}{{Kubota} \&
  {Done}}{2018}]{kubota:2018.soft.excess.comptonizing.layers}
{Kubota} A.,  {Done} C.,  2018, \mn@doi [\mnras] {10.1093/mnras/sty1890}, \href
  {https://ui.adsabs.harvard.edu/abs/2018MNRAS.480.1247K} {480, 1247}

\bibitem[\protect\citeauthoryear{{Lacki}, {Thompson}, {Quataert}, {Loeb}  \&
  {Waxman}}{{Lacki} et~al.}{2011}]{lacki:2011.cosmic.ray.sub.calorimetric}
{Lacki} B.~C.,  {Thompson} T.~A.,  {Quataert} E.,  {Loeb} A.,   {Waxman} E.,
  2011, \mn@doi [\apj] {10.1088/0004-637X/734/2/107}, \href
  {http://adsabs.harvard.edu/abs/2011ApJ...734..107L} {734, 107}

\bibitem[\protect\citeauthoryear{{Laor}}{{Laor}}{1991}]{laor:1991.blr.disk.line.profiles}
{Laor} A.,  1991, \mn@doi [\apj] {10.1086/170257}, \href
  {https://ui.adsabs.harvard.edu/abs/1991ApJ...376...90L} {376, 90}

\bibitem[\protect\citeauthoryear{{Laor} \& {Davis}}{{Laor} \&
  {Davis}}{2014}]{laor:2014.disk.winds.low.temp}
{Laor} A.,  {Davis} S.~W.,  2014, \mn@doi [\mnras] {10.1093/mnras/stt2408},
  \href {https://ui.adsabs.harvard.edu/abs/2014MNRAS.438.3024L} {438, 3024}

\bibitem[\protect\citeauthoryear{{Laor}, {Fiore}, {Elvis}, {Wilkes}  \&
  {McDowell}}{{Laor} et~al.}{1997}]{laor:warm.absorber}
{Laor} A.,  {Fiore} F.,  {Elvis} M.,  {Wilkes} B.~J.,   {McDowell} J.~C.,
  1997, \mn@doi [\apj] {10.1086/303696}, \href
  {http://adsabs.harvard.edu/cgi-bin/nph-bib_query?bibcode=1997ApJ...477...93L&db_key=AST}
  {477, 93}

\bibitem[\protect\citeauthoryear{{Laor}, {Barth}, {Ho}  \& {Filippenko}}{{Laor}
  et~al.}{2006}]{laor:2006.blr.could.be.smooth.disk.not.clumpy.but.must.be.turb}
{Laor} A.,  {Barth} A.~J.,  {Ho} L.~C.,   {Filippenko} A.~V.,  2006, \mn@doi
  [\apj] {10.1086/497908}, \href
  {https://ui.adsabs.harvard.edu/abs/2006ApJ...636...83L} {636, 83}

\bibitem[\protect\citeauthoryear{{Lawrence}}{{Lawrence}}{2012}]{lawrence:2012.big.blue.bump.agn.theory.problems.needs.reprocessing.and.scattering}
{Lawrence} A.,  2012, \mn@doi [\mnras] {10.1111/j.1365-2966.2012.20889.x},
  \href {https://ui.adsabs.harvard.edu/abs/2012MNRAS.423..451L} {423, 451}

\bibitem[\protect\citeauthoryear{{Lawrence} \& {Elvis}}{{Lawrence} \&
  {Elvis}}{1982}]{lawrence:1982.torus.alignment}
{Lawrence} A.,  {Elvis} M.,  1982, \mn@doi [\apj] {10.1086/159918}, \href
  {http://adsabs.harvard.edu/abs/1982ApJ...256..410L} {256, 410}

\bibitem[\protect\citeauthoryear{{Leighly}}{{Leighly}}{2004}]{leighly:2004.agn.winds}
{Leighly} K.~M.,  2004, \mn@doi [\apj] {10.1086/422089}, \href
  {https://ui.adsabs.harvard.edu/abs/2004ApJ...611..125L} {611, 125}

\bibitem[\protect\citeauthoryear{{Leighly}, {Choi}, {Eracleous}, {Terndrup},
  {Gallagher}  \& {Richards}}{{Leighly}
  et~al.}{2024}]{leighly:2024.agn.winds.stronger.line.driven.highedd.dust.driven.alledd}
{Leighly} K.~M.,  {Choi} H.,  {Eracleous} M.,  {Terndrup} D.~M.,  {Gallagher}
  S.~C.,   {Richards} G.~T.,  2024, \mn@doi [\apj] {10.3847/1538-4357/ad2f2a},
  \href {https://ui.adsabs.harvard.edu/abs/2024ApJ...966...87L} {966, 87}

\bibitem[\protect\citeauthoryear{{Liu} \& {Qiao}}{{Liu} \&
  {Qiao}}{2022}]{liu.qiao:2022.agn.acc.disk.review.w.focus.on.coronae.disk.states}
{Liu} B.~F.,  {Qiao} E.,  2022, \mn@doi [iScience]
  {10.1016/j.isci.2021.103544}, \href
  {https://ui.adsabs.harvard.edu/abs/2022iSci...25j3544L} {25, 103544}

\bibitem[\protect\citeauthoryear{{Loeb} \& {Laor}}{{Loeb} \&
  {Laor}}{1992}]{loeb.laor:1992.radiative.viscosities.in.accretion.disks}
{Loeb} A.,  {Laor} A.,  1992, \mn@doi [\apj] {10.1086/170857}, \href
  {https://ui.adsabs.harvard.edu/abs/1992ApJ...384..115L} {384, 115}

\bibitem[\protect\citeauthoryear{{Luo} \& {Shlosman}}{{Luo} \&
  {Shlosman}}{2024}]{luo:2024.magnetically.dominated.disk.like.our.zoomins.zoomin.on.first.supermassive.star.situation}
{Luo} Y.,  {Shlosman} I.,  2024, \mn@doi [\apj] {10.3847/1538-4357/ad7fec},
  \href {https://ui.adsabs.harvard.edu/abs/2024ApJ...976...85L} {976, 85}

\bibitem[\protect\citeauthoryear{{Luo} et~al.,}{{Luo}
  et~al.}{2015}]{luo:2015.thick.disk.agn.shielding.wind.unification}
{Luo} B.,  et~al., 2015, \mn@doi [\apj] {10.1088/0004-637X/805/2/122}, \href
  {https://ui.adsabs.harvard.edu/abs/2015ApJ...805..122L} {805, 122}

\bibitem[\protect\citeauthoryear{{Lyu} \& {Rieke}}{{Lyu} \&
  {Rieke}}{2022}]{lyu:2022.torus.variability.sed.models}
{Lyu} J.,  {Rieke} G.,  2022, \mn@doi [Universe] {10.3390/universe8060304},
  \href {https://ui.adsabs.harvard.edu/abs/2022Univ....8..304L} {8, 304}

\bibitem[\protect\citeauthoryear{{Magdziarz} \& {Zdziarski}}{{Magdziarz} \&
  {Zdziarski}}{1995}]{magdziarz.zdziarski.95:compton.reflection.model}
{Magdziarz} P.,  {Zdziarski} A.~A.,  1995, \mnras, \href
  {http://adsabs.harvard.edu/abs/1995MNRAS.273..837M} {273, 837}

\bibitem[\protect\citeauthoryear{{Marinucci}, {Tamborra}, {Bianchi},
  {Dov{\v{c}}iak}, {Matt}, {Middei}  \& {Tortosa}}{{Marinucci}
  et~al.}{2018}]{marinucci:2018.agn.coronae.review}
{Marinucci} A.,  {Tamborra} F.,  {Bianchi} S.,  {Dov{\v{c}}iak} M.,  {Matt} G.,
   {Middei} R.,   {Tortosa} A.,  2018, \mn@doi [Galaxies]
  {10.3390/galaxies6020044}, \href
  {https://ui.adsabs.harvard.edu/abs/2018Galax...6...44M} {6, 44}

\bibitem[\protect\citeauthoryear{{McConnell} \& {Ma}}{{McConnell} \&
  {Ma}}{2013}]{mcconnell:mbh.host.revisions}
{McConnell} N.~J.,  {Ma} C.-P.,  2013, \mn@doi [\apj]
  {10.1088/0004-637X/764/2/184}, \href
  {https://ui.adsabs.harvard.edu/abs/2013ApJ...764..184M} {764, 184}

\bibitem[\protect\citeauthoryear{{Meena} et~al.,}{{Meena}
  et~al.}{2022}]{meena:2022.strong.nlr.rad.pressure.driven.winds}
{Meena} B.,  et~al., 2022, arXiv e-prints, \href
  {https://ui.adsabs.harvard.edu/abs/2022arXiv221202513M} {p. arXiv:2212.02513}

\bibitem[\protect\citeauthoryear{{Mercedes-Feliz} et~al.,}{{Mercedes-Feliz}
  et~al.}{2023}]{mercedes.feliz:2023.agn.feedback.positive.negative}
{Mercedes-Feliz} J.,  et~al., 2023, \mn@doi [\mnras] {10.1093/mnras/stad2079},
  \href {https://ui.adsabs.harvard.edu/abs/2023MNRAS.tmp.2027M} {}

\bibitem[\protect\citeauthoryear{{Mercedes-Feliz} et~al.,}{{Mercedes-Feliz}
  et~al.}{2024}]{mercedes.feliz:2023.qso.feedback.fire.induced.clumps.in.gas.and.stars}
{Mercedes-Feliz} J.,  et~al., 2024, \mn@doi [\mnras] {10.1093/mnras/stae1021},
  \href {https://ui.adsabs.harvard.edu/abs/2024MNRAS.530.2795M} {530, 2795}

\bibitem[\protect\citeauthoryear{{Mihalas} \& {Mihalas}}{{Mihalas} \&
  {Mihalas}}{1984}]{mihalas:1984oup..book.....M}
{Mihalas} D.,  {Mihalas} B.~W.,  eds, 1984, {Foundations of radiation
  hydrodynamics}.
New York, Oxford University Press, 731 p.

\bibitem[\protect\citeauthoryear{{Mor}, {Netzer}  \& {Elitzur}}{{Mor}
  et~al.}{2009}]{mor:2009.torus.structure.from.fitting.obs}
{Mor} R.,  {Netzer} H.,   {Elitzur} M.,  2009, \mn@doi [\apj]
  {10.1088/0004-637X/705/1/298}, \href
  {http://adsabs.harvard.edu/abs/2009arXiv0907.1654M} {705, 298}

\bibitem[\protect\citeauthoryear{{Most} \& {Wang}}{{Most} \&
  {Wang}}{2024}]{most:2024.bh.circumbinary.acc.disk.decoupling.when.mad}
{Most} E.~R.,  {Wang} H.-Y.,  2024, \mn@doi [arXiv e-prints]
  {10.48550/arXiv.2410.23264}, \href
  {https://ui.adsabs.harvard.edu/abs/2024arXiv241023264M} {p. arXiv:2410.23264}

\bibitem[\protect\citeauthoryear{{Murray}, {Chiang}, {Grossman}  \&
  {Voit}}{{Murray} et~al.}{1995}]{murray:1995.acc.disk.rad.winds}
{Murray} N.,  {Chiang} J.,  {Grossman} S.~A.,   {Voit} G.~M.,  1995, \mn@doi
  [\apj] {10.1086/176238}, \href
  {http://adsabs.harvard.edu/abs/1995ApJ...451..498M} {451, 498}

\bibitem[\protect\citeauthoryear{{Naddaf}, {Czerny}  \& {Szczerba}}{{Naddaf}
  et~al.}{2021}]{naddaf:2021.blr.structure.from.failed.winds}
{Naddaf} M.-H.,  {Czerny} B.,   {Szczerba} R.,  2021, \mn@doi [\apj]
  {10.3847/1538-4357/ac139d}, \href
  {https://ui.adsabs.harvard.edu/abs/2021ApJ...920...30N} {920, 30}

\bibitem[\protect\citeauthoryear{{Naddaf}, {Czerny}  \&
  {Zaja{\v{c}}ek}}{{Naddaf} et~al.}{2022}]{naddaf:2022.frado.failed.wind.model}
{Naddaf} M.-H.,  {Czerny} B.,   {Zaja{\v{c}}ek} M.,  2022, \mn@doi [Dynamics]
  {10.3390/dynamics2030015}, \href
  {https://ui.adsabs.harvard.edu/abs/2022Dynam...2..295N} {2, 295}

\bibitem[\protect\citeauthoryear{{Narayan} \& {Yi}}{{Narayan} \&
  {Yi}}{1995a}]{narayan.yi.95:adaf.self.similarity.outflows}
{Narayan} R.,  {Yi} I.,  1995a, \mn@doi [\apj] {10.1086/175599}, \href
  {http://adsabs.harvard.edu/abs/1995ApJ...444..231N} {444, 231}

\bibitem[\protect\citeauthoryear{{Narayan} \& {Yi}}{{Narayan} \&
  {Yi}}{1995b}]{narayan.yi.95:adaf.lowmass.bhs}
{Narayan} R.,  {Yi} I.,  1995b, \mn@doi [\apj] {10.1086/176343}, \href
  {http://adsabs.harvard.edu/abs/1995ApJ...452..710N} {452, 710}

\bibitem[\protect\citeauthoryear{{Narayan}, {Mahadevan}  \&
  {Quataert}}{{Narayan} et~al.}{1998}]{narayan:bh.review.1998}
{Narayan} R.,  {Mahadevan} R.,   {Quataert} E.,  1998, in {M.~A.~Abramowicz,
  G.~Bjornsson, \& J.~E.~Pringle} ed., Theory of Black Hole Accretion Disks;
  Cambridge University Press. pp 148--+

\bibitem[\protect\citeauthoryear{{Narayanan}, {Groppi}, {Kulesa}  \&
  {Walker}}{{Narayanan} et~al.}{2005}]{narayanan:2005.co32.lirgs}
{Narayanan} D.,  {Groppi} C.~E.,  {Kulesa} C.~A.,   {Walker} C.~K.,  2005,
  \mn@doi [\apj] {10.1086/431171}, \href
  {http://adsabs.harvard.edu/abs/2005ApJ...630..269N} {630, 269}

\bibitem[\protect\citeauthoryear{{Narayanan}, {Krumholz}, {Ostriker}  \&
  {Hernquist}}{{Narayanan} et~al.}{2011}]{narayanan:2011.xco}
{Narayanan} D.,  {Krumholz} M.,  {Ostriker} E.~C.,   {Hernquist} L.,  2011,
  \mn@doi [\mnras] {10.1111/j.1365-2966.2011.19516.x}, \href
  {http://adsabs.harvard.edu/abs/2011MNRAS.418..664N} {418, 664}

\bibitem[\protect\citeauthoryear{{Netzer} et~al.,}{{Netzer}
  et~al.}{2003}]{netzer:2003.detailed.modeling.warm.absorber.parsec.scale.outflow.weak}
{Netzer} H.,  et~al., 2003, \mn@doi [\apj] {10.1086/379508}, \href
  {https://ui.adsabs.harvard.edu/abs/2003ApJ...599..933N} {599, 933}

\bibitem[\protect\citeauthoryear{{Nomura}, {Ohsuga}  \& {Done}}{{Nomura}
  et~al.}{2020}]{nomura:2020.line.driven.winds.agn.modest.massloss.rates}
{Nomura} M.,  {Ohsuga} K.,   {Done} C.,  2020, \mn@doi [\mnras]
  {10.1093/mnras/staa948}, \href
  {https://ui.adsabs.harvard.edu/abs/2020MNRAS.494.3616N} {494, 3616}

\bibitem[\protect\citeauthoryear{{Nomura}, {Omukai}  \& {Ohsuga}}{{Nomura}
  et~al.}{2021}]{nomura:2021.agn.wind.massloss.modest.strongly.sensitive.to.metallicity}
{Nomura} M.,  {Omukai} K.,   {Ohsuga} K.,  2021, \mn@doi [\mnras]
  {10.1093/mnras/stab2214}, \href
  {https://ui.adsabs.harvard.edu/abs/2021MNRAS.507..904N} {507, 904}

\bibitem[\protect\citeauthoryear{{Orr} et~al.,}{{Orr}
  et~al.}{2020}]{orr:2020.resolved.dispersions.sfrs.correlations}
{Orr} M.~E.,  et~al., 2020, \mn@doi [\mnras] {10.1093/mnras/staa1619}, \href
  {https://ui.adsabs.harvard.edu/abs/2020MNRAS.496.1620O} {496, 1620}

\bibitem[\protect\citeauthoryear{{Ostriker} \& {Shetty}}{{Ostriker} \&
  {Shetty}}{2011}]{ostriker.shetty:2011.turb.disk.selfreg.ks}
{Ostriker} E.~C.,  {Shetty} R.,  2011, \mn@doi [\apj]
  {10.1088/0004-637X/731/1/41}, \href
  {http://adsabs.harvard.edu/abs/2011ApJ...731...41O} {731, 41}

\bibitem[\protect\citeauthoryear{{Ostriker}, {Stone}  \& {Gammie}}{{Ostriker}
  et~al.}{2001}]{ostriker:2001.gmc.column.dist}
{Ostriker} E.~C.,  {Stone} J.~M.,   {Gammie} C.~F.,  2001, \mn@doi [\apj]
  {10.1086/318290}, \href {http://adsabs.harvard.edu/abs/2001ApJ...546..980O}
  {546, 980}

\bibitem[\protect\citeauthoryear{{Ouyed} \& {Pudritz}}{{Ouyed} \&
  {Pudritz}}{1999}]{ouyed:1999.numerical.sims.mhd.winds.vs.massloading}
{Ouyed} R.,  {Pudritz} R.~E.,  1999, \mn@doi [\mnras]
  {10.1046/j.1365-8711.1999.02828.x}, \href
  {https://ui.adsabs.harvard.edu/abs/1999MNRAS.309..233O} {309, 233}

\bibitem[\protect\citeauthoryear{{Paardekooper}}{{Paardekooper}}{2012}]{paardekooper:2012.stochastic.disk.frag}
{Paardekooper} S.-J.,  2012, \mn@doi [\mnras]
  {10.1111/j.1365-2966.2012.20553.x}, \href
  {http://adsabs.harvard.edu/abs/2012MNRAS.421.3286P} {421, 3286}

\bibitem[\protect\citeauthoryear{{Paczynski}}{{Paczynski}}{1978}]{paczynski:1978.selfgrav.disk}
{Paczynski} B.,  1978, \actaa, \href
  {https://ui.adsabs.harvard.edu/abs/1978AcA....28...91P} {28, 91}

\bibitem[\protect\citeauthoryear{{Paczy{\'n}sky} \& {Wiita}}{{Paczy{\'n}sky} \&
  {Wiita}}{1980}]{paczynsky.wiita:1980.slim.disk}
{Paczy{\'n}sky} B.,  {Wiita} P.~J.,  1980, \aap, \href
  {https://ui.adsabs.harvard.edu/abs/1980A&A....88...23P} {88, 23}

\bibitem[\protect\citeauthoryear{{Palit} et~al.,}{{Palit}
  et~al.}{2024}]{palit:2024.warm.covering.agn.corona.expands.with.luminosity.reprocesses.most.emission}
{Palit} B.,  et~al., 2024, \mn@doi [arXiv e-prints]
  {10.48550/arXiv.2406.14378}, \href
  {https://ui.adsabs.harvard.edu/abs/2024arXiv240614378P} {p. arXiv:2406.14378}

\bibitem[\protect\citeauthoryear{{Pariev}, {Blackman}  \& {Boldyrev}}{{Pariev}
  et~al.}{2003}]{pariev:2003.mag.dominated.disk.models}
{Pariev} V.~I.,  {Blackman} E.~G.,   {Boldyrev} S.~A.,  2003, \mn@doi [\aap]
  {10.1051/0004-6361:20030868}, \href
  {https://ui.adsabs.harvard.edu/abs/2003A&A...407..403P} {407, 403}

\bibitem[\protect\citeauthoryear{{Peterson}}{{Peterson}}{1997}]{peterson:1997.agn.book}
{Peterson} B.~M.,  1997, {An Introduction to Active Galactic Nuclei}.
Cambridge, New York Cambridge University Press

\bibitem[\protect\citeauthoryear{Peterson}{Peterson}{2006}]{Peterson2006:BLR.review}
Peterson B.,  2006, The Broad-Line Region in Active Galactic Nuclei.
Springer Berlin Heidelberg (Eds. Alloin, Danielle and Johnson, Rachel and Lira,
  Paulina), Berlin, Heidelberg, pp 77--100, \mn@doi{10.1007/3-540-34621-X_3},
  \url {https://doi.org/10.1007/3-540-34621-X_3}

\bibitem[\protect\citeauthoryear{{Petrucci}, {Ursini}, {De Rosa}, {Bianchi},
  {Cappi}, {Matt}, {Dadina}  \& {Malzac}}{{Petrucci}
  et~al.}{2018}]{petrucci:2018.agn.comptonizing.layer.properties}
{Petrucci} P.~O.,  {Ursini} F.,  {De Rosa} A.,  {Bianchi} S.,  {Cappi} M.,
  {Matt} G.,  {Dadina} M.,   {Malzac} J.,  2018, \mn@doi [\aap]
  {10.1051/0004-6361/201731580}, \href
  {https://ui.adsabs.harvard.edu/abs/2018A&A...611A..59P} {611, A59}

\bibitem[\protect\citeauthoryear{{Piran}}{{Piran}}{1978}]{piran:1978.thermal.viscous.accretion.disk.instability}
{Piran} T.,  1978, \mn@doi [\apj] {10.1086/156069}, \href
  {https://ui.adsabs.harvard.edu/abs/1978ApJ...221..652P} {221, 652}

\bibitem[\protect\citeauthoryear{{Pope} et~al.,}{{Pope}
  et~al.}{2008}]{pope:2008.pah.agn.dont.dominate.smgs}
{Pope} A.,  et~al., 2008, \mn@doi [\apj] {10.1086/527030}, \href
  {http://adsabs.harvard.edu/abs/2008ApJ...675.1171P} {675, 1171}

\bibitem[\protect\citeauthoryear{{Prieto} \& {Escala}}{{Prieto} \&
  {Escala}}{2016}]{prieto:2016.zoomin.sims.to.fewpc.hydro.cosmo.highz}
{Prieto} J.,  {Escala} A.,  2016, \mn@doi [\mnras] {10.1093/mnras/stw1285},
  \href {https://ui.adsabs.harvard.edu/abs/2016MNRAS.460.4018P} {460, 4018}

\bibitem[\protect\citeauthoryear{{Prieto}, {Escala}, {Volonteri}  \&
  {Dubois}}{{Prieto} et~al.}{2017}]{prieto:2017.zoomin.sims.agn.fueling.sne.fb}
{Prieto} J.,  {Escala} A.,  {Volonteri} M.,   {Dubois} Y.,  2017, \mn@doi
  [\apj] {10.3847/1538-4357/aa5be5}, \href
  {https://ui.adsabs.harvard.edu/abs/2017ApJ...836..216P} {836, 216}

\bibitem[\protect\citeauthoryear{{Querejeta} et~al.,}{{Querejeta}
  et~al.}{2016}]{querejeta:grav.torque.obs.m51}
{Querejeta} M.,  et~al., 2016, \mn@doi [\aap] {10.1051/0004-6361/201527536},
  \href {https://ui.adsabs.harvard.edu/abs/2016A&A...588A..33Q} {588, A33}

\bibitem[\protect\citeauthoryear{{Ramos Almeida} et~al.}{{Ramos Almeida}
  et~al.}{2009}]{ramosalmeida:pc.scale.torus.emission}
{Ramos Almeida} C.,  et~al., 2009, \mn@doi [\apj]
  {10.1088/0004-637X/702/2/1127}, \href
  {http://adsabs.harvard.edu/abs/2009arXiv0906.5368R} {702, 1127}

\bibitem[\protect\citeauthoryear{{Rees}}{{Rees}}{1987}]{rees:1987.mag.confined.blr}
{Rees} M.~J.,  1987, \mn@doi [\mnras] {10.1093/mnras/228.1.47P}, \href
  {https://ui.adsabs.harvard.edu/abs/1987MNRAS.228P..47R} {228, 47P}

\bibitem[\protect\citeauthoryear{{Ren}, {Sun}, {Wang}  \& {Cai}}{{Ren}
  et~al.}{2024}]{ren:2024.reverb.mapping.sizes.larger.than.expected}
{Ren} G.,  {Sun} M.,  {Wang} J.-X.,   {Cai} Z.-Y.,  2024, \mn@doi [\apj]
  {10.3847/1538-4357/ad3d53}, \href
  {https://ui.adsabs.harvard.edu/abs/2024ApJ...967...25R} {967, 25}

\bibitem[\protect\citeauthoryear{{Reynolds}}{{Reynolds}}{1997}]{reynolds:1997.xray.agn.warm.absorbers.launched.from.dusty.torus}
{Reynolds} C.~S.,  1997, \mn@doi [\mnras] {10.1093/mnras/286.3.513}, \href
  {https://ui.adsabs.harvard.edu/abs/1997MNRAS.286..513R} {286, 513}

\bibitem[\protect\citeauthoryear{{Rice}, {Lodato}  \& {Armitage}}{{Rice}
  et~al.}{2005}]{rice:2005.disk.frag.firstlook}
{Rice} W.~K.~M.,  {Lodato} G.,   {Armitage} P.~J.,  2005, \mn@doi [\mnras]
  {10.1111/j.1745-3933.2005.00105.x}, \href
  {http://adsabs.harvard.edu/abs/2005MNRAS.364L..56R} {364, L56}

\bibitem[\protect\citeauthoryear{{Rice}, {Martini}, {Greene}, {Pogge},
  {Shields}, {Mulchaey}  \& {Regan}}{{Rice} et~al.}{2006}]{rice:nlr.kinematics}
{Rice} M.~S.,  {Martini} P.,  {Greene} J.~E.,  {Pogge} R.~W.,  {Shields} J.~C.,
   {Mulchaey} J.~S.,   {Regan} M.~W.,  2006, \mn@doi [\apj] {10.1086/498091},
  \href {http://adsabs.harvard.edu/abs/2006ApJ...636..654R} {636, 654}

\bibitem[\protect\citeauthoryear{{Richards} et~al.}{{Richards}
  et~al.}{2006}]{richards:seds}
{Richards} G.~T.,  et~al., 2006, \mn@doi [\apjs] {10.1086/506525}, \href
  {http://adsabs.harvard.edu/cgi-bin/nph-bib_query?bibcode=2006ApJS..166..470R&db_key=AST}
  {166, 470}

\bibitem[\protect\citeauthoryear{{Riols} \& {Latter}}{{Riols} \&
  {Latter}}{2016}]{riols:2016.mhd.ppd.gravitoturb}
{Riols} A.,  {Latter} H.,  2016, \mn@doi [\mnras] {10.1093/mnras/stw1112},
  \href {https://ui.adsabs.harvard.edu/abs/2016MNRAS.460.2223R} {460, 2223}

\bibitem[\protect\citeauthoryear{{Riols} \& {Latter}}{{Riols} \&
  {Latter}}{2018}]{riols:2018.mhd.sims.ppd.gravitoturb.fx.on.mri}
{Riols} A.,  {Latter} H.,  2018, \mn@doi [\mnras] {10.1093/mnras/stx2455},
  \href {https://ui.adsabs.harvard.edu/abs/2018MNRAS.474.2212R} {474, 2212}

\bibitem[\protect\citeauthoryear{{Rowan-Robinson}, {Valtchanov}  \&
  {Nandra}}{{Rowan-Robinson}
  et~al.}{2009}]{rowanrobinson:xr.ir.comparison.of.torii}
{Rowan-Robinson} M.,  {Valtchanov} I.,   {Nandra} K.,  2009, \mn@doi [\mnras]
  {10.1111/j.1365-2966.2009.15094.x}, \href
  {http://adsabs.harvard.edu/abs/2009arXiv0905.4389R} {397, 1326}

\bibitem[\protect\citeauthoryear{{Rybicki} \& {Lightman}}{{Rybicki} \&
  {Lightman}}{1986}]{rybicki.lightman:1986.radiative.processes.book}
{Rybicki} G.~B.,  {Lightman} A.~P.,  1986, {Radiative Processes in
  Astrophysics},.
Wiley-VCH; Weinheim, Germany

\bibitem[\protect\citeauthoryear{{Sanders}, {Soifer}, {Elias}, {Neugebauer}  \&
  {Matthews}}{{Sanders} et~al.}{1988}]{sanders88:warm.ulirgs}
{Sanders} D.~B.,  {Soifer} B.~T.,  {Elias} J.~H.,  {Neugebauer} G.,
  {Matthews} K.,  1988, \mn@doi [\apjl] {10.1086/185155}, \href
  {http://adsabs.harvard.edu/cgi-bin/nph-bib_query?bibcode=1988ApJ...328L..35S&db_key=AST}
  {328, L35}

\bibitem[\protect\citeauthoryear{{Schmidt}}{{Schmidt}}{1963}]{schmidt:1963.qso.redshift}
{Schmidt} M.,  1963, \mn@doi [\nat] {10.1038/1971040a0}, \href
  {https://ui.adsabs.harvard.edu/abs/1963Natur.197.1040S} {197, 1040}

\bibitem[\protect\citeauthoryear{{Seifried}, {Pudritz}, {Banerjee}, {Duffin}
  \& {Klessen}}{{Seifried}
  et~al.}{2012}]{seifried:2012.mhd.outflow.disk.criterion}
{Seifried} D.,  {Pudritz} R.~E.,  {Banerjee} R.,  {Duffin} D.,   {Klessen}
  R.~S.,  2012, \mn@doi [\mnras] {10.1111/j.1365-2966.2012.20610.x}, \href
  {https://ui.adsabs.harvard.edu/abs/2012MNRAS.422..347S} {422, 347}

\bibitem[\protect\citeauthoryear{{Semenov}, {Henning}, {Helling}, {Ilgner}  \&
  {Sedlmayr}}{{Semenov} et~al.}{2003}]{semenov:2003.dust.opacities}
{Semenov} D.,  {Henning} T.,  {Helling} C.,  {Ilgner} M.,   {Sedlmayr} E.,
  2003, \mn@doi [\aap] {10.1051/0004-6361:20031279}, \href
  {http://adsabs.harvard.edu/abs/2003A%26A...410..611S} {410, 611}

\bibitem[\protect\citeauthoryear{{Shakura} \& {Sunyaev}}{{Shakura} \&
  {Sunyaev}}{1973}]{shakurasunyaev73}
{Shakura} N.~I.,  {Sunyaev} R.~A.,  1973, \aap, \href
  {http://adsabs.harvard.edu/cgi-bin/nph-bib_query?bibcode=1973A%26A....24..337S&db_key=AST}
  {24, 337}

\bibitem[\protect\citeauthoryear{{Shen}, {Hopkins}, {Faucher-Gigu{\`e}re},
  {Alexander}, {Richards}, {Ross}  \& {Hickox}}{{Shen}
  et~al.}{2020}]{shen:bolometric.qlf.update}
{Shen} X.,  {Hopkins} P.~F.,  {Faucher-Gigu{\`e}re} C.-A.,  {Alexander} D.~M.,
  {Richards} G.~T.,  {Ross} N.~P.,   {Hickox} R.~C.,  2020, \mn@doi [\mnras]
  {10.1093/mnras/staa1381}, \href
  {https://ui.adsabs.harvard.edu/abs/2020MNRAS.495.3252S} {495, 3252}

\bibitem[\protect\citeauthoryear{{Shi}, {Kremer}  \& {Hopkins}}{{Shi}
  et~al.}{2024a}]{shi:2024.imbh.growth.feedback.survey}
{Shi} Y.,  {Kremer} K.,   {Hopkins} P.~F.,  2024a, \mn@doi [arXiv e-prints]
  {10.48550/arXiv.2405.12164}, \href
  {https://ui.adsabs.harvard.edu/abs/2024arXiv240512164S} {p. arXiv:2405.12164}

\bibitem[\protect\citeauthoryear{{Shi}, {Kremer}  \& {Hopkins}}{{Shi}
  et~al.}{2024b}]{shi:2024.seed.to.smbh.case.study.subcluster.merging.pairing.fluxfrozen.disk}
{Shi} Y.,  {Kremer} K.,   {Hopkins} P.~F.,  2024b, \mn@doi [arXiv e-prints]
  {10.48550/arXiv.2405.17338}, \href
  {https://ui.adsabs.harvard.edu/abs/2024arXiv240517338S} {p. arXiv:2405.17338}

\bibitem[\protect\citeauthoryear{Shlosman, Frank  \& Begelman}{Shlosman
  et~al.}{1989}]{shlosman:bars.within.bars}
Shlosman I.,  Frank J.,   Begelman M.~C.,  1989, \mn@doi [Nature]
  {10.1038/338045a0}, 338, 45

\bibitem[\protect\citeauthoryear{{Simcoe}, {McLeod}, {Schachter}  \&
  {Elvis}}{{Simcoe} et~al.}{1997}]{simcoe:1997.agn.host.alignment}
{Simcoe} R.,  {McLeod} K.~K.,  {Schachter} J.,   {Elvis} M.,  1997, \mn@doi
  [\apj] {10.1086/304819}, \href
  {http://adsabs.harvard.edu/abs/1997ApJ...489..615S} {489, 615}

\bibitem[\protect\citeauthoryear{{Sirko} \& {Goodman}}{{Sirko} \&
  {Goodman}}{2003}]{sirko:qso.seds.from.selfgrav.disks}
{Sirko} E.,  {Goodman} J.,  2003, \mn@doi [\mnras]
  {10.1046/j.1365-8711.2003.06431.x}, \href
  {http://adsabs.harvard.edu/abs/2003MNRAS.341..501S} {341, 501}

\bibitem[\protect\citeauthoryear{{Soltan}}{{Soltan}}{1982}]{soltan82}
{Soltan} A.,  1982, \mnras, \href
  {http://adsabs.harvard.edu/cgi-bin/nph-bib_query?bibcode=1982MNRAS.200..115S&db_key=AST}
  {200, 115}

\bibitem[\protect\citeauthoryear{{Spitzer}}{{Spitzer}}{1962}]{spitzer:1962.ionized.gases.book}
{Spitzer} L.,  1962, {Physics of Fully Ionized Gases; New York: Interscience}.
New York: Interscience

\bibitem[\protect\citeauthoryear{{Spruit}}{{Spruit}}{1996}]{spruit:1996.mhd.disk.winds.review}
{Spruit} H.~C.,  1996, in {Wijers} R. A.~M.~J.,  {Davies} M.~B.,   {Tout}
  C.~A.,  eds,  NATO Advanced Study Institute (ASI) Series C Vol. 477,
  Evolutionary Processes in Binary Stars; Kluwer academic publishers.
  Evolutionary processes in binary stars. pp 249--286

\bibitem[\protect\citeauthoryear{{Squire}, {Quataert}  \& {Hopkins}}{{Squire}
  et~al.}{2024}]{squire:2024.mri.shearing.box.strongly.magnetized.different.beta.states}
{Squire} J.,  {Quataert} E.,   {Hopkins} P.~F.,  2024, \mn@doi [arXiv e-prints]
  {10.48550/arXiv.2409.05467}, \href
  {https://ui.adsabs.harvard.edu/abs/2024arXiv240905467S} {p. arXiv:2409.05467}

\bibitem[\protect\citeauthoryear{{Stark} \& {Carlson}}{{Stark} \&
  {Carlson}}{1984}]{starkcarlson:m82.nlr}
{Stark} A.~A.,  {Carlson} E.~R.,  1984, \mn@doi [\apj] {10.1086/161871}, \href
  {http://adsabs.harvard.edu/abs/1984ApJ...279..122S} {279, 122}

\bibitem[\protect\citeauthoryear{{Stevans}, {Shull}, {Danforth}  \&
  {Tilton}}{{Stevans}
  et~al.}{2014}]{stevans:2014.soft.excess.uv.slopes.corr.luminosity.agn}
{Stevans} M.~L.,  {Shull} J.~M.,  {Danforth} C.~W.,   {Tilton} E.~M.,  2014,
  \mn@doi [\apj] {10.1088/0004-637X/794/1/75}, \href
  {https://ui.adsabs.harvard.edu/abs/2014ApJ...794...75S} {794, 75}

\bibitem[\protect\citeauthoryear{{Tchekhovskoy}, {Narayan}  \&
  {McKinney}}{{Tchekhovskoy} et~al.}{2011}]{tchekhovskoy:2011.mad.disk.jets}
{Tchekhovskoy} A.,  {Narayan} R.,   {McKinney} J.~C.,  2011, \mn@doi [\mnras]
  {10.1111/j.1745-3933.2011.01147.x}, \href
  {https://ui.adsabs.harvard.edu/abs/2011MNRAS.418L..79T} {418, L79}

\bibitem[\protect\citeauthoryear{{Temple}, {Banerji}, {Hewett}, {Rankine}  \&
  {Richards}}{{Temple}
  et~al.}{2021a}]{temple:2021.agn.outflows.assoc.hot.dust.covering}
{Temple} M.~J.,  {Banerji} M.,  {Hewett} P.~C.,  {Rankine} A.~L.,   {Richards}
  G.~T.,  2021a, \mn@doi [\mnras] {10.1093/mnras/staa3842}, \href
  {https://ui.adsabs.harvard.edu/abs/2021MNRAS.501.3061T} {501, 3061}

\bibitem[\protect\citeauthoryear{{Temple}, {Ferland}, {Rankine}, {Chatzikos}
  \& {Hewett}}{{Temple}
  et~al.}{2021b}]{temple:2021.qso.spectra.solar.metallicity.not.supersolar.components.just.come.from.different.radii}
{Temple} M.~J.,  {Ferland} G.~J.,  {Rankine} A.~L.,  {Chatzikos} M.,   {Hewett}
  P.~C.,  2021b, \mn@doi [\mnras] {10.1093/mnras/stab1610}, \href
  {https://ui.adsabs.harvard.edu/abs/2021MNRAS.505.3247T} {505, 3247}

\bibitem[\protect\citeauthoryear{{Temple} et~al.,}{{Temple}
  et~al.}{2023}]{temple:2023.outflow.lines.wind.association.eddington.ratio}
{Temple} M.~J.,  et~al., 2023, \mn@doi [\mnras] {10.1093/mnras/stad1448}, \href
  {https://ui.adsabs.harvard.edu/abs/2023MNRAS.523..646T} {523, 646}

\bibitem[\protect\citeauthoryear{{Thompson}, {Quataert}  \&
  {Murray}}{{Thompson} et~al.}{2005}]{thompson:rad.pressure}
{Thompson} T.~A.,  {Quataert} E.,   {Murray} N.,  2005, \mn@doi [\apj]
  {10.1086/431923}, \href {http://adsabs.harvard.edu/abs/2005ApJ...630..167T}
  {630, 167}

\bibitem[\protect\citeauthoryear{{Tielens}}{{Tielens}}{2005}]{tielens:2005.book}
{Tielens} A.~G.~G.~M.,  2005, {The Physics and Chemistry of the Interstellar
  Medium}.
Cambridge, UK: Cambridge University Press

\bibitem[\protect\citeauthoryear{{Tombesi}, {Cappi}, {Reeves}, {Nemmen},
  {Braito}, {Gaspari}  \& {Reynolds}}{{Tombesi}
  et~al.}{2013}]{tombesi:2013.ufo.warm.absorbers.may.be.connected.through.outflow}
{Tombesi} F.,  {Cappi} M.,  {Reeves} J.~N.,  {Nemmen} R.~S.,  {Braito} V.,
  {Gaspari} M.,   {Reynolds} C.~S.,  2013, \mn@doi [\mnras]
  {10.1093/mnras/sts692}, \href
  {http://adsabs.harvard.edu/abs/2013MNRAS.430.1102T} {430, 1102}

\bibitem[\protect\citeauthoryear{{Torrey}, {Hopkins}, {Faucher-Gigu{\`e}re},
  {Vogelsberger}, {Quataert}, {Kere{\v s}}  \& {Murray}}{{Torrey}
  et~al.}{2017}]{torrey.2016:fire.galactic.nuclei.star.formation.instability}
{Torrey} P.,  {Hopkins} P.~F.,  {Faucher-Gigu{\`e}re} C.-A.,  {Vogelsberger}
  M.,  {Quataert} E.,  {Kere{\v s}} D.,   {Murray} N.,  2017, \mn@doi [\mnras]
  {10.1093/mnras/stx254}, \href
  {http://adsabs.harvard.edu/abs/2017MNRAS.467.2301T} {467, 2301}

\bibitem[\protect\citeauthoryear{{Tortosa} et~al.,}{{Tortosa}
  et~al.}{2022}]{tortosa:2022.hard.coronal.constraints.in.hyper.eddington.qsos}
{Tortosa} A.,  et~al., 2022, \mn@doi [\mnras] {10.1093/mnras/stab3152}, \href
  {https://ui.adsabs.harvard.edu/abs/2022MNRAS.509.3599T} {509, 3599}

\bibitem[\protect\citeauthoryear{{Trump}}{{Trump}}{2011}]{trump:2011.host.agn.morph.discussion}
{Trump} J.~R.,  2011, \mnras, in press, arXiv:1112.3970, \href
  {http://adsabs.harvard.edu/abs/2011arXiv1112.3970T} {}

\bibitem[\protect\citeauthoryear{{Trump} et~al.}{{Trump}
  et~al.}{2009}]{trump:lowl.agn.dilution}
{Trump} J.~R.,  et~al., 2009, \mn@doi [\apj] {10.1088/0004-637X/706/1/797},
  \href {http://adsabs.harvard.edu/abs/2009arXiv0910.2672T} {706, 797}

\bibitem[\protect\citeauthoryear{{Tumlinson}, {Peeples}  \& {Werk}}{{Tumlinson}
  et~al.}{2017}]{tumlinson:2017.cgm.review}
{Tumlinson} J.,  {Peeples} M.~S.,   {Werk} J.~K.,  2017, \mn@doi [\araa]
  {10.1146/annurev-astro-091916-055240}, \href
  {http://adsabs.harvard.edu/abs/2017ARA%26A..55..389T} {55, 389}

\bibitem[\protect\citeauthoryear{{Urry} \& {Padovani}}{{Urry} \&
  {Padovani}}{1995}]{urry:radio.unification.review}
{Urry} C.~M.,  {Padovani} P.,  1995, \mn@doi [\pasp] {10.1086/133630}, \href
  {http://adsabs.harvard.edu/abs/1995PASP..107..803U} {107, 803}

\bibitem[\protect\citeauthoryear{{Vanden Berk} et~al.}{{Vanden Berk}
  et~al.}{2001}]{vandenberk01:composite.qso.seds}
{Vanden Berk} D.~E.,  et~al., 2001, \mn@doi [\aj] {10.1086/321167}, \href
  {http://adsabs.harvard.edu/abs/2001AJ....122..549V} {122, 549}

\bibitem[\protect\citeauthoryear{{Veilleux} et~al.}{{Veilleux}
  et~al.}{2009}]{veilleux:ulirg.to.qso.sample.big.mdot.changes}
{Veilleux} S.,  et~al., 2009, \mn@doi [\apjs] {10.1088/0067-0049/182/2/628},
  \href {http://adsabs.harvard.edu/abs/2009ApJS..182..628V} {182, 628}

\bibitem[\protect\citeauthoryear{{Weingartner} \& {Draine}}{{Weingartner} \&
  {Draine}}{2001}]{weingartner:2001.dust.size.distrib}
{Weingartner} J.~C.,  {Draine} B.~T.,  2001, \mn@doi [\apj] {10.1086/318651},
  \href {http://adsabs.harvard.edu/abs/2001ApJ...548..296W} {548, 296}

\bibitem[\protect\citeauthoryear{{Wellons} et~al.,}{{Wellons}
  et~al.}{2023}]{wellons:2022.smbh.growth}
{Wellons} S.,  et~al., 2023, \mn@doi [\mnras] {10.1093/mnras/stad511}, \href
  {https://ui.adsabs.harvard.edu/abs/2023MNRAS.520.5394W} {520, 5394}

\bibitem[\protect\citeauthoryear{{Wilkins} \& {Gallo}}{{Wilkins} \&
  {Gallo}}{2015}]{wilkins:2015.patchy.corona.for.comptonizing.and.reprocessing}
{Wilkins} D.~R.,  {Gallo} L.~C.,  2015, \mn@doi [\mnras]
  {10.1093/mnras/stu2524}, \href
  {https://ui.adsabs.harvard.edu/abs/2015MNRAS.448..703W} {448, 703}

\bibitem[\protect\citeauthoryear{{Williamson}, {B{\"o}sch}  \&
  {H{\"o}nig}}{{Williamson}
  et~al.}{2022}]{williamson:2022.gizmo.rhd.psph.sims.binary.smbh.torii.radiation.reduces.grav.torques}
{Williamson} D.~J.,  {B{\"o}sch} L.~H.,   {H{\"o}nig} S.~F.,  2022, \mn@doi
  [\mnras] {10.1093/mnras/stab3792}, \href
  {https://ui.adsabs.harvard.edu/abs/2022MNRAS.510.5963W} {510, 5963}

\bibitem[\protect\citeauthoryear{{Wolfire}, {Hollenbach}, {McKee}, {Tielens}
  \& {Bakes}}{{Wolfire} et~al.}{1995}]{wolfire:1995.neutral.ism.phases}
{Wolfire} M.~G.,  {Hollenbach} D.,  {McKee} C.~F.,  {Tielens} A.~G.~G.~M.,
  {Bakes} E.~L.~O.,  1995, \mn@doi [\apj] {10.1086/175510}, \href
  {http://adsabs.harvard.edu/abs/1995ApJ...443..152W} {443, 152}

\bibitem[\protect\citeauthoryear{{Wolfire}, {McKee}, {Hollenbach}  \&
  {Tielens}}{{Wolfire} et~al.}{2003}]{wolfire.2003:neutral.atomic.cooling}
{Wolfire} M.~G.,  {McKee} C.~F.,  {Hollenbach} D.,   {Tielens} A.~G.~G.~M.,
  2003, \mn@doi [\apj] {10.1086/368016}, \href
  {http://adsabs.harvard.edu/abs/2003ApJ...587..278W} {587, 278}

\bibitem[\protect\citeauthoryear{{Woo} et~al.,}{{Woo}
  et~al.}{2023}]{woo:2023.reverb.map.updated.compilation.higher.lum}
{Woo} J.-H.,  et~al., 2023, \mn@doi [arXiv e-prints]
  {10.48550/arXiv.2311.15518}, \href
  {https://ui.adsabs.harvard.edu/abs/2023arXiv231115518W} {p. arXiv:2311.15518}

\bibitem[\protect\citeauthoryear{{Younger} et~al.,}{{Younger}
  et~al.}{2009a}]{younger:mm.obs.z2.ulirgs}
{Younger} J.~D.,  et~al., 2009a, \mn@doi [\mnras]
  {10.1111/j.1365-2966.2009.14455.x}, \href
  {http://adsabs.harvard.edu/abs/2009MNRAS.394.1685Y} {394, 1685}

\bibitem[\protect\citeauthoryear{{Younger}, {Hayward}, {Narayanan}, {Cox},
  {Hernquist}  \& {Jonsson}}{{Younger} et~al.}{2009b}]{younger:warm.ulirg.evol}
{Younger} J.~D.,  {Hayward} C.~C.,  {Narayanan} D.,  {Cox} T.~J.,  {Hernquist}
  L.,   {Jonsson} P.,  2009b, \mn@doi [\mnras]
  {10.1111/j.1745-3933.2009.00663.x}, \href
  {http://adsabs.harvard.edu/abs/2009MNRAS.396L..66Y} {396, L66}

\bibitem[\protect\citeauthoryear{{Yuan} \& {Narayan}}{{Yuan} \&
  {Narayan}}{2014}]{yuan:2014.hot.accretion.flows.review}
{Yuan} F.,  {Narayan} R.,  2014, \mn@doi [\araa]
  {10.1146/annurev-astro-082812-141003}, \href
  {https://ui.adsabs.harvard.edu/abs/2014ARA&A..52..529Y} {52, 529}

\bibitem[\protect\citeauthoryear{{Zhang}, {Soria}, {Zhang}, {Swartz}  \&
  {Liu}}{{Zhang} et~al.}{2009}]{zhang:2009.agn.vs.hubble.type}
{Zhang} W.~M.,  {Soria} R.,  {Zhang} S.~N.,  {Swartz} D.~A.,   {Liu} J.~F.,
  2009, \mn@doi [\apj] {10.1088/0004-637X/699/1/281}, \href
  {http://adsabs.harvard.edu/abs/2009ApJ...699..281Z} {699, 281}

\bibitem[\protect\citeauthoryear{{Zhang}, {Peng}  \& {Wang}}{{Zhang}
  et~al.}{2019}]{zhang:2019.new.cosmic.ray.compilation.vs.calorimetry.sub.calor}
{Zhang} Y.,  {Peng} F.-K.,   {Wang} X.-Y.,  2019, \mn@doi [\apj]
  {10.3847/1538-4357/ab0ae2}, \href
  {https://ui.adsabs.harvard.edu/abs/2019ApJ...874..173Z} {874, 173}

\bibitem[\protect\citeauthoryear{{Zhu}, {Bu}, {Yang}, {Yuan}  \& {Lin}}{{Zhu}
  et~al.}{2022}]{zhu:2022.idealized.wind.driving.sims.start.thick.configuration}
{Zhu} Y.,  {Bu} D.-F.,  {Yang} X.-H.,  {Yuan} F.,   {Lin} W.-B.,  2022, \mn@doi
  [\mnras] {10.1093/mnras/stac1015}, \href
  {https://ui.adsabs.harvard.edu/abs/2022MNRAS.513.1141Z} {513, 1141}

\bibitem[\protect\citeauthoryear{{Zier} \& {Springel}}{{Zier} \&
  {Springel}}{2022}]{zier:2022.gravitoturb.sims.validating.stochastic.frag}
{Zier} O.,  {Springel} V.,  2022, arXiv e-prints, \href
  {https://ui.adsabs.harvard.edu/abs/2022arXiv221202526Z} {p. arXiv:2212.02526}

\makeatother
\end{thebibliography}

\end{document}